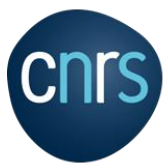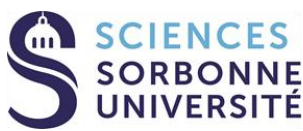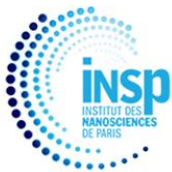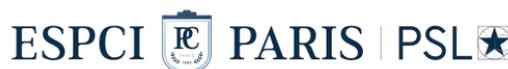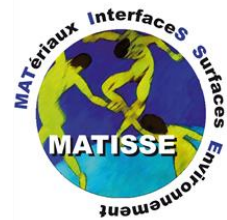

# THÈSE DE DOCTORAT DE SORBONNE UNIVERSITÉ

Spécialité : Physique et chimie des Matériaux

École doctorale 397

réalisée à l'Institut des Nanosciences de Paris (INSP)

présentée par Bertille MARTINEZ

pour obtenir le grade de docteur de SORBONNE UNIVERSITÉ

Sujet de la thèse

## ÉTUDE DES PROPRIÉTÉS OPTOÉLECTRONIQUES DE NANOCRISTAUX COLLOÏDAUX À FAIBLE BANDE INTERDITE : APPLICATION À LA DÉTECTION INFRAROUGE

Soutenue le 10 juillet 2019

devant le jury composé de

| | | |
|---|---|---|
| Laurence Ressier, Professeur des universités, INSA Toulouse | | Rapportrice |
| Bruno Masenelli, Professeur des universités, INSA Lyon | | Rapporteur |
| Alexa Courty, Professeur des universités, Sorbonne Université | | Examinatrice |
| Nicolas Péré-Laperne, Ingénieur recherche & développement, Lynred | | Examinateur |
| Emmanuelle Lacaze, Directrice de recherche, INSP | | Directrice de thèse |
| Emmanuel Lhuillier, Chargé de recherche, INSP | | Encadrant de thèse |

# Table des matières

# Notations et abréviations

Constantes physiques

| Symbole | Valeur | Description |
|---|---|---|
| e | $\approx 1,60 \times 10^{-19}$ C | Charge élémentaire |
| c | $299792458$ m.s$^{-1}$ | Célérité de la lumière dans le vide |
| h | $\approx 6,62 \times 10^{-34}$ J.s | Constante de Planck |
| $\hbar$ | $\approx 1,05 \times 10^{-34}$ J.s | Constante de Planck réduite |
| $k_B$ | $\approx 1,38 \times 10^{-23}$ J.K$^{-1}$ | Constante de Boltzmann |
| $m_0$ | $\approx 9,11 \times 10^{-31}$ kg | Masse de l'électron dans le vide |
| $\varepsilon_0$ | $\approx 8,85 \times 10^{-12}$ m$^{-3}$.kg$^{-1}$.s$^4$.A$^2$ | Permittivité du vide |

Notations

| Symbole | Description |
|---|---|
| R | Rayon du nanocristal |
| T | Température |
| $\lambda$ | Longueur d'onde |
| $E_F$ | Énergie de Fermi |
| $E_G$ | Énergie de bande interdite |
| S | Source |
| D | Drain |
| $V_{DS}$ | Tension drain-source aux bornes d'un échantillon |
| $I_{DS}$ | Courant drain-source circulant dans un échantillon |
| $V_{GS}$ | Tension grille-source appliquée à un échantillon dans le cas d'une mesure de transistor à effet de champ |
| $I_{noir}$ | Courant d'obscurité circulant dans un échantillon |
| $\mathcal{R}$ | Réponse (s'exprime en A/W) |
| D* | Détectivité (s'exprime en Jones ou cm.f$^{0,5}$.W$^{-1}$) |
| NEP | Puissance équivalente de bruit (*Noise Equivalent Power*) |
| $m_e$ | Masse effective de l'électron |
| $m_h$ | Masse effective du trou |
| $a_0$ | Rayon de Bohr |
| $\varepsilon_r$ | Permittivité relative |
| F | Champ électrique |

| Symbole | Description |
|---|---|
| $E_C$ | Énergie cinétique |
| $E_L$ | Énergie de liaison |
| $1S_h$ | Pour un nanocristal, niveau électronique issu de la bande de valence ayant la plus haute énergie |
| $1S_e$ | Pour un nanocristal, niveau électronique issu de la bande de conduction ayant la plus basse énergie |
| $1P_e$ | Pour un nanocristal, niveau électronique issu de la bande de conduction ayant la deuxième plus basse énergie |
| $1D_e$ | Pour un nanocristal, niveau électronique issu de la bande de conduction ayant la troisième plus basse énergie |
| WF | Travail de sortie (*Work Function*) |
| $E_u$ | Énergie d'Urbach |

Abréviations – composés chimiques

| Symbole | Description |
|---|---|
| Au | Or |
| $As_2S_3$ | Sulfure d'arsenic |
| AZO | Oxyde de zinc dopé à l'aluminium |
| BeSH | 1,4 benzenedithiol |
| BuSH | 1 butanethiol |
| $Cl^-$ | Ion chlorure |
| CdS | Sulfure de cadmium |
| CdSe | Séléniure de cadmium |
| CdTe | Tellure de cadmium |
| DDT | Dodécanethiol |
| DMF | N,N- diméthylformamide |
| ECS | Électrode au calomel saturé |
| EDT | 1,2 éthanedithiol |
| FTO | Oxyde d'étain dopé au fluor (*Fluor Tin Oxide*) |
| GZO | Oxyde de zinc dopé au gallium |
| $H_2O$ | Eau |
| $HgCl_2$ | Chlorure de mercure |
| HgSe | Séléniure de mercure |
| HgTe | Tellure de mercure |
| ITO | Oxyde d'étain dopé à l'indium (*Indium Tin Oxide*) |
| $LiClO_4$ | Perchlorate de lithium |
| Mg | Magnésium |

| Symbole | Description |
|---|---|
| MoO$_3$ | Oxyde de molybdène |
| MPOH | 1,2 mercaptoéthanol, ou mercaptoalcool |
| NMF | N-méthylformamide |
| O$_2$ | Dioxygène |
| OA | Acide oléique |
| OLA | Oleylamine |
| OSH | Octanethiol |
| PbS | Sulfure de plomb |
| PbSe | Séléniure de plomb |
| PEG | Polyéthylène glycol |
| PET | Polyéthylène téréphtalate |
| POM | Polyoxométallate |
| POM-SH | Polyoxométallate greffé avec des groupements thiols |
| QD | *Quantum Dot* (nanocristal) |
| S$^{2-}$ | Ion sulfure |
| SCN$^-$ | Ion thiocyanate |
| SiO$_2$ | Silice |
| TiO$_2$ | Oxyde de titane |
| ZnO | Oxyde de zinc |
| ZnS | Sulfure de zinc |

Abréviatons – autres

| Symbole | Description |
|---|---|
| resp. | respectivement |
| SWIR | Proche infrarouge (Short Wave Infrared Range) : domaine infrarouge pour lequel : $1\ \mu m < \lambda < 1,7\ \mu m$ |
| ESWIR | Proche infrarouge étendu (Extended Short Wave Infrared Range) : domaine infrarouge pour lequel : $1,7\ \mu m < \lambda < 2,5\ \mu m$ |
| MWIR | Moyen infrarouge (Mid Wave Infrared Range) : domaine infrarouge pour lequel : $3\ \mu m < \lambda < 5\ \mu m$ |
| LWIR | Infrarouge lointain (Long-Wave Infrared Range) : domaine infrarouge pour lequel : $8\ \mu m < \lambda < 12\ \mu m$ |
| UV | ultraviolets |
| u.a. | unités arbitraires |

# Remerciements

J'aimerais avant tout remercier mes rapporteurs, Bruno Masenelli de l'Institut des Nanosciences de Lyon (INL) et Laurence Ressier du Laboratoire de Physique et de Chimie des Nano-Objets (LPCNO) qui ont accepté de rapporter mes travaux de thèse. Je tiens également à remercier Nicolas Péré-Laperne de LYNRED et Alexa Courty du laboratoire MONARIS pour avoir accepté de faire partie de mon jury. J'ai beaucoup apprécié les discussions que j'ai pu avoir avec tous les membres du jury pendant la soutenance : vous vous êtes montrés très enthousiastes envers mes résultats de thèse et avez posé des questions très pertinentes et intéressantes sur tous les aspects de mon travail.

Bien évidemment, je souhaite remercier mon encadrant Emmanuel Lhuillier pour avoir monté ce projet de thèse qui m'a beaucoup plu. Je ne connaissais pas spécialement les *quantum dots*, le confinement quantique ni même quoi que ce soit sur la détection infrarouge. Mais j'ai tout de suite senti que je serai dans mon élément avec ce sujet très pluridisciplinaire qui mêlait synthèse des matériaux, caractérisations à la fois structurales, optiques et électroniques, pour aller jusqu'à la conception d'un dispositif de détection infrarouge. La maîtrise de toutes ces étapes m'a permis de m'approprier pleinement ce sujet de recherche, de comprendre au mieux les enjeux et d'avoir très vite un certain recul sur les expériences, et c'est une des grandes forces de ton groupe de recherche. J'aimerais également te remercier pour ton encadrement : j'ai l'impression que tu as très vite compris la manière dont je travaillais, tu m'as rapidement fait confiance et j'ai toujours pu avancer à mon rythme. En plus, grâce à ton don assez incroyable pour écrire et relire les articles ou les manuscrits, on peut facilement enchaîner les projets. Enfin, tu assures une très bonne ambiance au sein de ton groupe, notamment grâce à la traditionnelle bière du vendredi soir !

Merci également à Emmanuelle Lacaze pour avoir accepté d'être ma directrice de thèse. Même si nous ne nous voyions pas très souvent pour discuter des avancées de mon projet, tu as relu le manuscrit et pris le temps d'assister à ma répétition de soutenance en faisant des remarques pertinentes malgré ton emploi du temps parfois chargé. Merci également au labex Matisse qui a permis à cette thèse de voir le jour et de m'avoir financée. L'année de mon arrivée en thèse (2016) fut l'année de création des Comités de Suivi de Thèse à l'ED397. Merci donc à François Dubin, Gilles Hug et Riad Haidar pour avoir assisté à ces rencontres et avoir suivi l'évolution de mes travaux de thèse pendant ces 3 ans.

J'aimerais ensuite remercier les membres (actuels ou anciens) de l'équipe OCN : Clément Livache, Nicolas Goubet, Charlie Gréboval, Audrey Chu, Junling Qu, Julien Ramade, Prachi Rastogi. Ça a été un vrai plaisir de travailler avec vous ! Il y a un super esprit de collaboration dans ce groupe : tout le monde est prêt à donner un coup de main quand il y a une manip' un peu originale à faire. Merci aussi à Hervé Cruguel et Sébastien Royer, toujours prêts à venir aider.

Merci à tous les jeunes de l'INSP, doctorants, anciens doctorants, post-docs, anciens post-docs, stagiaires : Camille Lagoin, Violette Steinmetz, Dylan Amelot, Suzanne Dang, Sunniva Indrehus, Samuel Peillon, Alberto Curcella, Danilo Longo, Léo Bossard-Gianesini, Sophie Malaquin, Ronan Delalande, Lounès Lounis, Lise Picaut, Romain Anankine, Nathalie Bonnatout, Pierre-Yves Perrin, Francesco Bresciani, Amaury Triboulin, Louis Bondaz, Elisa Meriggio Helen Ibrahim, et tous les autres.

Je remercie également tous les membres de l'INSP qui ont facilité ces trois années au laboratoire, tant sur le plan technique qu'administratif : Valérie Guezo et toute l'équipe de gestion, Cécile Lefèbvre au

Pour terminer, j'aimerais remercier tous les amis de l'ESPCI que j'ai continué à voir de manière plus ou moins régulière pendant ces trois ans de thèse, en particulier les BH et affiliés qui sont restés à Paris et avec qui j'ai pu tester les différents restos (chinois, burgers, salades, sandwichs) du quartier autour de Jussieu. Une attention particulière pour mes fillots Elia Henry et Florian Benoît, ma copine de cross fit Caroline Venet (je suis chaud pour continuer l'année prochaine si tu veux !) et mes anciens colocs Reda Belbahri, Jean-Baptiste Thomazo, Domitille Le Cornec, Antoine Delhomme et Marie Boulez avec qui j'ai vécu deux belles années dans la plus chouette des maisons. Merci également à mon amie d'enfance, Tiphaine Maurel, installée à Paris depuis peu qui a pu écouter mes péripéties de thèse dans les meilleurs restaurants pas chers de Paris, bientôt les Petites Tables n'auront plus de secrets pour nous !

Je remercie évidemment mes parents qui m'ont toujours soutenue dans tous mes projets. Merci d'avoir fait le déplacement pour assister à ma soutenance et d'avoir essayé de comprendre ce sur quoi je travaillais. Merci également à ma petite sœur Florette Martinez qui commence sa thèse quand je la termine: tu vois ce qui t'attend à partir de l'année prochaine !

Enfin, merci à Quentin Magdelaine qui me supporte depuis plus de cinq ans maintenant. Tu es toujours là pour m'épauler, pour me calmer quand je m'énerve après mes manips qui ne fonctionnent pas, pour me remonter le moral. Je ne sais pas ce que je ferai sans toi, merci d'être à mes côtés.

# Introduction générale

Les nanomatériaux, c'est-à-dire les matériaux dont l'une des dimensions est inférieure à la centaine de nanomètres, suscitent un grand intérêt dans le domaine de la physico-chimie des matériaux. Leur petite taille leur permet d'acquérir des propriétés nouvelles que l'on ne retrouve pas dans les matériaux massifs.

Dans le cas des métaux par exemple, réduire la taille du matériau à quelques nanomètres permet de modifier l'interaction du métal avec la lumière. Les métaux massifs sont bien plus grands que la longueur d'onde de la lumière visible (400 – 800 nm). À l'inverse, les métaux nanométriques dont la taille est bien inférieure à la longueur d'onde de la lumière perçoivent un champ électromagnétique quasi uniforme à leur échelle. Cela entraîne l'apparition de propriétés optiques caractéristiques de l'échelle nanométrique. Ainsi, au XIX$^{ème}$ et au début du XX$^{ème}$ siècle, Michael Faraday (1) et Gustav Mie (2) ont été les premiers à attribuer la couleur rouge « rubis » de solutions d'or à la taille (< 100 nm) des particules d'or qui la constituaient.

Les semiconducteurs nanométriques peuvent également présenter de nouvelles propriétés par rapport aux semiconducteurs massifs. Lorsque la taille du semiconducteur est inférieure à la taille d'une paire électron-trou, ou exciton, des propriétés de confinement quantique apparaissent. Les propriétés optiques, optoélectroniques ainsi que les propriétés de transport deviennent alors dépendantes de la taille. Ainsi, le séléniure de cadmium (CdSe) et le phosphure d'indium (InP) qui émettent dans le rouge à l'état massif, peuvent émettre dans l'orange, le jaune, le vert et le bleu quand leur taille diminue.

La synthèse colloïdale, en solution, de nanocristaux de semiconducteurs ayant des propriétés optiques dans le visible tels que le séléniure de cadmium (CdSe), le sulfure de cadmium (CdS), le sulfure de zinc (ZnS) ou le phosphure d'indium (InP), est bien maîtrisée. Des dispositifs commerciaux à base de nanocristaux de CdSe ou InP sont déjà commercialisés, dans les écrans QLED pour les télévisions par *Samsung* par exemple. À présent, l'enjeu est de transférer cette technologie de nanocristaux colloïdaux de semiconducteurs dans des gammes d'énergies plus faibles. On peut par exemple penser à l'utilisation de tels nanocristaux pour faire de la détection infrarouge, d'autant que les technologies actuelles présentent souvent le défaut d'être soit chères, soit peu performantes.

C'est dans ce cadre que s'inscrivent mes travaux de doctorat. Les nanocristaux de tellure de mercure (HgTe) et de séléniure de mercure (HgSe) que j'étudie absorbent les photons infrarouges et présentent des propriétés de photoconduction. Les électrons et les trous générés sous illumination peuvent être collectés par des électrodes, et le courant électrique ainsi généré constitue le signal de détection infrarouge. L'objectif de mes travaux de thèse est d'améliorer notre connaissance et notre maîtrise des propriétés optoélectroniques de ces nanocristaux à différents degrés de confinement. Ces études nous permettront de proposer des systèmes de détection adaptés aux propriétés optiques et de transport des nanocristaux.

Mon doctorat a été réalisé à l'Institut des Nanosciences de Paris (INSP) à Sorbonne Université, en collaboration avec le Laboratoire d'Étude et Physique des Matériaux (LPEM) de l'ESPCI. Il a été financé par le Labex Matisse.

# Organisation du manuscrit

Le PREMIER CHAPITRE commence par un bref rappel sur les propriétés des semiconducteurs, notamment sur l'énergie de bande interdite et les différents moyens permettant de la modifier. À cette occasion, le confinement quantique est présenté en détail. Les matériaux utilisés pendant le doctorat sont ensuite introduits. Il s'agit des nanocristaux colloïdaux confinés de séléniure de mercure (HgSe) et de tellure de mercure (HgTe) qui présentent des propriétés optiques dans l'infrarouge et qui sont donc candidats potentiels pour être utilisés dans des détecteurs infrarouges. Les figures de mérite caractéristiques du domaine de la détection infrarouge qui permettent d'évaluer les performances des détecteurs, ainsi que les technologies actuelles sont présentées. La dernière partie de ce chapitre est consacrée au transport dans les films de nanocristaux, et aux propriétés de photoconduction de ces dispositifs.

Le DEUXIÈME CHAPITRE de ce manuscrit est dédié à la **structure électronique** des nanocristaux colloïdaux de HgSe et HgTe. La structure électronique renseigne notamment sur les transitions optiques accessibles et sur les énergies absolues des niveaux énergétiques des nanocristaux. Ces informations sont essentielles pour concevoir un dispositif de détection capable d'extraire efficacement les porteurs générés sous illumination. Les techniques expérimentales qui permettent de remonter à la structure électronique sont présentées, et l'accent est mis sur la photoémission que j'ai beaucoup utilisée pendant mon doctorat. L'influence de la taille des nanocristaux sur les structures électroniques de HgSe et HgTe est ensuite étudiée. Pour ces deux matériaux à faible bande interdite, la taille influe sur le niveau de dopage, c'est-à-dire sur l'énergie de Fermi par rapport aux niveaux électroniques. Dans le cas de HgSe, une transition semiconducteur-métal est même observée à mesure que la taille des cristaux augmente. Enfin, pour avoir une vision complète de la structure électronique, la distribution des pièges (qui modifient les énergies de transition et perturbent les propriétés de transport) est étudiée.

Le TROISIÈME CHAPITRE est consacré au **contrôle du dopage** dans les nanocristaux. La modification du dopage permet d'accéder à différentes transitions, notamment aux transitions intrabandes, au sein d'un nanocristal. Le dopage a également une influence sur le courant d'obscurité généré par les nanocristaux, courant qui est directement lié au bruit dans un détecteur infrarouge. Dans ce chapitre, plusieurs techniques permettant de contrôler le dopage des nanocristaux post-synthèse sont présentées. L'introduction de dipôles à la surface des nanocristaux qui modifient l'énergie des niveaux électroniques est étudiée. Une deuxième stratégie basée sur le greffage de molécules oxydantes de polyoxométallates à la surface des nanocristaux est également étudiée et un transfert d'électrons du nanocristal vers ces molécules oxydantes est démontré.

Enfin, le QUATRIÈME CHAPITRE présente **l'intégration des nanocristaux colloïdaux dans des systèmes de détection infrarouge** qui permettent d'extraire efficacement les porteurs créés sous illumination, tout en gardant un niveau de bruit faible. Des photodiodes à base de nanocristaux dans le visible et dans le proche infrarouge ont été développées dans la littérature dans le but de réaliser des cellules solaires. L'architecture de ces diodes nous sert donc de point de départ à l'élaboration d'un dispositif de détection infrarouge pour des longueurs d'onde plus élevées, autour de 2.5 µm, à base de nanocristaux de HgTe. Les travaux du chapitre 2 sur la structure électronique permettent d'élaborer une structure adaptée à cette gamme d'énergie. Ce chapitre sera également l'occasion d'aborder la question de la reconstruction d'une image infrarouge, dans l'idée d'intégrer les détecteurs infrarouges dans des caméras. Une matrice de 100 pixels fabriquée au laboratoire, couverte d'un film de nanocristaux colloïdaux de HgTe, permet de reconstruire le profil d'intensité d'un laser infrarouge.

# CHAPITRE 1
# Contexte et état de l'art

L'objectif de mon doctorat est de développer la technologie des nanocristaux colloïdaux pour les intégrer dans des détecteurs infrarouges à la fois performants et à moindre coût. Dans ce premier chapitre, je commencerai par rappeler les principales caractéristiques des semiconducteurs et plus spécifiquement sous la forme de nanocristaux colloïdaux. Je décrirai les différents domaines de la gamme infrarouge ainsi que les principales figures de mérite associées. J'exposerai ensuite les différentes technologies actuelles de détecteurs infrarouges et identifierai leurs avantages et leurs défauts. Je terminerai ce chapitre en présentant l'intégration des nanocristaux colloïdaux dans des dispositifs infrarouges et les paramètres importants à contrôler pour obtenir un détecteur fonctionnel.

## I.   Propriétés des semiconducteurs

### 1.   Structure de bandes d'un semiconducteur

#### a.   Bande de valence et bande de conduction

À l'inverse des atomes isolés qui ont des niveaux électroniques discrets, les propriétés électroniques des matériaux massifs, métaux, isolants et semiconducteurs, peuvent être décrites en utilisant la théorie des bandes. Ces bandes décrivent des intervalles d'énergies autorisées pour les électrons, séparés par des bandes dites « interdites ». Deux bandes jouent un rôle particulier dans le transport des électrons et des trous (Figure 1) : la **bande de valence**, qui est la bande complètement remplie la plus haute en énergie, et la **bande de conduction** qui est la première bande inoccupée.[1]

L'espacement entre la bande de valence et la bande de conduction permet de déterminer la nature du matériau : si elles sont séparées de plusieurs électronvolts (eV), le matériau est un isolant ; si elles sont séparées de l'ordre de 1 eV, le matériau est un semiconducteur ; si elles se touchent, on parle de semimétal et si elles se chevauchent, le matériau est un métal. L'énergie entre la bande de conduction et la bande de valence est appelée **énergie de bande interdite** ou *gap* ($E_G$).

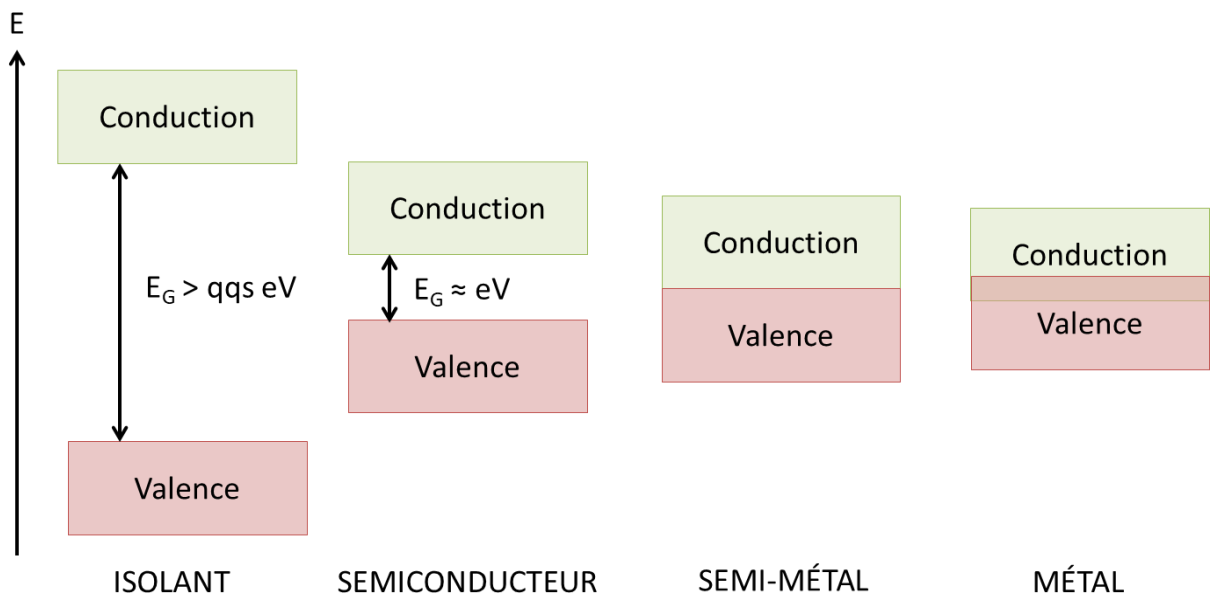

*Figure 1 : Schéma présentant les positions des bandes de conduction et de valence pour les métaux, les semimétaux, les semiconducteurs et les isolants.*

---

[1] Sauf dans le cas d'un métal où des électrons sont déjà présents dans la bande de conduction

Dans ce manuscrit, nous nous intéresserons plus précisément au cas des semimétaux et des semiconducteurs. Ces derniers sont présents dans la colonne IV du tableau périodique de Mendeleïev (Si, Ge), mais peuvent également être obtenus avec des composés isoélectroniques de la colonne IV tels que GaAs (III – V) ou CdSe (II-VI). Des composés IV- VI, tels que PbSe ou SnTe sont également des semiconducteurs.

Le faible écart énergétique entre la bande de valence et la bande de conduction permet à quelques électrons d'être promus dans la bande de conduction par excitation thermique.[2] Lorsque le semiconducteur est illuminé par des photons dont l'énergie est supérieure à celle de la bande interdite, les photons sont absorbés et des électrons passent dans la bande de conduction. Ces électrons peuvent :

-   soit être extraits en appliquant une tension : la conductivité change sous illumination, c'est le principe de la photoconduction ;
-   soit se recombiner avec les trous laissés dans la bande de valence : le semiconducteur peut alors réémettre des photons.

### b.  Diagramme de bandes d'un semiconducteur

L'énergie cinétique d'un électron libre est proportionnelle au carré du vecteur d'onde de l'électron et peut s'écrire :

$$E = \frac{\hbar^2 k^2}{2m_0} \tag{1.1}$$

Où E est l'énergie de l'électron, $\hbar$ est la constante de Planck réduite, k est le vecteur d'onde et $m_0$ est la masse de l'électron. Dans le cas où l'électron est dans un cristal, le potentiel du cristal vient renormaliser cette dépendance parabolique, mais elle reste conservée au centre de la zone de Brillouin (k = 0, point Γ), à condition d'utiliser la masse effective des porteurs, qui dépend de la courbure des bandes. On peut donc écrire, en prenant l'origine des énergies en haut de la bande de valence :

$$E_h = - \frac{\hbar^2 k^2}{2m_h^*} \; et \; E_e = \; E_G + \frac{\hbar^2 k^2}{2m_e^*} \tag{1.2}$$

Où $E_h$ (respectivement $E_e$) est l'énergie des états dans la bande de valence (resp. dans la bande de conduction), $\hbar$ est la constante de Planck réduite, k est le vecteur d'onde de l'électron, $m_h^*$ (resp. $m_e^*$) est la masse effective du trou dans la bande de valence (resp. de l'électron dans la bande de conduction) et $E_G$ est l'énergie de bande interdite.

Le vecteur d'onde k est le même dans l'expression de $E_h$ et $E_e$ (1.2) si le semiconducteur est à *gap* direct, c'est-à-dire que le maximum de la bande de valence se situe au même vecteur d'onde que le minimum de la bande de conduction (c'est le cas de CdSe, voir Figure 2). Dans le cas contraire, le semiconducteur est dit à *gap* indirect.

---

[2] La probabilité de présence des électrons dans la bande de conduction en fonction de la température peut être calculée en utilisant la statistique de Fermi – Dirac : $f(E) = \frac{1}{1+\exp(\frac{E-\mu}{k_B T})}$, où f(E) est la probabilité de trouver un électron à l'énergie E, μ est le potentiel chimique des électrons, $k_B$ est la constante de Boltzmann et T est la température.

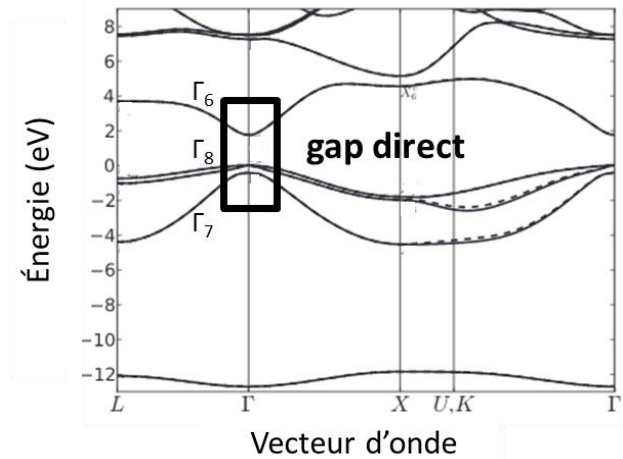

*Figure 2 : Diagramme de bandes de CdSe (3, 4). Les lettres en abscisses correspondent à plusieurs valeurs particulières du vecteur d'onde. En particulier, le point Γ correspond au point k = 0 dans les trois directions de l'espace réciproque.*

Lors de l'étude de la structure de bandes d'un semiconducteur, l'attention est focalisée sur quatre bandes: la bande $\Gamma_6$, qui est souvent la bande de conduction, les deux bandes $\Gamma_8$, dégénérées au point Γ, jouant souvent le rôle de bande de valence, et la bande $\Gamma_7$ légèrement décalée vers les basses énergies[3] par rapport à $\Gamma_8$. Dans la plupart des semiconducteurs, l'énergie de bande interdite $E_G$ est donc égale à la différence d'énergie $E_{\Gamma 6} - E_{\Gamma 8}$.

Dans la suite de ce manuscrit, je garderai l'écriture $E_G$ pour décrire l'écart en énergie entre la bande de conduction et la bande de valence, et préciserai la valeur de $E_{\Gamma 6} - E_{\Gamma 8}$ lorsque celle-ci sera différente.

## 2. Accordabilité de l'énergie de bande interdite

Les propriétés optiques et électriques des semiconducteurs dépendent fortement de l'énergie de bande interdite, dont certaines sont présentées dans le Tableau 1. HgTe a une structure de bande inversée (la structure de bandes sera décrite plus en détails dans la suite de ce manuscrit, p29), $E_G$ est donc différent de $E_{\Gamma 6} - E_{\Gamma 8}$. Ce matériau est un semi-métal, son énergie de bande interdite est donc nulle, $E_{\Gamma 6} - E_{\Gamma 8}$ est négatif et vaut – 0,3 eV.

---

[3] La différence $\Gamma_8 - \Gamma_7$ est le couplage spin orbite.

*Tableau 1 : Énergies de bande interdite de quelques semiconducteurs classiques. Les lettres I et D dans la ligne « Nature gap » correspondent au caractère indirect et direct respectivement (5).*

| Groupe | IV | | III - V | | | | II - VI | | | |
|---|---|---|---|---|---|---|---|---|---|---|
| **Nom** | Si | Ge | InP | GaAs | InSb | InAs | CdSe | CdTe | ZnS | HgTe |
| $E_G$ **(eV)** | 1,12 | 0,66 | 1,27 | 1,42 | 0,17 | 0,36 | 1,70 | 1,56 | 3,68 | 0 |
| **Nature gap** | I | I | D | D | D | D | D | D | D | D |

Dans cette partie, je présenterai plusieurs paramètres permettant de modifier l'énergie de bande interdite : le dopage, les alliages, la température et le confinement quantique.

### a. Dopage

Le dopage des semiconducteurs consiste à introduire des dopants dans le cristal qui vont apporter un excès d'électrons (dopage n) ou de trous (dopage p) dans le système. Par exemple, dans le silicium, introduire des atomes de phosphore (colonne V) revient à introduire un excès d'électrons tandis qu'introduire des atomes de bore (colonne III) revient à retirer des électrons donc à introduire un excès de trous. Ces dopants entraînent l'apparition de niveaux dans la bande interdite, près de la bande de conduction pour un dopage n, près de la bande de valence pour un dopage p (Figure 3).

Cette technique ne permet pas à proprement parler d'accorder l'énergie de bande interdite, puisque les énergies de la bande de valence et la bande de conduction ne sont pas modifiées, mais elle introduit un nouveau niveau dans la bande interdite qui rend possible des transitions de plus faibles énergies.

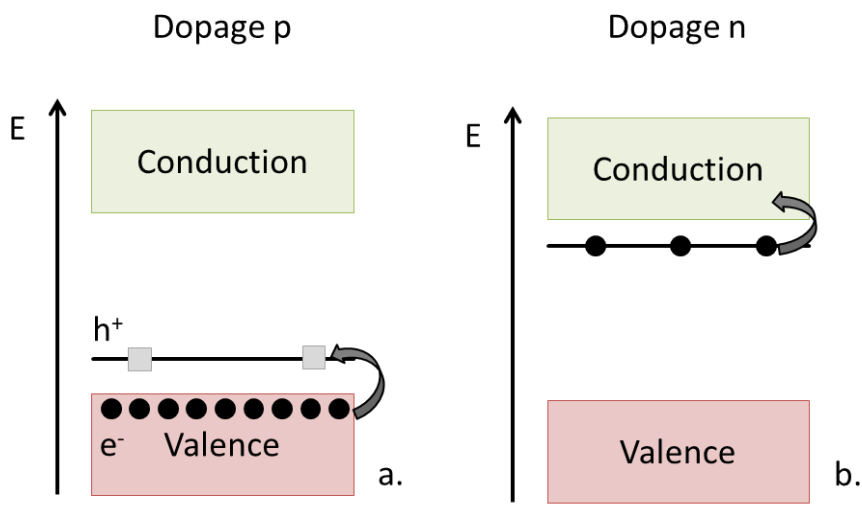

*Figure 3 : Schéma de principe d'un dopage p (a) ou n (b) dans un semiconducteur. Les carrés gris représentent les trous, les disques noirs représentent les électrons.*

Un paramètre important à déterminer et à contrôler lors du dopage de semiconducteurs est l'énergie du **niveau de Fermi**, qui correspond à l'énergie maximale des électrons à température nulle. Pour un semiconducteur intrinsèque (non dopé), le niveau de Fermi se situe au milieu de la bande interdite. Pour un semiconducteur dopé n (resp. p), le niveau de Fermi se rapproche de la bande de conduction (resp. de valence).

### b. Alliages

Pour ajuster précisément l'énergie de bande interdite, une possibilité est d'utiliser un alliage de deux semiconducteurs. L'énergie de bande interdite de l'alliage sera comprise entre les énergies des bandes interdites des deux semiconducteurs seuls (*6*).

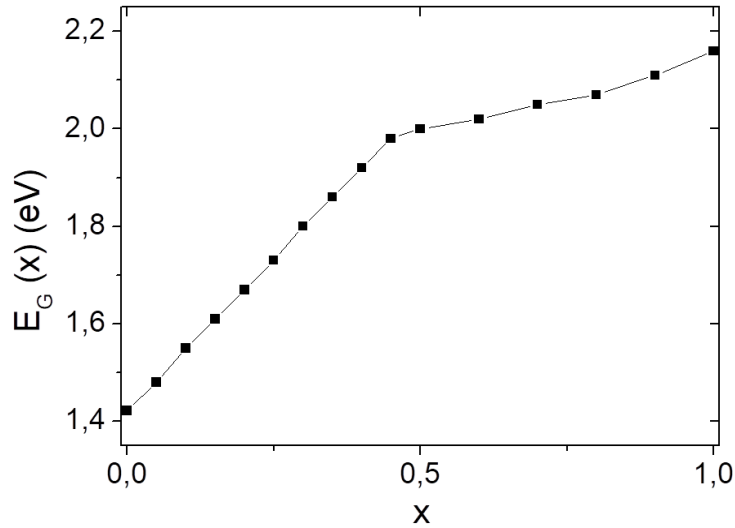

*Figure 4 : Évolution de l'énergie de la bande interdite $E_G$ avec x dans l'alliage $Al_xGa_{1-x}As$ (7).*

Les alliages sont réalisés entre deux semiconducteurs appartenant au même groupe : AlAs avec GaAs pour les III-V par exemple. Sur la Figure 4 sont présentées les énergies de bande interdite (du *gap* direct) de l'alliage $Al_xGa_{1-x}As$ pour différentes valeurs de x.

### c. Température

L'énergie de bande interdite d'un semiconducteur dépend de la température selon la loi empirique de Varshni (*8*) :

$$E_G(T) = E_G(0\,K) - \frac{\alpha T^2}{T + \beta} \qquad (1.3)$$

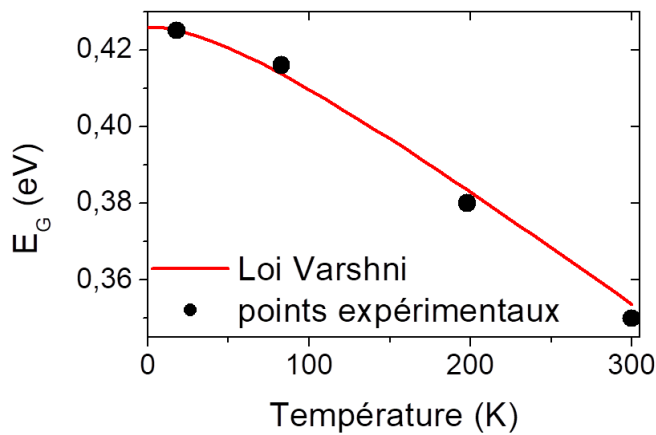

*Figure 5 : Dépendance de l'énergie de bande interdite du gap direct dans InAs en fonction de la température. Les points noirs représentent les résultats expérimentaux, la courbe rouge représente la dépendance empirique prévue par la loi de Varshni.*

Pour la plupart des semiconducteurs, α est positif : la bande interdite est donc plus large à basses températures qu'à hautes températures. Ainsi, pour InAs, on a α = 3,158 × 10⁻⁴ eV/K, β = 93 K et $E_G$ (0 K) = 0,426 eV (voir Figure 5). Le comportement est inversé dans le cas de HgTe, dont le coefficient α est négatif.

### d. Confinement quantique

Le confinement quantique est un phénomène qui intervient dans les semiconducteurs lorsqu'au moins une de ses dimensions devient inférieure au rayon de Bohr ($a_0$), qui est la distance moyenne entre l'électron et le trou générés après absorption d'un photon. Le modèle de Bohr donne l'expression analytique du rayon de Bohr :

$$a_0 = \frac{h^2 \varepsilon_0 \varepsilon_r}{m^* e^2 \pi}, avec\ m^* = \frac{1}{\frac{1}{m_e^*} + \frac{1}{m_h^*}} \qquad (1.4)$$

Où $a_0$ est le rayon de Bohr, h est la constante de Planck, $\varepsilon_0$ est la permittivité diélectrique du vide, $\varepsilon_r$ est la constante diélectrique relative du matériau, $m_e^*$ est la masse effective de l'électron, $m_h^*$ est la masse effective du trou et e est la charge élémentaire. Le rayon de Bohr est généralement compris entre 2 et 50 nm : il est de 6 nm pour CdSe (*9*) et de 40 nm pour HgTe (*10, 11*) par exemple.

Pour un semiconducteur confiné dans une direction (puits quantique), par exemple la direction z, les vecteurs d'onde de l'espace réciproque dans la direction z ($k_z$) ne peuvent prendre que des valeurs discrètes valant $n\pi/a$, où n est un entier relatif non nul et a est la taille du cristal dans la direction de confinement (voir Figure 6a). L'énergie de bande interdite correspond donc à l'écart entre la bande de conduction et la bande de valence à k = $k_z$ = $\pi/a$ et non plus à k = 0.

Ce confinement entraîne également l'apparition de sous-bandes dans les directions non confinées comme illustré sur la Figure 6. Si des électrons sont présents dans la bande de conduction, ces transitions apparues avec le confinement quantique seront accessibles : on parle alors de transitions inter sous-bandes ou intrabandes.

Changer la taille d'un semiconducteur en dessous du rayon de Bohr permet donc de contrôler finement l'énergie de la bande interdite. On notera que l'énergie obtenue est nécessairement supérieure à l'énergie du semiconducteur massif.

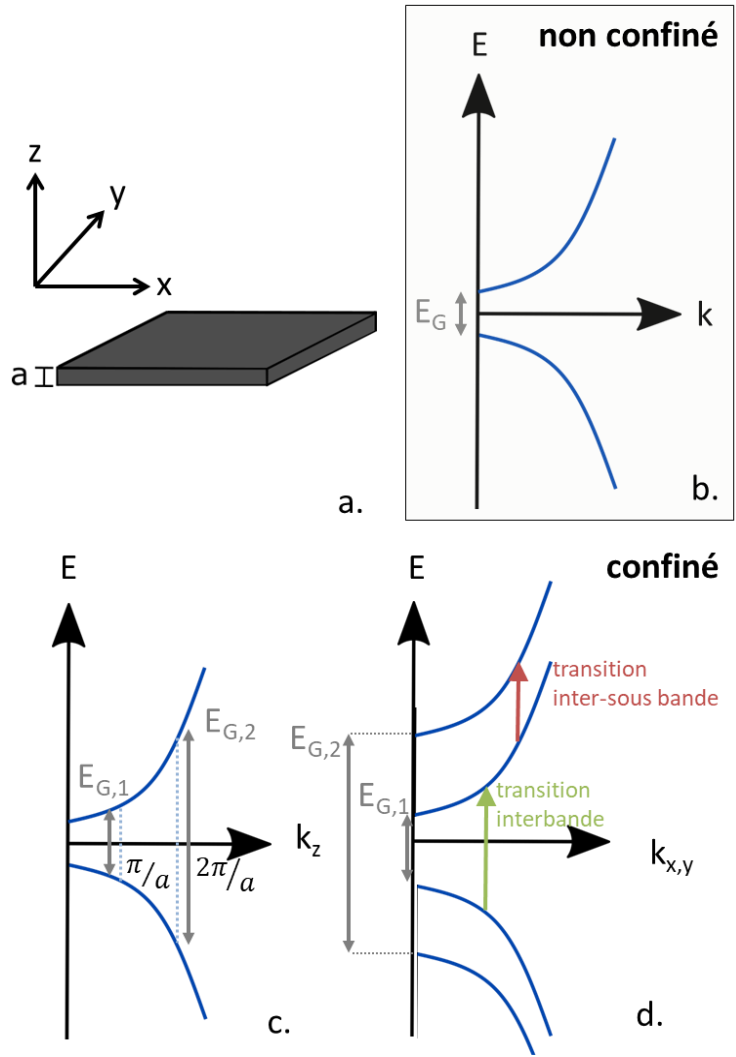

*Figure 6 : (a) Schéma d'un semiconducteur dont la taille selon la dimension z est plus petite que selon les autres dimensions ; (b) diagramme de bandes dans le cas non confiné où a > a₀. Dans le cas confiné a < a₀, le diagramme de bandes est modifié : selon la direction z, seules des valeurs discrètes de k sont autorisées (c) ; selon les directions x et y, on note l'apparition de sous bandes issues du confinement en z (d).*

## II. Nanocristaux colloïdaux à base de semiconducteurs

Les nanocristaux colloïdaux de semiconducteurs sont des matériaux synthétisables en laboratoire, relativement peu coûteux, et dont l'énergie de bande interdite peut être ajustée en changeant la taille (donc leur confinement quantique). Dans cette partie, je commencerai par détailler les propriétés électroniques de ces matériaux en revenant sur le confinement quantique, cette fois à trois dimensions avant de présenter les nanocristaux les plus matures dans différentes gammes du spectre électromagnétique. Je terminerai par la présentation de la synthèse par voie colloïdale.

### 1. Confinement quantique des nanocristaux

Dans la partie précédente, j'ai présenté le confinement quantique dans une direction de l'espace. Dans le cas des nanocristaux, le confinement se fait dans les trois directions de l'espace : on parle alors de boîte quantique (ou *quantum dot, QD*). Avec ces matériaux, il n'y a plus de dispersion possible dans aucune direction (*12*). Les niveaux électroniques sont discrets, et l'écartement entre les niveaux est fixé par les valeurs de k autorisées par le confinement quantique, comme présenté sur la Figure 7.

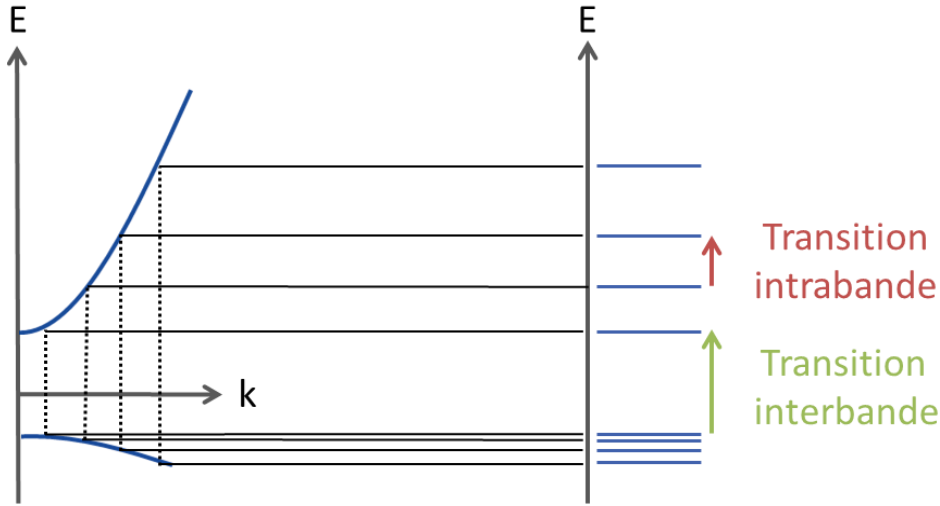

*Figure 7 : Schéma représentant les différents niveaux d'énergie d'un nanocristal colloïdal. Comme dans le cas des systèmes que j'étudie, la masse effective des trous est plus importante que celle des électrons : les niveaux correspondant à la bande de valence sont donc plus rapprochés que ceux correspondant à la bande de conduction.*

Dans le cas d'un nanocristal sphérique, les valeurs de k autorisées ne sont plus des multiples de $\pi/a$ où a est la taille du nanocristal, mais sont liées aux racines de la fonction de Bessel sphérique (*12*) par la formule :

$$k = \frac{\emptyset_{n,l}}{R} \qquad (1.5)$$

Où k est le vecteur d'onde, $\emptyset_{n,l}$ est la n-ième racine de la L-ième fonction de Bessel ($J_L$) et R est le rayon du nanocristal.

*Tableau 2 : Calcul des premières valeurs de $\emptyset_{n,L}$ et noms des niveaux correspondants*

| (n,L) | (1,0) | (1,1) | (1,2) | (2,0) |
|-------|-------|-------|-------|-------|
| $\Phi_{n,L}$ | π | 4,49 | 5,76 | 2π |
| Nom niveau | 1S | 1P | 1D | 2D |

Au premier ordre, l'espacement entre les niveaux peut être calculé en supposant une relation parabolique entre l'énergie et le vecteur d'onde dans le cas du matériau massif. En prenant l'origine des énergies au niveau du haut de la bande de valence, on a alors

$$E_{k,h} \approx \frac{\hbar^2 k^2}{2m_h^*} \; et \; E_{k,e} \approx E_G + \frac{\hbar^2 k^2}{2m_e^*} \qquad (1.6)$$

Où $E_{k,h}$ (resp. $E_{k,e}$) est l'énergie d'un niveau provenant de la bande de valence (resp. bande de conduction) associé au vecteur d'onde k, k est le vecteur d'onde quantifié selon la formule définie en (1.6), $\hbar$ est la constante de Planck réduite, $m_h^*$ est la masse effective du trou, $m_e^*$ est la masse effective de l'électron et $E_G$ est l'énergie de bande interdite du matériau massif.

Bien qu'il n'y ait pas de bandes dans les nanocristaux, on utilise les termes « interbande » pour désigner une transition entre un niveau issu de la bande de valence et un niveau issu de la bande de conduction, et « intrabande » pour désigner une transition entre deux niveaux issus d'une même bande (conduction ou valence), voir Figure 7. La plus petite énergie de transition interbande correspond donc à l'écart en énergie entre le dernier niveau de la bande de valence $1S_h$ et le premier niveau de la bande de conduction $1S_e$.[4] Elle vaut :

$$1S_e - 1S_h \approx E_G + \frac{\hbar^2 \pi^2}{2m_h^* R^2} + \frac{\hbar^2 \pi^2}{2m_e^* R^2} + E_c \qquad (1.7)$$

Où R est le rayon du nanocristal et $E_c$ l'énergie d'interaction coulombienne entre le trou et l'électron, également appelée énergie de liaison de l'exciton. Ainsi, pour la transition interbande de plus basse énergie $1S_e - 1S_h$, le confinement quantique ajoute l'énergie $\frac{\hbar^2 \pi^2}{2m_h^* R^2} + \frac{\hbar^2 \pi^2}{2m_e^* R^2} + E_c$ à l'énergie de bande interdite du matériau massif.

## 2. Quels matériaux pour quelle application ?

La synthèse colloïdale est particulièrement mature pour les nanocristaux de semiconducteurs des colonnes II-VI. Les nanocristaux à base de cadmium ou de zinc sont bien adaptés pour fabriquer des émetteurs de lumières ; ceux à base de mercure ou de plomb semblent prometteurs pour les applications de détection infrarouge.

### a. Dans le visible

Les nanocristaux II-VI tels que CdSe, CdS, ZnS …, ont été très étudiés pendant les trente dernières années (*13*). Les paramètres de synthèse permettent d'obtenir des matériaux de tailles et de formes variées, comme présenté sur la Figure 8.

Les nanocristaux ont un ratio surface/volume important. Les atomes de surface apportent des défauts car ils ont une coordination incomplète, ce qui se traduit par l'apparition d'états de pièges dans la bande interdite. Pour utiliser des nanocristaux colloïdaux comme émetteurs de lumière, la présence de pièges dans la bande interdite qui peuvent favoriser les transitions non radiatives limite le rendement quantique de fluorescence.

---

[4] Les indices « h » indiquent que le niveau est issu de la bande de valence, les indices « e » que le niveau est issu de la bande de conduction

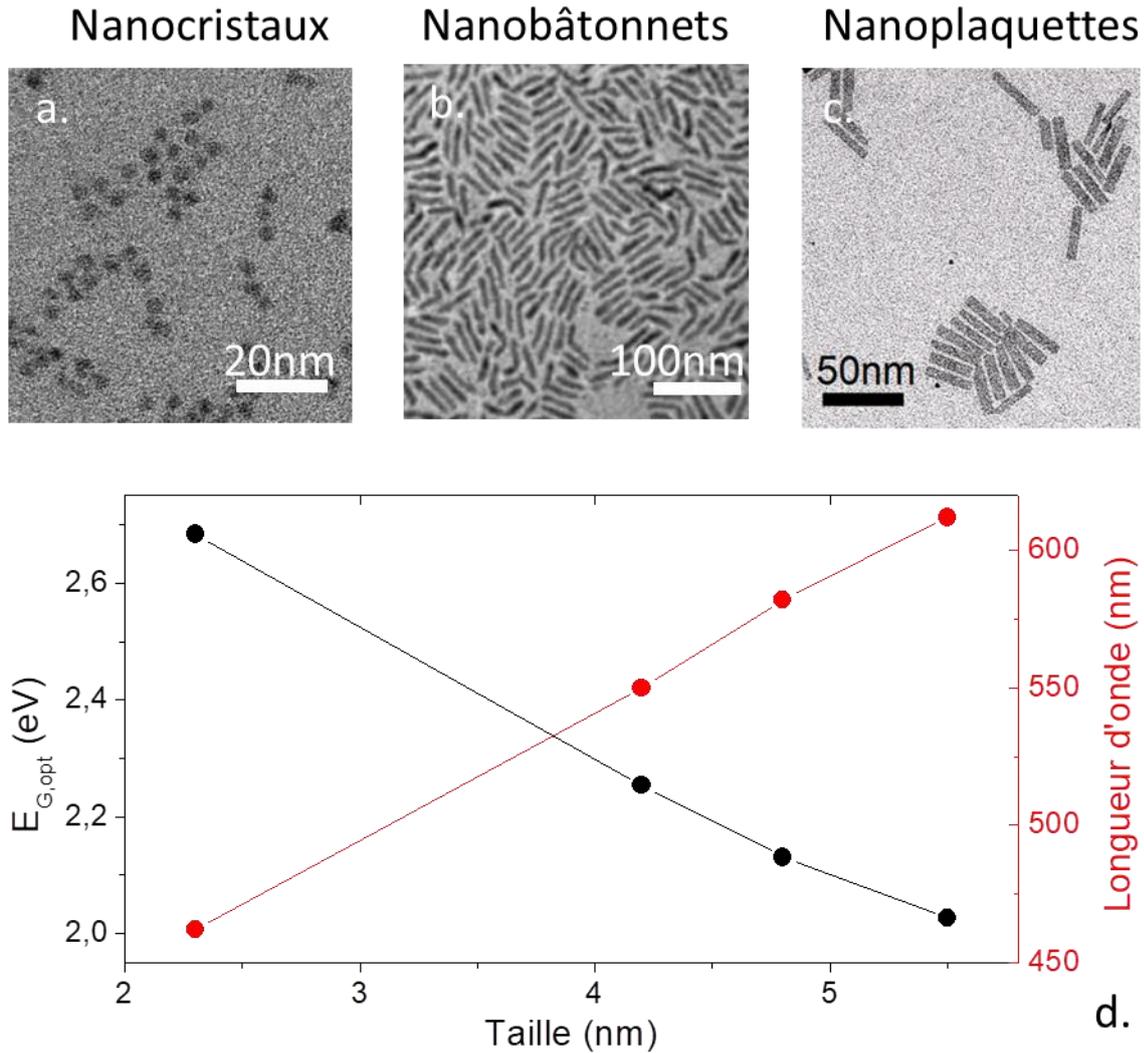

*Figure 8 : Les nanostructures de CdSe peuvent prendre plusieurs formes : des nanocristaux 0D (a), des nanobâtonnets 1D (b) et des nanoplaquettes 2D (c). La taille des nanocristaux est également bien contrôlable et permet de balayer une grande partie du spectre visible. (d) Énergie de la transition optique $E_{G,opt}$ et longueur d'onde correspondante pour différentes rayons de nanocristaux sphériques de CdSe (14).*

Des structures plus complexes, telles que des hétérostructures cœur – coquille basées sur la croissance d'une coque sur un nanocristal fabriqué au préalable, ont vu le jour. Elles permettent de confiner les électrons et les trous loin de la surface pour éviter le piégeage et donc augmenter le rendement de l'émission (Figure 9). Les synthèses de nanocristaux de CdSe/ZnS (15) sont robustes et matures, et ces matériaux sont actuellement utilisés dans des dispositifs grand public tels que les télévisions QLED.

Au-delà des technologies historiques II-VI, des nanocristaux III-V, comme InP (*16, 17*), ou à base de perovskites (*18*) sont également très étudiés dans la littérature pour leur application en tant qu'émetteurs de lumière.

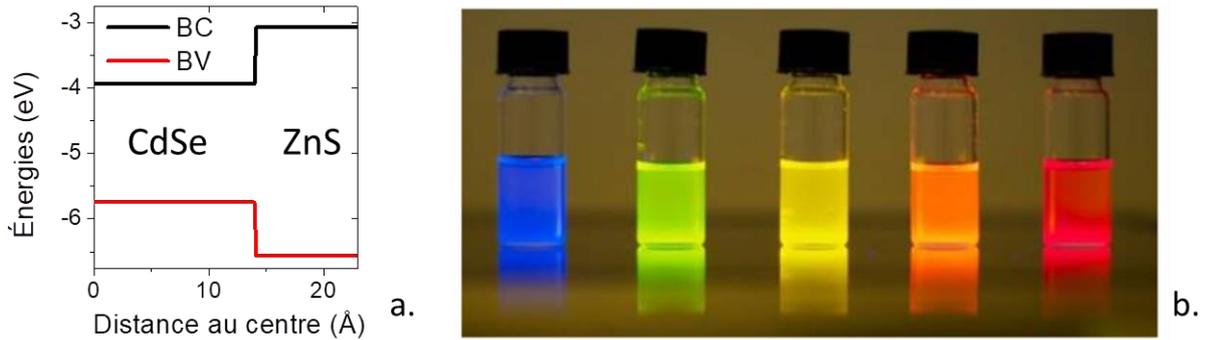

*Figure 9 : (a) Alignement de bandes entre CdSe et ZnS. Cet alignement est un alignement de type I, il confine donc les électrons et les trous dans le matériau central, loin de la surface. (b) Photo de nanocristaux de CdSe/Zns de différentes tailles. Les bleus sont fortement confinés (≈ 2 nm) tandis que les rouges ont la taille du rayon de Bohr (≈ 6 nm) - les synthèses de nanocristaux ont été réalisées par Nexdot© ainsi que la photo.*

### b. Dans l'infrarouge

L'infrarouge est la gamme du spectre électromagnétique qui s'étend entre 800 nm et quelques millimètres. À l'intérieur de cette gamme, on peut distinguer plusieurs domaines qui mènent à des applications différentes. Je commencerai donc par présenter ces différentes applications avant d'indiquer quels types de nanocristaux peuvent être utilisés.

### i. Présentation de la gamme infrarouge

La gamme de longueurs d'onde d'intérêt pour mon travail de thèse se situe entre le proche et le moyen infrarouge, soit entre 1 et 5 μm. Pour faciliter la compréhension du lecteur, une correspondance entre les différentes unités souvent utilisées dans ce domaine est présentée sur la Figure 10.

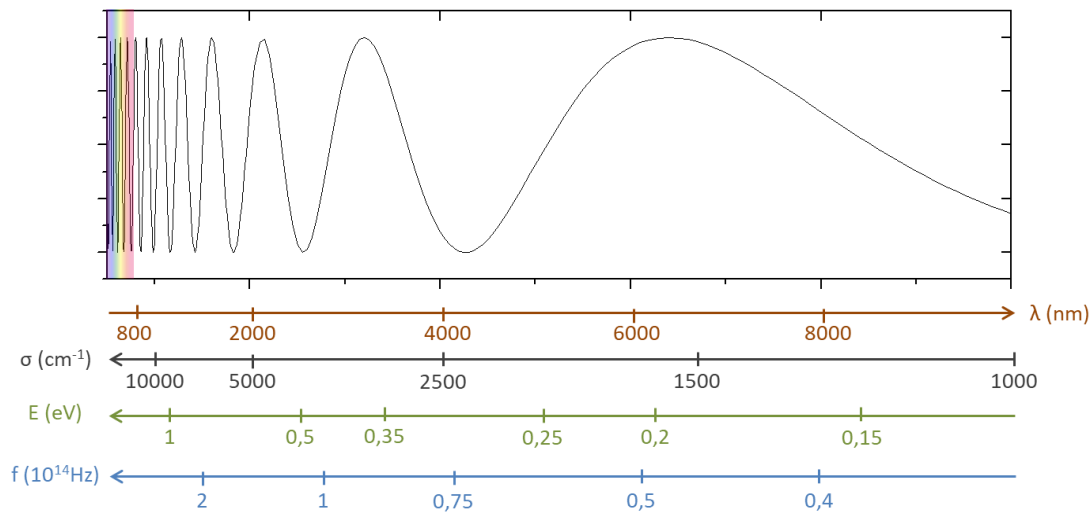

*Figure 10 : Illustration des différents domaines de l'infrarouge et correspondance entre les différentes unités (longueur d'onde en orange, nombre d'onde en noir, énergie en vert, fréquence en bleu)*

Entre 0,8 et 10 μm, on distingue quatre catégories menant à des applications différentes :

- le NIR (*near infrared* ou proche infrarouge) entre 0,8 et 1 μm ;
- le SWIR (*short wave infrared* ou infrarouge à ondes courtes) entre 1 et 2,5 μm ;
- le MWIR (*mid wave infrared* ou infrarouge à ondes moyennes) entre 3 et 5 μm ;

- le LWIR (*long wave infrared* ou infrarouge à grandes ondes) entre 8 et 12 μm.

Entre 5 et 8 μm, l'atmosphère absorbe les photons infrarouges, comme présenté sur la Figure 11. Il n'est donc pas possible d'envisager d'applications d'imagerie dans cette gamme.

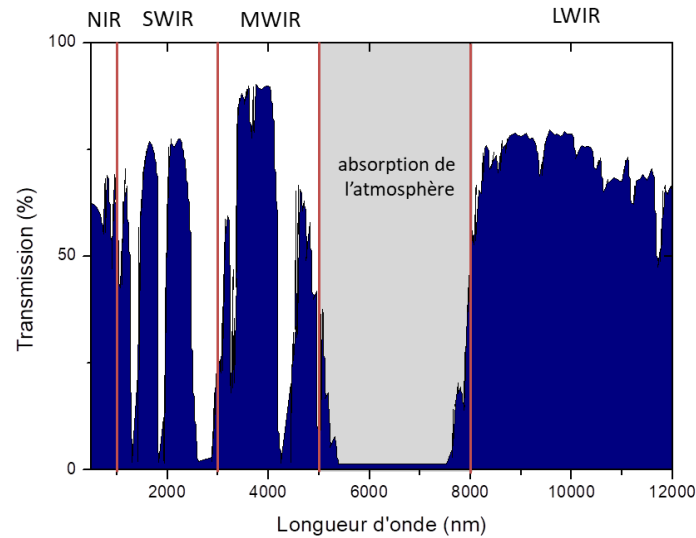

*Figure 11 : Transmission des infrarouges dans l'atmosphère à différentes longueurs d'ondes. Entre 5 et 8 μm, la transmission des photons infrarouges dans l'atmosphère est presque nulle.*

**NIR : 0,8 – 1 μm**

Dans cette gamme de longueurs d'onde, les infrarouges peuvent être utilisés pour imager les tissus, car ces derniers sont relativement transparents entre 700 et 900 nm (l'eau absorbe peu dans cette gamme).

Si le détecteur NIR est capable d'absorber les plus grandes énergies (dans le cas d'un semiconducteur intrinsèque par exemple), il peut être utilisé pour le photovoltaïque puisqu'il absorbe tous les photons visibles ainsi qu'une partie des photons infrarouges émis par le soleil.

**SWIR : 1 – 2,5 μm**

Cette gamme de longueurs d'onde est typiquement utilisée pour faire de l'imagerie active, c'est-à-dire avec une source de lumière telle qu'un laser infrarouge, les photons émis par les étoiles ou l'illumination nocturne. Les applications sont multiples : surveillance, détermination de défauts ou contrefaçons (les détecteurs SWIR ne sont pas sensibles à la couleur des objets et peuvent donc être plus sensibles aux défauts), tri, télécommunications, technologies LIDAR (*Light Detection And RAnging*, semblable au RADAR mais avec un signal lumineux)…

En progressant vers le moyen infrarouge, la réflexion des photons infrarouges sur les surfaces devient moins efficace car de nombreux matériaux absorbent dans cette gamme de longueurs d'onde et les objets deviennent leur propre source de lumière (voir section suivante).

**MWIR 3 – 5 μm et LWIR 8 – 12 μm**

Les gammes du MWIR et du LWIR permettent de faire de l'imagerie thermique, donc de faire une cartographie de la température d'une scène. L'imagerie thermique repose sur le fait que les objets deviennent leur propre source de lumière et rayonnent des photons dont le nombre et la longueur d'onde dépendent de la température, selon la loi de Planck :

$$L° = \frac{2hc^2n^3}{\lambda^5} \frac{1}{\exp\left(\frac{hc}{\lambda k_B T}\right) - 1} \tag{1.8}$$

Où L° est la luminance énergétique spectrale par unité de longueur d'onde (en W.m⁻³.sr⁻¹), h est la constante de Planck, c est la célérité de la lumière dans le vide, n est l'indice du milieu, λ est la longueur d'onde du photon émis et T est la température. L'évolution de L° en fonction de la température du corps et de la longueur d'onde est présentée sur la Figure 12. Plus source est chaude, plus elle émet de photons et plus la longueur d'onde de ces derniers est faible.

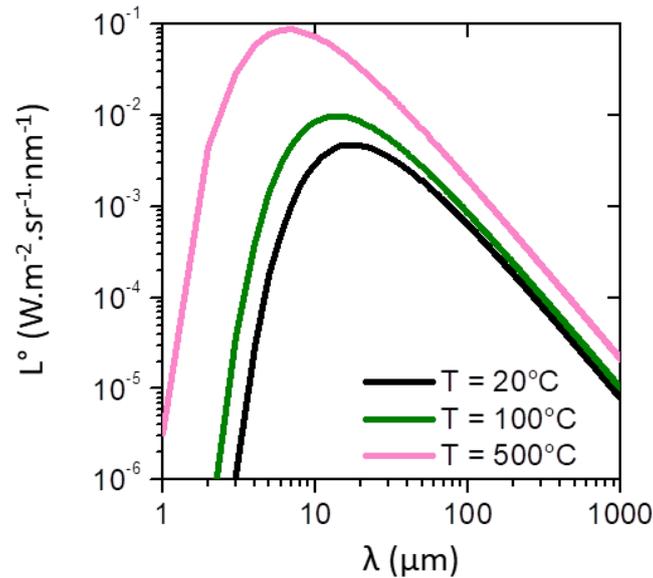

*Figure 12 : Évolution de la luminance énergétique spectrale par unité de longueur d'onde en fonction de la longueur d'onde et à différentes températures.*

À des températures proches de la température ambiante, les objets ou corps émettent donc des photons dans les gammes MWIR et LWIR. L'imagerie thermique permet notamment de faire de la surveillance ou de la détection d'objets chauds. Elle peut aussi être utilisée pour détecter des gaz qui absorbent dans la gamme MWIR – LWIR.

Pour que la transition $1S_e - 1S_h$ soit dans l'infrarouge, il faut que l'énergie de bande interdite du matériau massif $E_G$ soit faible. Les nanocristaux de PbS et PbSe, dont les énergies de bande interdite valent respectivement 0,41 eV et 0,27 eV [19] sont utilisés dans le proche infrarouge (< 2 µm) [20], notamment pour réaliser des cellules photovoltaïques ou dans le domaine des télécommunications. Pour utiliser ces matériaux dans les gammes SWIR et MWIR, il faudrait utiliser des nanocristaux peu confinés, dont la taille est proche du rayon de Bohr ($a_{PbS}$ = 18 nm [21]). Or, plus la taille des nanocristaux est grande, plus leur synthèse est difficile à contrôler et moins la solution de nanocristaux est stable colloïdalement (voir partie 3 « Principe de la synthèse des nanocristaux par voie colloïdale »).

Une solution alternative est d'utiliser des nanocristaux dont l'énergie de bande interdite du matériau massif est nulle : on ne parle alors plus de semiconducteurs mais de semimétaux. J'ai choisi de focaliser mon attention sur deux d'entre eux : le séléniure de mercure (HgSe) et le tellure de mercure (HgTe), qui font partie de la famille des chalcogénures de mercure.

### ii. Semimétaux de chalcogénures de mercure HgSe et HgTe

Les chalcogénures de mercure HgSe et HgTe présentent tous deux une structure de bande inversée. La bande $\Gamma_6$ qui joue classiquement le rôle de bande de conduction dans les semiconducteurs est en dessous des bandes $\Gamma_8$. $E_{\Gamma_6\text{-}\Gamma_8}$ est donc négatif.

Dans ces matériaux, les bandes de conduction et de valence sont les deux bandes $\Gamma_8$ qui se touchent au centre de la zone de Brillouin à l'état massif (Figure 13a). Le niveau de Fermi se situe entre ces deux bandes.

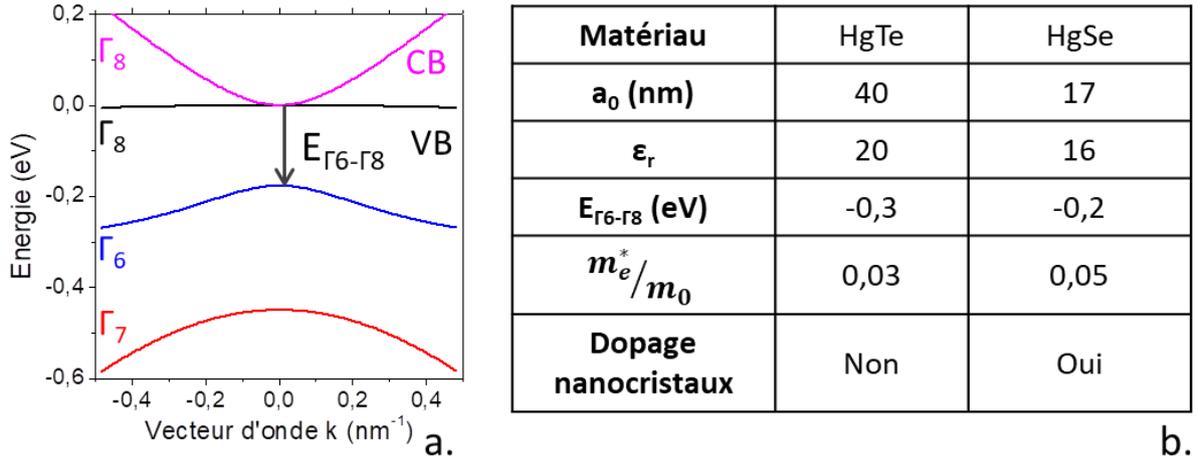

| Matériau | HgTe | HgSe |
|---|---|---|
| $a_0$ (nm) | 40 | 17 |
| $\varepsilon_r$ | 20 | 16 |
| $E_{\Gamma_6\text{-}\Gamma_8}$ (eV) | -0,3 | -0,2 |
| $m_e^*/m_0$ | 0,03 | 0,05 |
| Dopage nanocristaux | Non | Oui |

Figure 13 : (a) Structure de bande de HgSe, celle de HgTe étant très similaire. Le niveau de Fermi sert de référence à 0 eV. (b) Paramètres utiles pour caractériser les chalcogénures de mercure : $a_0$ est le rayon de Bohr, $\varepsilon_r$ est la permittivité relative statique, $E_G$ est l'écart en énergie entre les bandes $\Gamma_8$ et $\Gamma_6$ à k = 0, $m_e^*/m_0$ est le ratio de la masse effective de l'électron sur la masse de l'électron libre. La dernière ligne « dopage » se réfère aux nanocristaux confinés de HgTe et HgSe, et indique si ces derniers présentent ou non des électrons dans un ou plusieurs niveau(x) issu(s) de la bande de conduction (22, 23).

Dans un nanocristal confiné de semimétal, l'énergie de bande interdite du matériau massif est nulle. L'énergie de la transition interbande vaut environ $\frac{\hbar^2\pi^2}{2m_h^*R^2} + \frac{\hbar^2\pi^2}{2m_e^*R^2}$. La bande de valence dans les semi-métaux de chalcogénures de mercure est moins dispersive que la bande de conduction, ce qui revient à dire que $m_h^*/m_e^* \gg 1$. On peut donc simplifier l'écriture de la transition interbande (1.7) en :

$$1S_e - 1S_h \approx \frac{\hbar^2\pi^2}{2m_e^*R^2} \qquad (1.9)$$

Ainsi, pour un nanocristal de HgTe de 10 nm de diamètre, la transition interbande aura lieu vers 375 meV, soit 3,3 µm.

Pour atteindre des transitions de faibles énergies, il est aussi possible d'utiliser les transitions intrabandes, c'est-à-dire entre sous-niveaux issus d'une même bande. Pour ce faire, il faut que des électrons soient déjà présents dans un ou plusieurs niveau(x) issu(s) de la bande de conduction, c'est-à-dire que les nanocristaux soient dopés. C'est le cas des nanocristaux de HgSe (voir Figure 13b).

L'espacement entre les sous-niveaux issus de la bande de conduction est souvent plus faible que la transition interbande. Utiliser des transitions intrabandes permet donc d'avoir des transitions à basses énergies avec des petites particules, donc faciles à synthétiser. En contrepartie, les électrons présents dans la bande de conduction vont engendrer du courant d'obscurité et donc du bruit dans le détecteur infrarouge.

### 3. Principe de la synthèse des nanocristaux par voie colloïdale

#### a. Particularité des nanocristaux colloïdaux : les ligands

Les ligands sont des molécules qui complexent les nanocristaux pendant la synthèse. Ils sont généralement constitués d'une longue chaîne organique et d'une tête ayant une affinité pour les atomes de surface, comme une fonction amine, acide, thiol, phosphine ...

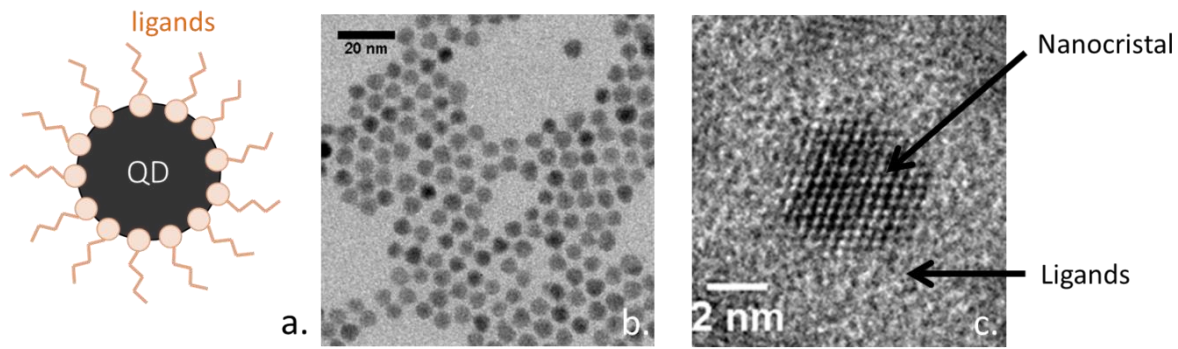

*Figure 14 : (a) Schéma d'un nanocristal (ou quantum dot, QD) dont la surface est recouverte de ligands ; (b) image de microscopie électronique à transmission de nanocristaux colloïdaux de HgSe, les nanocristaux ne sont pas en contact les uns avec les autres à cause de la présence de ligands ; (c) image haute résolution de microscopie électronique d'un nanocristal de HgSe : la partie cristallisée apparaissant en foncé au centre correspond au nanocristal HgSe, la partie claire à l'extérieur correspond aux ligands amorphes.*

Le rôle des ligands est triple. Ils permettent d'assurer la stabilité colloïdale des particules dans le milieu réactionnel grâce à leur longue chaîne organique soluble dans le solvant apolaire utilisé. Ils passivent la surface des cristaux, en assurant la coordination des atomes de surface et limitent l'arrivée des précurseurs à la surface grâce à leur encombrement stérique, ce qui permet un meilleur contrôle de la croissance du nanocristal.

Pour arrêter la croissance, on injecte généralement un large excès de ligands ayant une excellente affinité avec les atomes de surface. Cela permet de diminuer la réactivité du nanocristal à la fin de la synthèse. Ces ligands permettent d'assurer une stabilité colloïdale dans le temps à la solution de nanocristaux.

#### b. Protocole de synthèse des nanocristaux colloïdaux de semiconducteurs

L'atout majeur des nanocristaux colloïdaux pour la fabrication de détecteurs infrarouges est leur simplicité de production par rapport aux techniques concurrentes. La synthèse colloïdale, sous atmosphère inerte, se fait en quatre étapes, comme présenté sur la Figure 15. Le précurseur métallique est d'abord dissous dans un solvant à une température proche de 100 °C sous vide. Une fois le dioxygène, l'eau et les sous-produits dus à une éventuelle réaction du précurseur évacués, la solution est placée sous atmosphère argon à la température réactionnelle voulue.

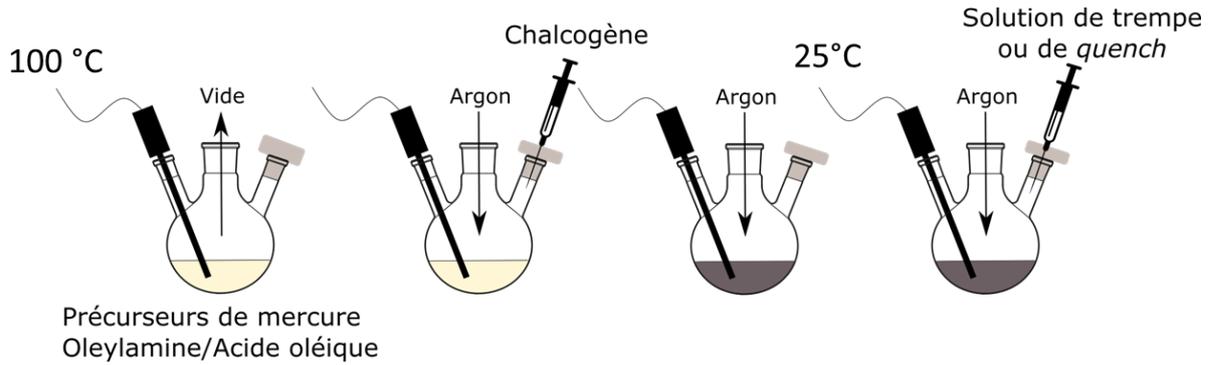

*Figure 15 : Schéma présentant les différentes étapes de synthèse de nanocristaux : exemple appliqué à la synthèse de chalcogénures de mercure.*

Ensuite, une solution contenant l'anion est très rapidement injectée dans le ballon afin que la croissance des nanocristaux soit la plus monodisperse possible. En effet, si les nanocristaux obtenus ne font pas tous la même taille, ils n'auront pas le même confinement et présenteront donc des transitions à des énergies différentes.

À la fin de la réaction, une solution de trempe (ou de *quench*), contenant entre autres un grand excès de ligands, est injectée dans le ballon et la température du milieu réactionnel est rapidement diminuée jusqu'à température ambiante pour éviter que la réaction ne se poursuive. Plus la réaction est longue et plus la température de réaction est élevée, plus les nanocristaux obtenus seront larges (Figure 16).

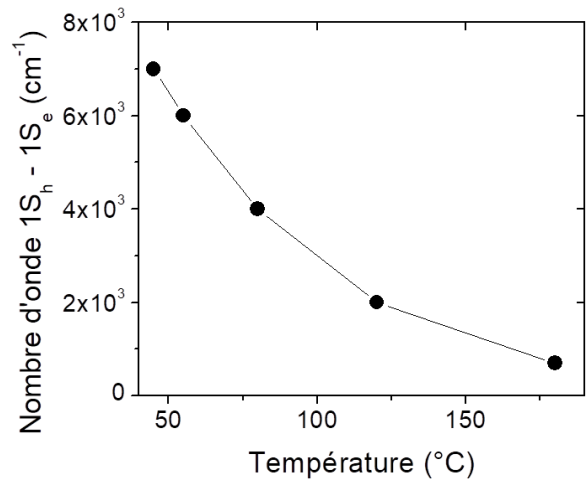

*Figure 16 : Influence de la température pendant la synthèse de nanocristaux de tellure de mercure sur l'énergie de la transition interbande.(24)*

La solution obtenue est ensuite lavée, c'est-à-dire que l'on retire tous les précurseurs n'ayant pas réagi et les ligands « libres » qui ne passivent pas la surface des nanocristaux. Cette étape se fait en ajoutant des mauvais solvants, tels que l'éthanol ou l'acétone, à la solution de nanocristaux et en centrifugeant. Les nanocristaux recouverts de ligands peuvent alors être séparés du reste. Ils sont ensuite dispersés dans un bon solvant tel que l'hexane, le toluène, le chloroforme ou le chlorobenzène.

### c. Techniques de caractérisation des nanocristaux colloïdaux post-synthèse

Après la synthèse, les propriétés spectroscopiques et structurales des nanocristaux sont contrôlées.

### i. Propriétés spectroscopiques

Pour mesurer l'énergie de la transition interbande et/ou de la transition intrabande, on utilise la spectrométrie infrarouge à transformée de Fourier (*Fourier Transform Infrared Spectroscopy* ou *FTIR*).

Dans le cas de nanocristaux non dopés comme HgTe, le niveau électronique peuplé le plus haut en énergie est le dernier niveau issu de la bande de valence $1S_h$. La transition de plus basse énergie observable est donc la transition interbande $1S_h - 1S_e$ (Figure 17a et b). On remarque que l'absorption ne diminue que très peu aux énergies (donc aux nombres d'onde) légèrement supérieures à celle de la transition interbande. L'absorption ré-augmente ensuite, aux nombres d'ondes supérieurs à 5000 cm⁻¹ (Figure 17b). La bande de valence de HgTe étant peu dispersive, les niveaux issus de cette bande sont très proches en énergie (Figure 13a et Figure 17a). Les transitions $1P_h - 1S_e$, $1D_h - 1S_e$ se font donc à des énergies légèrement supérieures à l'énergie de la transition $1S_h - 1S_e$. Dans le cas d'un matériau quasi intrinsèque, c'est-à-dire non dopé, le signal observé en absorption infrarouge est donc un seuil (ou *cut-off*) au-delà duquel toutes les énergies sont absorbées par le matériau.

Dans le cas de nanocristaux dopés comme HgSe, le niveau électronique peuplé le plus haut en énergie est un niveau issu de la bande de conduction (souvent il s'agit de $1S_e$). La transition $1S_e - 1P_e$ peut donc être observée (Figure 17c). Les niveaux issus de la bande de conduction étant beaucoup plus séparés en énergie que ceux issus de la bande de valence, les photons ayant une énergie supérieure à celle de cette transition ne peuvent pas être absorbés : le signal observé est donc un pic. La largeur du pic dépend de la polydispersité de la solution : plus la taille des cristaux est uniforme, plus le pic sera fin.

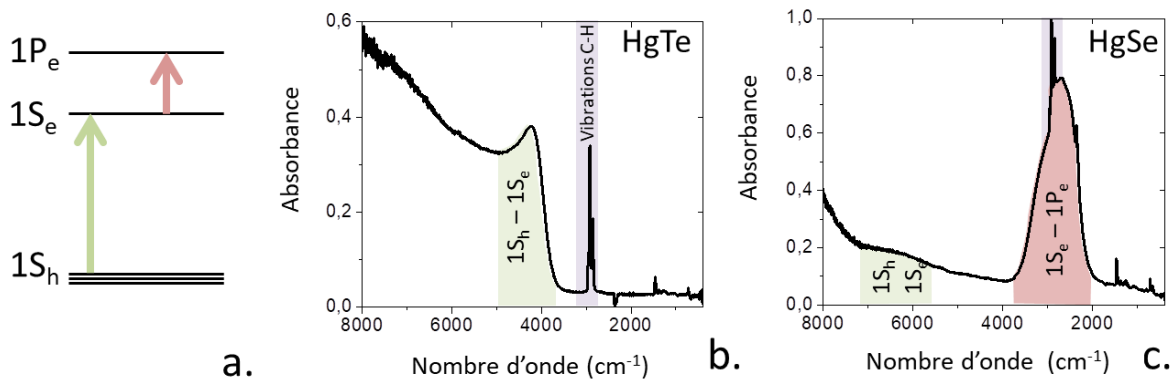

*Figure 17 : (a) Schéma représentant les transitions interbande (en vert) et intrabande (en rouge) dans un nanocristal confiné ; (b) spectre d'absorption infrarouge d'un nanocristal de HgTe ; (c) spectre d'absorption infrarouge d'un nanocristal de HgSe. Les vibrations C-H indiquées en violet proviennent des chaînes aliphatiques des ligands à la surface des nanocristaux.*

Dans la suite de ce manuscrit, les nanocristaux de chalcogénures de mercure seront décrits par leur nom (HgTe ou HgSe) suivi du nombre d'onde correspondant à la transition observée. Ainsi, pour les nanocristaux de la Figure 17b, on parlera de HgTe 4200 cm⁻¹.

### ii. Propriétés structurales

La taille et la forme des nanocristaux sont contrôlées par microscopie électronique à transmission (*Transmission Electronic Microscopy* ou *TEM*), comme présenté sur la Figure 18. HgSe donne généralement des nanocristaux sphériques (Figure 18a) tandis que HgTe donne des nanocristaux en forme de tétraèdres ou de tétrapodes (Figure 18b). La polydispersité des nanocristaux peut également être évaluée grâce à cette technique. Pour les nanocristaux sphériques de HgSe, elle est de l'ordre de 5-6 %, pour les nanocristaux de HgTe ayant une forme plus complexe, la polydispersité est de l'ordre de 10 %.

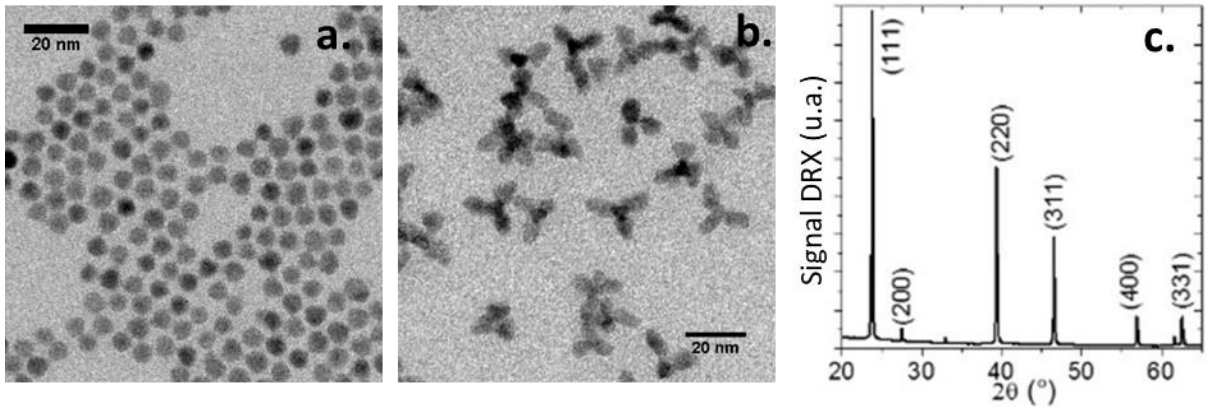

*Figure 18 : (a) Cliché TEM de nanocristaux de HgSe ; (b) cliché TEM de nanocristaux de HgTe, signal de diffraction des rayons X obtenu sur des nanocristaux de HgTe ; (c) diagramme de diffraction de nanocristaux de HgTe.*

La phase cristalline des nanocristaux est contrôlée par diffraction des rayons X (DRX). Cette technique permet également d'estimer la taille des cristallites en fonction de la largeur des pics via la formule de Scherrer :

$$D = \frac{k.\lambda}{H.\cos(\theta)} \qquad (1.10)$$

Où D est le diamètre de la particule, λ est la longueur d'onde incidente utilisée pour la mesure de DRX, H est la largeur à mi-hauteur du pic étudié et θ est le demi-angle de ce pic.

### III. Introduction à la détection infrarouge et présentation des technologies actuelles

Les nanocristaux de séléniure et de tellure de mercure présentent des propriétés optoélectroniques dans l'infrarouge, qui les rendent donc potentiellement intéressants pour des applications de détection infrarouge.

Dans cette partie, je commencerai par introduire les figures de mérite les plus courantes des détecteurs infrarouges et expliquerai comment les déterminer. Je présenterai ensuite les technologies de détection infrarouge actuelles.

#### 1. Caractérisation des détecteurs infrarouges : quelques figures de mérite usuelles

Contrairement aux dispositifs photovoltaïques pour lesquels seuls l'efficacité quantique, c'est-à-dire le nombre de porteurs (électrons ou trous) générés par l'absorption d'un photon, et le prix relatif à la puissance récoltée sont à optimiser, les détecteurs infrarouges doivent répondre à de nombreux critères. Dans le but de réaliser une caméra infrarouge par exemple, la fréquence d'acquisition des images doit permettre de suivre la scène en temps réel ; le rapport signal sur bruit doit permettre de distinguer l'objet à imager de son environnement…

Dans cette partie, j'expliciterai les principaux critères, ou figures de mérite, qui permettent de quantifier les performances des détecteurs infrarouges. Les plus utilisés sont les suivants : la réponse, la détectivité, la résolution thermique (*Noise Equivalent Temperature Difference* ou NETD), le temps de réponse et la température de fonctionnement.

##### a. Réponse

La réponse correspond au ratio entre le signal électrique mesuré en sortie et la puissance incidente :

$$\mathcal{R} = \frac{I_{lum} - I_{noir}}{P_{inc}} \qquad (1.11)$$

Où $\mathcal{R}$ est la réponse (en A/W), $I_{lum}$ est le courant mesuré sous illumination (en A), $I_{noir}$ est le courant d'obscurité (en A) et $P_{inc}$ est la puissance optique incidente (en W).

La réponse caractérise la capacité du détecteur à convertir les photons infrarouges (ou l'énergie infrarouge) en signal électrique.

##### b. Bruit et détectivité

Les sources de bruit dans un détecteur infrarouge peuvent être multiples. Comme pour tout composant électronique, il faut tenir compte du bruit thermique lié à la résistance électrique du dispositif $R_\Omega$, aussi appelé bruit de *Johnson-Nyquist*, et du bruit dû au passage du courant I, appelé bruit de *Schottky* ou bruit de grenaille. Ces deux composantes peuvent s'écrire :

$$S_I^2 = \frac{4k_bT}{R_\Omega} + 2e|I| \qquad (1.12)$$

Où $S_I$ est la densité spectrale de bruit (en A), $k_B$ est la constante de Boltzmann, T est la température de fonctionnement du dispositif (en K), $R_\Omega$ est la résistance électrique du détecteur infrarouge (en Ω), e est la charge élémentaire et I est l'intensité passant à travers le dispositif dans l'obscurité (en A).

À ces deux sources de bruit peut s'ajouter un bruit inversement proportionnel à la fréquence (on parle de bruit en 1/f). Il peut provenir d'une fluctuation du nombre de porteurs (électrons ou trous) (*25*) ou d'une fluctuation de la mobilité des porteurs (*26*). Dans un détecteur à base de nanocristaux colloïdaux, une fluctuation de mobilité peut être observée lors des sauts tunnels effectués par les porteurs pour passer d'un nanocristal à un autre et on retrouve souvent cette composante en 1/f (*27*, *28*). Dans un film de nanocristaux colloïdaux, le bruit s'écrit :

$$S_I^2 = \frac{4k_bT}{R_\Omega} + 2e|I| + \frac{\alpha I^2}{f.N}$$ (1.13)

Où α est un paramètre ajustable appelé constante de Hooge et N est la quantité de porteurs mobiles dans le système.

La **détectivité** est une mesure du rapport signal sur bruit. Elle s'exprime grâce à la formule suivante :

$$D^* = \frac{\sqrt{A}\sqrt{\Delta f}}{NEP}$$ (1.14)

Où D* est la détectivité (en Jones, ou cm. Hz$^{1/2}$. W$^{-1}$), A est la surface du détecteur (en cm²), Δf est la bande de fréquence sur laquelle est effectuée la mesure (en Hz) et NEP est la puissance équivalente de bruit (*Noise Equivalent Power*, en W). Cette dernière est définie comme étant la puissance donnant un rapport signal sur bruit de un pour une bande passante de sortie de 1 Hz. La détectivité sera d'autant meilleure que la NEP est faible.

La NEP est reliée à la réponse selon la formule :

$$NEP = \frac{S_I\sqrt{\Delta f}}{\mathcal{R}}$$ (1.15)

Où S$_I$ est la densité spectrale de bruit (en A.Hz$^{-0,5}$), qui peut se calculer comme : $S_I = \frac{FFT\,(I_{noir})}{\sqrt{\Delta f}}$. La détectivité peut donc finalement s'écrire :

$$D^* = \frac{\sqrt{A}.\mathcal{R}}{S_I}$$ (1.16)

### c. Résolution thermique et NETD

La NETD (*Noise Equivalent Temperature Difference*, ou température équivalente de bruit) est une figure de mérite qui renseigne sur la plus petite différence de température mesurable avec un détecteur infrarouge (on parle aussi de résolution thermique). Elle se calcule en mesurant le signal reçu pour un même objet à deux températures différentes, selon la formule :

$$NETD = \frac{\Delta T}{\Delta I_{signal}\big/I_{bruit}}$$ (1.17)

Où NETD est la température équivalente de bruit (en K), ΔT est la différence de température entre les deux signaux collectés (en K), ΔI$_{signal}$ est la différence de courant mesurée entre les deux signaux collectés (en A) et I$_{bruit}$ est le courant correspondant au niveau de bruit du détecteur (en A).

La NETD se situe aux alentours de 20 mK pour les technologies quantiques actuelles.

### d. Température de fonctionnement

Comme mentionné dans la partie précédente, le bruit du détecteur provient en partie du bruit thermique lié à sa résistance. Par conséquent, diminuer la température de fonctionnement du détecteur peut permettre d'augmenter le rapport signal sur bruit. Notamment, pour la détection dans la gamme LWIR où les énergies en question sont de l'ordre de 10 à 100 meV, le froid est nécessaire.

Cependant, ajouter un système de refroidissement engendre nécessairement des coûts supplémentaires. Pour la détection dans les gammes MWIR et SWIR, il y'a donc un compromis à trouver pour la température de fonctionnement, pour maximiser le rapport signal sur bruit tout en limitant l'augmentation des coûts de fonctionnement.

### e. Temps de réponse du capteur

Le temps de réponse du détecteur infrarouge est un paramètre essentiel puisqu'il renseigne sur la capacité à acquérir des images à haute fréquence, ce qui est primordial pour des applications de type LIDAR ou détection d'objets chauds sur fond froid. Il se mesure généralement en utilisant une source de type « créneau » ou « impulsion » et en étudiant la réponse du détecteur à des changements rapides du signal.

## 2. Technologies actuelles

Les technologies actuellement commercialisées pour la détection infrarouge se répartissent en deux catégories principales : les détecteurs thermiques, qui sont sensibles au flux d'énergie infrarouge, et les détecteurs quantiques, à base de semiconducteurs, qui sont sensibles aux photons infrarouges.

Dans cette partie, je décrirai les principes de fonctionnement des principales technologies ainsi que les avantages et les inconvénients qu'elles présentent.

### a. Détecteurs thermiques

Les détecteurs thermiques sont sensibles au flux d'énergie et donc à la température. Le matériau actif qui les constitue voit ses caractéristiques physiques modifiées quand la température varie.

Ainsi, la technologie la plus utilisée de détecteurs thermiques, le bolomètre, est basée sur la mesure de la résistance en fonction de la température. Pour cela, on peut utiliser des matériaux ayant une transition de phase à une température proche de la température ambiante. C'est le cas de l'oxyde de vanadium (*29*), qui passe d'un comportement isolant à métallique à une température autour de 60 °C : sa résistance peut alors varier de plusieurs ordres de grandeurs. Le silicium amorphe est également souvent utilisé pour sa facilité d'intégration dans des dispositifs CMOS (*Complementary Metal Oxide Semiconductor*). Utilisés sous la forme de matrices de pixels, chacun constitué d'un bolomètre, les microbolomètres permettent de recréer une image thermique.

D'autres technologies, basées sur la dépendance en température d'autres propriétés physiques telles que la polarisation (on parle alors de pyromètres) ou la tension de seuil existent également mais sont moins répandues.

Les détecteurs thermiques présentent l'avantage d'être utilisable à température ambiante et d'être peu coûteux (environ 100 $). Cependant, leur temps de réponse est limité (de l'ordre de la milliseconde) à

cause de phénomènes de rémanence thermique, et leur détectivité est limitée à quelques $10^9$ Jones (*30*).

### b. Semiconducteurs à faible énergie de bande interdite

Les détecteurs quantiques sont basés sur l'absorption d'un photon par un semiconducteur, générant une paire électron-trou. En appliquant une tension aux bornes du semiconducteur, l'électron et le trou peuvent être séparés et migrer vers une des électrodes, ce qui génère un courant.

Afin de détecter des photons infrarouges avec un semiconducteur, il est nécessaire que le photon à absorber ait une énergie supérieure à celle de la bande interdite du semiconducteur. Ainsi, comme présenté sur la Figure 19, Si et GaSb ne peuvent être utilisés que dans la gamme NIR ($E_G \approx 1$ eV), Ge et InAs dans la gamme SWIR ($1\ eV < E_G < 0.5$ eV).

Pour ajuster précisément l'énergie de bande interdite du détecteur, deux options sont possibles : fabriquer des alliages ternaires de semiconducteurs ou ajouter des dopants.

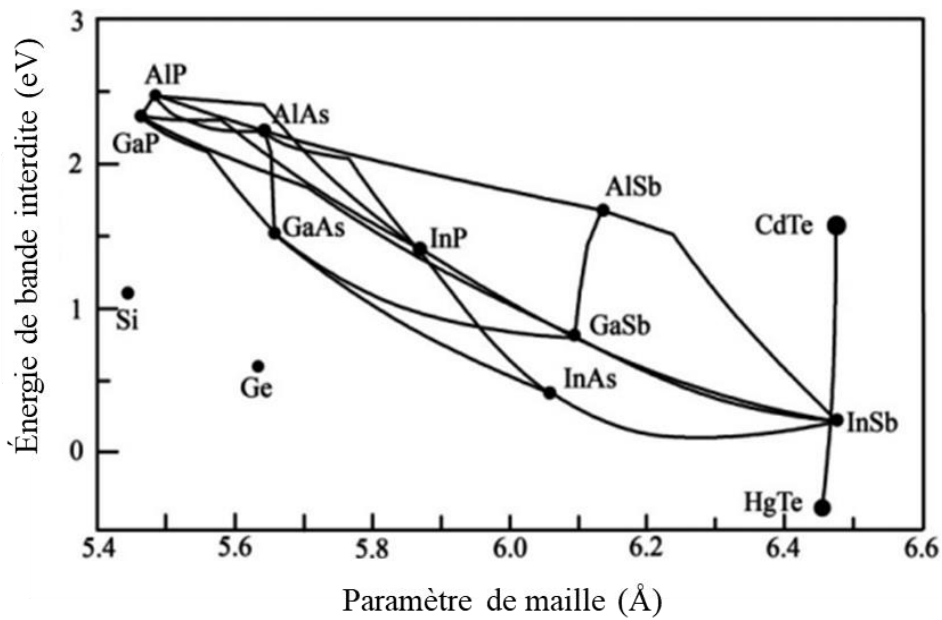

*Figure 19 : Caractéristiques (énergies de bande interdite et paramètres de maille) des principaux semiconducteurs utilisés à faible énergie de bande interdite.*

#### i. Alliages ternaires de semiconducteurs

Il est possible de fabriquer des alliages de semiconducteurs (ternaires). Ces derniers sont présentés sur la Figure 19 par les lignes noires. Ainsi, $In_{0.53}Ga_{0.47}As$ a une bande interdite de l'ordre de 0.75 eV et est communément utilisé dans la gamme SWIR. Pour les bandes interdites plus faibles, le ternaire $Hg_xCd_{1-x}Te$ permet d'explorer toutes les énergies entre 0 (HgTe pur) et 1,5 eV (CdTe pur).

Les technologies à base de semiconducteurs à faible énergie de bande interdite présentent de bonnes détectivités ($10^{12}$ à $10^{13}$ Jones pour InGaAs dans le SWIR, $10^{11}$ à $10^{12}$ Jones pour HgCdTe dans le MWIR/LWIR) (*30*) et des temps de réponse courts, adaptés aux applications de type caméras infrarouges hautes performances.

En revanche, les méthodes de croissance (épitaxie par jet moléculaire : MBE, ou épitaxie en phase vapeur aux organométalliques : MOCVD) sont lentes et demandent des bâtis de dépôt très sophistiqués, typiquement sous ultravide. Les semiconducteurs doivent être déposés sur des substrats permettant une croissance épitaxiale avec un nombre limité de dislocations. Ainsi, $In_xGa_{1-x}As$ est souvent utilisé avec la stœchiométrie x = 0,53 pour être compatible avec un substrat InP. HgCdTe ne voit pas son paramètre de maille changer avec la stœchiométrie ($a_{HgTe}$ = 6,46 Å, $a_{CdTe}$ = 6,48 Å) (**31**), mais le seul substrat compatible avec sa croissance est CdZnTe, dont le prix est d'environ 1000 \$/cm². Enfin, la composition de l'alliage doit être parfaitement maîtrisée pour contrôler l'énergie de bande interdite, et certaines compositions sont difficiles à obtenir : ainsi HgCdTe n'est pas utilisé au-delà de 12 µm car les matériaux trop riches en mercure tendent à être moins uniformes (**30**).

Il faut également noter que, pour les détecteurs utilisés dans les gammes MWIR et LWIR, les températures de fonctionnement de ces technologies sont basses : autour de 120-180 K pour le MWIR, 90 K pour le LWIR, ce qui augmente encore leur coût d'utilisation, leur poids et limite leur durée de vie.

### ii. Semiconducteurs extrinsèques

Les semiconducteurs extrinsèques les plus utilisés pour la détection infrarouge sont à base de silicium dopé. Il existe également des détecteurs à base de germanium dopé mais ceux-ci sont moins répandus, notamment car les dopants sont plus solubles dans le silicium et que des détecteurs avec une meilleure résolution spatiale peuvent être fabriqués avec le silicium.

Le dopant peut être le gallium, l'indium, l'antimoine ou l'arsenic, ce qui permet de détecter des longueurs d'onde allant jusqu'à 25 µm. En augmentant la quantité de dopants jusqu'à l'apparition d'une « bande d'impureté » (*block impurity band*), la détection infrarouge peut être poussée jusqu'à 100 voire 200 µm (**30**).

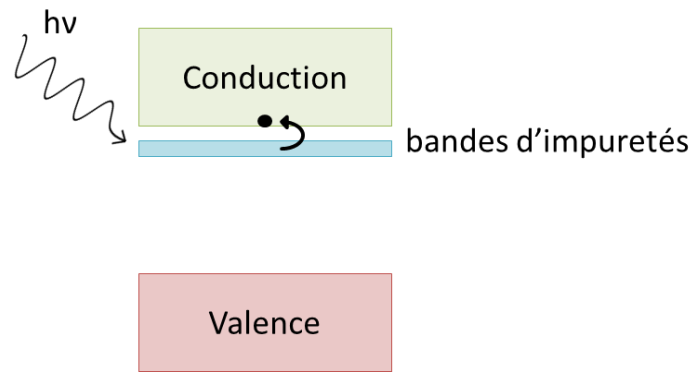

*Figure 20 : Schéma d'un semiconducteur extrinsèque dont le dopage a entraîné l'apparition d'une bande d'impuretés près de la bande de conduction.*

La température de fonctionnement de ces détecteurs est très faible, entre 4 et 10 K, ce qui réduit fortement les applications possibles.

### c. Détecteurs à base d'hétérostructures

Dans la partie précédente, nous avons vu qu'il était possible de modifier la composition du matériau actif pour obtenir des propriétés d'absorption dans l'infrarouge, mais que cela pouvait entraîner des problèmes de miscibilité ou de solvabilité. En utilisant le confinement quantique, les détecteurs à base de puits quantiques permettent de faire varier les propriétés optiques en modifiant l'épaisseur de la couche, et non plus sa composition.

### i. Photodétecteurs infrarouges à puits quantiques (QWIP)

La structure la plus utilisée pour les photodétecteurs infrarouges à puits quantiques est Al$_x$Ga$_{1-x}$As/ GaAs, où GaAs est le puits quantique et Al$_x$Ga$_{1-x}$As la barrière.

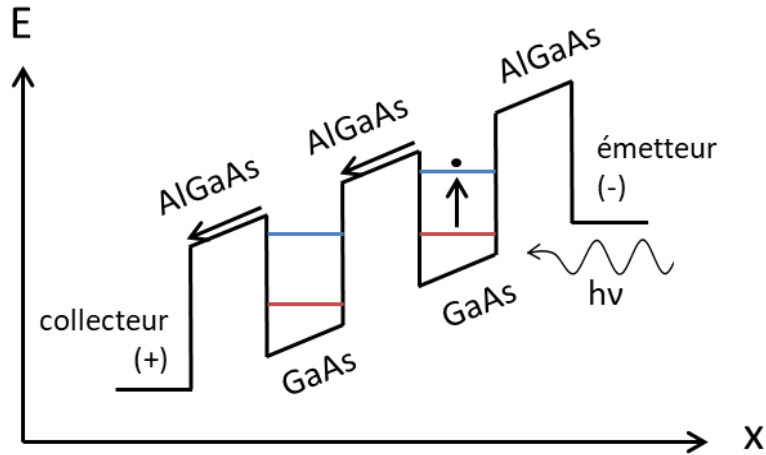

*Figure 21 : Schéma représentant le fonctionnement d'un photodétecteur infrarouge à puits quantiques (ou QWIP) à base de GaAs/Al$_x$Ga$_{1-x}$As. x correspond à la direction de croissance du QWIP*

Le principe de fonctionnement est le suivant : les couches de GaAs, qui constituent les puits quantiques, sont dopées avec du silicium afin que des électrons soient présents dans le premier niveau électronique (ou la première sous bande) de la bande de conduction. Ce niveau est piégé dans le puits. Après absorption d'un photon infrarouge, une transition intrabande promeut un électron dans un second niveau, proche en énergie du continuum. En appliquant une tension aux bornes du QWIP, les électrons du second niveau peuvent passer dans le continuum et être détectés au niveau du collecteur (Figure 21). Les épaisseurs des couches étant inférieures au rayon de Bohr, ces matériaux sont confinés dans la direction de croissance du QWIP. Il est alors possible de modifier la position des niveaux en changeant leur épaisseur : la hauteur de la barrière peut être ajustée pour que le continuum et le second niveau soient résonants.

Ces photodétecteurs sont opérés à basse température (autour de 77 K), et ne peuvent pas absorber les photons à incidence normale. Ces photons doivent donc être déviés grâce à un réseau gravé à la surface du pixel. Tout comme les semiconducteurs à faible énergie de bande interdite, ils sont fabriqués dans des bâtis d'épitaxie.

### ii. Super réseaux

Comme les QWIPs, les super-réseaux sont construits par empilement de couches confinées, mais ne sont pas dopés : les transitions en jeu sont des transitions interbandes. La combinaison GaSb/InAs est la plus connue : l'alignement entre les deux matériaux est de type II, ce qui donne naissance à deux minibandes comme présenté sur la Figure 22. En modifiant le pas du réseau, on peut contrôler l'espacement entre les deux minibandes et donc la longueur d'onde d'absorption.

Ces détecteurs fonctionnent également à basse température, autour de 100 K, mais ils peuvent absorber des photons à incidence normale. Cette technologie permet également d'atteindre des détectivités autour de $10^{10}$ Jones dans le LWIR. En revanche, ils doivent comme les QWIPs et les semiconducteurs à faible énergie de bande interdite, être préparés dans des bâtis d'épitaxie, ce qui augmente le coût de production, et supporter un système de refroidissement.

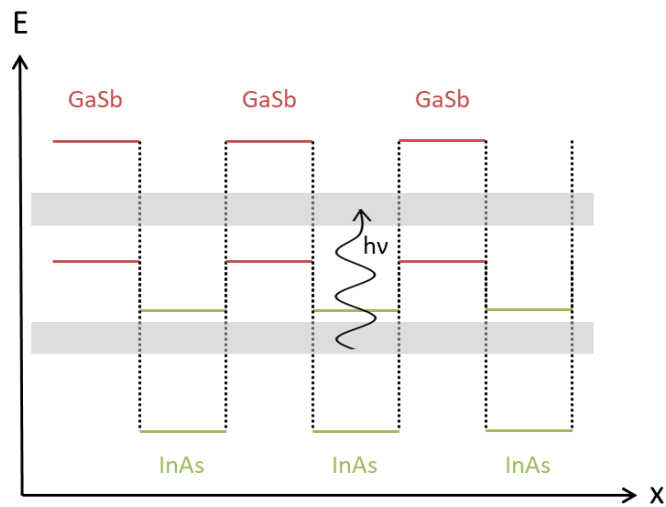

*Figure 22 : Schéma représentant un super réseau de GaSb (niveaux électroniques en rouge)/InAs (niveaux électroniques en vert).  Des minibandes (en gris) dues à l'alignement de type II entre les deux matériaux apparaissent.*

### d. Bilan

Les détecteurs infrarouges commercialisés peuvent se classer selon deux catégories : les détecteurs thermiques, utilisables à température ambiante et peu coûteux mais lents ; et les détecteurs quantiques qui présentent des bonnes détectivités et des temps de réponse courts mais très coûteux de par leur fabrication et leur utilisation à froid.

De par leur simplicité de fabrication, les nanocristaux colloïdaux à faible bande interdite (HgTe et HgSe) semblent être une alternative peu coûteuse aux détecteurs infrarouges quantiques classiques qui présentent aujourd'hui les meilleures performances. Dans la partie suivante, je présenterai donc comment ces nanocristaux peuvent être intégrés dans un dispositif de type détecteur. J'indiquerai ensuite les axes principaux de recherche que j'ai suivis pendant mon doctorat pour améliorer la compréhension des propriétés électroniques des films de nanocristaux et ainsi réaliser un détecteur infrarouge optimisé.

## IV.    Des nanocristaux colloïdaux au photodétecteur

Cette partie est consacrée à l'intégration des nanocristaux dans des systèmes de détection infrarouge. Les nanocristaux sont déposés sous forme de films qui doivent vérifier plusieurs propriétés comme une bonne conductivité des porteurs (électrons ou trous) et une bonne mobilité.

### 1.    Films de nanocristaux photoconducteurs

Pour être intégrés dans des détecteurs, les nanocristaux colloïdaux sont déposés sous forme de film sur des substrats contenant une ou plusieurs électrodes. Contrairement aux technologies quantiques actuelles, une large gamme de substrats peut être utilisée puisqu'il n'est pas nécessaire qu'il y ait un accord de paramètre de maille entre le substrat et le nanocristal de chalcogénure de mercure. En fonction du solvant de la solution de nanocristaux, plusieurs méthodes de dépôt sont possibles : l'enduction centrifuge (ou *spin-coating*), l'enduction par trempage-retrait (ou *dip-coating*), le dépôt par spray …

Au sein du film, la conduction des porteurs se fait par saut tunnel (ou *hopping*) d'un nanocristal à l'autre jusqu'à l'électrode. Cependant, les ligands longs qui permettaient la stabilité colloïdale des nanocristaux en solution jouent le rôle de barrière tunnel entre les nanocristaux et rendent le film isolant. Il est donc nécessaire de les échanger pour rendre le film photoconducteur, voir Figure 23 pour un schéma de principe.

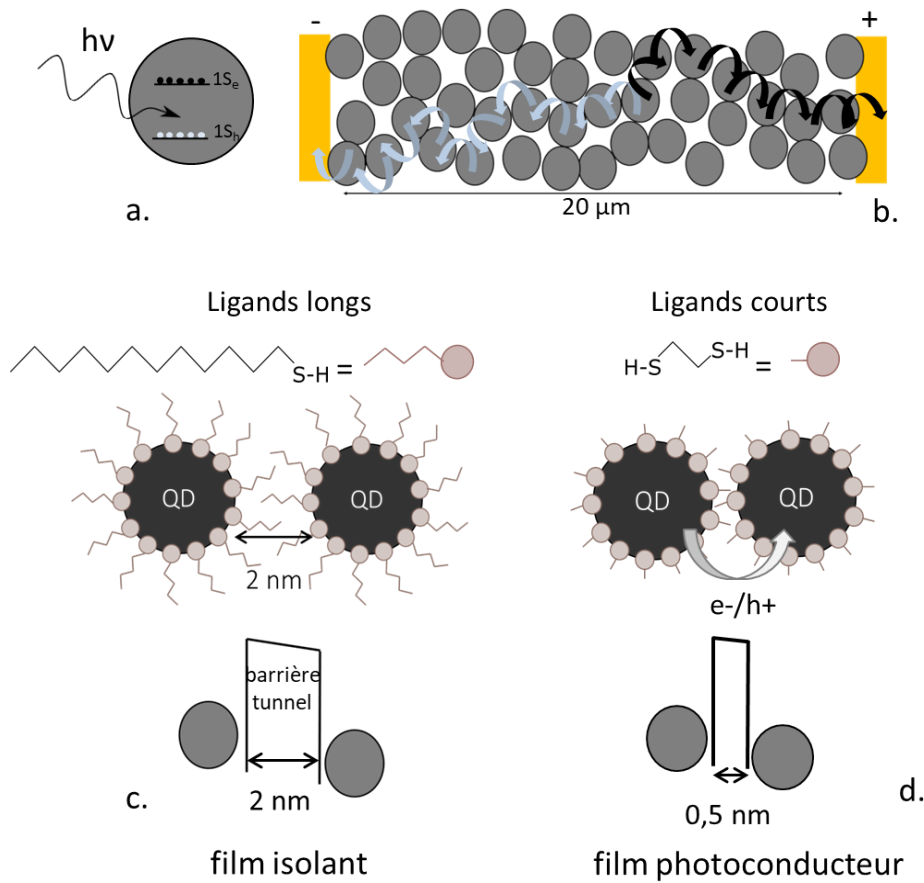

*Figure 23 : (a) Schéma représentant la promotion d'électrons dans le niveau $1S_e$ et de trous dans le niveau $1S_h$ après absorption d'un photon infrarouge ; (b) schéma d'un film de nanocristaux entre deux électrodes : les électrons (en noir) rejoignent par saut tunnel l'électrode +, les trous (en bleu clair) l'électrode - ; (c) schéma de la*

*barrière tunnel due à la présence de ligands longs dodécanethiol (ou DDT) dans l'exemple, sur la surface des nanocristaux ; (d) schéma représentant l'amincissement de la barrière tunnel avec le changement de ligands (ethanedithiol ou EDT dans l'exemple).*

L'échange de ligands peut se faire de deux manières : directement sur le film ou en solution.

### a. Échange de ligands sur film

Pour réaliser un échange de ligands sur film, une fine couche de nanocristaux est déposée sur un substrat. Une fois le solvant évaporé, le substrat est plongé dans un mauvais solvant pour les nanocristaux, typiquement de l'éthanol, contenant un large excès de nouveaux ligands (1% en masse). Les ligands courts en excès pénètrent et diffusent dans la couche de nanocristaux facilement, tandis que les ligands longs (ayant une diffusion moins efficace) ont une dynamique de décrochage – réadsorption sur la surface plus lente. De plus, les ligands courts étant en excès, l'échange de ligands se fait en leur faveur. Après une minute d'échange, le substrat est retiré de la solution de ligands et rincé dans une solution d'éthanol pour retirer les ligands libres (longs et courts) restés dans le film. Cette technique a l'avantage de pouvoir être utilisée avec de nombreux ligands, à partir du moment où ils peuvent être dissouts dans la solution d'échange.

Pour que la diffusion des ligands courts soit efficace dans le film, il est nécessaire que la couche de nanocristaux soit assez fine, de l'ordre de 20 à 40 nm. Pour obtenir un film échangé en ligands plus épais, l'opération doit être répétée plusieurs fois : après l'échange de ligands, l'échantillon est séché et une nouvelle couche de nanocristaux est déposée à la surface. On réalise donc un échange de ligands couche par couche. Toutefois, il est difficile d'augmenter l'épaisseur du film de nanocristaux au-delà de 200 nm avec cette technique, le film devenant moins résistant mécaniquement.

Cette technique d'échange de ligands est partielle (70 – 80%). L'échange de ligands peut être suivi par absorption infrarouge : les pics correspondants aux liaisons -CH caractéristiques des ligands doivent diminuer en intensité. On peut également mesurer la baisse de la résistance du matériau, amenée par la réduction de la barrière tunnel entre les nanoparticules, pendant l'échange de ligands

Enfin, le remplacement des ligands longs par des ligands courts entraînant le rapprochement des nanocristaux entre eux, des craquelures peuvent apparaître au sein du film et réduire la mobilité des porteurs. Pour limiter cet effet, un dépôt multi-couches peut être effectué.

### b. Échange de ligands en solution

Afin d'obtenir un échange de ligands complet, ce dernier peut être réalisé en solution par la méthode de « transfert de phases » (*32*). Cette technique, présentée sur la Figure 24 fonctionne de la manière suivante : la solution obtenue après la synthèse est généralement constituée de nanocristaux de chalcogénures de mercure (HgX) recouverts de ligands longs, souvent du dodécanethiol (DDT) dans un solvant apolaire. On ajoute à cette solution une solution d'ions courts comme des sulfures ($S^{2-}$) dissouts dans un solvant polaire : formamide (FA), N-méthylformamide (NMF) ou N,N diméthylformamide (DMF). Si les deux solvants ne sont pas miscibles[5] : deux phases apparaissent.

---

[5] Si le solvant apolaire utilisé n'est pas l'hexane, la séparation de phases n'apparaît pas. Pour la forcer, il suffit de rajouter le même volume d'hexane que de solution de ligands dans le solvant polaire.

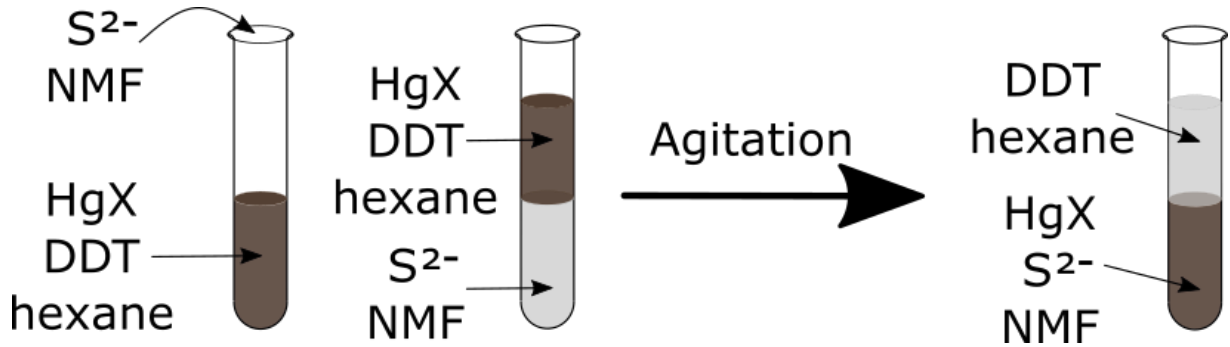

*Figure 24 : Schéma de principe de l'échange de ligands par transfert de phases.*

En agitant vigoureusement la solution, la surface de contact entre les deux phases augmente. Les ligands courts présents dans la phase polaire peuvent alors s'échanger sur la surface des nanocristaux, les entraînant ainsi dans la phase polaire. L'échange est terminé quand le solvant apolaire est translucide.

Cette technique assure un échange de ligands quasi-complet (aucune trace de liaison C-H n'est observée par spectroscopie infrarouge) mais présente l'inconvénient d'utiliser des solvants polaires avec des températures d'ébullition élevées qui rendent le dépôt plus difficile. De plus, elle ne peut pas être utilisée sur de trop faibles quantités de nanocristaux (au moins une centaine de milligrammes est nécessaire).

### c. Observation de la photoconduction

Pour observer la photoconduction, les nanocristaux doivent être déposés sur un substrat, entre deux électrodes, que l'on appellera drain (D) et source (S) dans la suite du manuscrit. Ces électrodes appliqueront un potentiel qui permettra de dissocier les paires électron-trou et d'extraire les porteurs, donc de mesurer un courant (voir Figure 23b). Pour plus d'informations sur la fabrication des électrodes, se reporter à l'Annexe 2.

Dans l'obscurité, seuls les porteurs générés par activation thermique participent à la conduction. Sous illumination, des paires électrons-trous sont générées par l'absorption d'un photon, et ces porteurs participeront à la conduction. Le courant mesuré sera donc plus important sous illumination que dans l'obscurité (Figure 25).

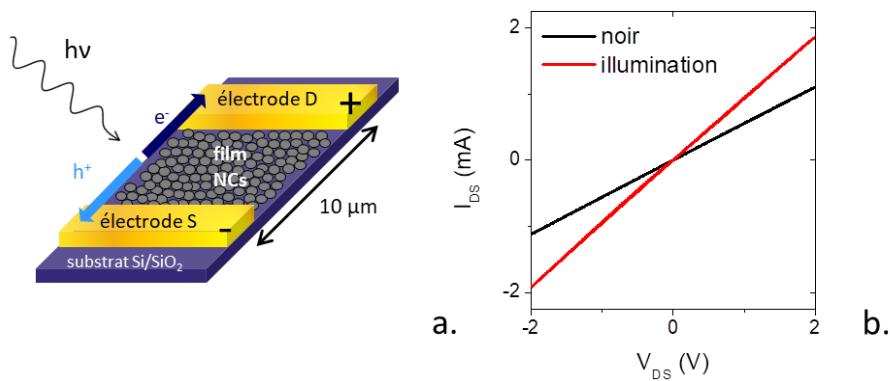

*Figure 25 : (a) Schéma de principe d'une mesure de réponse d'un film de nanocristaux photoconducteurs. Les électrodes D et S correspondent au drain et à la source respectivement. (b) Mesure de l'intensité mesurée aux bornes d'un film de nanocristaux de HgTe en fonction de la tension appliquée. Sous illumination (rouge), la conductivité est meilleure que dans l'obscurité (noir).*

## 2. Objets d'étude du doctorat

La synthèse de nanocristaux de chalcogénures de mercure a atteint une certaine maturité, et la photoconduction de films à base de ces nanocristaux a été démontrée (*33*). En revanche, de nombreuses études restent à faire avant d'obtenir un détecteur infrarouge performant à base de ces nanocristaux. Dans cette partie, je présenterai les trois grands axes que j'ai suivis pendant mon doctorat pour approfondir la connaissance de ce type de matériaux : la détermination absolue de la structure électronique, le contrôle du dopage, et l'élaboration de structures de détecteurs adaptées à HgTe pour la détection dans le SWIR.

### a. Connaissance précise de la structure électronique

La connaissance de la structure électronique est indispensable pour prévoir les propriétés optoélectroniques des nanocristaux colloïdaux. D'un point de vue applicatif, elle permet d'élaborer des détecteurs bien adaptés au matériau actif, notamment pour éviter les résistances de contact entre le matériau et les électrodes. Pendant mon doctorat, j'ai donc travaillé à déterminer les énergies absolues des différents niveaux, c'est-à-dire par rapport au niveau de Fermi, en fonction du confinement quantique. Ces études permettent de déterminer la nature des porteurs majoritaires (électrons ou trous) ainsi que d'identifier les métaux les plus adaptés pour jouer le rôle d'électrodes. Elles permettent également de quantifier le niveau de dopage dans les nanocristaux qui, lorsqu'il devient supérieur à quelques électrons par nanocristal, entraîne l'émergence de comportements collectifs de type plasmoniques.

À ces travaux s'ajoute l'étude de la densité de pièges dans la bande interdite, qui peuvent engendrer du bruit dans le système et détériorer les performances des détecteurs infrarouges. Les résultats concernant la structure électronique des nanocristaux de chalcogénures de mercure seront présentés dans le chapitre 2 de ce manuscrit.

### b. Contrôle du dopage

Le contrôle du dopage dans les nanocristaux colloïdaux est un enjeu important. Le niveau de dopage régit les transitions autorisées et interdites dans les nanocristaux. Il est également lié au niveau de bruit dans un système de détection. Or, ce contrôle du dopage a été pendant longtemps difficile à mettre en place. Par exemple, lors de l'ajout d'atomes dopants pendant la synthèse, des processus d'auto-purification des nanocristaux (*34*) peuvent entraîner l'exclusion de ces atomes dopants et donc empêcher la modification du niveau de dopage.

Dans le chapitre 3, j'explorerai plusieurs méthodes permettant de contrôler le dopage des nanocristaux colloïdaux. En particulier, je présenterai comment l'utilisation de ligands ayant un fort moment dipolaire ou un fort pouvoir oxydant permet de modifier le niveau de dopage de nanocristaux de séléniure de mercure.

### c. Fabrication d'un détecteur infrarouge optimisé

Dans le dernier chapitre de ce manuscrit, je présenterai les stratégies qui peuvent être adoptées pour réaliser un détecteur infrarouge adapté aux nanocristaux colloïdaux. Je présenterai une technique d'échange de ligands permettant de déposer des films épais, donc absorbant efficacement la lumière, ainsi qu'une architecture de détecteur infrarouge qui tient compte de la position des niveaux pour optimiser la récupération des porteurs. Je finirai en proposant un dispositif multipixels, une preuve de concept pour imager une source laser avec un détecteur à base de nanocristaux de tellure de mercure.

# CHAPITRE 2
# Détermination de la structure électronique des nanocristaux colloïdaux de chalcogénures de mercure

**Publications associées à ces travaux :**

- B. Martinez *et al.*, HgSe self-doped nanocrystals as a platform to investigate the effects of vanishing confinement, *ACS Applied Materials and Interfaces*, **9**, 41, 36173-36180 (2017)
- B. Martinez *et al.*, Probing charge carrier dynamics to unveil the role of surface ligands in HgTe narrow band gap nanocrystals, *the Journal of Physical Chemistry C*, **122**, 1, 859-865 (2018)
- N. Goubet *et al.*, Terahertz HgTe nanocrystals: beyond confinement, *Journal of American Chemical Society*, **140**, 15, 5033-5036 (2018)

**Mots clés :** structure électronique, niveau de Fermi, dopage, photoémission, interbande, intrabande, plasmonique, transition semiconducteur-métal, énergie d'Urbach, photocourant transitoire

**Techniques expérimentales :**

- Électrochimie sur film de nanocristaux
- Transistors à effet de champ
- Spectroscopie infrarouge
- Photoémission
- Mesures de photocourant transitoire

Avant de pouvoir intégrer un film de nanocristaux colloïdaux dans un détecteur infrarouge, il est nécessaire de maîtriser ses propriétés optoélectroniques, en particulier de connaître les énergies des différents niveaux électroniques (voir Figure 26a). En plus des énergies des transitions optiques, qui peuvent être obtenues par spectroscopie optique, l'énergie du niveau de Fermi renseignera sur le niveau de dopage du cristal, et donc sur sa capacité à transporter préférentiellement les électrons ou les trous. Dans le cas où il se situe au-dessus du premier niveau de conduction $1S_e$, il renseigne sur le fait que le cristal est dopé et permet d'estimer quantitativement le niveau de dopage.

La valeur du travail de sortie des nanocristaux est également essentielle dans la démarche d'intégration des nanocristaux dans un composant. Elle permet de connaître les énergies des niveaux électroniques par rapport à l'énergie du vide. Pour faciliter l'extraction des porteurs et avoir des contacts ohmiques, il faut choisir des électrodes dont les travaux de sortie sont compatibles avec les énergies des niveaux du nanocristal, c'est-à-dire pour lesquels les barrières Schottky pour les électrons et les sont minimales (voir Figure 26b).

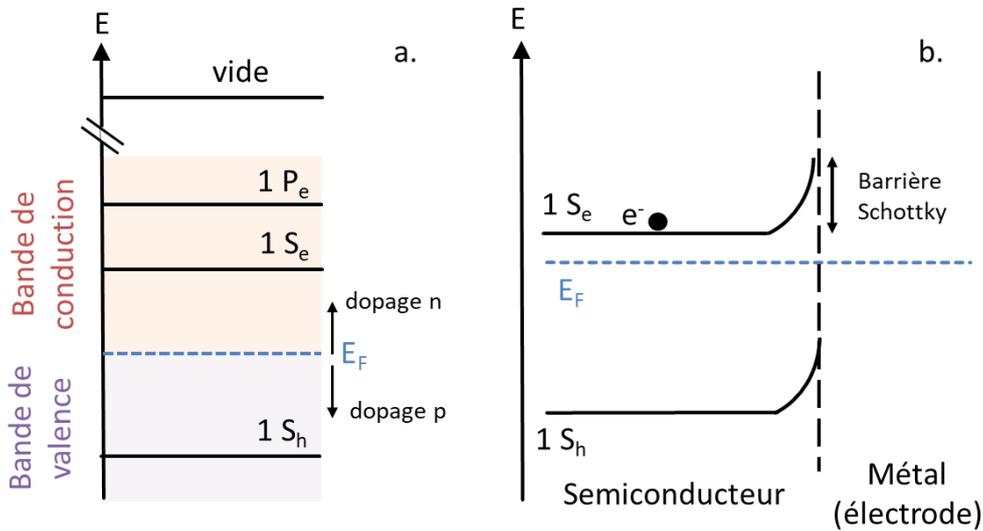

*Figure 26 : (a) Schéma représentant la structure électronique d'un nanocristal ; (b) schéma présentant l'alignement de bandes d'un film de nanocristaux entre deux électrodes : une permettant de récolter les trous (en jaune) et une permettant de récolter les électrons (en gris).*

La structure de bande inversée des chalcogénures de mercure, additionnée au confinement quantique et à la dépendance de l'énergie des niveaux avec la chimie de surface, rend difficile la connaissance *a priori* de la structure électronique. Or, pour concevoir les dispositifs de détection infrarouge efficaces, la connaissance du spectre électronique complet, c'est-à-dire les énergies des niveaux électroniques ainsi que les pièges dans la bande interdite, est indispensable.

Pour connaître les énergies des états, il est nécessaire de les mesurer expérimentalement. Dans ce chapitre, je commencerai donc par présenter certaines techniques expérimentales permettant de remonter à l'énergie des niveaux électroniques des nanocristaux, en insistant sur la photoémission que j'utilise pour connaître l'énergie de Fermi et le travail de sortie. J'utiliserai ensuite ces techniques de reconstruction de la structure électronique sur des nanocristaux de tellure de mercure (HgTe) et de séléniure de mercure (HgSe) à différents degrés de confinements.

HgTe et HgSe sont des semimétaux à l'état massif. Une des spécificités des nanocristaux de HgSe et HgTe est donc d'être à la limite entre semiconducteur et métal, selon qu'ils sont plus ou moins confinés. Les techniques que j'ai mises en place pour reconstruire la structure électronique de ces matériaux permettent de révéler la transition semiconducteur-métal dans les nanocristaux de HgSe à différents degrés de confinement quantique.

La dernière partie de ce chapitre sera consacrée à la détermination de la distribution des pièges dans la bande interdite des nanocristaux. Je présenterai une technique expérimentale originale basée sur le photo-transport transitoire permettant de mesurer l'énergie d'Urbach, caractéristique de la répartition de ces pièges.

## I.    Techniques de détermination de la structure électronique

Reconstruire le spectre électronique des matériaux n'est pas un sujet récent. Plusieurs techniques sont bien développées dans la littérature pour remonter au spectre électronique. Dans cette partie, je présenterai l'électrochimie, le transistor à effet de champ qui est une spécialité de notre équipe à l'Institut des Nanosciences de Paris, et la photoémission.

### 1.    Mesure des énergies des niveaux électroniques par électrochimie

L'électrochimie est un outil permettant de modifier le remplissage des différents niveaux (*35–38*) ou autrement dit, l'énergie du niveau de Fermi dans le milieu. Cette mesure se fait en utilisant un système à quatre électrodes : deux électrodes de travail $WE_1$ et $WE_2$, une contre électrode CE et une électrode de référence RE (Figure 27). Le film de nanocristaux à étudier est déposé entre les deux électrodes de travail et l'ensemble est placé dans un électrolyte, c'est-à-dire une matrice liquide contenant des ions mobiles (*38*).

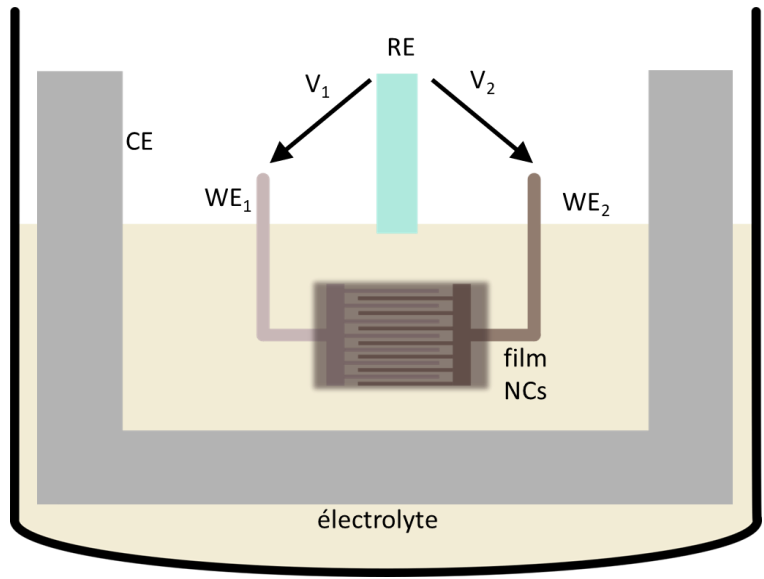

*Figure 27 : Schéma de cellule électrochimique. Le film de nanocristaux est déposé entre les électrodes interdigitées $WE_1$ et $WE_2$. L'ensemble est plongé dans un électrolyte liquide.*

La conductivité d'un film de nanocristaux dépend de l'énergie du niveau de Fermi. S'il se situe au milieu de la bande interdite, c'est-à-dire si les nanocristaux sont intrinsèques, elle est minimale. S'il se rapproche du niveau $1S_e$, les électrons pourront plus facilement y accéder par activation thermique et la conductivité sera meilleure que dans le régime intrinsèque. De même avec les trous si le niveau de Fermi se rapproche du niveau $1S_h$.

Pour modifier l'énergie du niveau de Fermi dans les nanocristaux, une tension est appliquée entre l'électrode de référence et les électrodes de travail. Grâce aux ions présents dans l'électrolyte, cette tension permet d'induire un champ électrique F à la surface des nanocristaux et donc de courber les niveaux électroniques, comme présenté sur la Figure 28 :

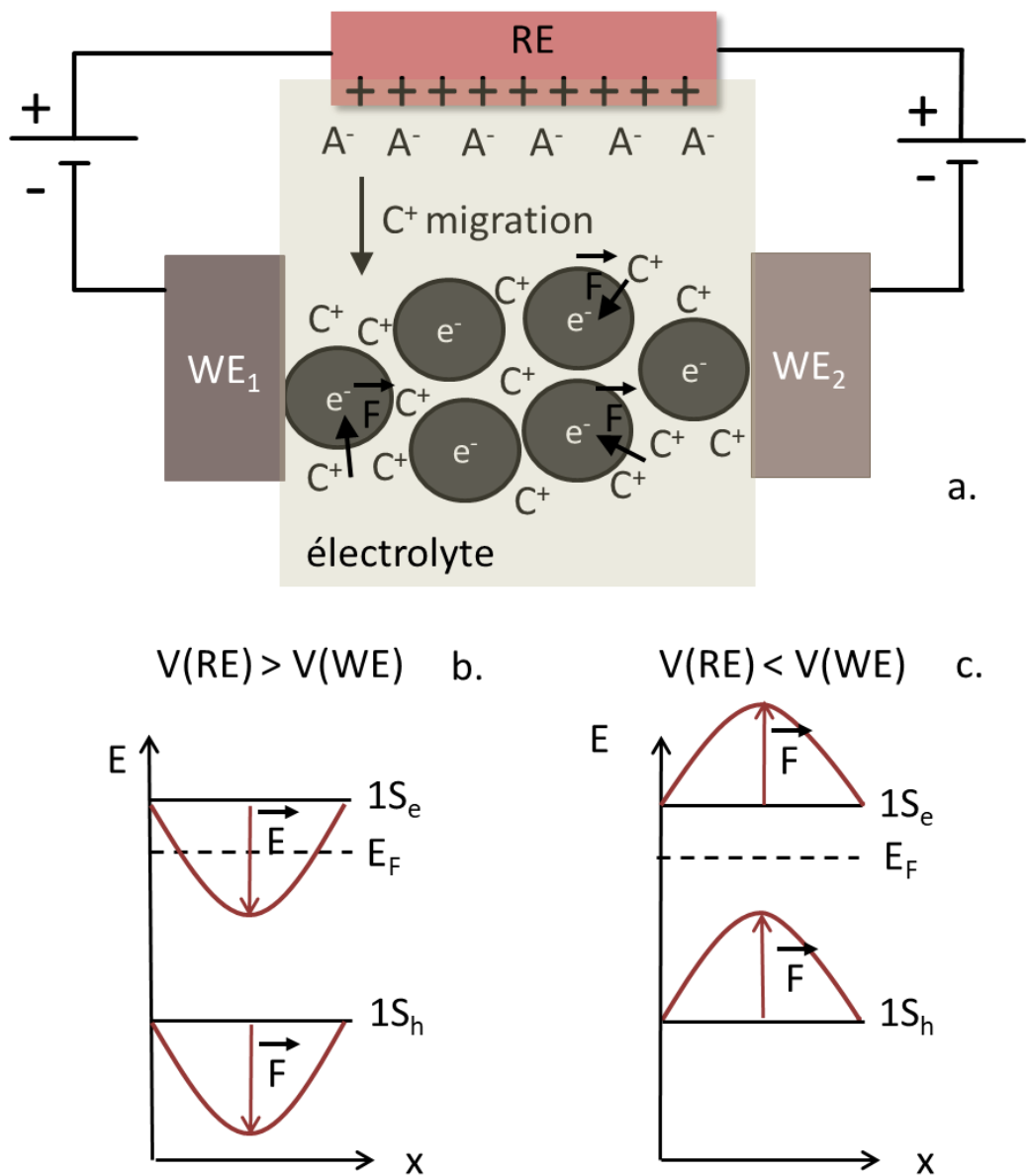

*Figure 28 : (a) Schéma représentant la migration des ions de l'électrolyte lorsqu'une tension positive est appliquée à l'électrode de référence par rapport aux électrodes de travail. Les anions (A⁻) sont attirés par les charges positives sur l'électrode de référence tandis que les cations (C⁺) migrent à travers le film. (b) Représentation de l'énergie des niveaux électroniques dans le nanocristal sans champ électrique (en noir) et avec (en rouge) sous application d'une tension de grille positive. (c) Représentation de l'énergie des niveaux électroniques dans le nanocristal sans champ électrique (en noir) et avec (en rouge) sous application d'une tension de grille négative. L'axe des abscisses des figures b et c correspond à l'axe séparant les deux électrodes WE₁ et WE₂.*

Les deux électrodes de travail $WE_1$ et $WE_2$ se voient appliquer par rapport à l'électrode de référence les potentiels $V_1$ et $V_2$ respectivement. La différence $V_2 - V_1$ est fixée à quelques millivolts de telle sorte à ce qu'un courant puisse circuler dans le film de nanocristaux. Les courants sont mesurés entre chacune des électrodes de travail et la contre électrode ($I_1$ entre WE$_1$ et CE, $I_2$ entre WE$_2$ et CE). On peut ensuite déterminer la conductance G :

$$G = \frac{I_1 - I_2}{2(V_1 - V_2)} \tag{2.1}$$

Comme décrit sur la Figure 28b et la Figure 28c, faire varier les potentiels $V_1$ et $V_2$ revient à faire varier la position du niveau de Fermi dans le système et donc à changer le remplissage des différents niveaux.

L'évolution de la conductance en fonction du potentiel appliqué présente des maxima locaux suivis de minima locaux. Ces valeurs caractéristiques de conductance, appelées blocages de Pauli, témoignent d'une énergie particulière du niveau de Fermi. En effet, un maximum local est atteint lorsqu'un niveau électronique est demi-rempli : le niveau de Fermi est à la même énergie que le niveau électronique en question (*39*). Le minimum local est atteint lorsque ce niveau est totalement rempli. La conductivité ré-augmente ensuite, les électrons étant conduits via le niveau suivant. Sur la Figure 29a, les maxima locaux de conductance correspondent au demi-remplissage du niveau 1S$_e$ (en rouge), et au demi-remplissage du niveau 1P$_e$ (en bleu). Les potentiels correspondants sont mesurés par rapport à une électrode standard, (l'électrode au calomel saturé, ECS, par exemple), et permettent donc de positionner les niveaux 1S$_e$ et 1P$_e$ par rapport au potentiel de l'ECS, comme présenté sur la Figure 29b. Le travail de sortie de l'ECS (WF$_{ECS}$) est fixe et vaut 4,68 ± 0,02 eV : la position des niveaux par rapport au niveau du vide peut donc se calculer facilement. Dans le cas du niveau 1S$_e$, on a :

$$1S_e/vide = WF_{ECS} - 1S_e/ECS \tag{2.2}$$

Où 1S$_e$/vide est l'écart en énergie entre le niveau 1S$_e$ et le niveau du vide, WF$_{ECS}$ est le travail de sortie de l'ECS et 1S$_e$/ECS est l'énergie du niveau 1S$_e$ par rapport à l'ECS. Le niveau de Fermi initial du film est le potentiel de repos (sans circulation de courant ni potentiel appliqué à l'électrode de référence). Il est mesuré par rapport à l'ECS et peut donc également être recalculé par rapport au niveau du vide (Figure 29b).

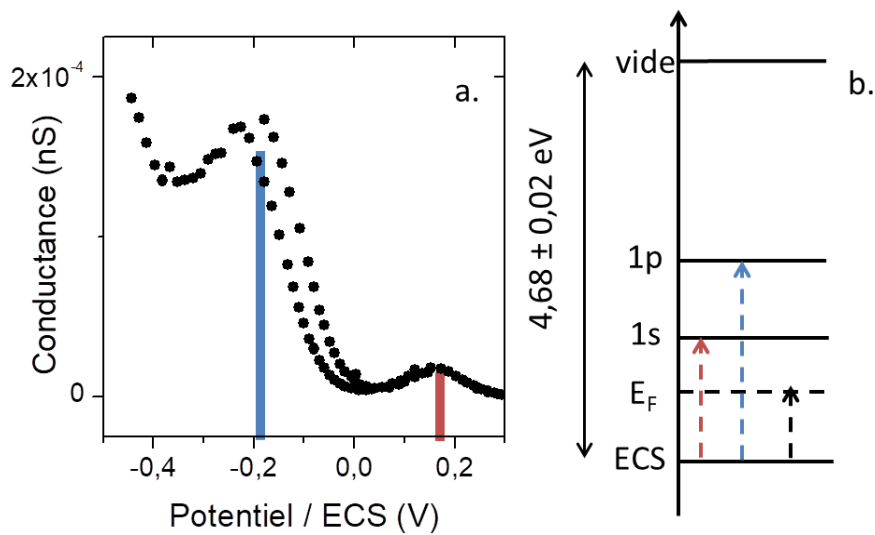

*Figure 29 : Données issues de l'article « Reversible Electrochemistry of Mercury Chalcogenide Colloidal Quantum Dot Films » de M. Chen et P. Guyot-Sionnest en 2017 (38) (a) Mesure de la conductance en fonction du potentiel de l'électrode de référence. Les potentiels pour lesquels la conductance est maximale correspondent aux*

*énergies des niveaux électroniques des nanocristaux. (b) Reconstruction de la structure électronique de nanocristaux de HgTe par cette technique. Le niveau de Fermi est celui mesuré à l'équilibre.*

La méthode électrochimique permet de déterminer l'intégralité de la structure électronique mais présente quelques inconvénients. Notamment, la mesure est faite dans un électrolyte qui peut limiter la gamme de potentiels accessibles. De plus, cet électrolyte autour du film de nanocristaux est un environnement diélectrique très différent (liquide et ions) de celui du dispositif en fonctionnement.

Dans notre groupe, nous préférons utiliser deux autres techniques. (i) Les transistors à effet de champ, dont le principe est similaire à celui de l'électrochimie : ils sont beaucoup plus simples à mettre en place mais ne donnent que des informations qualitatives sur l'énergie du niveau de Fermi. (ii) La photoémission qui, combinée à l'absorption infrarouge, permet de reconstruire l'intégralité de la structure électronique.

### 2. Mesures de transistor à effet de champ

#### a. Fonctionnement d'un transistor à effet de champ

Le transistor à effet de champ se construit en déposant un film de nanocristaux entre deux électrodes, la source (S) et le drain (D), et en ajoutant une capacité, ou grille, notée G. De même que pour l'électrochimie, appliquer une tension de grille positive permet d'injecter des charges positives dans le film de nanocristaux. Cela a pour effet de courber les niveaux énergétiques vers les basses énergies et donc de rapprocher le niveau $1S_e$ du niveau de Fermi. À l'inverse, une tension de grille négative courbera les niveaux énergétiques vers les hautes énergies et rapprochera le niveau $1S_h$ du niveau de Fermi (voir Figure 28b et c).

Dans le cas de nanocristaux dopés n, c'est-à-dire quand le niveau de Fermi est dans la moitié supérieure de la bande interdite, ce sont les électrons qui assurent la conduction au sein du film. Appliquer une tension de grille positive permettra de rapprocher les niveaux $1S_e$ et $E_F$, donc d'augmenter le nombre d'électrons thermiquement activés et ainsi la conductivité. Au contraire, une tension de grille négative éloignera le niveau de Fermi du niveau $1S_e$ ce qui diminuera la conductivité. Dans le cas de nanocristaux dopés p, on observe le phénomène inverse, et la conductivité augmente quand on applique une tension de grille négative.

À cause de l'absence d'électrode de référence, les mesures de transistors à effet de champ ne permettent pas de déterminer quantitativement l'énergie du niveau de Fermi avec la tension de grille, et donc ne permettent pas de déterminer quantitativement les énergies des niveaux. Cependant, en fonction de l'allure de la courbe de transfert obtenue, nous pourrons déterminer si le film de nanocristaux est n ou p en fonction de l'évolution de la conductivité avec la tension de grille.

#### b. Grille diélectrique ou grille électrolytique

Conventionnellement, les grilles utilisées pour réaliser des transistors à effet de champ sont les grilles diélectriques et les grilles électrolytiques.

Le déplacement des niveaux électroniques par rapport au niveau de Fermi sera d'autant plus important que la quantité de charges injectées dans le film q sera grande. La quantité de charges injectées est proportionnelle à la capacité et à la tension de grille (voir équation 2.3). Pour obtenir un transistor à effet de champ efficace, permettant de sonder de grands intervalles d'énergie, il faut donc maximiser la capacité et la tension de la grille.

$$q = C.V_{GS} \tag{2.3}$$

Les **grilles diélectriques** sont des solides isolants ayant une grande permittivité relative. Leur capacité se calcule selon la formule :

$$C = \frac{\varepsilon_0 \varepsilon_r S}{d} \tag{2.4}$$

Où C est la capacité, $\varepsilon_0$ est la permittivité diélectrique du vide, $\varepsilon_r$ est la permittivité relative de la grille, S sa surface et d son épaisseur. Ainsi, la capacité surfacique d'une grille de silice (SiO$_2$) de 300 nm d'épaisseur vaut 12 nF/cm². La capacité sera maximale pour un matériau à forte constante diélectrique (voir Tableau 3) et de faible épaisseur.

*Tableau 3 : Permittivités relatives $\varepsilon_r$ de quelques isolants (40)*

| Matériau | $\varepsilon_r$ |
|:---:|:---:|
| SiO$_2$ | 3.9 |
| Si$_3$N$_4$ | 7.5 |
| Al$_2$O$_3$ | 9 |
| HfO$_2$ | 25 |

La faible épaisseur de la couche de matériau diélectrique peut être incompatible avec la tension de grille appliquée : plus l'épaisseur est faible, plus le champ électrique est important dans la couche, et donc plus le risque de claquage du matériau augmente.

L'épaisseur du canal (*i.e.* film de nanocristaux à sonder) est également à prendre en compte dans le cas de l'utilisation de grilles diélectriques. En effet, comme présenté sur la Figure 30a, les charges injectées dans le film restent localisées près de la surface de la grille diélectrique, là où le champ est appliqué. Dans le cas d'un film de nanocristaux épais (> 50 nm), une partie du film ne sera pas affectée par la tension de grille : l'effet de grille sera donc plus difficile à observer.

Les **grilles électrolytiques** sont composées d'une matrice (liquide, polymère ou solide) contenant des ions mobiles, voir Figure 30b. Dans le cas des nanocristaux, les transistors électrolytiques à base d'une matrice de polymères de polyéthylène glycol (PEG) contenant des ions Li$^+$ et ClO$_4^-$ sont couramment utilisés. L'intérêt de ce matériau par rapport aux grilles diélectriques est leur forte capacité (> µF/cm²), qui ne dépend pas de l'épaisseur de la grille. En effet, les ions étant mobiles, ils peuvent migrer à l'intérieur de la couche de nanocristaux et appliquer un champ électrique directement à la surface de chaque nanocristal. La distance sur laquelle le champ électrique est appliqué est donc la double couche électrique, ou longueur de Debye, de l'ordre du nanomètre :

$$C = \frac{\varepsilon_0 \varepsilon_r S}{d_{Debye}} \tag{2.5}$$

Où $d_{Debye}$ est l'épaisseur de la double couche électrostatique à la surface des nanocristaux. Les grilles électrolytiques permettent d'explorer une large gamme de densités de porteurs en appliquant des tensions de grilles faibles. C'est donc ce type de grille que j'ai choisi d'utiliser pour déterminer les propriétés de transport des films de nanocristaux.

Le transistor à effet de champ électrolytique est un outil bien maîtrisé dans l'équipe dans laquelle j'ai travaillé pendant mon doctorat (*41, 42*). En plus de sa capacité élevée, il est utilisable à l'air, et permet, grâce à la diffusion des ions, de sonder des films épais. Cette dernière propriété nous permet de tester des films absorbants, dont l'épaisseur se rapproche de celle que nous utilisons pour mesurer la photoréponse des nanocristaux. Nous pouvons donc réaliser des phototransistors, c'est-à-dire étudier la photoréponse en fonction de l'énergie de Fermi dans les nanocristaux (*43, 44*).

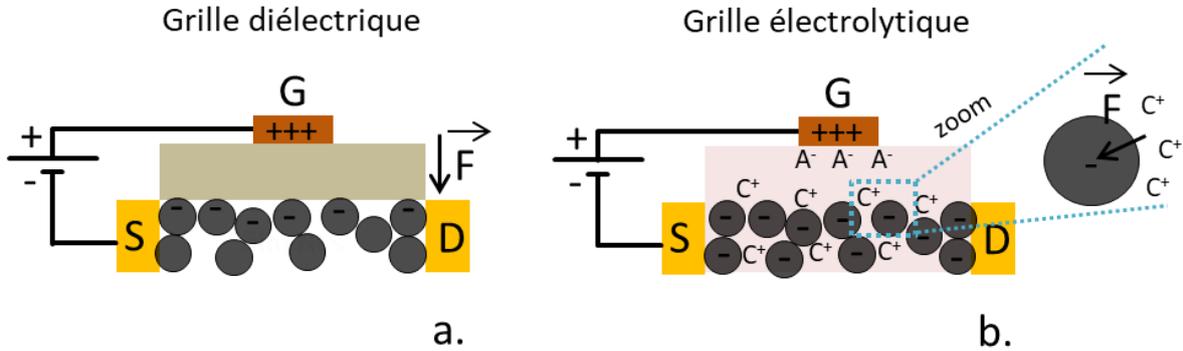

*Figure 30 : (a) Schéma d'un transistor diélectrique. Dans ce cas, le champ électrique s'applique à travers l'épaisseur de la grille diélectrique. (b) Schéma d'un transistor électrolytique. $A^-$ correspond aux anions de l'électrolyte, $C^+$ aux cations. Dans ce cas, le champ électrique s'applique sur l'épaisseur de la couche de Debye autour du nanocristal.*

### c. Application du transistor à effet de champ sur des films de nanocristaux de chalcogénures de mercure

Les courbes de transfert, c'est-à-dire la mesure de l'intensité du courant circulant dans le canal en fonction de la tension de grille appliquée, ont été mesurées pour des films de nanocristaux de chalcogénures de mercure (Figure 31). Pour les nanocristaux de tellure de mercure (HgTe), l'évolution de la conductivité en fonction de la tension de grille varie avec le confinement quantique du nanocristal.[6] Les HgTe 6000 cm$^{-1}$ sont dopés p (Figure 31a), ce qui correspond à un niveau de Fermi dans la moitié inférieure de la bande interdite (Figure 31b). Les HgTe 4000 cm$^{-1}$ sont ambipolaires, conduisant les trous et les électrons (Figure 31c). Cela correspond à un niveau de Fermi au milieu de la bande interdite (Figure 31d). Les HgTe 2000 cm$^{-1}$ sont dopés n (Figure 31e), ce qui correspond à un niveau de Fermi dans la moitié supérieure de la bande interdite (Figure 31f) (*11, 45*).

Pour les nanocristaux de séléniure de mercure (HgSe), la courbe de transfert présente un maximum local suivi d'un minimum local, non observés pour les nanocristaux de HgTe. Ces caractéristiques sont les mêmes que celles que j'ai décrites pour les mesures d'électrochimie (voir p49). Le maximum local correspond à la superposition des niveaux de Fermi et $1S_e$ (niveau de Fermi demi-rempli). Le minimum local est atteint lorsque le niveau $1S_e$ est totalement rempli. Comme présenté précédemment, les nanocristaux de HgSe présentent déjà des électrons dans les niveaux de conduction après la synthèse. Le niveau de Fermi est donc proche du niveau 1Se. L'origine de ce dopage sera discutée dans le chapitre 3.

---

[6] Pour rappel, dans la suite du manuscrit les nanocristaux de chalcogénures de mercure seront identifiés par leur nom suivi du nombre d'onde correspondant à la transition mesurée par absorption infrarouge. Ex : HgTe 6000 cm$^{-1}$

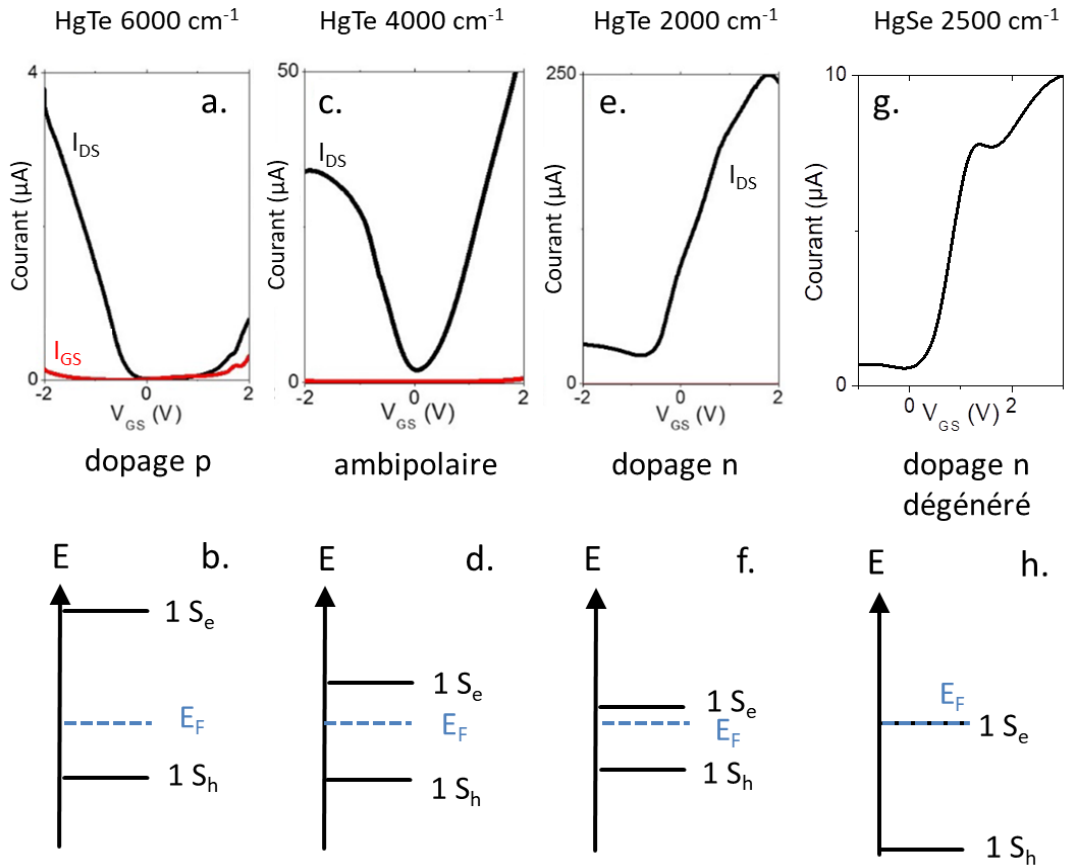

*Figure 31 : Courbes de transfert réalisées sur des films de nanocristaux de HgTe 6000 cm⁻¹ (a), HgTe 4000 cm⁻¹ (c), HgTe 2000 cm⁻¹ (e) et HgSe 2500 cm⁻¹ (g). Le caractère n, ambipolaire ou p permet de connaître approximativement l'énergie du niveau de Fermi dans la bande interdite. On peut donc reconstruire les structures électroniques de ces matériaux : HgTe 6000 cm⁻¹ (b), HgTe 4000 cm⁻¹ (d) HgTe 2000 cm⁻¹ (f), HgSe 2500 cm⁻¹ (h).*

Le transistor à effet de champ permet donc d'obtenir une estimation de l'écart entre l'énergie du niveau de Fermi et les énergies des niveaux électroniques des nanocristaux. Il n'est cependant pas suffisant pour déterminer de manière quantitative cet écart sans passer par l'ajout d'une quatrième électrode de référence dans l'électrolyte. Pour des mesures plus quantitatives, nous pouvons utiliser la photoémission.

### 3. Structure électronique déterminée par photoémission des rayons X (ou XPS)

La photoémission permet de déterminer avec précision l'énergie du niveau de Fermi et le travail de sortie. Combinée avec l'absorption infrarouge, elle permet de remonter à la structure électronique des nanocristaux. Cette technique est une méthode traditionnelle pour déterminer l'énergie absolue des niveaux électroniques dans les semiconducteurs épitaxiés.

#### a. Présentation de la technique

Les mesures de photoémission des rayons X consistent à envoyer des photons énergétiques, entre 20 eV et quelques keV, sur un matériau pour en extraire des électrons de cœur ou de valence, comme présenté sur la Figure 32. Un analyseur à la sortie de l'échantillon permet de quantifier le nombre de ces photoélectrons extraits en fonction de leur énergie cinétique, reliée à l'énergie de liaison du niveau dont est extrait l'électron par :

$$E_C = h\nu - E_L \qquad\qquad (2.6)$$

Où $E_C$ est l'énergie cinétique de l'électron émis, $h\nu$ est l'énergie du photon initial et $E_L$ est l'énergie de liaison du niveau dont provient l'électron par rapport au niveau de Fermi. En réalité, une correction $C_{EF}$ doit être ajoutée pour que le niveau de Fermi ait une énergie de liaison de 0 eV au niveau du détecteur (Figure 32). Cette technique permet ainsi de remonter aux énergies de tous les niveaux de cœur d'un matériau et celle de la bande de valence.

L'échantillon est déposé sur un substrat conducteur, dans notre cas un substrat de silice recouvert d'or, et introduit dans un bâti ultravide via des sas à des pressions différentes : le sas d'entrée est à quelques $10^{-3}$ mbar, la chambre de préparation est à quelques $10^{-7}$ mbar, tandis que la pression de la chambre d'analyse est inférieure à quelques $10^{-9}$ mbar. La source de photons peut être une source X, comme la raie Kα d'une anode d'aluminium, ou un rayonnement synchrotron monochromaté. Dans le cas du faisceau synchrotron, comme celui de la ligne TEMPO du synchrotron SOLEIL que j'ai utilisé pendant mon doctorat, l'énergie de photons $h\nu_{exp}$ n'est pas connue avec une grande précision (± 0,3 eV), il est donc nécessaire de la mesurer expérimentalement.

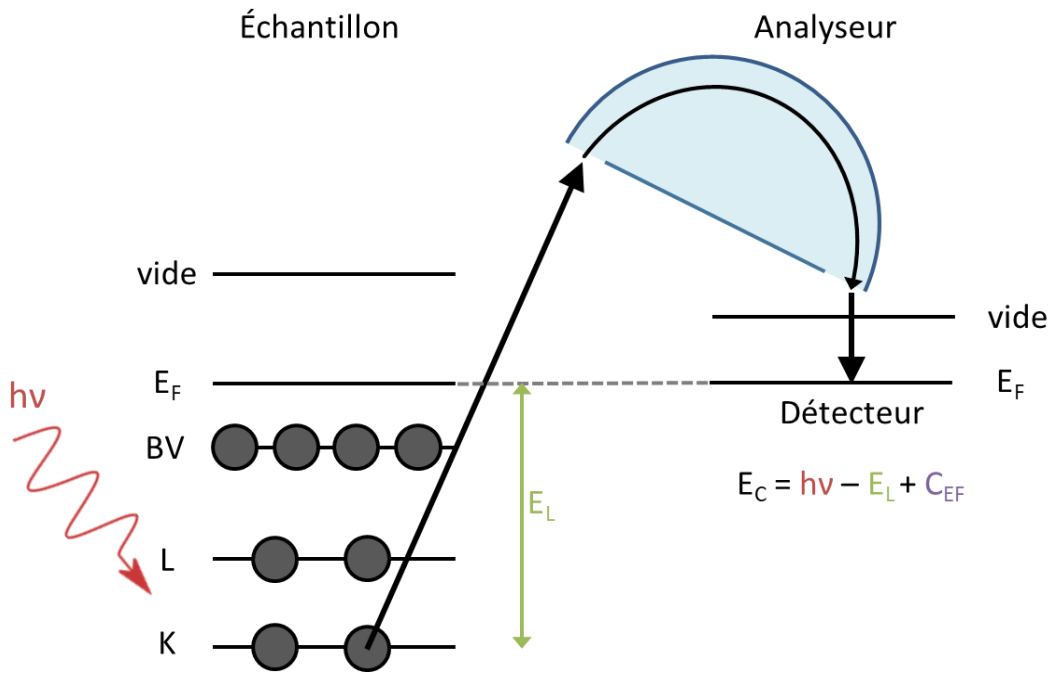

*Figure 32 : Schéma de principe de la spectroscopie par photoémission de rayons X. Les niveaux K et L représentent des niveaux de cœur, BV et $E_F$ sont les abréviations de bande de valence et niveau de Fermi respectivement ; $E_C$ et $E_L$ les abréviations d'énergie cinétique et énergie de liaison respectivement ; $C_{EF}$ l'abréviation de correction du niveau de Fermi.*

Avant de pouvoir déterminer avec précisions les différentes énergies de liaison du matériau, l'énergie expérimentale de photon $h\nu_{exp}$ et la correction du niveau de Fermi $C_{EF}$ doivent être mesurées.

### b. Mesure de l'énergie de photons $h\nu_{exp}$

L'énergie des photons incidents se détermine précisément en mesurant les niveaux de cœur d'un matériau, par exemple un cristal d'or monocristallin. Un monochromateur permet de sélectionner une énergie de photons $h\nu_{exp}$, ainsi que ses harmoniques ($2h\nu_{exp}$, $3h\nu_{exp}$ …).

Pour déterminer l'énergie de photons, il faut sélectionner un niveau de cœur connu, par exemple le niveau 4f de l'or (les énergies de liaison des niveaux de cœur de tous les atomes sont tabulées, on les retrouve notamment dans le *X-ray Data Booklet (46)*). Ce niveau est un doublet dû au couplage spin orbite et ses deux composantes sont appelées $4f_{7/2}$ et $4f_{5/2}$. Les photoélectrons éjectés du niveau $4f_{7/2}$ (d'énergie de liaison $E_{L\ 4f7/2}$, 84,0 eV) par des photons ayant une énergie $h\nu_{exp}$ auront donc l'énergie cinétique $E_{C,1}$ :

$$E_{C,1} = h\nu_{exp} - E_{L\ 4f\ 7/2} + C_{EF} \qquad (2.7)$$

Les photoélectrons éjectés du niveau $4f_{7/2}$ par des photons ayant une énergie $2h\nu_{exp}$ auront l'énergie cinétique $E_{C,2}$ :

$$E_{C,2} = 2h\nu_{exp} - E_{L\ 4f\ 7/2} + C_{EF} \qquad (2.8)$$

Ainsi, en mesurant ces deux énergies cinétiques et en faisant la différence entre les deux, on retrouve la valeur de l'énergie de photon $h\nu_{exp}$ (voir Figure 33a).

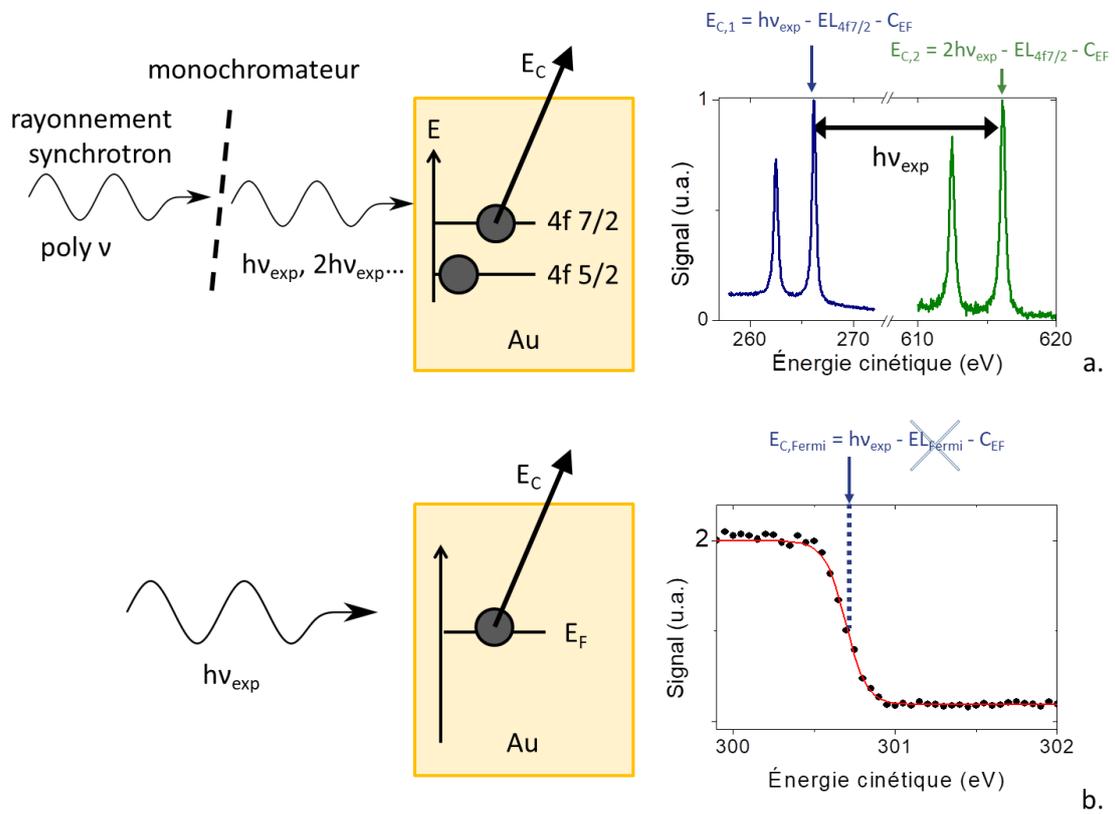

*Figure 33 : (a) Schéma de principe de la mesure de l'énergie incidente $h\nu_{exp}$ ; (b) schéma de principe de la mesure de la correction de l'analyseur $C_{EF}$. La position du niveau de Fermi est déterminée en faisant une modélisation de la courbe par la statistique de Fermi-Dirac.*

### c. Mesure de la correction du niveau de Fermi $C_{EF}$

Pour calibrer le niveau de Fermi à une énergie de liaison de 0 eV, il peut être nécessaire de réaliser une calibration. Pour cela, on utilise un échantillon d'or monocristallin sur lequel on mesure l'énergie cinétique du dernier niveau occupé : pour l'or, il s'agit du niveau de Fermi. On a donc :

$$E_{C,Fermi} = h\nu_{exp} - E_{L,Fermi} + C_{EF} \qquad (2.9)$$

Où $E_{C,Fermi}$ est l'énergie cinétique des photoélectrons éjectés du niveau de Fermi, $E_{L,Fermi}$ est l'énergie de liaison du niveau de Fermi de l'or et $C_{EF}$ la correction à appliquer au niveau de Fermi recherchée. Par définition, l'énergie de liaison du niveau de Fermi de l'or est nulle, on peut donc déterminer $C_{EF}$ en mesurant $E_{C,Fermi}$ et en utilisant la valeur de l'énergie de photons mesurée au paragraphe b. Le principe de la mesure est présenté sur la Figure 33b.

### d. Détermination des niveaux de cœur, de la bande de valence et du travail de sortie

Une fois les paramètres expérimentaux calibrés, la photoémission peut être utilisée pour mesurer les niveaux de cœur et l'énergie de la bande de valence des matériaux. À la fin de cette partie, je montrerai comment la photoémission permet également de mesurer le travail de sortie d'un matériau.

#### i. Niveaux de cœur

Les valeurs des énergies de liaison des niveaux de cœur par rapport au niveau de Fermi de tous les atomes sont tabulées dans la littérature (*46*). Par conséquent, si tous les paramètres expérimentaux (énergie de photons, correction du niveau de Fermi) ont été calibrés auparavant, il est facile de repérer les différents éléments présents dans le matériau grâce à leur énergie de liaison.[7] Mesurer les niveaux de cœur permet donc en premier lieu de confirmer la composition de l'échantillon, et de vérifier que ce dernier n'est pas contaminé par d'autres éléments. Les niveaux de cœur d'un film de nanocristaux de HgSe, sondés avec une énergie de photons de 350 eV, sont présentés sur la Figure 34a.

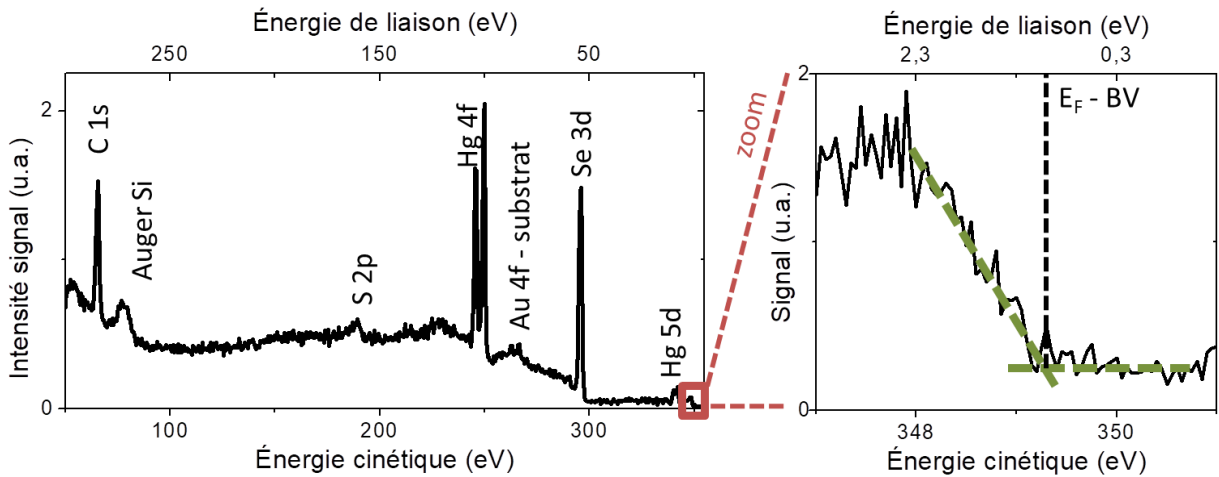

*Figure 34 : Signal de photoémission en fonction de l'énergie cinétique des électrons éjectés. L'énergie de liaison est calculée avec la formule $E_L = h\nu_{exp} - E_C - C_{EF}$. L'insert est un zoom du spectre de photoémission sur les basses énergies de liaison, donc sur la bande de valence (BV). L'énergie de liaison de la bande de valence se mesure au pied de la marche de plus faible énergie de liaison.*

On note qu'en plus des niveaux de cœur, des électrons Auger peuvent également être détectés par photoémission : lorsque des électrons de cœur viennent remplir les lacunes créées par la photoémission, leur énergie peut être transférée à un électron voisin qui sera à son tour éjecté du matériau (voir Figure 34a).

---

[7] Seuls les niveaux de cœur ayant une énergie de liaison inférieure à celle de l'énergie des photons peuvent être sondés. Les électrons provenant de niveaux plus profonds ne peuvent pas être extraits.

Les niveaux de cœur tabulés dans la littérature sont ceux d'atomes, au degré d'oxydation 0. Or, les énergies des niveaux de cœur dépendent du degré d'oxydation de l'élément : si celui-ci est positif, donc si l'élément est pauvre en électrons, les énergies de liaison seront plus élevées. En revanche, si le degré d'oxydation est négatif, l'élément est riche en électrons et ses énergies de liaisons seront plus faibles.

Ainsi, pour des nanocristaux de HgSe, nous pouvons étudier les niveaux de cœur du mercure, notamment les pics 4f (dans la littérature, $E_{L,4f7/2}$ = 99,9 eV), comme présenté sur la Figure 35. Ces pics sont asymétriques, ce qui témoigne de la présence de plusieurs composantes 4f au sein de ce niveau de cœur : après modélisation, le niveau 4f du mercure dans ces nanocristaux est la somme de deux composantes. Dans les nanocristaux de séléniure de mercure, deux environnements sont possibles pour le mercure : celui-ci peut être lié au sélénium à l'intérieur du cristal, ou au soufre s'il est en surface car les nanocristaux sont recouverts de ligands thiols. Le sélénium étant moins électronégatif que le soufre, les atomes de mercure liés à un atome de sélénium auront une énergie de liaison plus basse (violet sur la Figure 35) que ceux liés à un atome de soufre (rose sur la Figure 35).

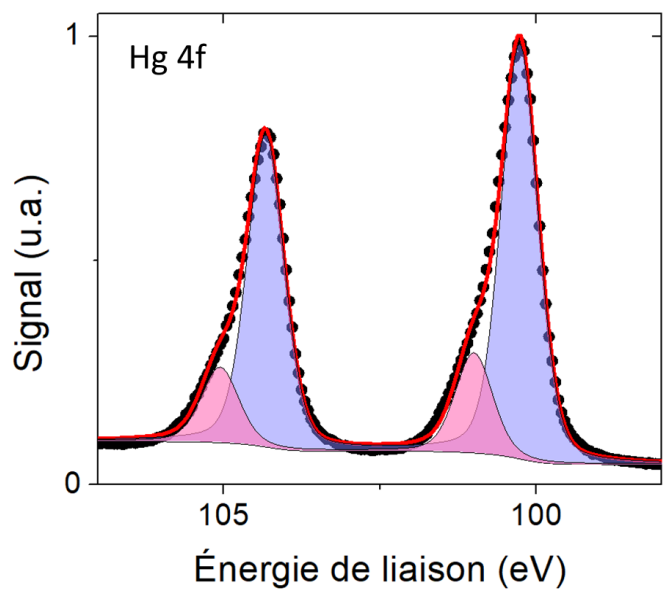

*Figure 35 : Niveaux de cœur Hg 4f obtenus pour les nanocristaux de 5,8 nm. L'énergie de photons vaut 600 eV. Les points expérimentaux sont représentés par les cercles noirs, les deux pics rose et violets sont les deux composantes du signal. La somme de ces deux composantes est représentée par la ligne rouge, qui suit parfaitement les points expérimentaux.*

### ii. Bande de valence

Comme pour les niveaux de cœur, la photoémission permet de mesurer l'énergie de liaison de la bande de valence par rapport au niveau de Fermi. La bande de valence est le niveau dont l'énergie de liaison est la plus faible (voir Figure 34).

### iii. Travail de sortie

Le travail de sortie n'intervient pas dans l'expression de l'énergie cinétique des photoélectrons récoltés par l'analyseur mais peut tout de même se déterminer par photoémission.

La barrière énergétique à franchir pour un photoélectron afin d'être éjecté de l'échantillon vaut $E_L$ + WF, où $E_L$ est l'énergie de liaison du niveau de cœur dont provient le photoélectron et WF le travail de sortie du matériau. Par conséquent, les photoélectrons étant expulsés avec une énergie cinétique nulle du matériau vérifient :

$$h\nu_{exp} - E_L - WF = 0 \qquad (2.10)$$

Ces électrons ayant une très faible énergie cinétique sont des électrons secondaires. Ce sont des électrons qui ont subi des collisions qui ont fait diminuer leur énergie cinétique.

Le détecteur est paramétré pour collecter les électrons au niveau de Fermi. Par conséquent, au niveau du détecteur, l'énergie cinétique des électrons a augmenté d'une valeur égale au travail de sortie des nanocristaux par rapport au niveau du vide. L'énergie cinétique minimale que l'on peut mesurer au niveau du détecteur est donc égale au travail de sortie. On parle aussi de *cut-off* des électrons secondaires. Un schéma de principe et un exemple de mesure de travail de sortie (WF) sont présentés sur la Figure 36.

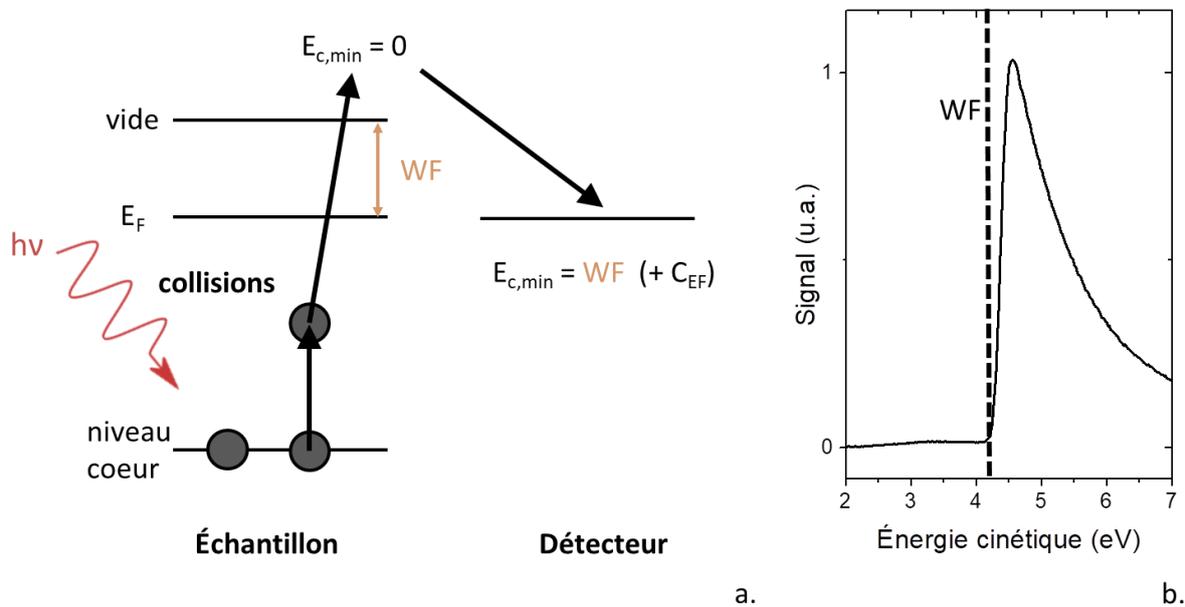

*Figure 36 : (a) Schéma de principe de mesure du travail de sortie à partir de la collecte d'électrons secondaires. (b) Exemple de mesure de travail de sortie sur des nanocristaux de HgSe.*

## II.    Énergies absolues des niveaux de chalcogénures de mercure en fonction de leur taille

Dans cette partie, nous allons utiliser la photoémission pour reconstruire la structure électronique de nanocristaux de HgTe et HgSe de différentes tailles, donc de différents confinements quantiques.

### 1.    Étude sur HgTe

Des cristaux de cinq tailles différents ont donc été synthétisés par Nicolas Goubet, post doctorant dans notre groupe, selon la méthode de Keuleyan (*47*). Les détails des protocoles de synthèse sont fournis dans l'Annexe 1. Leurs tailles ont été mesurées par microscopie électronique et les énergies des transitions interbandes optiques mesurées par absorption infrarouge, voir le Tableau 4. Pour obtenir l'énergie $1S_h - 1S_e$, il faut ajouter à la valeur de la transition optique l'énergie de liaison de l'exciton $E_{LE}$ :

$$E_{LE} = 1,8 \ \frac{e^2}{4\pi\varepsilon_0\varepsilon_r R} \hspace{2cm} (2.11)$$

Où R est le rayon du nanocristal. L'énergie de liaison de l'exciton est généralement comprise entre 10 et 50 meV pour les nanocristaux colloïdaux confinés dans les trois directions de l'espace.

*Tableau 4 : Caractéristiques des 5 tailles de nanocristaux de HgTe dont la structure électronique a été mesurée. La ligne dopage a été déterminée via des mesures par transistor à effet de champ : P correspond aux films de nanocristaux dopés p, A correspond aux films de nanocristaux ambipolaires, N correspond aux films de nanocristaux dopés n.*

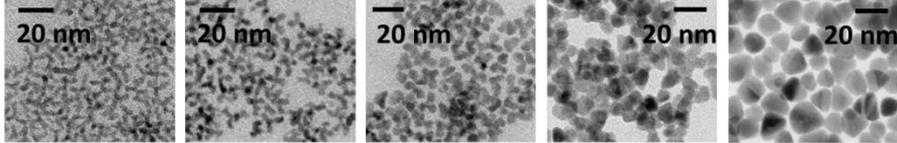

| | | | | | |
|---|---|---|---|---|---|
| Rayon (nm) | 2,5 | 3 | 4 | 6 | 12 |
| $E_{G,opt}$ (eV) | 0,89 | 0,76 | 0,51 | 0,37 | 0,09 |
| $E_{LE}$ (meV) | 51 | 43 | 32 | 21 | 10 |
| $E_G$ (eV) | 0,94 | 0,80 | 0,54 | 0,39 | 0,10 |
| Dopage | P | P | A | A | N |

Ces nanocristaux ont ensuite été déposés sous forme de film et échangés avec des ligands éthanedithiol (EDT). Des mesures de transistor à effet de champ ont été effectuées pour déterminer le type de dopage : n, p ou ambipolaire. Les nanocristaux de tellure de mercure les plus confinés sont dopés p, les moins confinés sont dopés n.

Ces mesures sont confirmées par photoémission. Sur la Figure 37, les structures électroniques obtenues par photoémission sont reportées et on observe que le niveau de Fermi se situe dans la moitié inférieure de la bande interdite pour les nanocristaux d'énergie interbande élevée (donc les plus petits) ; qu'il se situe dans la moitié supérieure de la bande interdite pour les nanocristaux d'énergie interbande plus faible (0,54 et 0,39 eV), et résonant avec le niveau $1S_e$ pour les nanocristaux les moins confinés ($E_G$ = 100 meV) (*24*).

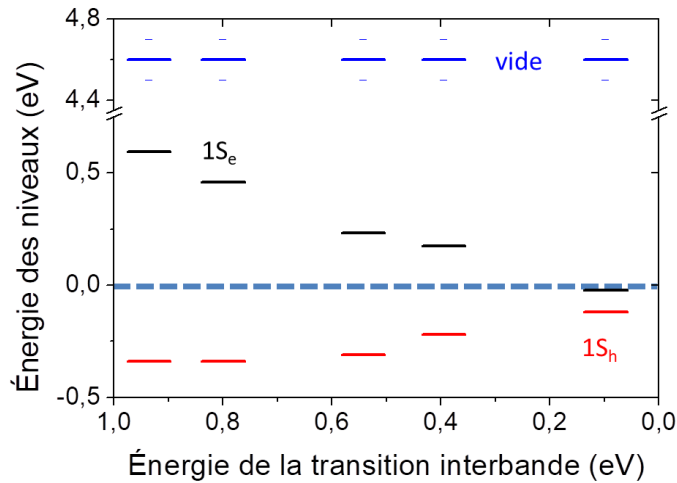

*Figure 37 : Énergie des niveaux électroniques ($1S_h$ en rouge, $1S_e$ en noir, niveau du vide en bleu, $E_F$) de nanocristaux de HgTe à différents confinements quantiques. Le niveau de Fermi est pris comme référence, à 0 eV. Il est représenté par un trait pointillé bleu clair.*

Le travail de sortie des nanocristaux est également déterminé par photoémission : le niveau du vide ne dépend pas du confinement quantique dans les nanocristaux de tellure de mercure, et vaut environ 4,7 eV. Ce travail de sortie est donc une propriété caractéristique de la surface HgTe – EDT.

La reconstruction de la structure électronique pour les différentes tailles de nanocristaux de HgTe est primordiale pour construire un dispositif de détection infrarouge : **je réutiliserai ces résultats dans le chapitre 4 afin de construire une diode à base de nanocristaux de HgTe.**

2. **Étude sur HgSe**

Les nanocristaux de séléniure de mercure sont dopés à la fin de la synthèse, et montrent une transition intrabande par spectroscopie infrarouge. Dans cette partie, nous allons étudier l'influence du confinement quantique sur le niveau de dopage de ces nanocristaux et donc, sur les transitions accessibles. En augmentant la taille des cristaux, le niveau de Fermi remonte dans les niveaux de conduction, modifiant ainsi le dopage de 1 à 2 électrons par nanocristal à une vingtaine d'électrons par nanocristal.

a. **Gamme couverte par les HgSe en fonction du confinement quantique**

Plusieurs tailles de nanocristaux de séléniure de mercure ont été synthétisées en laboratoire : 4,5 nm ± 0,5 nm, 5,8 nm ± 0,8 nm et 17 nm ± 3 nm de diamètre (voir Figure 38a, b et c). L'énergie de la transition intrabande dépend fortement de la taille des nanocristaux, passant de 3,5 µm pour les nanocristaux les plus petits à 10 µm pour les nanocristaux les plus gros, ce qui confirme la présence de confinement quantique pour ces nanocristaux ($a_{HgSe} \approx 17$ nm). Pour des nanocristaux encore plus larges, il est possible d'atteindre des transitions à une longueur d'onde de 25 µm (*22*). Au moment de la publication de mon article (*22*), ces nanocristaux synthétisés par Nicolas Goubet (Laboratoire de Physique et d'Étude des Matériaux, LPEM, de l'ESPCI) étaient les nanocristaux de HgSe les plus larges jamais synthétisés par voie colloïdale.

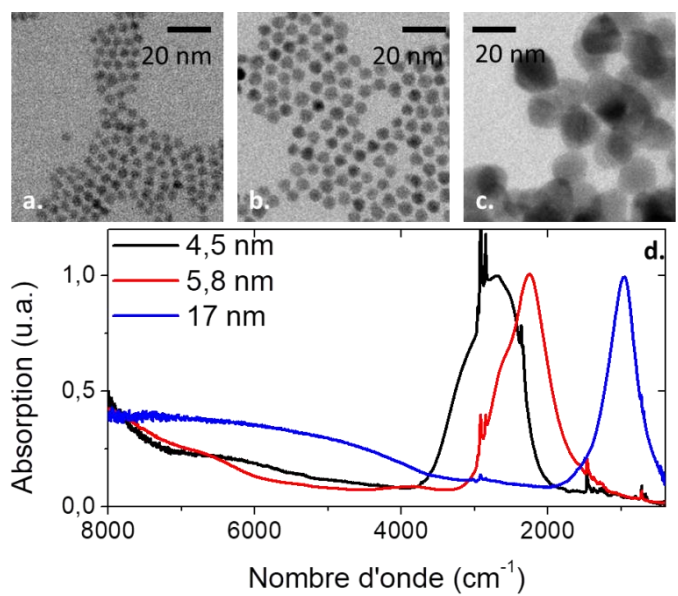

*Figure 38 : Présentation des différentes tailles de matériaux utilisés. (a) Cliché TEM des nanocristaux de 4,5 nm ; (b) cliché TEM des nanocristaux de 5,8 nm ; (c) cliché TEM des nanocristaux de 17 nm ; (d) spectres FTIR des trois tailles de nanocristaux synthétisés (4,5 nm en noir, 5,8 nm en rouge, 17 nm en bleu). Les ligands à la surface des nanocristaux sont des dodécanethiols (ou DDT).*

### b. Détermination de la structure électronique à différents confinements quantiques

#### i. Mesure des transitions interbandes $1S_h – 1S_e$

Avant de réaliser des études de photoémission pour déterminer l'énergie du niveau $1S_h$ par rapport au niveau de Fermi, il faut mesurer les énergies des transitions interbandes et intrabandes des trois tailles de nanocristaux. Or, comme présenté sur la Figure 38d, la transition interbande $1S_e – 1S_h$ n'est pratiquement pas visible pour aucune des tailles. En effet, cette transition n'est autorisée que dans le cas où le niveau $1S_e$ n'est pas rempli, c'est-à-dire lorsque le niveau de Fermi a une énergie inférieure au niveau $1S_e$, comme présenté sur la Figure 39a. Dans le cas contraire, elle est interdite.

Pour déterminer l'énergie de cette transition, nous avons décidé de diminuer le niveau de dopage des nanocristaux de séléniure de mercure en échangeant les ligands à la surface. Nous avons vu au début de ce chapitre, dans la partie sur les transistors à effet de champ, que la présence de dipôles à la surface du nanocristal pouvait courber les niveaux énergétiques et ainsi modifier leur énergie par rapport au niveau de Fermi. Les ligands à la surface du nanocristal peuvent ainsi introduire un dipôle, plus ou moins intense selon leur nature.[8] Ainsi, le dodécanethiol à un dipôle quasi nul, tandis que les ions sulfures $S^{2-}$ ont un dipôle de l'ordre de 1 Debye (*48*) : les nanocristaux couverts de ligands $S^{2-}$ seront donc moins dopés que les nanocristaux couverts de dodécanethiols. En abaissant le niveau de Fermi, certaines transitions auparavant interdites deviennent ré-autorisées, et peuvent être mesurées par spectroscopie infrarouge, voir Figure 39b, c et d.

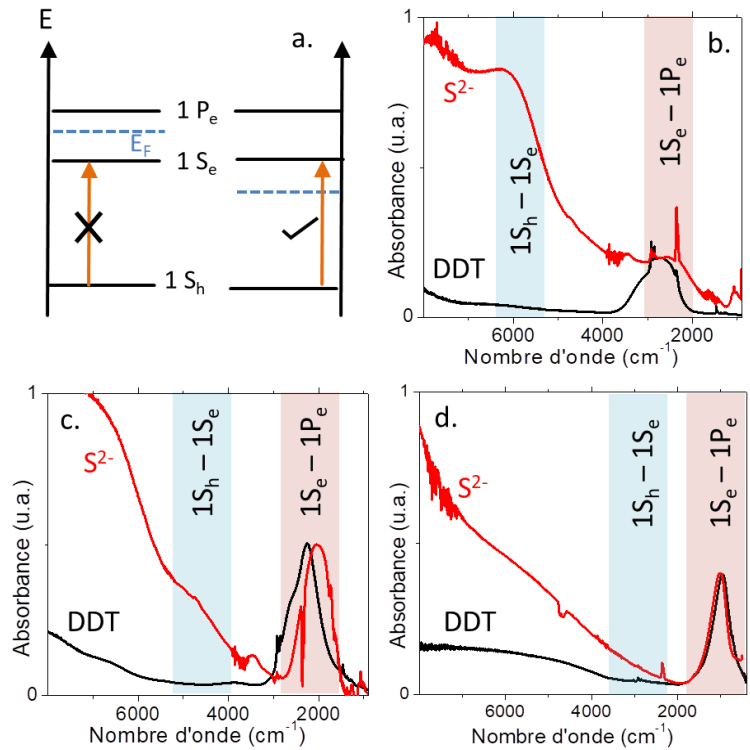

*Figure 39 : (a) Schéma de structures électroniques pour lesquelles la transition interbande, en orange, est interdite (à gauche) ou autorisée (à droite) en fonction de la position du niveau de Fermi. Spectres d'absorption infrarouge pour deux ligands ($S^{2-}$ en rouge, DDT en noir) et trois tailles de nanocristaux différents : 4,5 nm (b),*

---

*5,8 nm (c) et 17 nm (d). Sur chaque spectre infrarouge, la transition interbande est indiquée par un rectangle bleu et la transition intrabande par un rectangle rouge.*

L'énergie de la transition 1Se – 1Sh peut être obtenue en additionnant l'énergie de la transition optique mesurée par spectroscopie infrarouge à l'énergie de liaison de l'exciton (voir équation 2.11).

### ii. Calcul de l'énergie du niveau 1D$_e$

L'énergie du niveau 1D$_e$ (Tableau 2, p23) se calcule via un modèle k.p à trois bandes (*38, 49, 50*). Dans le cas de HgSe, les trois bandes en question sont les deux bandes $\Gamma_8$ qui jouent le rôle de bandes de valence et de conduction, et la bande $\Gamma_6$ (la structure de bandes est rappelée sur la Figure 40).

En supposant que la bande de valence $\Gamma_8$ est non-dispersive, on peut écrire pour la bande de conduction $\Gamma_8$:

$$E_{\Gamma 8} = \frac{E_{\Gamma 6 - \Gamma 8}}{2} + \sqrt{\frac{E_{\Gamma 6 - \Gamma 8}^2}{4} + \frac{E_p \hbar^2 k^2}{3 m_0}} \qquad (2.12)$$

Où E$_{\Gamma 8}$ est l'énergie de la bande de conduction $\Gamma_8$, E$_{\Gamma 6\text{-}\Gamma 8}$ est l'écart en énergie à k = 0 entre les bandes $\Gamma_6$ et $\Gamma_8$, E$_p$ est un paramètre appelé énergie de Kane, $\hbar$ est la constante de Planck réduite, k est le vecteur d'onde, et m$_0$ est la masse de l'électron libre.

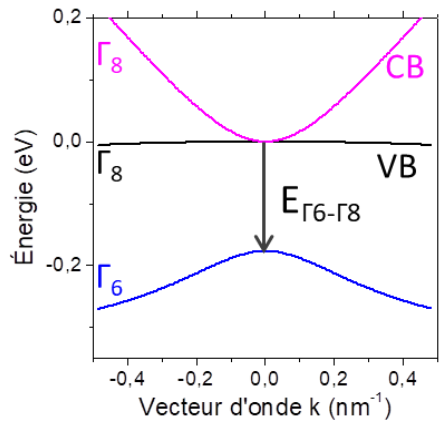

*Figure 40 : Diagramme de bandes de HgSe*

Les seuls paramètres de cette équation sont E$_{\Gamma 6\text{-}\Gamma 8}$ et E$_p$ que l'on peut déterminer par simulation. En ajoutant les résultats de la littérature à nos données pour les valeurs des transitions intrabandes dans HgSe à différentes tailles(*49, 51*), j'obtiens que E$_{\Gamma 6\text{-}\Gamma 8}$ vaut -0,2 eV et E$_p$ vaut 10,3 eV pour HgSe.

Les valeurs des transitions 1D$_e$ – 1P$_e$ sont ensuite obtenues en utilisant les valeurs de k correspondant à ces niveaux que j'ai rappelées dans le chapitre 1 (p23), soit 4,49/R pour 1P$_e$ et 5,76/R pour 1D$_e$.

### iii. Détermination de la structure électronique de nanocristaux de HgSe à différents confinements quantiques

En combinant mesures de photoémission pour les énergies des niveaux 1S$_h$ et du vide, spectroscopie infrarouge pour les énergies des niveaux 1S$_e$ et 1P$_e$, et modélisation k.p pour l'énergie du niveau 1D$_e$, je peux reconstruire la structure électronique des nanocristaux de HgSe entourés de ligands dodécanethiol, comme présenté sur la Figure 41. Le niveau de Fermi est au-dessus du niveau 1S$_e$ dans le cas des nanocristaux de 4,5 nm ($E_{inter} = 1S_h - 1S_e = 780$ meV), tandis qu'il passe au-dessus du niveau $1D_e$ pour les nanocristaux de 17 nm ($E_{inter} = 460$ meV).

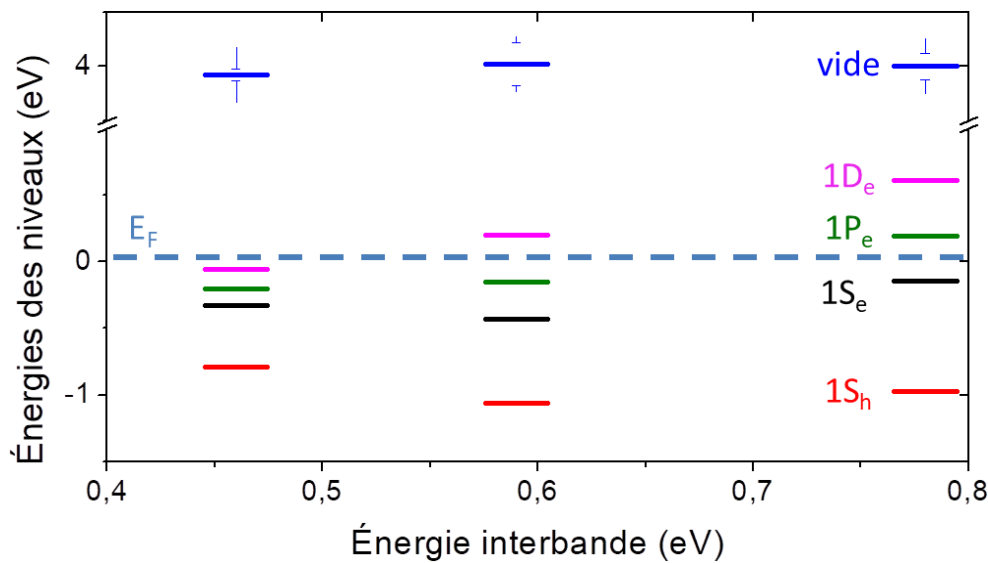

*Figure 41 : Structures électroniques obtenues pour les trois tailles de matériaux. Les niveaux en rouge correspondent au niveau $1S_h$, en noir $1S_e$, en vert $1P_e$, en rose $1D_e$ et en bleu le niveau du vide. Le trait bleu clair en pointillés à 0 eV correspond à l'énergie du niveau de Fermi. Pour les trois tailles étudiées, le ligand utilisé pour mesurer l'énergie du niveau $1S_h$ par rapport au niveau de Fermi est le dodécanéthiol.*

Dans le cas des nanocristaux les moins confinés, l'énergie du niveau de Fermi est supérieure à celle du niveau $1D_e$, les transitions $1S_h - 1S_e$ et $1S_e - 1P_e$ sont normalement interdites. Les valeurs des énergies de transitions que nous avons mesurées grâce à la Figure 39 sont donc moins précises que pour les nanocristaux les plus confinés.

Ces mesures ont été réalisées avec différents ligands : $S^{2-}$ mais aussi l'éthanedithiol (EDT) et le sulfure d'arsenic ($As_2S_3$). Je les ai regroupés dans un digramme de phase représentant le dopage (représenté par l'énergie du niveau de Fermi) en fonction de l'énergie de confinement. Ce diagramme de phase est présenté sur la Figure 42. Plusieurs zones se distinguent sur ce diagramme :

- si le niveau de Fermi est situé dans la moitié inférieure de la bande interdite ($1S_h < E_F < E_G/2$), le matériau est dopé p ;
- s'il est situé dans la moitié supérieure de la bande interdite ($E_G/2 < E_F < 1S_e$), le matériau est dopé n ;
- s'il est entre les niveaux $1S_e$ et $1P_e$, 2 électrons sont présents dans le niveau $1S_e$ ;
- s'il est entre les niveaux $1P_e$ et $1D_e$, 6 électrons présents dans le niveau $1P_e$ s'ajoutent aux 2 électrons du niveau $1S_e$ : 8 électrons par nanocristal ;
- s'il est au-dessus du niveau $1D_e$, 10 électrons du niveau $1D_e$ s'ajoutent aux 8 déjà présents dans les niveaux inférieurs : 18 électrons par nanocristal.

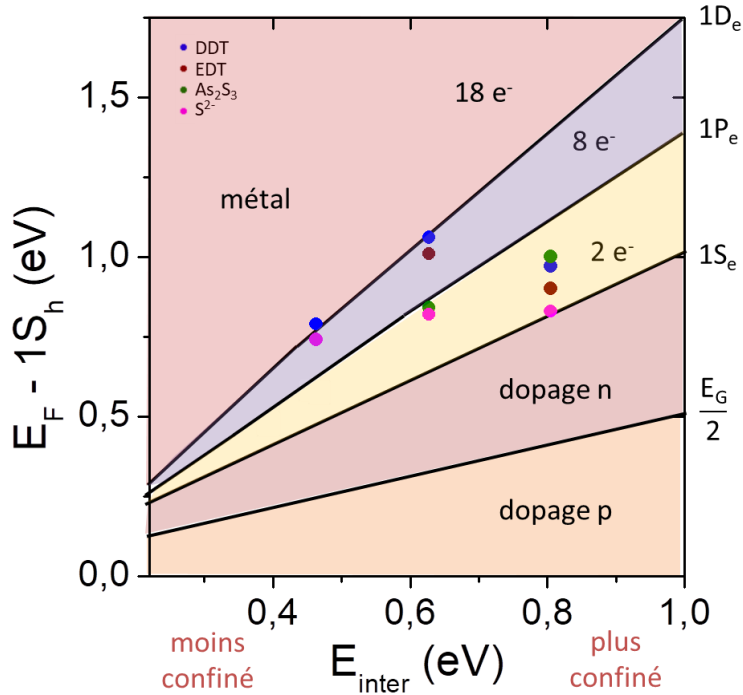

*Figure 42 : Diagramme de phases de nanocristaux de HgSe. L'axe des abscisses correspond à l'énergie de la transition interbande $1S_h - 1S_e$, l'axe des ordonnées correspond à l'écart en énergie entre le niveau $1S_h$ et le niveau de Fermi.*

Ce diagramme de phases nous indique que lorsque le confinement quantique diminue, le nombre d'électrons par nanocristal augmente : il passe de 2 pour les nanocristaux les plus confinés à 18 pour les nanocristaux les moins confinés. **Les nanocristaux acquièrent donc un caractère de plus en plus métallique quand le confinement quantique diminue.** Des transitions semiconducteur-métal ont déjà été étudiées dans la littérature (*52, 53*), dans les nanocristaux de silicium dopés par exemple (*54*). Dans cet exemple, c'était le niveau de dopage qui était modulé. En revanche, dans le cas des nanocristaux de séléniure de mercure, c'est le confinement quantique qui est modifié et qui donne lieu à la transition semiconducteur-métal.

### 3. Étude de la transition semiconducteur-métal dans les nanocristaux de séléniure de mercure en fonction du confinement quantique

Deux techniques complémentaires nous permettent de valider l'observation de la transition semiconducteur-métal avec le confinement quantique. La première est l'absorption infrarouge, qui nous permettra d'étudier le caractère intrabande ou plasmonique des transitions. La seconde est une étude du transport qui montre une dépendance différente de la conductivité en fonction de la température pour des nanocristaux de semiconducteurs et des nanocristaux métalliques.

### a. Étude de la transition semi-conducteur métal par absorption infrarouge

Dans le cas où les nanocristaux sont faiblement dopés (1 à 2 électron(s) par nanocristal), ces derniers peuvent être décrits par la physique des semiconducteurs que j'ai décrite dans le chapitre 1 et dans le début de ce chapitre 2. En revanche, lorsque le niveau de dopage est plus important, des comportements collectifs tels que des plasmons peuvent apparaître. C'est ce qui est observé sur des nanocristaux d'oxydes dopés tels que $In_2O_3$ dopés par Sn (ITO) (*55*) ou ZnO dopés par Al (AZO) (*56*) par exemple. La transition intrabande que nous observons dans le cas de nanocristaux de HgSe confinés à faibles niveaux de dopage devrait donc changer de nature et devenir plasmonique lorsque les nanocristaux deviennent plus riches en électrons.

En premier lieu, nous pouvons vérifier que la transition observée pour les nanocristaux les plus larges (17 nm de diamètre) à 970 cm$^{-1}$ (Figure 38d et Figure 39d) est cohérente avec un comportement plasmonique. En effet, grâce à la structure électronique reconstruite à la partie précédente, nous pouvons estimer qu'il y'a environ 18 électrons par nanocristal : un film de ces nanocristaux a donc une densité de porteurs de l'ordre de $4,5 \times 10^{18}$ cm$^{-3}$. Si la transition observée est plasmonique, il est aisé de remonter à la densité de porteurs en utilisant les formules (2.13) :

$$w_{pic} = \sqrt{\frac{w_p^2}{2\varepsilon_{env} + \varepsilon_\infty}} \ avec \ w_p = \sqrt{\frac{ne^2}{\varepsilon_0 \varepsilon_r m_e^*}} \qquad (2.13)$$

Où $w_{pic}$ est la fréquence du pic observé par spectroscopie infrarouge, $\varepsilon_{env}$ est la constante diélectrique de l'environnement des cristaux, $\varepsilon_\infty$ est la constante diélectrique du nanocristal à hautes fréquences, n est la densité de porteurs et $m_e^*$ est la masse effective des électrons, qui vaut 0,05m$_0$ pour HgSe. Appliqué à la transition à 970 cm$^{-1}$ observée pour nanocristaux de séléniure de mercure de 17 nm (Figure 38d et Figure 39d), ce modèle donne une valeur de densité de porteurs de $6,3 \times 10^{18}$ cm$^{-3}$, du même ordre de grandeur que ce qui peut être estimé en déterminant le niveau de dopage via la structure électronique. La transition observée est donc cohérente avec un comportement plasmonique.

Pour sonder la nature intrabande ou plasmonique de cette transition en fonction du confinement quantique, la spectroscopie infrarouge à différentes températures est un bon outil. En effet, il a été observé que la transition intrabande de nanocristaux faiblement dopés de séléniure de mercure se décalait vers les plus hautes énergies (*blueshift*) lorsque la température diminue (*51, 57*), à cause d'interactions électrons-phonons. En revanche, l'énergie de la transition plasmonique ne dépend que de la densité de porteurs dans le film (voir 2.13), qui ne varie quasiment pas avec la température.

Pour réaliser les spectres d'absorption infrarouge de films de nanocristaux à différentes températures, nous avons collaboré avec Ricardo Lobo du laboratoire de Physique et d'Étude des Matériaux, LPEM, de l'ESPCI. Les spectres d'absorption infrarouge entre 300 et 4 K pour les nanocristaux de séléniure de mercure les plus confinés (4,5 nm de diamètre) sont présentés sur la Figure 43a. La transition interbande, vers 5500 cm$^{-1}$, se décale vers les plus basses énergies quand la température diminue : ce comportement, à l'opposé de ce qui est observé pour la plupart des semiconducteurs qui voient leur énergie interbande augmentée à basse température, est typique des chalcogénures de mercure (*50*). Le pic intrabande/plasmonique vers 2800 cm$^{-1}$ se décale quant à lui vers les plus hautes énergies lorsque la température diminue. Ce déplacement est quantifié sur la Figure 43b et vaut 200 cm$^{-1}$ (entre 300 et 4 K). Pour les nanocristaux moins confinés, on observe également un déplacement vers les plus grandes énergies du pic intrabande/plasmonique lorsque la température est diminuée, mais l'amplitude de ce déplacement décroit pour les confinements quantiques plus faibles (voir Figure 43b). Ces observations semblent confirmer une transition intrabande pour les nanocristaux confinés, et une transition plus plasmonique pour les nanocristaux moins confinés. Autrement dit, on observe un comportement type semiconducteur pour les nanocristaux confinés, et un comportement métallique pour les nanocristaux moins confinés.

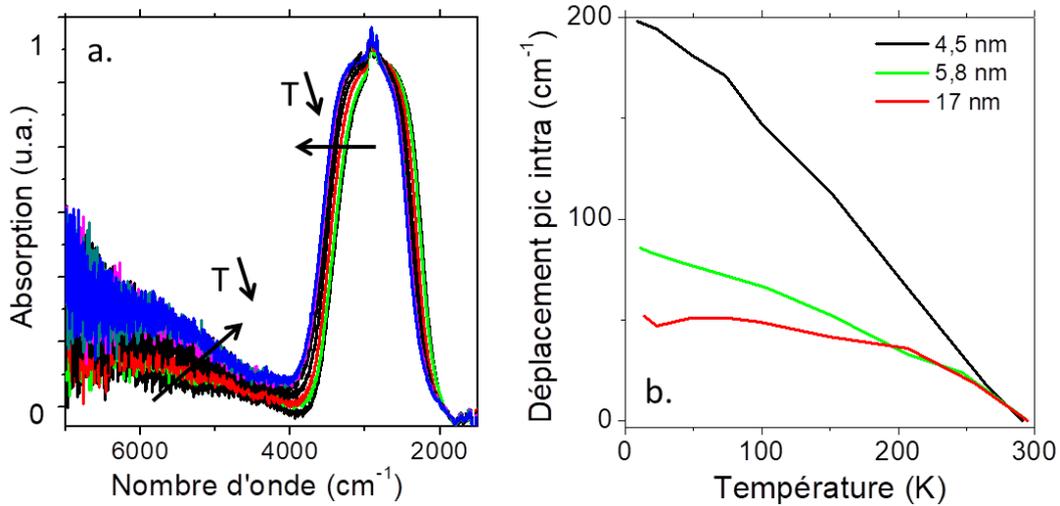

Figure 43 : (a) Spectre d'absorption infrarouge d'un film de nanocristaux de HgSe (4,5 nm de diamètre, fort confinement) à différentes températures, entre 300 K et 4 K. Le pic vers 2800 cm$^{-1}$ se déplace vers les plus grandes énergies lorsque la température diminue. (b) Déplacement du pic intrabande/plasmonique vers les plus grandes énergies en fonction de la température pour différentes tailles de nanocristaux : 4,5 nm en noir, 5,8 nm en vert et 17 nm en rouge.

L'évolution continue de la nature de la transition, d'intrabande à plasmonique, sur des nanocristaux confinés a été étudiée sur d'autres matériaux dans la littérature comme ZnO (*58*), HgS et HgS/CdS (*59*)). Les auteurs observent une évolution entre une transition intrabande pour des nanocristaux faiblement dopés (1 à 2 électron(s) par nanocristal) confinés à une transition plasmonique pour des nanocristaux moins confinés et plus dopés.

Il est important de noter que, grâce à leur faible densité de porteurs, plus faible que celles des oxydes dopés, les nanocristaux peu confinés de semimétaux permettent d'accéder à des transitions plasmoniques à de grandes longueurs d'onde, dans le moyen voire dans le lointain infrarouge, comme présenté sur la Figure 44.

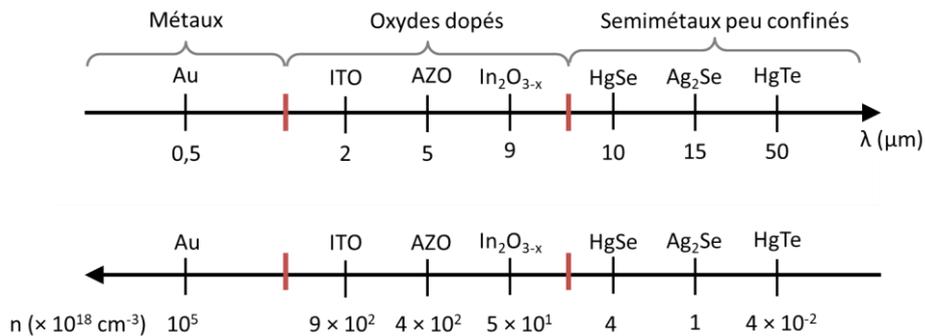

Figure 44 : Longueurs d'onde de transitions plasmoniques observées sur des nanocristaux de métaux, d'oxydes dopés ou de semimétaux peu confinés, et densités volumiques de porteurs (n) correspondantes.

### b. Étude de la transition semi-conducteur métal par mesure de transport à basse température

Le transport par saut tunnel entre nanoparticules dépend fortement de leur nature, semiconducteur ou métal. À des températures proches de la température ambiante, le transport se fait par saut tunnel d'un nanocristal à son voisin le plus proche, pour que la barrière tunnel à franchir soit la plus fine possible. On parle alors de conduction par voisin le plus proche, ou *Nearest Neighbour Hopping* en anglais, et la dépendance du courant avec la température suit une loi de type Arrhénius. À plus basses températures, il est possible d'observer des comportements sur des distances plus longues, on parle alors de sauts tunnels à distance variable, ou *Variable Range Hopping* (VRH) en anglais.

Dans le cas du VRH, la nature du nanocristal joue un rôle important. Ainsi, entre nanocristaux de semiconducteurs, le paramètre limitant le transport est la disponibilité d'états à l'arrivée du saut, c'est-à-dire la densité d'états. Dans les métaux, les niveaux sont très proches les uns des autres, le paramètre limitant est alors l'énergie de chargement (ou *charging energy*) : les nanoparticules métalliques étant déjà riches en électrons, l'énergie de chargement est l'énergie à payer pour ajouter un électron supplémentaire à la nanoparticule (Figure 45).

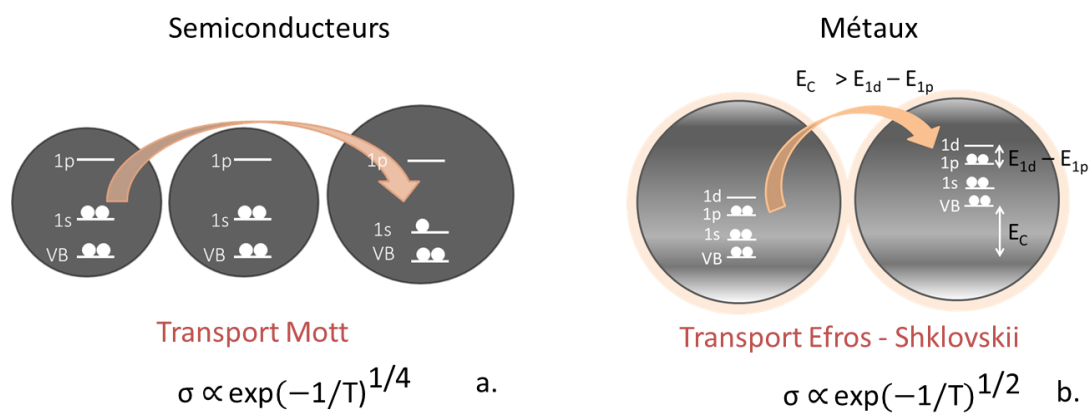

*Figure 45 : Schéma présentant le Variable Range Hopping (VRH) entre nanoparticules de semiconducteurs (transport limité par la densité d'états) (a) et entre nanoparticules métalliques (transport limité par l'énergie de chargement) (b).*

Ces deux types de comportements, décrits dans la littérature (*60, 61*), présentent des dépendances en température différentes. En effet, si le transport est limité par la densité d'états, on parle de régime de Mott et la conductivité est proportionnelle à $\exp(T)^{-0,25}$ où T est la température ; si le transport est limité par l'énergie de chargement, on parle de transport d'Efros-Shklovskii et la conductivité est proportionnelle à $\exp(T)^{-0,5}$, comme présenté sur la Figure 45. La dépendance de la résistance en fonction de la température est donc mesurée sur des films de nanocristaux. Ces mesures ont été effectuées par nos collaborateurs de l'Institut de Physique et de Chimie des Matériaux de Strasbourg, IPCMS : Louis Donald Notemgnou Mouafo et Jean-François Dayen. Les résultats obtenus pour les nanocristaux de 4,5 nm (2 électrons par nanocristal) et ceux de 17 nm (18 électrons par nanocristal) sont présentés sur la Figure 46.

Expérimentalement, on observe qu'à basse température, la loi d'Arrhénius n'est plus vérifiée (Figure 46a). La dépendance en $T^{-1/4}$ pour les nanoparticules les plus confinées (Figure 46b) confirme le transport de Mott, tandis que celle en $T^{-1/2}$ pour les nanoparticules moins confinées confirme le transport d'Efros-Shklovskii (Figure 46c). Avec la diminution du confinement quantique, ces mesures de transport valident la transition semiconducteur-métal observée par reconstruction de la structure électronique et l'absorption infrarouge à basse température.

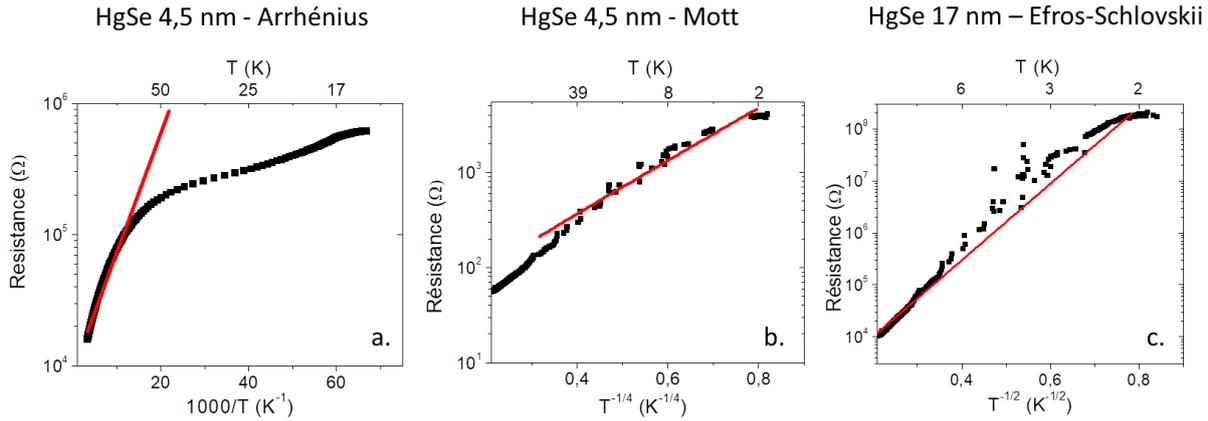

*Figure 46 : Évolution de la résistance ($R_\Omega$) de films de nanocristaux de HgSe en fonction de la température (a) $R_\Omega$ = f(1000/T) (Arrhénius) pour les nanocristaux faiblement dopés (2 électrons par nanocristal, 4,5 nm de diamètre) ; (b) $R_\Omega$ = f($T^{-1/4}$) (Mott) pour les nanocristaux faiblement dopés (2 électrons par nanocristal, 4,5 nm de diamètre) ; (c) $R_\Omega$ = f($T^{-1/2}$) (Efros-Shklovskii) pour les nanocristaux fortement dopés (18 électrons par nanocristal, 17 nm de diamètre).*

On notera qu'à l'échelle du film et non plus de la particule, la transition semiconducteur-métal n'est pas observée : en effet, la résistance diminue quand la température augmente dans les deux cas, ce qui est caractéristique d'un semiconducteur. Les particules étant faiblement couplées entre elles, le transport ne peut se faire que par saut tunnel, ce qui empêche l'apparition d'un comportement métallique à l'échelle du film.

## III. Localisation des pièges dans la bande interdite

Pour donner une image plus complète de la structure électronique, je vais à présent aborder la question des pièges dans la bande interdite. Ces pièges ont une influence sur les propriétés de transport des nanocristaux, en particulier dans les diodes que je présenterai au chapitre 4, il est donc important de déterminer leurs caractéristiques.

Dans cette partie, je commencerai donc par présenter les origines potentielles des pièges et aborderai comment les propriétés de transport au sein d'un film de nanocristaux peuvent être modifiées par la présence de pièges. J'introduirai ensuite la notion d'énergie d'Urbach, qui caractérise la répartition des pièges dans la bande interdite et présenterai le montage expérimental original qui nous permet de la mesurer. Enfin, je conclurai en mesurant la densité de pièges dans la bande interdite induite par différents ligands, afin de déterminer lesquels sont les plus adaptés pour des applications de détection infrarouge.

### 1. Origine des pièges dans la bande interdite

La présence de pièges dans la bande interdite peut avoir plusieurs origines. La première est la polydispersité des nanocristaux. En effet, les nanocristaux les plus larges sont moins confinés et leur énergie de bande interdite est plus faible. Ils peuvent donc jouer le rôle de pièges au sein du film de nanocristaux, comme présenté sur la Figure 47. Sous illumination, ces larges nanocristaux seront rapidement remplis et ne participeront plus au transport, en particulier à basse température.

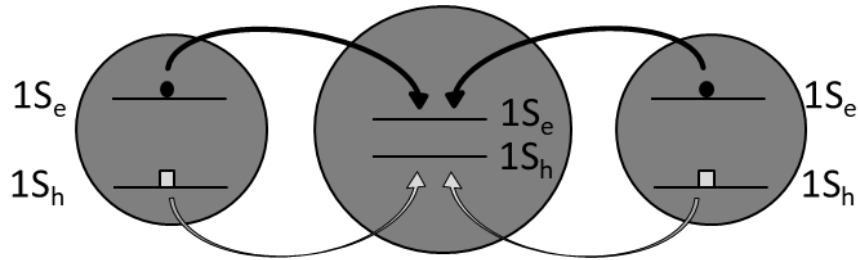

*Figure 47 : Schéma présentant le rôle des nanocristaux plus larges, donc moins confinés, dans le piégeage des porteurs.*

Les atomes de surface ayant une coordination incomplète peuvent également introduire des pièges dans la bande interdite. Un des rôles essentiels des ligands est de s'hybrider avec les atomes de mercure en excès pour augmenter l'énergie des états de pièges au-dessus du niveau $1S_e$. À la fin de cette partie, j'étudierai donc la distribution des pièges dans la bande interdite de nanocristaux de tellure de mercure pour différents ligands afin de déterminer lesquels sont les plus efficaces pour passiver ces états de piège (p73).

### 2. Influence des pièges sur les propriétés de transport d'un film de nanocristaux

La présence de pièges dans la bande interdite vient perturber le transport des charges dans un film de nanocristaux de plusieurs manières : ils abaissent l'énergie de la transition interbande, augmentent le nombre de porteurs thermiquement activés, et augmentent le temps de vie des porteurs.

### a. Influence des pièges sur l'énergie de bande interdite

Les pièges affectent la valeur de la transition interbande, puisque la plus faible énergie accessible devient alors l'énergie entre les niveaux de pièges et non plus l'énergie entre les niveaux $1S_h$ et $1S_e$ (Figure 48).

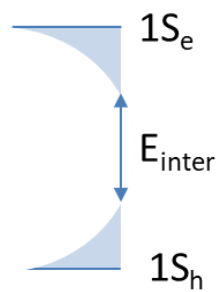

*Figure 48 : Rôle des pièges sur l'énergie de transition interbande*

En plus du décalage de la transition interbande vers les basses énergies, les pièges viennent modifier la distribution de la densité d'états autour de la bande interdite, qui devient moins abrupte. En l'occurrence, sur les photodiodes à base de nanocristaux, la tension de circuit ouvert (*Open Circuit Voltage*, $V_{OC}$) des diodes à base de nanocristaux, caractéristique de la bonne séparation des porteurs, est très affectée par la présence de pièges (*62, 63*).

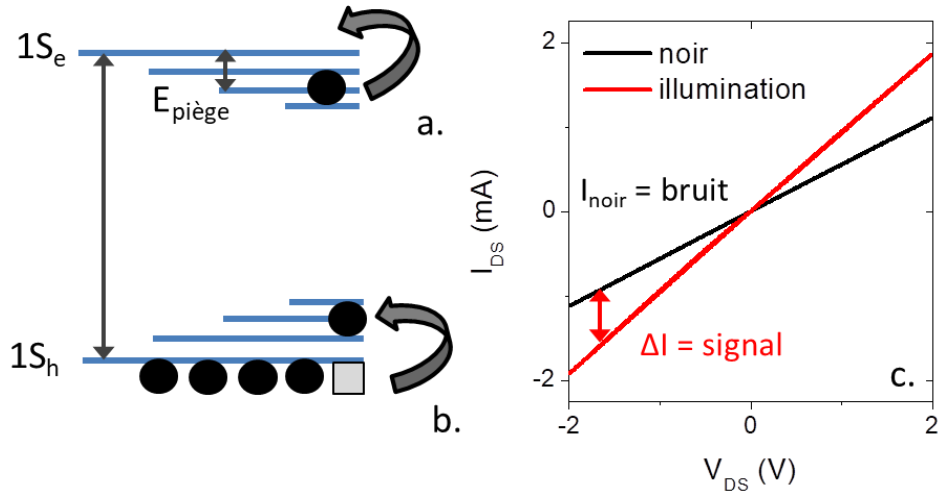

*Figure 49 : (a) Schéma représentant un électron (cercle noir) thermiquement activé par la présence de pièges sous le niveau 1S$_e$ ; (b) trou (carré gris) thermiquement activé par la présence de pièges au-dessus de niveau 1S$_h$ ; (c) courbes courant-tension d'un détecteur à base de nanocristaux de HgTe dans l'obscurité (noir) et sous illumination (rouge).*

De plus, selon le même mécanisme que celui décrit dans le chapitre 1 (p19), les pièges dans la bande interdite entraînent une diminution de l'énergie nécessaire pour introduire un électron dans le niveau 1S$_e$ (Figure 49a) ou un trou dans le niveau 1S$_h$ (Figure 49b). Si des pièges sont présents, la conductivité dans le noir est augmentée : la barrière énergétique à franchir (ou énergie d'activation) ne vaut plus 1S$_e$ − 1S$_h$ mais E$_{piège}$ (Figure 49a). Or, le courant d'obscurité induit du bruit : il vient s'ajouter au courant généré par l'illumination que nous voulons mesurer pour faire fonctionner le détecteur infrarouge (voir Figure 49c).

### b. Influence des pièges sur la recombinaison

Dans un film de nanocristaux photoconducteur, lorsqu'une paire électron-trou est générée, les porteurs peuvent soit se recombiner, soit rejoindre les électrodes par sauts tunnel successifs (dans le cas où une tension est appliquée). Dans le cas de la détection, c'est l'extraction des porteurs aux électrodes qui doit être favorisée.

Si les porteurs majoritaires sont les électrons (le niveau de Fermi des nanocristaux se situe dans la moitié supérieure de la bande interdite), le courant est détecté quand l'électron arrive à rejoindre l'électrode chargée positivement, le drain (D). Dès qu'un électron est collecté au niveau du drain, la source injecte un électron dans le film pour conserver la neutralité électrique. S'il existe des pièges près du niveau 1S$_h$, les trous (les porteurs minoritaires) ne peuvent pas se recombiner et s'accumulent dans le film. Tant que les trous sont piégés, les électrons peuvent circuler dans le canal. On définit le gain de photoconduction G selon l'expression :

$$G = \frac{\tau}{\tau_{transit}} \; avec \; \tau_{transit} = \frac{L}{\mu F} \qquad (2.14)$$

Où τ est le temps de vie du porteur minoritaire (dans notre exemple le trou), L est la distance entre les électrodes, μ est la mobilité des électrons et F est le champ électrique appliqué. Si le temps de transit, c'est-à-dire le temps nécessaire pour qu'un électron traverse le canal est inférieure au temps de vie du trou, le gain de photoconduction est supérieur à un. Cela implique que la génération d'une paire électron-trou entraîne la collecte de plusieurs électrons et augmente donc le signal mesuré.

En revanche, plus le temps de vie du porteur minoritaire est long, plus le système mettra du temps à revenir à l'état *off*, c'est-à-dire à retrouver la conductivité qu'il avait dans le noir. Les porteurs générés sous illumination doivent se recombiner pour que le courant diminue, et cette recombinaison est limitée par le temps de piégeage. Les pièges peuvent donc permettre d'augmenter le signal reçu sous illumination mais augmentent aussi le temps de réponse du dispositif.

### 3. Étude de la densité de pièges dans la bande interdite sur des nanocristaux de tellure de mercure

Dans cette étude, je me concentrerai sur les nanocristaux de tellure de mercure ayant une énergie de transition interbande de 4000 cm$^{-1}$ que je réutiliserai dans le chapitre 4. Ces matériaux sont ambipolaires, c'est-à-dire qu'ils conduisent aussi bien les électrons que les trous. Par conséquent, nous ne pourrons pas distinguer les pièges près de la bande de valence de ceux situés près de la bande de conduction.

#### a. Énergie d'Urbach

La densité de pièges dans la bande interdite est souvent modélisée par une queue d'Urbach décroissant exponentiellement dans la bande interdite comme présenté sur la Figure 50a. La largeur caractéristique de cette distribution de pièges est appelée énergie d'Urbach ($E_u$).

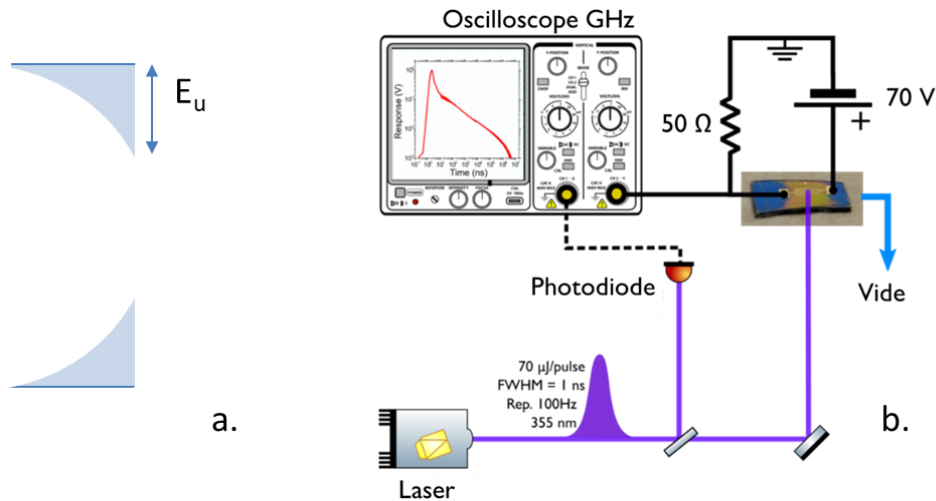

*Figure 50 : (a) Schéma représentant les densités de piège décroissant exponentiellement dans la bande interdite. La largeur caractéristique de ces queues d'Urbach est appelée énergie d'Urbach et est notée $E_u$. Dans le cas de HgTe ambipolaire, nous supposerons que la dispersion des pièges est la même près de la bande de conduction que près de la bande de valence ; (b) montage expérimental permettant de mesurer l'énergie d'Urbach.*

Il est possible de déterminer cette énergie d'Urbach optiquement en illuminant un film de nanocristaux avec une énergie inférieure à celle de la bande interdite et d'observer l'absorption due aux pièges uniquement.(*64*) Cette méthode présente toutefois l'inconvénient de mesurer des signaux de très faible amplitude.

L'énergie d'Urbach peut également se mesurer de manière indirecte en étudiant la réponse d'un film de nanocristaux à une impulsion lumineuse. On parle alors de mesure de **photocourant transitoire**. Sous illumination, l'intensité du courant augmente. À la fin de l'impulsion lumineuse, le photocourant décroît et cette décroissance peut se faire en plusieurs phases. Dans le cas des nanocristaux de HgTe, on observe d'abord une décroissance rapide exponentielle, pendant une phase durant une à deux nanoseconde(s) (voir Figure 51a). La durée de cette phase est du même ordre de grandeur que le temps de *hopping* ($\tau_{hop}$), c'est-à-dire le temps caractéristique pour qu'un porteur effectue un saut tunnel entre deux nanocristaux. Ce temps de *hopping* se calcule grâce à la formule :

$$\tau_{hop} = \frac{2eR^2}{3\mu k_b T} \qquad (2.15)$$

Dans le cas de nanocristaux de HgTe, le temps de *hopping* est de l'ordre de 2 à 3 ns. La diminution du photocourant pendant cette première phase est donc nécessairement due à un processus intra-nanocristal. Il s'agit de la recombinaison des porteurs générés par l'illumination au sein d'un nanocristal, et au piégeage de certains dans les niveaux de pièges dans la bande interdite (Figure 51b).

La décroissance du photocourant dans HgTe suit ensuite une loi de puissance pendant un temps beaucoup plus long, pouvant aller jusqu'à la milliseconde. Pendant cette phase, les porteurs restants entrent dans un régime de piégeage – dé-piégeage. Par conséquent, leur mobilité diminue et donc le photocourant également (*65–67*). Le photocourant sera d'autant plus important que le temps de vie radiatif sera important, et on pourra conserver une dynamique rapide si la queue d'Urbach n'apparaît que tardivement, c'est-à-dire une fois que le photocourant est redevenu suffisamment faible (< 10 % du photocourant maximal).

L'exposant de la loi de puissance est directement lié à l'énergie d'Urbach par la relation (*65*) :

$$E_u = \frac{k_B T}{b+1} \qquad (2.16)$$

où $E_u$ est l'énergie d'Urbach et b est l'exposant de la loi de puissance. Si b = -0,27, comme sur la Figure 51a, alors $E_u$ = 34 meV.

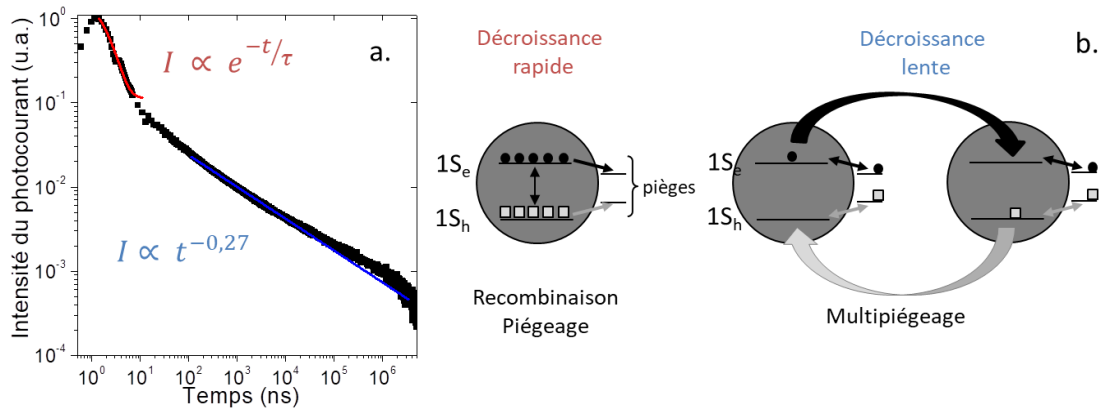

*Figure 51 : (a) Exemple de réponse sous impulsion lumineuse d'un film de nanocristaux de tellure de mercure, ligands EDT ($E_G$ = 4000 cm$^{-1}$). La source est un laser 355 nm, pulsé à 100 Hz, dont les impulsions font 1 ns. (b) schéma présentant les différents mécanismes de décroissance du courant après l'impulsion. Les disques noirs représentent des électrons, les carrés gris représentent des trous.*

Pour réaliser cette mesure expérimentalement, une tension drain source de 70 V est appliquée sur l'échantillon. Cette tension élevée permet d'augmenter la valeur du photocourant et donc d'éviter que la mesure soit limitée par la résolution de l'oscilloscope, et n'introduit pas de phénomènes non désirés dans le film de nanocristaux pendant la durée de la mesure. La mesure du photocourant après l'illumination se fait en utilisant un système à large bande passante (GHz) afin de décrire la réponse sur une gamme allant de la nanoseconde à la milliseconde. Enfin, pour minimiser les réflexions sur le substrat qui induisent des réponses décalées dans le temps, des électrodes d'ITO sur polyéthylène téréphtalate (PET) sont utilisées. Les détails concernant la fabrication et le *design* de ces électrodes sont fournis dans l'Annexe 2.

### b. Influence de la chimie de surface sur les pièges

À cause de leur coordination insuffisante, les atomes à la surface des nanocristaux peuvent générer des niveaux de piège dans la bande interdite. Les ligands ont pour rôle de passiver ces états de surface pour éviter ces pièges. Dans cette partie, je vais donc étudier l'influence de différents ligands sur l'énergie d'Urbach de nanocristaux de tellure de mercure.

### i. Ligands utilisés

Dans cette partie, j'étudie les principaux ligands utilisés pour réaliser du transport dans les films de nanocristaux colloïdaux (*32, 68–70*). Parmi eux se trouvent :

- des sulfures : $S^{2-}$, $As_2S_3$, $SCN^-$ ;
- des longs thiols : butanethiol (BuSH) et octanethiol (OSH) ;
- des dithiols courts : éthanedithiol (EDT), 1,4-benzenedithiol (BeSH) ;
- des halogénures : $Cl^-$.

Le soufre a une grande affinité avec le mercure, tandis que les halogénures sont de plus en plus utilisés dans la littérature sur les nanocristaux pour passiver les pièges et augmenter la conductivité.

Pour tous les ligands étudiés, l'échange de ligands sera réalisé sur film, selon la procédure décrite dans le chapitre 1 (p42). Pour $S^{2-}$, un échange de ligands en solution est également effectué. Les nanocristaux échangés sont dispersés dans la N-méthylformamide (NMF) et déposés par dépôt de goutte sur les électrodes. Pour que le solvant s'évapore, le film est chauffé à 100 °C sur une plaque chauffante pendant quelques minutes.

### ii. Mesure des énergies d'Urbach

Les énergies d'Urbach mesurées pour les différentes chimies de surface sont présentées sur la Figure 52. Toutes les mesures ont été effectuées à température ambiante, l'énergie d'activation thermique des porteurs $k_B T$ vaut donc 25 meV. Pour tous les ligands étudiés, les énergies d'Urbach sont légèrement supérieures à $k_B T$, dans la gamme [35 – 50] meV.

Les thiols longs (octanethiol et butanethiol) passivent mal les pièges dans la bande interdite : ils présentent des énergies d'Urbach supérieures à 40 meV. Ces ligands étant longs, leur diffusion au sein du film de nanocristal pendant l'échange de ligands est assez lente. Par conséquent, ils couvrent moins bien la surface que des ligands plus petits.

Le film réalisé par échange de ligands $S^{2-}$ en solution donne également une énergie d'Urbach élevée (50 meV). La technique de dépôt de la solution échangée est probablement à l'origine de l'apparition des pièges. En effet, les nanocristaux de chalcogénures de mercure sont très sensibles au chauffage. Les exposer à une température de 100 °C pendant plusieurs minutes (alors que leur synthèse se fait à 120 °C pendant trois minutes) entraîne nécessairement l'apparition de défauts, comme la percolation de nanocristaux entre eux. Les nanocristaux sont plus polydisperses, et les plus larges induisent des pièges dans le film, comme présenté en début de partie (p68).

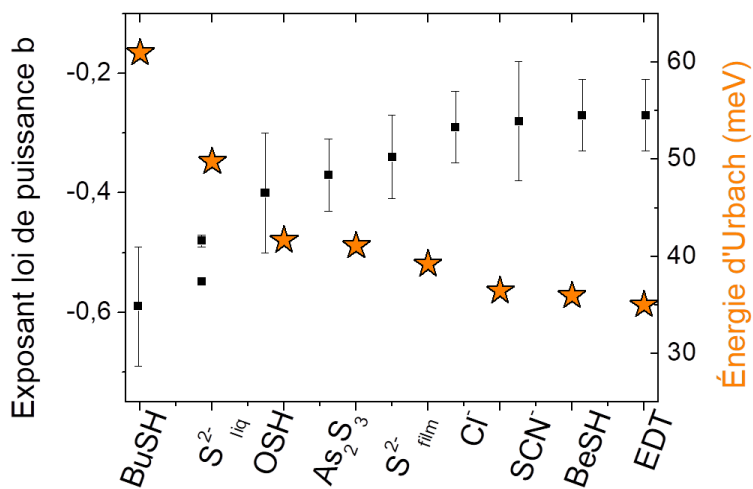

*Figure 52 : Détermination des énergies d'Urbach pour différents ligands sur des nanocristaux de tellure de mercure HgTe 4000 cm⁻¹.*

Les ligands sulfures, petits et complètement inorganiques, sont les plus intéressants pour maximiser la mobilité des porteurs au sein des films de nanocristaux (*42, 45*). Cependant, ils ne passivent pas suffisamment les pièges de la bande interdite. Au contraire, les dithiols courts EDT et BeSH qui ont une mobilité inférieure, sont les ligands qui passivent au mieux les pièges avec une énergie d'Urbach autour de 35 meV. Pour réaliser un dispositif à base de nanocristaux, un compromis doit être trouvé entre mobilité et passivation des pièges. Notamment, dans le cas de dispositifs à base de diodes, la passivation des pièges devra être privilégiée.

## IV. Conclusions et enjeux futurs

Dans ce chapitre, j'ai démontré qu'il était important de connaître la structure électronique des nanocristaux de chalcogénures de mercure afin de concevoir les dispositifs de détection infrarouge les plus performants. En combinant mesures de transport (transistors à effet de champs), spectroscopie optique, photoémission et modélisation k.p, j'ai pu obtenir avec précision les alignements de bande et les niveaux de dopage de nanocristaux de tellure de mercure et de séléniure de mercure.

Dans le cas des nanocristaux de séléniure de mercure, une transition semiconducteur-métal a été mise en évidence via la reconstruction de la structure électronique et a été confirmée par des mesures complémentaires d'absorption infrarouge et de transport à différentes températures. Dans le cas de HgSe, cette transition semiconducteur-métal ne se fait pas en injectant des porteurs, comme c'est souvent le cas dans la littérature, mais en modifiant le confinement quantique.

Enfin, j'ai montré que la détermination des énergies des niveaux électroniques pouvait être complétée par la mesure de la distribution de pièges dans la bande interdite. Pour cela, j'ai utilisé une méthode indirecte de mesure de l'énergie d'Urbach via la mesure de photocourant transitoire sur les films de nanocristaux colloïdaux. J'ai ainsi pu montrer que les dithiols courts permettaient la meilleure passivation des pièges.

Ces résultats seront particulièrement utiles pour la conception de dispositifs infrarouges performants. Ils seront entre autres utilisés dans le chapitre 4, lorsque je présenterai une diode fonctionnant dans le SWIR étendu (2,5 µm).

# CHAPITRE 3
# Contrôle du dopage dans les nanocristaux colloïdaux

**Publications associées à ces travaux :**

- B. Martinez *et al.*, HgSe self-doped nanocrystals as a platform to investigate the effects of vanishing confinement, *ACS Applied Materials and Interfaces*, **9**, 41, 36173-36180 (2017)
- B. Martinez *et al.*, Polyoxometalate as control agent for the doping in HgSe self-doped nanocrystals, *the Journal of Physical Chemistry C*, **122**, 46, 26680-26685 (2018)

**Mots clés :** dopage, niveau de Fermi, dipôles de surface, travail de sortie, oxydo-réduction, polyoxométallates

**Techniques expérimentales utilisées :**

- Spectroscopie infrarouge
- Photoémission
- Transistors à effet de champ

Dans le chapitre 1, j'ai montré que modifier la taille, donc le confinement quantique, permettait d'ajuster finement la transition interbande dans un nanocristal et donc, de choisir la gamme d'énergie absorbée. Pour obtenir des transitions dans le LWIR, les nanocristaux de tellure de mercure doivent être larges (leur diamètre doit être supérieur à une vingtaine de nanomètres). Leur croissance est donc plus difficile à contrôler (la dispersion en taille s'élargit) et leur stabilité colloïdale se détériore.

Un autre moyen de modifier la gamme d'énergie absorbée est le contrôle du dopage, donc de l'énergie de Fermi des nanocristaux. Par exemple, comme nous l'avons vu dans le chapitre 2 pour les nanocristaux de séléniure de mercure, lorsque le niveau de Fermi est au-dessus du niveau $1S_e$, la transition interbande est interdite et c'est la transition intrabande $1S_e - 1P_e$ qui devient autorisée.

Dans ce chapitre, je vais présenter différentes stratégies pour contrôler le niveau de dopage. Certaines approches peuvent être implémentées pendant la synthèse : il s'agit de l'ajout de dopants (*71*) ou du contrôle de la taille. D'autres méthodes peuvent s'appliquer après la synthèse des nanocristaux : il peut s'agir de méthodes physiques pour injecter des électrons dans les cristaux, ou de méthodes chimiques basées sur les ligands. Ces ligands jouent le rôle de dipôles et induisent un champ électrique qui diminue le dopage des nanocristaux. Dans ce chapitre, je mettrai en évidence la modification du remplissage des niveaux électroniques de HgSe en faisant varier la force du dipôle à la surface.

Je montrerai également que les ligands peuvent jouer le rôle d'oxydant ou de réducteur et donc injecter ou capter les électrons présents dans les nanocristaux. J'ai notamment utilisé les polyoxométallates (ou POMs) qui ont un fort pouvoir oxydant et captent les électrons, ce qui diminue l'énergie de Fermi des nanocristaux. Dans ce chapitre, je décrirai une série d'expériences qui démontrent le transfert d'électrons des nanocristaux vers les POMs et quantifierai la modification de dopage. Je montrerai que celle-ci peut être particulièrement efficace et modifier le niveau de dopage sur deux ordres de grandeur.

## I. Contrôle du dopage des nanocristaux pendant la synthèse

Dans cette partie, je présenterai plusieurs stratégies permettant de contrôler le niveau de dopage pendant la synthèse, en particulier l'ajout de dopants et la synthèse de nanocristaux non-stœchiométriques. Dans le cas des nanocristaux de semimétaux confinés, comme les chalcogénures de mercure que j'utilise pendant ma thèse, la taille des cristaux peut également avoir une influence sur le niveau de dopage. En effet, comme je l'ai montré dans le chapitre 2, en changeant la taille des cristaux, le dopage des nanocristaux de HgSe peut passer de 2 à 18 électrons par nanocristal.

### 1. Introduction de dopants

Pour contrôler le niveau de dopage de nanocristaux colloïdaux, une première approche peut être d'introduire des dopants. Lors des vingt dernières années, cette approche a pourtant été difficile à mettre en place. Des processus d'auto-purification des nanocristaux (*34*) pendant la synthèse entraînent l'exclusion des défauts, donc des dopants. Toutefois, des progrès pendant la synthèse des nanocristaux ont été réalisés et ont notamment permis l'intégration de dopants Mn dans des nanocristaux de CdSe afin de conférer des propriétés magnétiques (*72*). Pour modifier les propriétés optiques, l'intégration de dopants Cu ou Ag dans des nanocristaux de CdSe a également été démontrée (*71*, *73*, *74*).

Enfin, dans les nanocristaux d'oxydes, l'introduction d'impuretés non-isovalentes permet d'obtenir des charges libres, électrons ou trous et induit un dopage, comme c'est le cas dans les matériaux massifs (voir Chapitre 1, dopage des semiconducteurs, p19). Pour des oxydes dopés comme l'oxyde d'indium dopé avec de l'étain (ITO) ou l'oxyde de zinc dopé avec du gallium (GZO), l'influence de la concentration en dopants sur le nombre de charges libres dans les nanocristaux a été étudiée dans la littérature et quelques résultats sont présentés sur le Tableau 5 (*55, 75*).

*Tableau 5 : Évolution du nombre d'électrons par nanocristal (Nb e⁻ / nanocristal) en fonction de la concentration en dopants pour des nanocristaux d'ITO et de GZO.*

| Matériau | Concentration dopant | Nb e⁻ / nanocristal |
|---|---|---|
| ITO (dopage par Sn) | 1,7 % | 200 |
| | 5 % | 600 |
| | 10 % | 900 |
| GZO (dopage par Ga) | 3 % | 160 |
| | 6 % | 460 |
| | 9 % | 640 |

Les niveaux de dopage présentés dans le Tableau 5 sont très élevés, > 100 électrons par nanocristal. Intégrés dans des dispositifs de détection infrarouge, ils conduisent donc à des temps de réponse et à des niveaux de bruit trop importants (*55*). Dans le cadre de ma thèse, je cherche donc à limiter le dopage à quelques électrons par nanocristal.

### 2. Synthèse de nanocristaux non-stœchiométriques

Une autre stratégie pourrait être de changer le ratio des réactifs pour introduire une non-stœchiométrie dans les nanocristaux. Dans les nanocristaux d'oxyde d'indium, des vacances d'oxygène peuvent induire un dopage de l'ordre de quelques dizaines d'électrons par particule (*55*). Pour les nanocristaux de chalcogénure de mercure, cette stratégie est difficile à mettre en place car le ratio des réactifs n'a pas d'influence sur la stœchiométrie finale des nanocristaux, qui est toujours légèrement riche en cations (*24*).

### 3. Influence de la taille

Dans le cas de nanocristaux de semimétaux comme ceux que j'utilise pendant ma thèse, la taille des nanocristaux, donc le confinement quantique, influe sur le niveau de dopage. Au chapitre 2, nous avons pu voir l'influence du confinement quantique sur le dopage de HgSe. On retrouve également cette tendance sur les nanocristaux de $Ag_2Se$ (*76*) et HgTe (*11*) (voir Tableau 6).

Les nanocristaux présentés dans le Tableau 6 peuvent donc, en fonction de leur diamètre, présenter des niveaux de dopage plus ou moins importants sans que l'on ait introduit d'atomes dopants pendant la synthèse. Le contrôle du dopage avec la taille des nanocristaux est cependant difficile à mettre en place, car les particules les plus larges ne sont pas stables colloïdalement et sont donc très difficiles à manipuler et à déposer sous forme de film.

*Tableau 6 : Évolution du nombre d'électrons par nanocristal (Nb e⁻ / nanocristal) en fonction de la taille des nanocristaux, donc à différents degrés de confinement quantique.*

| Matériau | Diamètre (nm) | Nb e⁻ / nanocristal |
|----------|:-------------:|:-------------------:|
| HgSe | 4,5 | 1 - 2 |
| | 17 | 10 - 18 |
| Ag₂Se | 5 | 1 - 2 |
| | 27 | 20 - 25 |
| HgTe | 8 | 0,03 |
| | 25 | 1 - 2 |
| | 200 | 150 |

**Origine du dopage dans les nanocristaux de semimétaux**

Lorsque le niveau de Fermi se situe dans la bande interdite, le nombre d'électrons par nanocristal est très inférieur à 1, et les électrons présents dans les niveaux $1S_e$ et supérieurs ont été promus par excitation thermique. Leur nombre baisse donc lorsqu'on diminue la température. C'est le cas par exemple des nanocristaux de HgTe de 8 nm de diamètre présentés sur le Tableau 6.

En revanche, lorsque le nombre d'électrons par cristal est supérieur à 1, le niveau de Fermi est au-dessus du niveau $1S_e$. Dans ce cas, le nombre d'électrons par cristal diminue peu avec la température (c'est le cas de HgSe et Ag₂Se). Ces électrons ne proviennent pas du niveau $1S_h$ par excitation thermique, ce sont des électrons en excès présents dans les niveaux de conduction: les nanocristaux sont **auto-dopés** pendant leur synthèse.

Revenons sur l'origine du dopage dans les nanocristaux auto-dopés en prenant l'exemple de HgSe. À l'état massif , le travail de sortie de HgSe est élevé et vaut 6 eV (*77*). Pour les nanocristaux de HgSe, il est mesuré à 4,5 ± 0,3 eV par photoémission (*22, 78*). La combinaison de (i) un travail de sortie élevé, (ii) un excès de mercure à la surface qui induit un dopage de type n (*42, 49*), (iii) une faible énergie de bande interdite peut engendrer la situation présentée dans le rectangle gris sur la Figure 53. Le niveau $1S_e$ passe en dessous du potentiel du couple O₂/H₂O, ce qui conduit à une réduction du nanocristal par l'eau présente dans l'environnement : $2\ QD + H_2O \rightarrow 2\ QD^- + 0,5\ O_2 + 2\ H^+$. **La forme stable du nanocristal devient alors sa forme négativement chargée, ou dopée**. Le niveau $1S_e$ passe sous le niveau de Fermi.

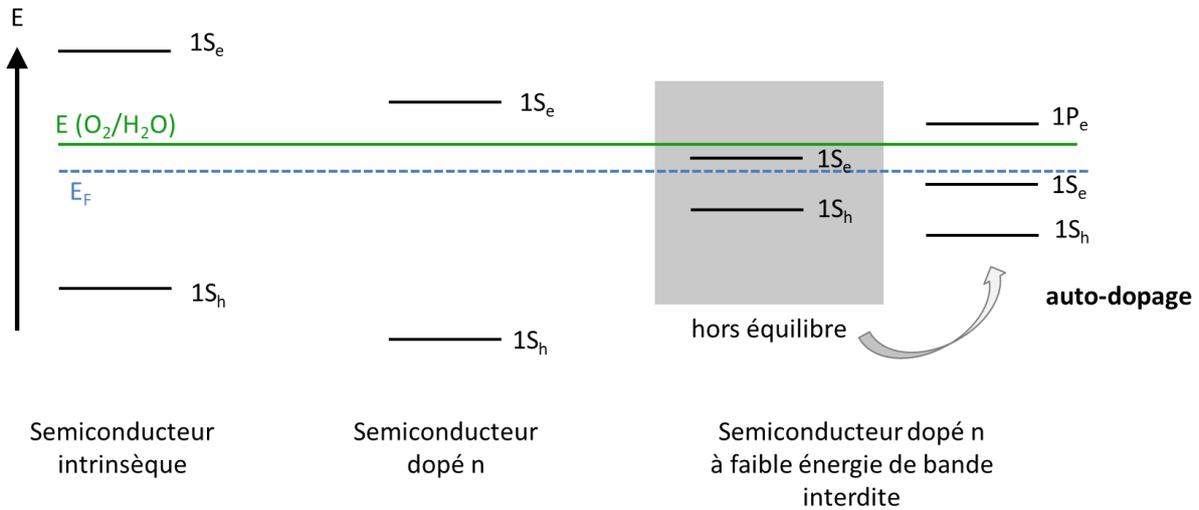

*Figure 53 : Schéma expliquant le mécanisme d'auto-dopage observé sur les nanocristaux de séléniure de mercure.*

## II. Modification du dopage post-synthèse – rôle des ligands

Le dopage des nanocristaux peut également être modifié après la synthèse. Une possibilité est d'utiliser des méthodes dites physiques, comme les transistors à effet de champ. Par effet de grille, l'énergie du niveau de Fermi peut être modifiée par rapport aux niveaux électroniques (voir chapitre 2). Cela permet donc de contrôler le niveau de dopage en fonction de la tension de grille appliquée.

Un autre moyen d'induire un dopage dans un film de nanocristaux est le pompage optique. Grâce à cette technique, le groupe de Philippe Guyot-Sionnest a observé des transitions intrabandes dans des nanocristaux de CdSe (*79*). Pour utiliser cette méthode dans un détecteur infrarouge, le dopage doit être permanent. Il faut donc saturer le niveau $1S_e$ en continu en utilisant un laser puissant (kW/cm²) et focalisé sur le film de nanocristaux.

Pour introduire un dopage permanent, sans utiliser ni tension électrique ni source de lumière en continu, on peut utiliser une approche chimique basée sur les ligands. Comme je l'ai brièvement introduit dans le chapitre 2, la présence de ligands à la surface des nanocristaux peut induire un dipôle, et ce dipôle dépend de la nature du ligand. Ainsi, entre le DDT et les ions sulfures $S^{2-}$, la différence de dipôle est de l'ordre de 1 Debye (*48*). Selon la force de leur dipôle, les ligands attirent plus ou moins les charges positives vers la surface du cristal tandis que les charges négatives restent dans le volume. Il se crée donc un champ électrique au sein du cristal qui vient courber les niveaux. Si le ligand utilisé est le DDT, le champ électrique est quasi inexistant et les niveaux restent plats, si au contraire on utilise des ions sulfures $S^{2-}$, les niveaux seront courbés vers les énergies plus faibles au centre du cristal (voir Figure 54).

Brown *et al.* (*80, 81*) ont travaillé sur les dipôles induits par les ligands à la surface de nanocristaux de sulfure de plomb (PbS). Ils ont montré un décalage du niveau de Fermi par rapport aux niveaux électroniques de l'ordre de 1 eV en changeant l'amplitude du dipôle. Ces résultats sont utilisés pour la construction de jonctions p-n pour les cellules solaires à base de PbS : lorsque les nanocristaux sont entourés par EDT ils présentent un dopage p, tandis qu'ils présentent un dopage n lorsqu'ils sont entourés par $NH_4I$ (*82*).

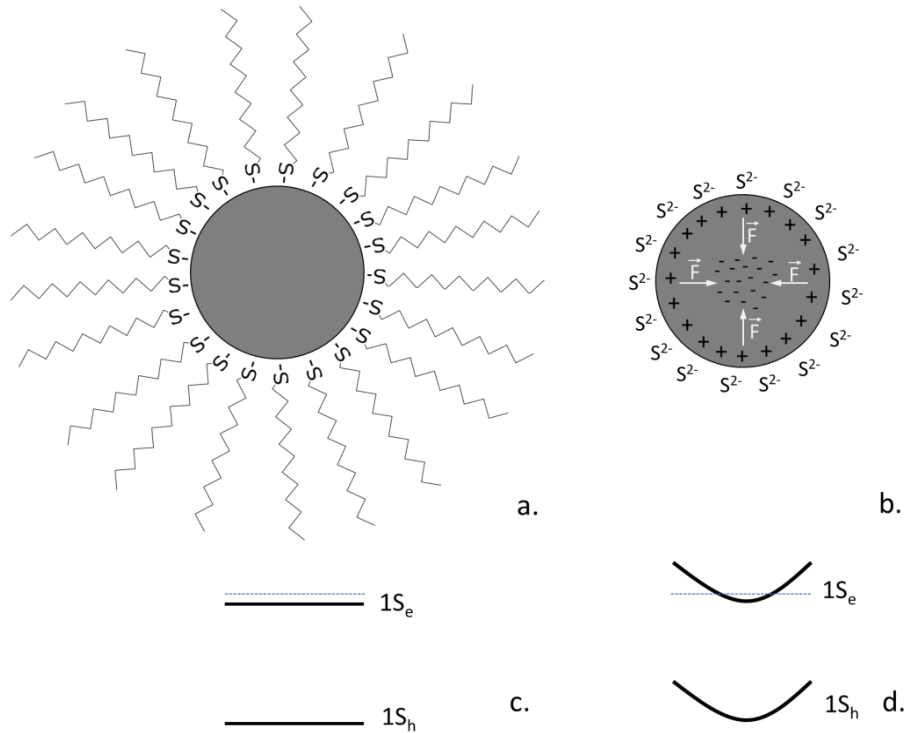

*Figure 54 : (a) Schéma d'un nanocristal entouré par des ligands DDT ; (b) schéma d'un nanocristal entouré par des ligands S²⁻ ; (c) courbure quasi-inexistante des niveaux dans le cas de ligands DDT ; (d) courbure des niveaux vers les énergies plus basses dans le cas de ligands S²⁻.*

Sur les nanocristaux dopés à la fin de la synthèse, tels que HgS et HgSe, l'effet du dipôle à la surface est encore plus flagrant (*48, 83*). En effet, comme le niveau de Fermi se situe très près du niveau $1S_e$ la transition intrabande $1S_e – 1P_e$ est autorisée. En appliquant un dipôle, le niveau de Fermi voit son énergie diminuée : le niveau $1S_e$ se vide et la transition intrabande devient interdite, tandis que la transition interbande devient autorisée (Figure 55). La variation du dopage obtenue en passant de ligands DDT à S²⁻ sur des nanocristaux de HgSe a été quantifiée par Robin *et al.* (*48*). Dans le cas du DDT, il y'a environ 2 électrons par nanocristal tandis qu'il n'y a que 0,2 électron par nanocristal dans le cas des ions S²⁻.

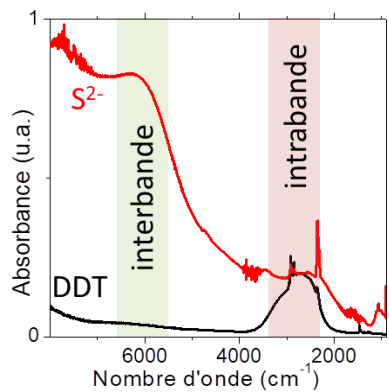

*Figure 55 : Spectres infrarouges de nanocristaux de HgSe pour deux ligands différents : le DDT (en noir), ayant un faible dipôle et pour lequel seule la transition intrabande apparaît ; les ions S²⁻, ayant un plus fort dipôle et pour lesquels la transition interbande devient autorisée.*

Pour vérifier l'effet des ligands sur les nanocristaux de HgSe, je vais reconstruire la structure électronique par les techniques que j'ai présentées au chapitre et m'intéresser en particulier à l'évolution du travail de sortie en fonction des ligands. En effet, le dipôle appliqué à la surface se traduit par une augmentation de l'énergie du niveau de vide par rapport au niveau de Fermi $\Delta E_{vide}$ : (*80, 84, 85*)

$$\Delta E_{vide} = -N \frac{\mu_{ligand}}{\varepsilon_0 \varepsilon_{ligand}} \tag{3.1}$$

Où N est la densité de surface de dipôles à la surface des nanocristaux, $\mu_{ligand}$ est l'amplitude du dipôle à la surface des nanocristaux moyennée et $\varepsilon_{ligand}$ est la constante diélectrique des ligands, qui est de l'ordre de 2 (*86*).

Pour des nanocristaux de 6 nm de diamètre, on obtient les résultats présentés sur la Figure 56.

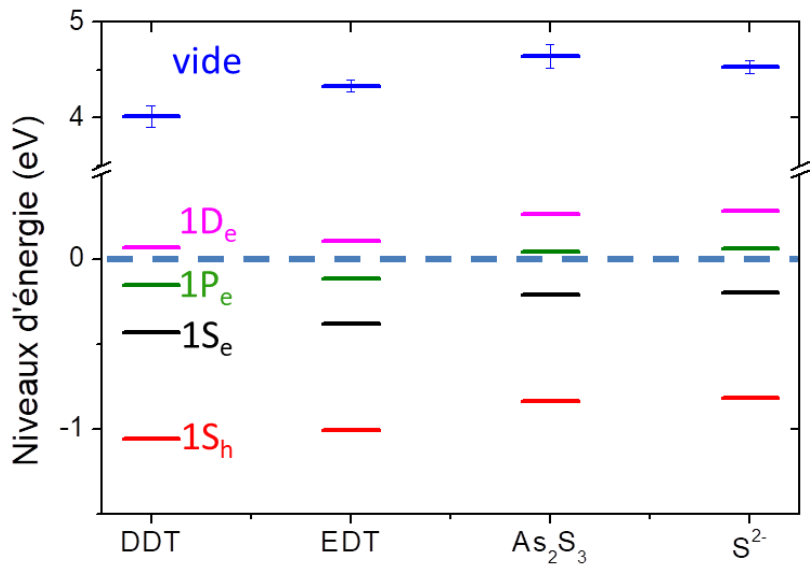

*Figure 56 : Structure électronique reconstruite par photoémission de nanocristaux de HgSe de 6 nm de diamètre pour 4 chimies de surface différentes (DDT, EDT, As$_2$S$_3$ et S$^{2-}$)*

D'après Robin et al. (*48*), les dipôles μ des chimies de surface utilisées vérifient : μ (S$^{2-}$) > μ (As$_2$S$_3$) > μ (EDT) > μ (DDT). On vérifie bien que :

-   le niveau de Fermi est plus profond dans la bande de conduction dans le cas où le dipôle est faible ; pour le DDT, on a 1P$_e$ < E$_F$ < 1D$_e$ tandis que pour S$^{2-}$ on a 1S$_e$ < E$_F$ < 1P$_e$.
-   le travail de sortie augmente avec le dipôle, comme prévu par l'équation (3.1).

Dans cette partie, j'ai présenté plusieurs techniques permettant d'ajuster le dopage. J'ai montré qu'il était possible d'ajuster le dopage après la synthèse des nanocristaux, et que ce dopage était réalisable par échange de ligands : une technique simple à mettre en œuvre. Cependant, les modifications de dopage rendues possibles par cette technique sont relativement faibles, de l'ordre de 1 à 2 électron(s) par nanocristal.

Dans la suite, je présenterai une technique, basée sur des ligands électroattracteurs, permettant de contrôler le dopage sur des gammes plus importantes que ce que nous avons observé avec les dipôles induits par les ligands.

### III.    Utilisation de ligands oxydants : les polyoxométallates (POMs)

Pour modifier le dopage des nanocristaux, une idée est d'utiliser des ligands à la surface capables d'échanger des électrons avec le nanocristal via une équation d'oxydoréduction, tels que les polyoxométallates, ou POMs.

#### 1.    Présentation des POMs

Les POMs sont des clusters à base de complexes oxo de métaux de transitions. Ces métaux sont à leur plus haut degré d'oxydation. Dans la suite de ce chapitre, nous utilisons des polyoxotungstates, donc à base de tungstène, au degré d'oxydation +VI. Leur structure est de type Keggin $[PW_{11}O_{39}]^{7-}$, et leur synthèse est décrite dans la littérature (*87, 88*). Ces polyoxotungstates sont fournis par Florence Volatron de l'équipe E-POM de l'Institut Parisien de Chimie Moléculaire (IPCM) de Sorbonne Université.

Ces matériaux sont utilisés pour leurs propriétés oxydantes. Greffés sur d'autres matériaux, leur caractère électroattracteur induit des transferts d'électrons du matériau vers les POMs. Cela a notamment été observé sur le graphène (*89*) ou sur des électrodes en carbone vitreux (*90*).

Pendant mon doctorat, j'ai donc étudié le greffage de ces POMs sur des nanocristaux de HgSe dans le but d'observer un transfert d'électrons des nanocristaux vers les POMs et donc une modification du dopage.

#### 2.    Diminution de dopage dans les nanocristaux de HgSe par les POMs

Pour cette étude, j'utilise des nanocristaux de 5 ± 0,7 nm de diamètre, dont les propriétés (taille, absorption et structure électronique) ont été reportées sur la Figure 57.

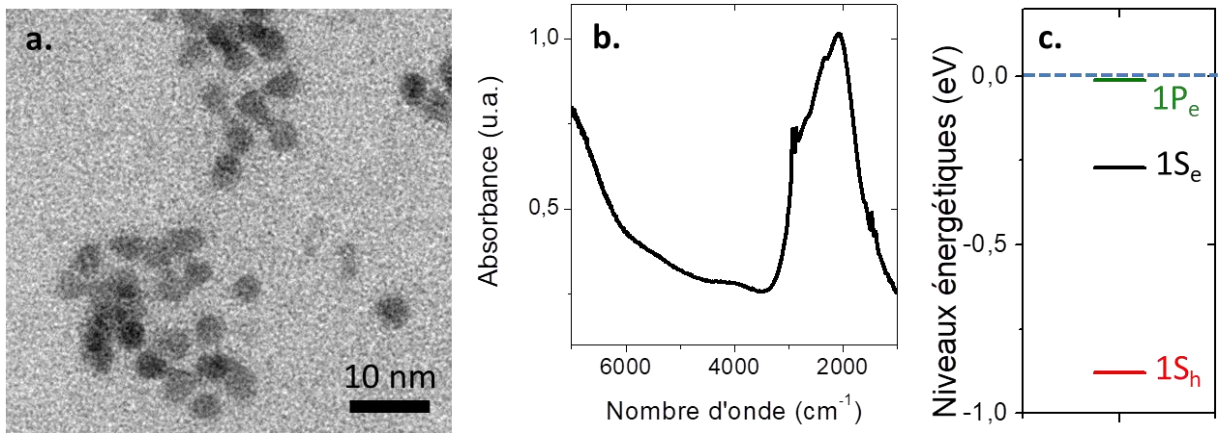

*Figure 57 : (a) Cliché TEM des nanocristaux de HgSe obtenus à la fin de la synthèse ; (b) spectre d'absorption infrarouge des nanocristaux infrarouges. La transition intrabande est observée à 2080 cm$^{-1}$ ; (c) structure électronique reconstruite des nanocristaux à la fin de la synthèse.*

#### a.    Greffage des POMs à la surface de nanocristaux de HgSe

Les POMs $[PW_{11}O_{39}]^{7-}$ sont des clusters anioniques, très solubles dans des solvants polaires comme la N,N diméthylformamide (DMF). À la fin de leur synthèse, les nanocristaux de HgSe sont couverts par le DDT et sont solubles dans des solvants apolaires, tels que l'hexane ou le toluène.

Afin de greffer les POMs à la surface, il faut retirer le DDT. Le triéthyloxonium tétrafluoroborate ($BF_4^-$) permet de retirer les ligands de surface, laissant alors le nanocristal nu, chargé positivement à la surface grâce aux cations en excès (*91*). Les nanocristaux peuvent donc ensuite être re-dispersés dans un solvant polaire. Cette technique fonctionne bien sur des nanocristaux de CdSe ou d'oxydes de fer $Fe_2O_3$ couverts d'oleylamine, mais elle ne permet pas de retirer le DDT de la surface des nanocristaux de HgSe. Cela est dû à la très grande affinité du mercure avec le soufre.

### i. Préparation des nanocristaux sans DDT

Il est donc nécessaire de modifier la synthèse des nanocristaux de HgSe pour mettre un autre ligand à la surface. Cette synthèse se faisant dans un milieu acide oléique – oleylamine (OA – OLA), n'ajouter que du toluène froid pour stopper la réaction permet d'obtenir des nanocristaux couverts de ligands OA – OLA.[9] Les propriétés de ces nanocristaux telles que la taille, la forme, ou les transitions optiques ne sont pas impactées par ce changement de ligands et restent celles présentées sur la Figure 57. En revanche, la stabilité colloïdale est bien moins bonne (< 1 jour). Les ligands OA – OLA sont plus faciles à retirer de la surface avec le triéthyloxonium tétrafluoroborate que le DDT. On obtient des nanocristaux nus sur lesquels il sera plus facile de greffer des POMs.

Pour greffer les POMs à la surface des nanocristaux, une solution de POMs dans la DMF est préparée. La concentration de cette solution est adaptée pour obtenir un ratio « nanocristaux : POMs » de 1 : 10 environ. Les nanocristaux sont introduits dans la solution de POMs, mais ne se solubilisent pas dans le solvant, même après agitation de quelques minutes. Étant donné la très bonne solubilité des POMs dans la DMF, cette « non-solubilisation » témoigne de l'échec du greffage des POMs à la surface des nanocristaux.

Afin de forcer le greffage des POMs, il peut être judicieux de les fonctionnaliser avec un ligand ayant une forte affinité avec les mercures à la surface des nanocristaux, comme des thiols.

### ii. Utilisation de POMs fonctionnalisées par des thiols

Les POMs sont donc fonctionnalisées par deux ligands thiols : dans la suite du chapitre, on utilisera l'écriture POM-SH. Pour éviter de créer une épaisseur isolante entre le nanocristal et le POM, nous utilisons une chaîne aliphatique courte, de trois carbones, pour implanter le thiol à la surface du POM. Les POM-SH peuvent être représentés par le schéma de la Figure 58. La synthèse de ces POM-SH a été réalisée par Florence Volatron de l'IPCM.

Ces POM-SH se greffent plus facilement aux nanocristaux nus de HgSe et permettent la re-dispersion de ces nanocristaux dans la DMF. La stabilité colloïdale de ces nanocristaux est très bonne (> 1 semaine) et après plusieurs mois, les nanocristaux tombés au fond du tube peuvent toujours être bien re-dispersés dans le solvant.

---

[9] Les différentes étapes d'une synthèse colloïdale sont décrites dans le chapitre 1 (p26), et les protocoles plus détaillés sont fournis dans l'Annexe 1 (p124).

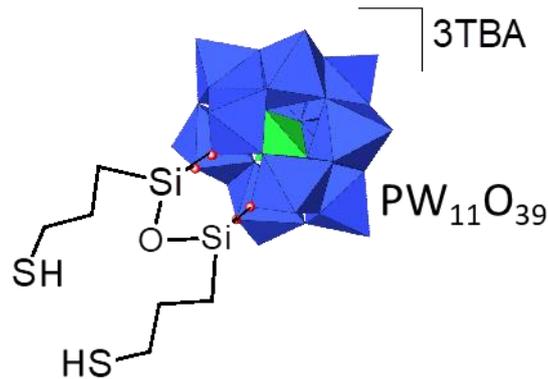

*Figure 58 : Schéma des POMs fonctionnalisés par des fonctions thiols. La partie en vert correspond aux atomes de métaux de transition (W) du POM, la partie bleue aux oxygènes.*

Une fois le greffage réalisé, nous pouvons vérifier l'impact des POM-SH sur la densité de porteurs.

### b. Étude de la modification de dopage induite par les POM-SH

Pour étudier la modification de dopage induite par les POM-SH sur les nanocristaux de HgSe, je vais utiliser plusieurs techniques : la reconstruction de la structure électronique, l'étude des niveaux de cœur du tungstène par photoémission et le transport via des mesures de transistor à effet de champ.

#### i. Reconstruction de la structure électronique

Une analyse qualitative du transfert des électrons des nanocristaux vers les POM-SH est réalisée par spectroscopie infrarouge. Les spectres d'absorption de deux films de nanocristaux, l'un ayant des ligands OA – OLA et l'autre ayant des ligands POM-SH, sont présentés sur la Figure 59. Sur ces spectres, nous observons que la transition intrabande obtenue pour les ligands OA – OLA est beaucoup moins prononcée pour les ligands POM-SH. Au contraire, la transition interbande qui était interdite pour les ligands OA – OLA devient autorisée et donne un signal d'absorption dans le cas des ligands POM-SH. Le niveau de Fermi qui était quasiment résonant avec le niveau $1P_e$ dans le cas OA – OLA passe donc en dessous du niveau $1S_e$ dans le cas POM-SH.

En utilisant les méthodes décrites dans le chapitre 2, la structure électronique des nanocristaux de HgSe couverts soit par OA – OLA soit par POM-SH, est reconstruite. On obtient les résultats présentés sur la Figure 59b qui confirment ce qui a été observé par spectroscopie infrarouge : le niveau de Fermi se situe dans la bande interdite après diminution du dopage par les POM-SH.

Nous pouvons estimer le nombre d'électrons par nanocristal pour les deux ligands. Dans le cas des ligands OA – OLA, le niveau de Fermi est pratiquement résonant avec le niveau $1P_e$. Le niveau $1S_e$ est donc plein (2 électrons) et le niveau $1P_e$ demi-rempli (3 électrons), ce qui donne 5 électrons par nanocristal. La formule (3.2) permet de calculer le nombre de porteurs thermiquement activés en fonction de la différence d'énergie entre le niveau $1S_e$ et le niveau de Fermi lorsque ce dernier est dans la bande interdite.

$$n = 2 \exp\left(-\frac{E_{1Se} - E_F}{k_B T}\right) \tag{3.2}$$

Où n est le nombre d'électrons par nanocristal. Dans le cas des ligands POM-SH, on trouve 0,03 électron par nanocristal. **Cette diminution du dopage est la plus importante jamais obtenue par échange de ligands sur des nanocristaux auto-dopés.**

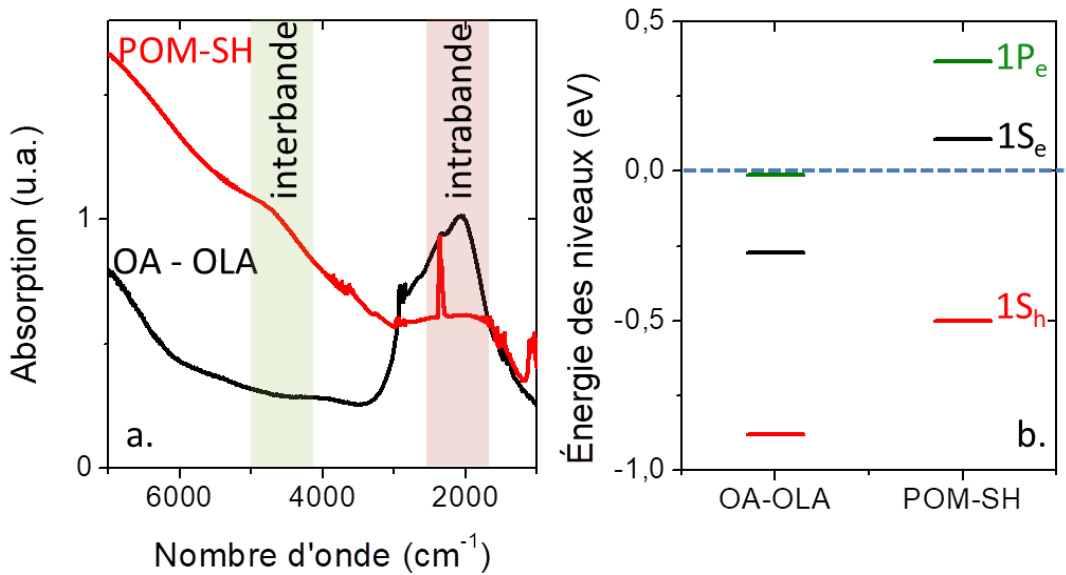

Figure 59 : (a) Spectres d'absorption infrarouge de nanocristaux HgSe avec deux types de ligands : OA – OLA en noir, POM-SH en rouge ; (b) Structure électronique des nanocristaux avec les deux types de ligands différents.

### ii.     Étude du degré d'oxydation des niveaux de cœur

Afin de m'assurer que la diminution du dopage est effectivement liée à une réaction d'oxydo-réduction entre les POM-SH et les nanocristaux, j'ai étudié les niveaux de cœur du tungstène sur un film de nanocristaux de HgSe couverts de POM-SH. En effet, les propriétés oxydantes des POM-SH proviennent du haut degré d'oxydation du tungstène (+VI) : si la diminution du dopage a été obtenue par transfert d'électron du nanocristal vers les POM-SH, le degré d'oxydation du tungstène doit diminuer.

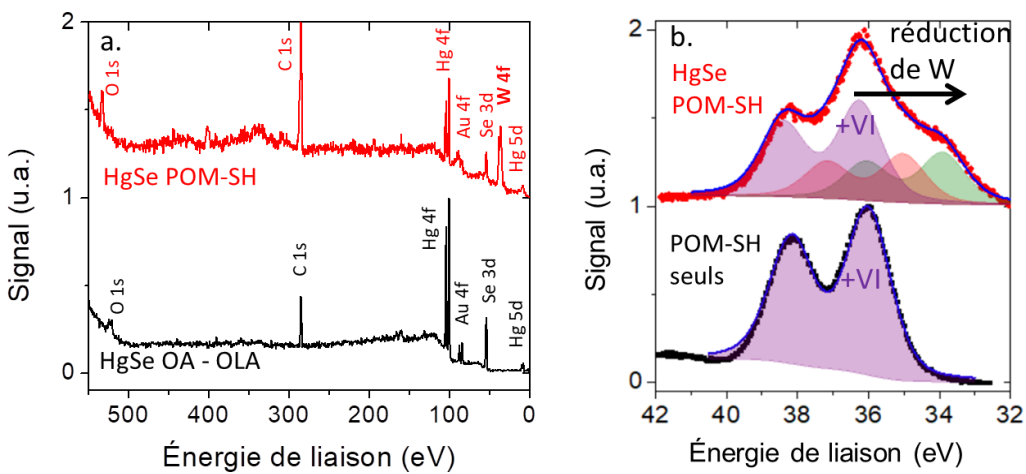

Figure 60 : (a) Niveaux de cœur de nanocristaux de HgSe pour différents ligands : POM-SH en rouge, OA – OLA en noir. (b) Zoom sur le niveau de cœur 4f du tungstène : sur un film de POM-SH seuls en noir et sur un film de nanocristaux de HgSe ayant des POM-SH greffés à la surface.

Les niveaux de cœur entre 0 et 550 eV de nanocristaux HgSe avec les deux chimies de surface, OA – OLA et POM-SH, ont été enregistrés (Figure 60a). Le tungstène, caractéristique des POM-SH est bien présent dans le spectre de photoémission de HgSe POM-SH (niveau W 4f à 36 eV), ce qui confirme que le greffage a bien fonctionné.

En zoomant sur ce pic 4f du tungstène (Figure 60b), on remarque que ce dernier présente plusieurs composantes. À 36,0 eV d'énergie de liaison, on retrouve la composante correspondant au degré d'oxydation +VI donc non réduit (en violet sur la Figure 60b) : c'est d'ailleurs la seule composante que l'on retrouve lors de l'analyse par photoémission d'un film de POM-SH seuls. On retrouve également deux composantes à des énergies de liaison plus faibles (35,1 et 33,9 eV) qui correspondent donc à des degrés d'oxydation plus faibles (+V et potentiellement +IV) (*92, 93*). Ces composantes témoignent de la réduction du tungstène dans le cas où les POM-SH sont greffés à la surface des nanocristaux de HgSe et donc du transfert d'électrons des nanocristaux vers les POM-SH.

### iii. Étude du transport

Les propriétés de transport des films de nanocristaux HgSe avec et sans POM-SH sont sondées via des mesures de transistor à effet de champ. Les ligands POM-SH sont assez gros ($\approx$ 1 nm de long) mais permettent tout de même d'obtenir des films conducteurs. En revanche, la chimie de surface OA – OLA donne des films de nanocristaux isolants. Pour pouvoir sonder les propriétés de transport de ces nanocristaux, un échange de ligands sur film est donc effectué pour remplacer les ligands OA – OLA par des ligands EDT. L'EDT a un dipôle assez faible (*48*) et permet de ne pas modifier le niveau de dopage des nanocristaux par rapport aux ligands OA – OLA. Les courbes de transfert obtenues sont présentées sur la Figure 61.

La chimie de surface EDT confirme le dopage n des nanocristaux puisque le courant dans le canal augmente continûment avec la tension de grille. En revanche, un comportement ambipolaire est observé dans le cas de la chimie de surface POM-SH, ce qui confirme que le niveau de Fermi se situe cette fois dans la bande interdite des nanocristaux de HgSe.

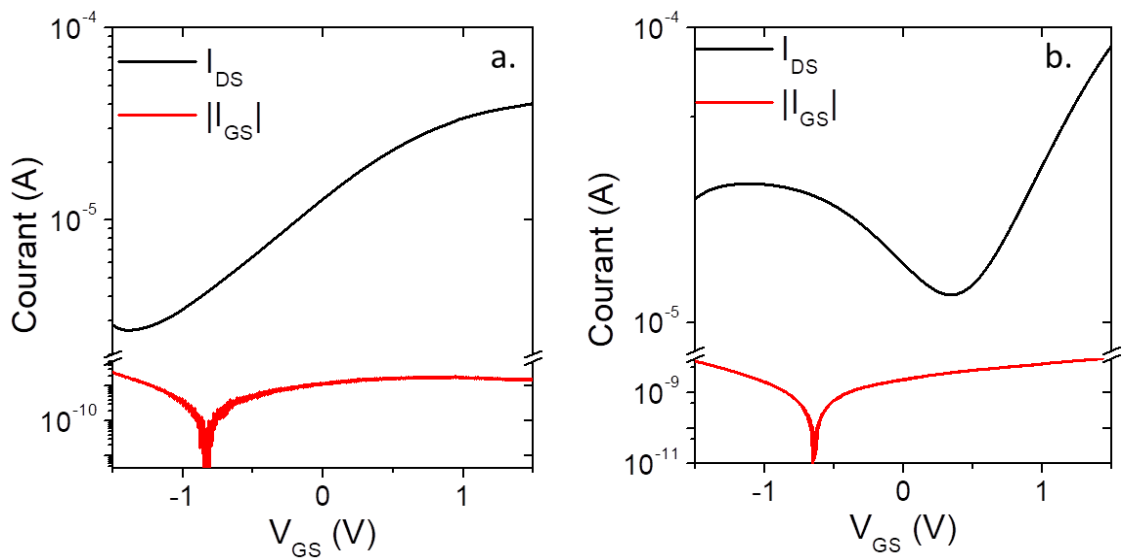

*Figure 61 : Courbes de transfert obtenues pour des nanocristaux de HgSe avec les deux chimies de surface : (a) EDT, (b) POM-SH. Les tensions drain-source sont fixées à 5 mV dans le cas d'EDT, à 100 mV dans le cas de POM-SH.*

### IV.    Conclusions et enjeux futurs

L'échange de ligands est une technique prometteuse dans le contrôle du dopage des nanocristaux. Elle est facile à mettre en œuvre et, contrairement aux méthodes physiques de contrôle du dopage, elle est compatible avec l'intégration des nanocristaux dans des dispositifs de détection. Les ligands permettent soit de modifier le dopage en appliquant un champ électrique à la surface des nanocristaux par effet dipolaire, soit de réaliser un transfert d'électrons des nanocristaux vers les ligands. J'ai montré qu'en utilisant des ligands POM-SH à la surface de nanocristaux de HgSe, un transfert de 5 électrons du nanocristal vers les ligands permettait de dé-doper ces nanocristaux. Le niveau de Fermi, résonant avec le niveau $1P_e$ à la fin de la synthèse, passe dans la bande interdite après greffage des POM-SH.

Le transfert d'électrons induits par les ligands POM-SH est prometteur pour obtenir des nanocristaux de type p. Ces derniers sont assez rares : on retrouve les nanocristaux HgTe fortement confinés ($1S_h - 1S_e > 6000$ cm$^{-1}$, voir les résultats sur la photoémission des nanocristaux de HgTe du chapitre 2, p58) qui de par leur petite taille et leur forte réactivité ne sont stables que quelques semaines en solution. Parmi les nanocristaux à plus grandes bandes interdites, on retrouve également PbS avec des ligands EDT (*82*) et CdTe (*94*). En utilisant la chimie de surface POM-SH, nous pouvons diminuer le dopage des matériaux auparavant dopés n ou ambipolaires. Par exemple, les HgTe 4000 cm$^{-1}$ sont ambipolaires lorsque leur chimie de surface est EDT (Figure 62a) et deviennent p après greffage de ligands POM-SH (Figure 62b).

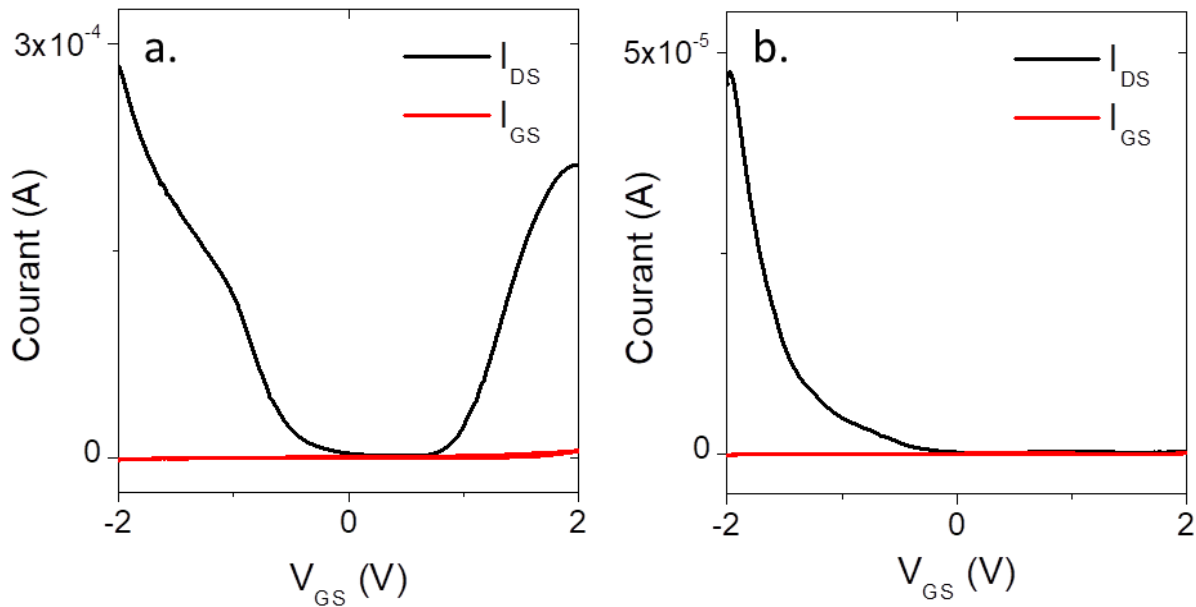

*Figure 62 : Courbes de transfert obtenues pour des HgTe 4000 cm$^{-1}$, dont la chimie de surface post-synthèse est OA – OLA. Après un échange EDT (a), le film de nanocristaux présente un comportement ambipolaire. Après un échange POM-SH (b), le film de nanocristaux présente un comportement de type p.*

Dans ce chapitre, j'ai surtout présenté des moyens de diminuer le niveau de dopage des nanocristaux. Pour certaines applications, il peut être intéressant d'augmenter le niveau de dopage, ce qui est plus difficile à réaliser post-synthèse. Le groupe de Philippe Guyot-Sionnest a proposé d'exposer des nanocristaux de CdSe à des réactifs réducteurs comme le biphényl de sodium (*95*) : un dopage n est observé mais ce dernier n'est stable que pendant environ 5 heures. Le groupe de Viktor Klimov a montré qu'un dopage de nanocristaux de PbSe et PbS était possible en introduisant un agent réducteur, le cobaltocène, à la surface des nanocristaux (*96*). Cependant, ce dopage n'est effectif que sur une quinzaine d'heures. Dans ces deux cas, le dopage obtenu n'est pas stable et n'est pas observable sur film. En particulier, aucune mesure de transistor à effet de champ sur film n'a été répertoriée dans la littérature.

Enfin, il est important de noter que plus le dopage est important, plus le nombre d'électrons participant à la conduction dans le noir est important. Cela a pour effet d'augmenter le niveau de bruit dans le système. De plus, l'utilisation de nanocristaux dopés dans des détecteurs infrarouges conduit souvent à des temps de réponse lents (*48, 51*). Pour réduire ces défauts, l'utilisation de systèmes cœur-coquille HgSe-HgTe peut être une solution. Les transitions intrabandes dans HgSe sont combinées à la dynamique rapide observée dans des films de nanocristaux de HgTe ce qui permet une diminution du temps de vie et une amélioration de la détectivité (*97*).

# CHAPITRE 4
# Intégration des nanocristaux dans des dispositifs à géométrie complexe

**Publications associées à ces travaux :**

- B. Martinez *et al.*, HgTe nanocrystal inks for extended short wave infrared detection, *Advanced Optical Materials*, 1900348 (2019) ;
- A. Jagtap *et al.*, Short wave infrared devices based on HgTe nanocrystals with air stable performances, *the Journal of Physical Chemistry C*, **122**, 23, 14979-14985 (2018);
- A. Jagtap *et al.*, Design of a unipolar barrier for nanocrystal-based short-wave infrared photodiode, *ACS Photonics*, **5**, 11, 4569-4576 (2018);
- C. Livache *et al.*, A random colloidal quantum dot infrared photodetector and its use for intraband detection, *Nature Communications*, **10**, 1, 2125 (2019).

**Mots clés :** diode, alignement de bandes, E-SWIR, HgTe, dispositif multipixels

**Techniques expérimentales utilisées :**

- Échange de ligands par transfert de phase
- Préparation de dispositifs à géométrie verticale
- Mesures de réponse et détectivité
- Mesures de photocourant
- Lithographie optique sur verre pour la fabrication d'électrodes

Dans les chapitres précédents, j'ai montré comment les propriétés électroniques des nanocristaux de chalcogénures de mercure (énergies des niveaux, dopage) pouvaient être déterminées et contrôlées. L'étape suivante pour intégrer ces matériaux dans un dispositif de détection infrarouge est de concevoir un détecteur adapté, qui permettra de maximiser le courant collecté sous illumination (réponse) ainsi que le rapport signal sur bruit (détectivité). Une architecture intéressante pour réaliser de tels détecteurs est la géométrie **photovoltaïque**, ou **diode**. En effet, avec cette géométrie, le courant d'obscurité est très faible à 0 V tandis que la réponse peut atteindre quelques mA/W dans la même gamme de tensions. Dans ce chapitre, je commencerai donc par expliquer le fonctionnement d'une cellule photovoltaïque et présenterai les diodes à base de nanocristaux dans l'infrarouge qui ont déjà été proposées dans la littérature.

Pour optimiser la réponse des détecteurs infrarouges photovoltaïques, une stratégie est d'augmenter l'épaisseur de la couche absorbante pour absorber plus de photons et donc générer plus de courant. En effet, pour des nanocristaux de HgTe à 2,5 µm, une couche d'épaisseur l = 200 nm n'absorbe que $1 - 10^{-\alpha l} = 17$ % de la lumière incidente, avec $\alpha$ le coefficient d'absorption des nanocristaux valant $3000 \pm 1$ cm$^{-1}$. Je présenterai une méthode de préparation des nanocristaux sous forme d'encre qui permet de déposer des couches plus épaisses de nanocristaux, de l'ordre de 500 nm, et qui simplifie la fabrication du détecteur.

J'utiliserai ces encres dans la gamme SWIR, en particulier dans le SWIR étendu (ESWIR). Dans cette gamme de longueurs d'onde, on trouve encore peu de détecteurs infrarouges à base de nanocristaux ayant de bonnes performances. Dans ce chapitre, je proposerai et testerai une structure de diode dans le SWIR étendu en utilisant des encres à base de HgTe.

Enfin, dans le but de construire une caméra et donc, et donc d'être capable de reconstruire une image, il est nécessaire de réaliser des systèmes à plusieurs pixels. Pour terminer ce chapitre, je présenterai la matrice de 100 pixels que j'ai développée et qui nous permet d'imager le profil d'un faisceau laser dans l'infrarouge.

## I.  Pourquoi utiliser des dispositifs photovoltaïques pour la détection infrarouge ?

### 1.  Comparaison dispositifs photoconducteurs et photovoltaïques

Les dispositifs photoconducteurs, que j'ai présentés au chapitre 1 et dont un schéma est rappelé sur la Figure 63, sont des dispositifs simples : deux électrodes de même nature (en or par exemple) sont déposées sur un substrat et un film de nanocristaux est déposé sur ces électrodes.

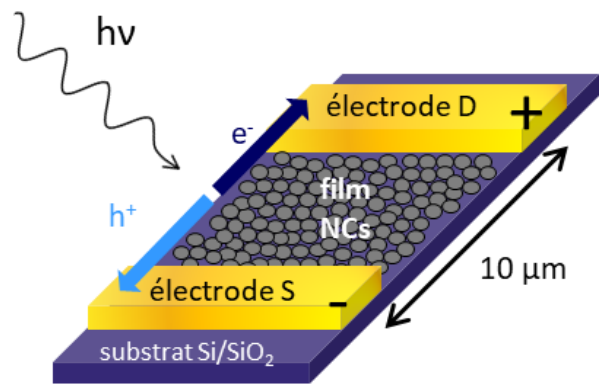

*Figure 63 : Schéma d'un dispositif photoconducteur à base de nanocristaux.*

Il est possible de schématiser le système comme suit : Au/nanocristaux/Au. C'est un système symétrique avec une caractéristique courant-tension symétrique (Figure 64a). Les dispositifs photoconducteurs à base de HgTe dans le SWIR permettent d'atteindre des réponses de l'ordre de quelques centaines de mA/W, mais pour de grandes tensions drain-source (de l'ordre de 10 V) (*33, 98, 99*). Les niveaux de bruit engendrés sont donc trop importants et les détectivités obtenues à température ambiante sont faibles (< $10^8$ Jones).

Pour garder une réponse importante tout en minimisant le niveau de bruit, une solution est d'utiliser une structure de détecteur photovoltaïque. Les détecteurs photovoltaïques présentent des caractéristiques courant-tension asymétriques comme celles présentées sur la Figure 64b. Ils présentent deux modes de fonctionnement : le sens bloquant ($V_{DS} < 0,5$ V sur la Figure 64b) pour lequel le courant d'obscurité est quasi nul, et le sens passant ($V_{DS} > 0,5$ V sur la Figure 64b).

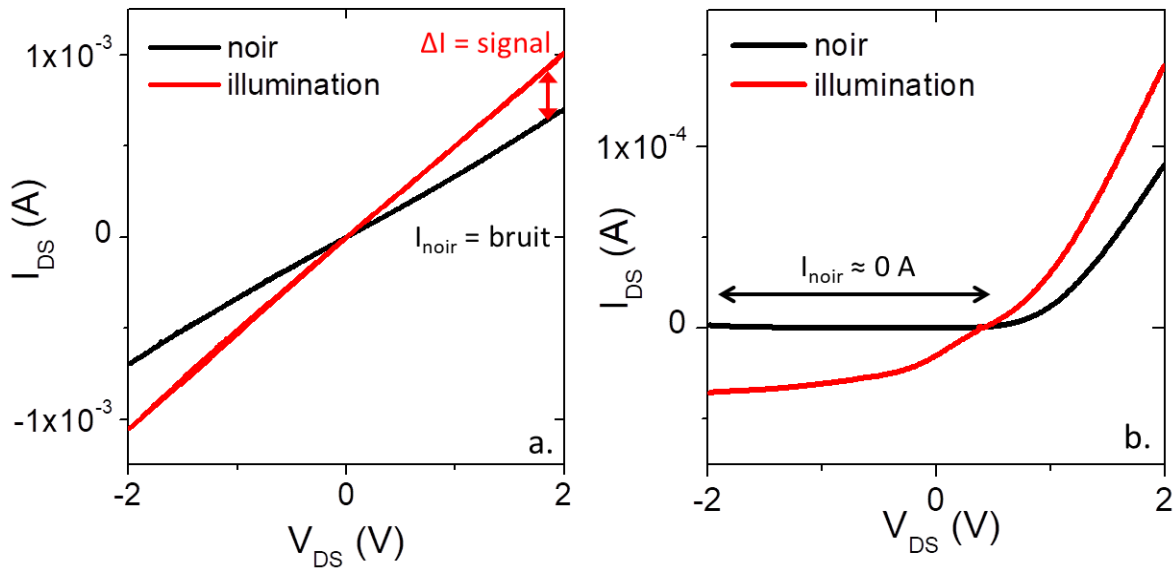

*Figure 64 : (a) Caractéristique courant-tension d'un dispositif photoconducteur ; (b) caractéristique courant-tension d'un dispositif photovoltaïque.*

Pour obtenir une caractéristique courant-tension asymétrique, il est nécessaire que la géométrie du détecteur soit asymétrique, par exemple en introduisant deux électrodes ayant des travaux de sortie différents. Cette asymétrie permet de séparer les charges créées sous illumination, sans appliquer de tension externe. Grâce à cette propriété, on peut observer sur la Figure 64b que le courant sous illumination est non nul près de 0 V, tandis que le courant d'obscurité est très faible. Dans une gamme de tensions près de 0 V, la réponse sera donc importante, tandis que le bruit sera faible.

Une des premières architectures proposées pour les diodes à base de chalcogénures de plomb est une diode Schottky dont la structure est présentée sur la Figure 65a (*100, 101*). De par leur énergie de bande interdite dans le proche infrarouge, les nanocristaux de chalcogénure de plomb sont pressentis pour être utilisés dans des cellules solaires. Dans cet exemple, les électrodes utilisées sont l'oxyde d'indium dopé avec de l'étain (ITO) et le magnésium (Mg), elles vérifient $WF_{ITO} > WF_{Mg}$. Lorsque les matériaux sont empilés, les niveaux de Fermi s'alignent, ce qui entraîne la courbure des bandes comme présenté sur la Figure 65b. Il est alors plus simple d'injecter les électrons par l'électrode de Mg que par l'électrode d'ITO : autrement dit, quand $V_{Mg-ITO}$ est négative, $I_{noir}$ est très faible (Figure 65c).

L'asymétrie de la structure permet d'extraire les porteurs générés sous illumination sans appliquer de tension (Figure 65d) : les trous rejoignent facilement l'électrode d'ITO et les électrons rejoignent l'électrode de Mg. Cela génère donc un photocourant à 0 V, tout en conservant un courant d'obscurité très faible (Figure 65e).

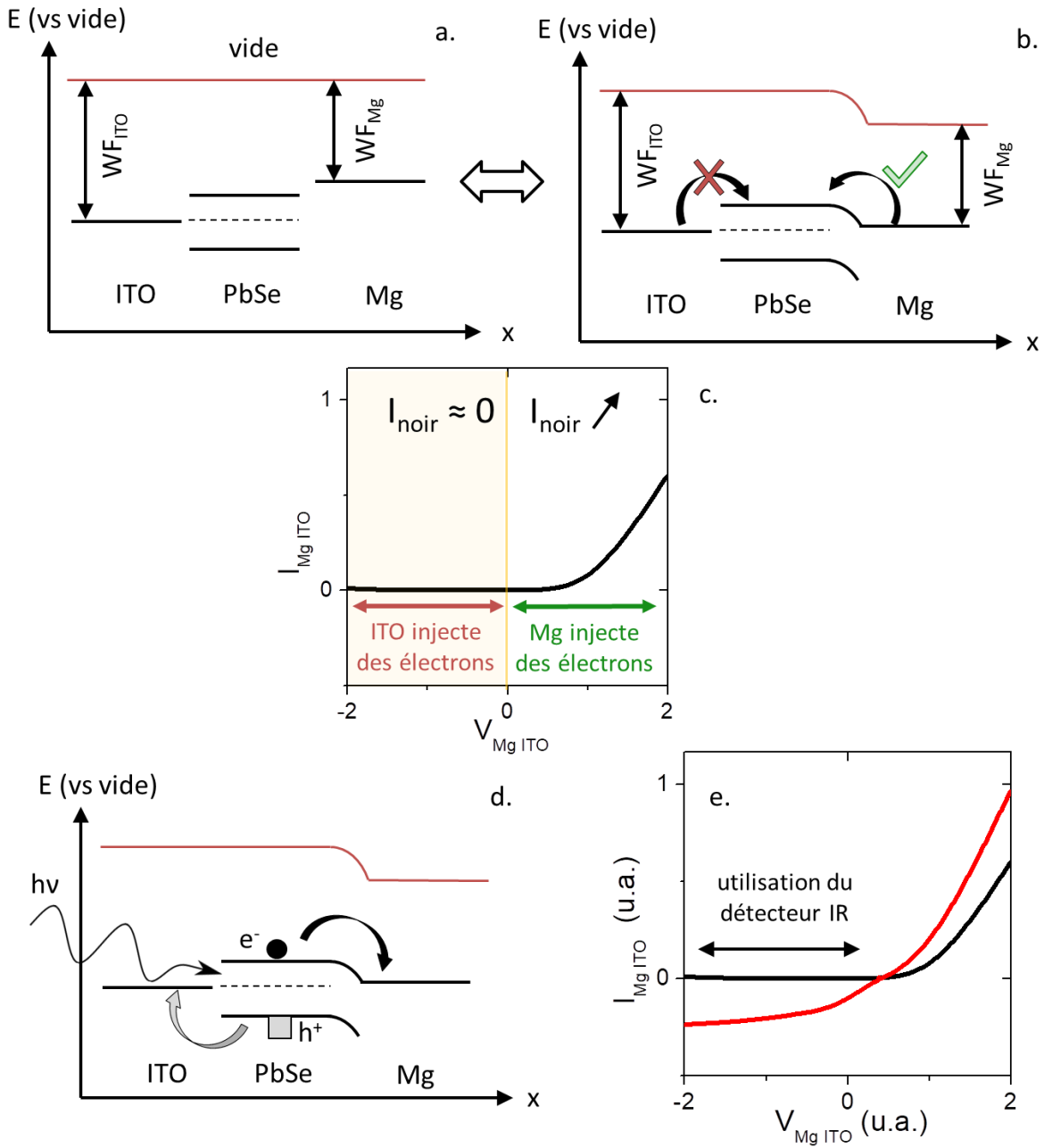

*Figure 65 : (a) Schéma de structure d'une diode à base de PbS et déplacements possibles des électrons au sein de la structure dans le noir; (b) alignement des niveaux dans le semiconducteur ; (c) allure de la caractéristique courant-tension dans le noir de cette structure ; (d) schéma de structure de la diode à base de PbS et déplacements possibles des électrons et photoélectrons au sein de la structure ; (e) allure de la caractéristique courant-tension dans le noir et sous illumination de cette structure. Un dispositif photovoltaïque s'utilise sur la gamme de tension indiquée par la flèche noire.*

Pour réaliser ces structures, la stratégie retenue est de préparer les dispositifs en géométrie verticale. Dans la Figure 65a, b et d, x représente l'épaisseur du dispositif et les couches d'ITO, PbSe et Mg ont été déposées les unes après les autres. La structure est réalisée en empilant les différents matériaux. À noter que la lumière doit pouvoir atteindre la couche de matériau actif. Dans le cas de la structure décrite dans la Figure 65d, la structure est illuminée avec une source visible et ITO est transparent dans cette gamme de longueurs d'onde. En utilisant une structure verticale dans laquelle les matériaux sont empilés les uns sur les autres, il est impératif que chaque couche soit bien homogène et ne présente ni rugosité, ni trou, ni craquelure au risque d'avoir des courts-circuits.

Pour optimiser le photocourant et réduire le courant d'obscurité, des structures de diode plus élaborées ont été réalisées.

### 2. État de l'art sur les diodes à base de nanocristaux

Les diodes les plus développées dans la littérature sont celles dans le visible et dans le proche infrarouge (< 1 µm), car elles peuvent être utilisées comme matériau photoconducteur dans les cellules solaires (*63, 101–107*).

### a. Cellules solaires

Pour optimiser la structure des dispositifs photovoltaïques, des couches peuvent être ajoutées. Ce sont soit des couches de transport, pour favoriser l'extraction des photo-porteurs, soit des couches de blocage, pour limiter le courant d'obscurité.

Parmi les matériaux qui suscitent le plus d'intérêt pour la réalisation de cellules solaires, on retrouve les pérovskites, y compris sous la forme de nanocristaux (*18, 108*). En particulier, les pérovskites inorganiques ($CsPbX_3$) présentent une meilleure stabilité dans le temps que leurs homologues hybrides. Une des structures les plus utilisées pour les cellules solaires à base de pérovskites est celle décrite sur la Figure 66a. Les couches de $TiO_2$ et Spiro permettent d'améliorer le transport des photoélectrons et des phototrous respectivement (*109–111*). De par leur absorption dans le proche infrarouge, les nanocristaux de PbS et PbSe sont également utilisés pour réaliser des cellules solaires (*68, 112, 113*).

Dans le but de favoriser l'extraction des porteurs, des efforts ont été réalisés pour passer d'une diode Schottky à une diode à jonction p-n. Pour cela, il faut que le matériau actif puisse être dopé p ou n, sans modifier l'énergie de la transition. Des développements ont été réalisés sur PbS, notamment par le groupe de Moungi Bawendi (*82*). Ils ont montré qu'en modifiant les ligands, le dopage pouvait être modifié sur des nanocristaux de PbS : le dopage est de type n avec des ligands halogénures, il est de type p avec des ligands éthanedithiol (EDT).

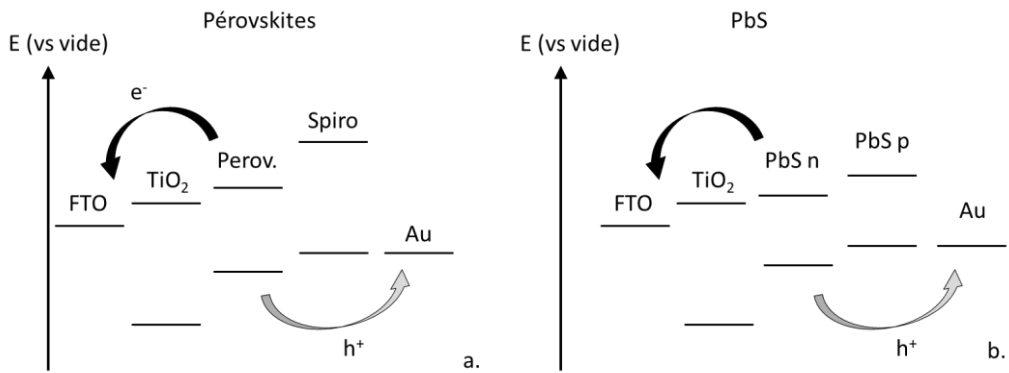

*Figure 66 : (a) Alignement de bandes utilisé pour les cellules solaires à base de pérovskites inorganiques ; (b) alignement de bandes utilisés pour les cellules solaires à base de nanocristaux de PbS.*

L'efficacité quantique externe (EQE), ratio du photocourant mesuré sur le nombre de photons incidents, atteint les 50 % pour les cellules solaires de PbS les plus optimisées.

### b. Détecteurs infrarouges à base de chalcogénures de mercure dans le ESWIR

La gamme ESWIR s'étend entre 1,7 et 2,5 µm (500 – 730 meV, ou encore 4000 – 6000 cm$^{-1}$). Cette gamme de longueurs d'onde est difficilement accessible aux nanocristaux de chalcogénures de plomb. En effet, des nanocristaux larges doivent être synthétisés (*114*), et ces larges nanocristaux ne sont pas stables colloïdalement. Leur dépôt peut entraîner des courts-circuits dans la diode. En ce qui concerne les détecteurs commerciaux, ceux à base d'InGaAs sont les plus souvent utilisés dans la gamme SWIR, mais la plupart ne fonctionne qu'à des longueurs d'onde inférieures à 1,7 µm. Pour atteindre la gamme ESWIR, une croissance du matériau sous contraintes est nécessaire, ce qui rend le dispositif plus compliqué à fabriquer et donc plus cher.

Pour obtenir des dispositifs photovoltaïques dans l'ESWIR, il faut adapter la stratégie utilisée pour les pérovskites et les nanocristaux de chalcogénures de plomb à des nanocristaux de plus faible énergie de bande interdite comme les chalcogénures de mercure. Avant le début de mon doctorat, la réalisation de tels dispositifs à base de chalcogénures de mercure était encore marginale. Le groupe de Philippe Guyot-Sionnest avait publié une diode à base de HgTe dans le MWIR qui présentait de bons résultats en terme de réponse (quelques 50 mA/W) et de détectivité (10$^{10}$ Jones). Cependant, ce détecteur fonctionnait à basse température (140 K) (*115*).

Amardeep Jagtap, post-doctorant dans notre groupe, a donc travaillé à l'élaboration d'une diode dans l'ESWIR, à base de nanocristaux de HgTe à 4000 cm$^{-1}$. Pour cela, il a repris une structure très similaire à celles utilisées pour des diodes à base de nanocristaux de chalcogénures de plomb, en remplaçant le matériau actif par des nanocristaux de HgTe 4000 cm$^{-1}$.

Pour la fabrication, un substrat commercial d'ITO sur verre est utilisé. Une solution colloïdale de particules de TiO$_2$ anatase est ensuite déposée par *spin-coating*. Un recuit de ces particules permet de donner une structure compacte aux particules de TiO$_2$. Les nanocristaux de HgTe sont ensuite déposés par *spin-coating* dans une boîte à gants sous atmosphère d'azote. En effet, sous forme de film, les nanocristaux de HgTe voient leurs performances diminuer après quelques jours s'ils sont conservés à l'air, notamment le courant d'obscurité augmente.[10] L'électrode d'or est ensuite déposée par évaporation, tout en conservant l'échantillon en atmosphère inerte. Enfin, des polymères protecteurs (PMMA et PVA) sont déposés à la surface du dispositif pour éviter la dégradation des films de nanocristaux par l'eau et l'oxygène présents dans l'air.

La diode finale est donc la suivante : ITO/TiO$_2$/HgTe 4000 cm$^{-1}$/Au. Les alignements des niveaux sont présentés sur la Figure 67. Sous illumination, les électrons rejoignent l'ITO tandis que les trous rejoignent l'or. La barrière de TiO$_2$ bloque le courant d'obscurité de trous, comme c'est le cas dans les cellules solaires à base de PbS ou de pérovskites. Aucune barrière ne permet de filtrer le courant d'obscurité dû aux électrons. Cette première génération de diode présente de faibles performances : la réponse à 0 V est de l'ordre de quelques µA/W, et la détectivité atteint 10$^7$ Jones (à 0 V et à 1 kHz).

---

[10] Des détails sur la dégradation des performances des films de nanocristaux de tellure de mercure seront fournis plus tard dans ce chapitre (p100).

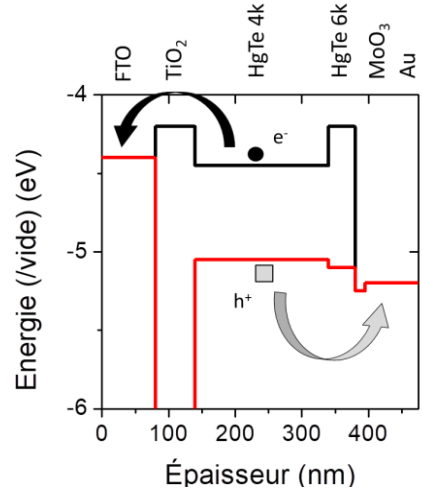

*Figure 67 : Architecture de la première génération de diodes ESWIR à base de HgTe réalisée dans notre équipe en 2018. La structure utilisée est la suivante : ITO / TiO$_2$ / HgTe 4k / Au. I$_{noir}$ correspond au courant d'obscurité, I$_{IR}$ au courant sous illumination infrarouge.*

Afin d'augmenter les performances de ces diodes, plusieurs modifications ont été apportées. (i) Pour améliorer l'extraction des photoélectrons, l'ITO a été remplacé par le FTO qui a un travail de sortie moins élevé (4,4 eV contre 4,7 pour l'ITO). (ii) Une couche d'oxyde de molybdène a été ajoutée pour favoriser le transport des trous jusqu'à l'électrode d'or. (iii) Enfin, une barrière unipolaire a été ajoutée. Cette barrière doit permettre d'extraire les trous vers l'électrode d'or et empêcher l'injection d'électrons de l'électrode d'or vers la diode, ce qui augmenterait le courant d'obscurité. Pour réaliser cette barrière, les nanocristaux de HgTe 6000 cm$^{-1}$ sont de bons candidats. En effet, j'ai démontré dans le chapitre 2 qu'ils vérifiaient les critères suivants :

- Le niveau 1S$_h$ des nanocristaux HgTe 6k a une énergie presque résonante avec celle du niveau 1S$_h$ des nanocristaux HgTe 4k (voir Figure 68a); le transport des trous vers l'électrode d'or ne sera donc pas affecté.
- Le niveau 1S$_e$ des nanocristaux HgTe 6k a une énergie supérieure au niveau 1S$_e$ des HgTe 4k ; les électrons ne peuvent pas rejoindre l'électrode d'or, et le courant d'obscurité provenant de l'électrode d'or est bloqué.

Les alignements des niveaux correspondants aux matériaux des différentes couches de la diode sont fournis sur la Figure 68b. Pour les deux couches de HgTe (4k et 6k), les ligands utilisés sont l'éthanedithiol (EDT), et l'échange de ligands est réalisé sur film.

Cette diode fonctionne à température ambiante et est stable pendant plusieurs semaines à l'air grâce à la présence des polymères protecteurs. La détectivité obtenue à température ambiante est de $3 \times 10^8$ Jones, mais l'amplitude de la réponse est assez faible, de l'ordre de la centaine de µA/W à 0 V. Cette faible réponse est due à plusieurs effets :

- La couche de TiO$_2$, qui joue le rôle de barrière pour les trous, était bien adaptée pour les systèmes à plus grandes énergies de bande interdite telles que les nanocristaux de PbS/PbSe ou de pérovskites (Figure 66). Or, dans le cas de HgTe 4000 cm$^{-1}$, elle peut également filtrer le photocourant dû aux électrons (Figure 68b).
- L'épaisseur de nanocristaux, de l'ordre de 200 nm, ne permet d'absorber que quelques pourcents de la lumière incidente (1-10$^{-\alpha l}$ = 17%).

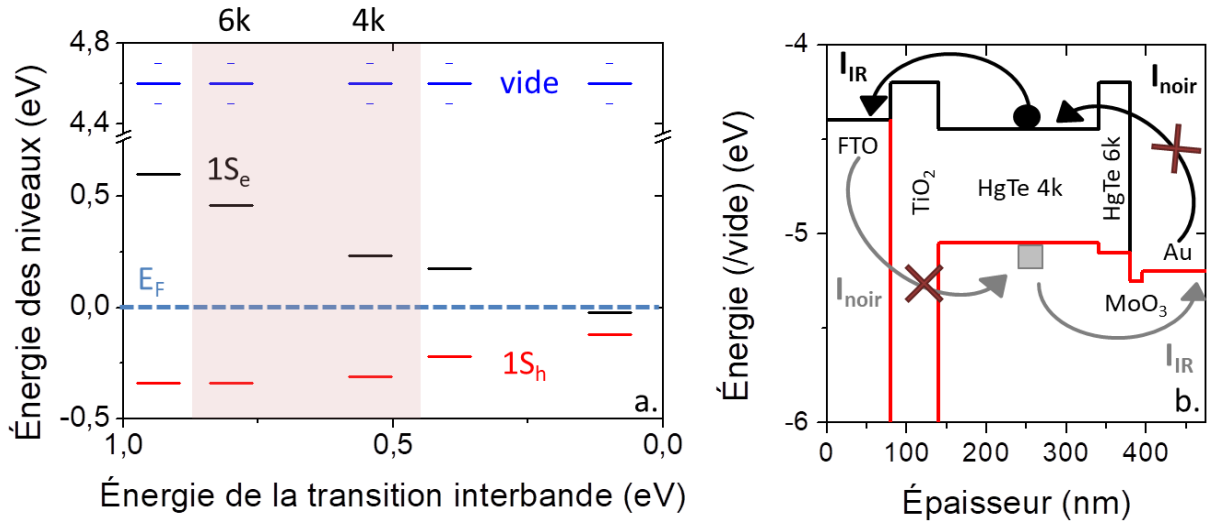

*Figure 68 : (a) Rappel des niveaux énergétiques des nanocristaux de HgTe de différentes tailles donc de différents degrés de confinement. Les niveaux des nanocristaux HgTe 6000 cm⁻¹ (6k) et 4000 cm⁻¹ (4k) ont été surlignés en rouge sur la figure. (b) Alignement des niveaux entre les matériaux de la deuxième génération de diode ESWIR réalisée au laboratoire. Les courants d'obscurité de trous et d'électrons sont filtrés par les barrières de $TiO_2$ et HgTe 6k respectivement.*

Pour augmenter la réponse de notre détecteur, je propose de retirer la couche de $TiO_2$ et d'utiliser une couche de nanocristaux de HgTe 4k plus épaisse. Pour réaliser ce deuxième point, j'ai développé une encre pour les nanocristaux de HgTe qui permet de s'affranchir du dépôt multicouche du matériau actif ainsi que d'obtenir des épaisseurs de l'ordre de 500 nm.

## II.   Utilisation d'encres

Les encres sont des solutions de nanocristaux dont les ligands longs nécessaires à la synthèse ont été échangés par des ligands compatibles avec le transport.  Le terme « encre » est utilisé puisque la solution est déjà prête à l'emploi. Une fois déposé, le film peut directement être utilisé, l'échange de ligands ayant déjà été effectué en solution.

L'intérêt de cette technique d'échange est de simplifier le dépôt du film de nanocristaux. Les échanges de ligands sur film peuvent être longs à réaliser. Il faut que la couche de nanocristaux soit suffisamment fine pour que les ligands diffusent à l'intérieur : pour construire un film épais qui absorbera plus de photons, il est donc nécessaire de répéter la procédure « dépôt de film – échange de ligands » plusieurs fois. Ces échanges présentent un risque de dégradation du film de nanocristaux : délamination du film pour les épaisseurs trop importantes, rayure du film pendant les nombreuses étapes de transfert entre les différentes solutions... Il est donc difficile d'obtenir des films épais qualitatifs. L'échange de ligands sur film présente également l'inconvénient de créer des micro-craquelures pendant l'échange : les nanocristaux se rapprochent lorsque les ligands à la surface sont plus petits, ce qui conduit à des craquelures dans le film. Dans le cas de dispositifs photovoltaïques en géométrie verticale, des craquelures peuvent conduire à des courts-circuits et rendre le détecteur inutilisable.

Dans le cas d'une encre, il est possible de réaliser un unique dépôt épais et sans micro-craquelures dues à l'échange de ligands, puisque l'échange a déjà été effectué en solution (un schéma rappelant le principe de la procédure d'échange est présenté sur la Figure 69). Il faut cependant parvenir à réaliser un film homogène à base d'une solution dont le solvant (polaire) est difficile à évaporer.

Les encres à base de nanocristaux ont été introduites par le groupe de Dmitri Talapin en 2009 dans le but d'améliorer le couplage entre nanocristaux (CdSe) sur film (*116*) et donc d'augmenter la mobilité. Une grande variété de ligands peut être utilisée pour ces échanges en solution : sulfures, séléniures, tellures, amidures et hydroxydes (*32*). Quelques années plus tard, le groupe d'Edward Sargent a montré que les ligands oleylamine à la surface des nanocristaux de PbS pouvaient être échangés en solution et être remplacés par des ligands 1-thioglycérol, et a démontré leur intégration dans un système de détection. Cependant, le solvant utilisé pour solubiliser les nanocristaux échangés étant le diméthylsulfoxyde (DMSO, $T_{ébullition}$ = 190 °C), le séchage du film est difficile à contrôler : les films sont utilisables après un séchage de 6 heures à 35 °C sous atmosphère inerte. Depuis 2014, les encres pour les gammes allant du visible au SWIR ont été développées et leur dépôt dans des systèmes de détection, types cellules solaires, a été amélioré (*117–122*).

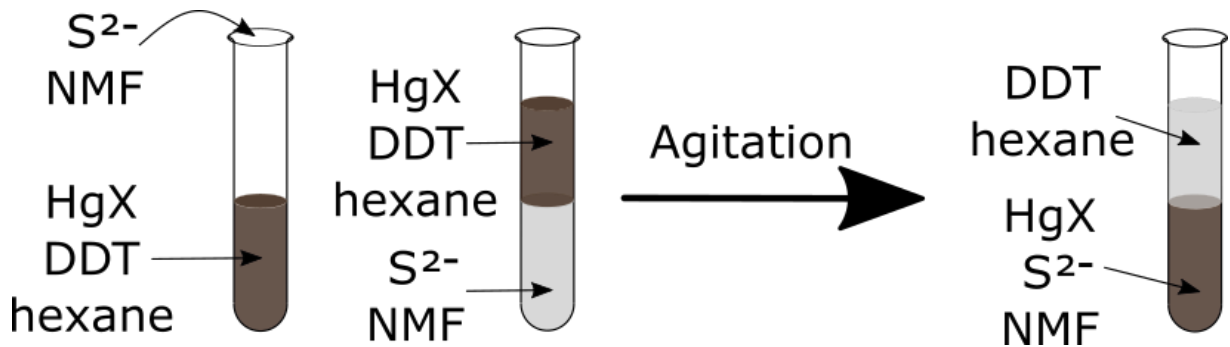

*Figure 69 : Schéma présentant la technique d'échange de ligands par transfert de phases. Dans cet exemple utilisant les nanocristaux de chalcogénures de mercure, le ligand long est le DDT, que l'on veut échanger par un ligand $S^{2-}$. Le solvant apolaire est l'hexane, le solvant polaire est la NMF.*

Dans les gammes de longueurs d'onde plus élevées, le dépôt de films de nanocristaux à base d'encre, d'une qualité suffisante pour être intégrés dans des systèmes de détection infrarouge à géométrie verticale, n'est pas encore optimal. Pour les nanocristaux de chalcogénures de mercure, les ligands et solvants permettant de réaliser des encres doivent être étudiés. Pendant ma thèse, j'ai développé une encre à base de mercaptoalcool (MPOH) et d'ions chlorures dans la *N,N*- diméthylformamide (DMF), qui permet de réaliser des films de nanocristaux de HgTe 4000 cm$^{-1}$ d'une épaisseur de l'ordre de 500 nm.

## 1. Optimisation de la formulation de l'encre

L'encre à base de nanocristaux doit permettre de réaliser des films très homogènes et épais, sans détériorer les propriétés optoélectroniques des nanocristaux de HgTe 4000 cm$^{-1}$. Dans cette partie, je vais optimiser la formulation de l'encre en jouant sur le solvant, les ligands et la concentration.

### a. Choix du solvant

Pour réaliser l'encre de nanocristaux de HgTe, j'ai sélectionné un solvant qui réunissait les critères suivants :

- excellente stabilité colloïdale des nanocristaux échangés ;
- température d'ébullition la moins élevée possible ;
- possibilité de réaliser des encres concentrées.

La stabilité colloïdale permet d'assurer une bonne homogénéité du film. Une faible température d'ébullition permet de réaliser un dépôt sans chauffer. Le fait de pouvoir obtenir des encres concentrées nous permettra de faire des couches épaisses en un seul dépôt. Les différents solvants testés et la stabilité des encres obtenues sont présentés dans le Tableau 7.

*Tableau 7 : Températures d'ébullition des différents solvants testés pour la formulation de l'encre et stabilité des encres obtenues. - - correspond à une agrégation totale des nanocristaux dans le solvant, - à une agrégation partielle, + à une solubilisation totale des nanocristaux à une concentration < 50 mg/mL, ++ à une solubilisation totale des nanocristaux pour des concentrations supérieures > 50 mg/mL.*

| Solvant | $T_{ébullition}$ (°C) | Stabilité |
|---|---|---|
| Propylamine | 48 | -- |
| Butylamine | 78 | - |
| Acétonitrile | 82 | -- |
| 2,6 difluoropyridine | 120 | + |
| Butylacétate | 126 | -- |
| DMF | 153 | ++ |
| NMF | 182 | ++ |
| Ethylène glycol | 198 | ++ |

Le solvant correspondant au mieux à ces trois critères est donc la DMF.

### b. Choix des ligands

Dans sa diode, Amardeep Jagtap utilisait des ligands EDT autour des nanocristaux de HgTe (*24*, *123*). Comme je l'ai montré dans le chapitre 2, ces ligands permettent de passiver les pièges dans la bande interdite. Cependant, de par leur encombrement stérique, ils ne permettent pas de passiver tous les mercures de surface et certaines facettes peuvent ne pas être recouvertes. Il est donc nécessaire de coupler les ligands thiols avec un autre type de ligands. Les ions halogénures ont montré de bonnes performances en termes de mobilité (*68*, *124*), de stabilité dans le temps et de passivation des pièges sur les films de nanocristaux (*42*, *125*). Cette passivation des pièges est un paramètre très important dans des cellules solaires à base de nanocristaux, puisque la présence de pièges peut drastiquement diminuer les performances (*62*). Je décide donc d'ajouter des sels de chlorure de mercure HgCl$_2$ à la formulation de l'encre (*126*).

Les ligands EDT présentent deux groupements thiols « -SH », très réactifs avec les sels de mercure, formant un précipité blanc dans la solution de nanocristaux. Il est donc difficile d'établir une formulation d'encre à base d'EDT et de sels de chlorure de mercure. Pour éviter ce phénomène tout en conservant une bonne passivation des pièges, je remplace l'EDT par le 2 mercaptoéthanol (mercaptoalcool, ou MPOH). Dans le mercaptoalcool, un des deux groupements « -SH » est remplacé par un groupement hydroxy « -OH ». Ce nouveau ligand compatible avec les sels de chlorure de mercure ne modifie pas l'environnement à la surface des nanocristaux et devrait permettre de passiver correctement les pièges. Les formules topologiques de ces deux ligands sont rappelées sur la Figure 70.

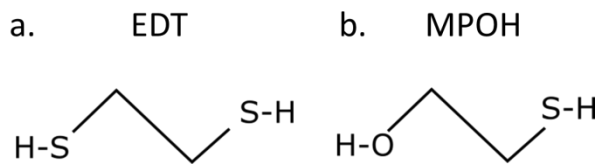

*Figure 70 : Formules topologiques des ligands 1,2 éthanedithiol (EDT) (a), et du 2 mercaptoéthanol (MPOH) (b)*

Ces ligands (MPOH et $HgCl_2$) ne modifient pas les propriétés optiques des nanocristaux. Les spectres infrarouges des nanocristaux avant et après échange de ligands sont présentés sur la Figure 71. Sur ces spectres, seuls les ligands (groupement alcool de MPOH, liaisons C-H de MPOH et de DDT) et le solvant (DMF) donnent lieu à des différences d'absorption entre les deux solutions étudiées.

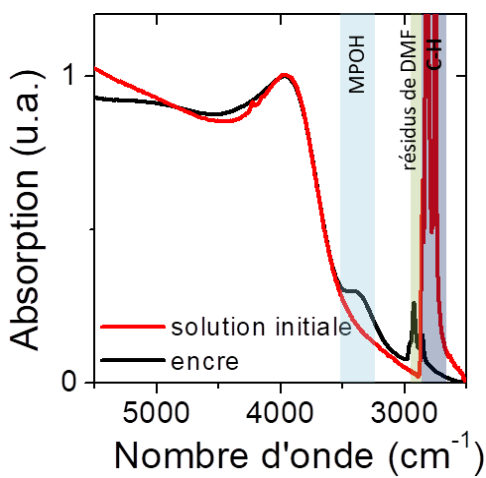

*Figure 71 : Spectres infrarouges en configuration ATR de la solution initiale de HgTe 4000 cm$^{-1}$ (solvant : toluène, ligands : DDT) en rouge, et de l'encre de HgTe 4000 cm$^{-1}$ (solvant DMF, ligands : MPOH et HgCl$_2$) en noir.*

Pour valider ce choix de ligands, il est nécessaire de vérifier les propriétés de photoconduction d'un film de nanocristaux avec cette chimie de surface. Pour cela, les nanocristaux doivent être déposés sous forme de film, puis des mesures de transport doivent être effectuées.

### c. Protocole de préparation de l'encre et paramètres de dépôt

Les nanocristaux de HgTe 4000 cm$^{-1}$, après la synthèse, sont dispersés dans le toluène, à une concentration de 50 mg/mL. Les ligands sont des dodécanethiols (DDT). La solution d'échange de ligands contient 5 mg de $HgCl_2$ et 100 μL de MPOH par mL de DMF.

### i. Protocole de préparation de l'encre

Le protocole de préparation de l'encre s'appuie sur la méthode d'échange de ligands par transfert de phases rappelée plus tôt dans ce chapitre (p97). Pour réaliser l'encre, un mélange équivolumique entre la solution de nanocristaux et la solution d'échange est réalisé. Le toluène et la DMF sont miscibles, il faut donc ajouter de l'hexane pour forcer la séparation de phases. Une fois les phases séparées, on constate que les nanocristaux ont migré dans la phase polaire, ce qui confirme que l'échange de ligands a été réalisé. La phase apolaire est alors retirée et de l'hexane est ajouté à la solution. Cette étape permet de retirer le maximum de ligands DDT libres qui sont restés dans la phase polaire. Ces derniers vont alors migrer dans la phase hexane. La phase hexane est alors retirée et l'étape est répétée deux fois.

Les ligands HgCl$_2$ et MPOH en excès doivent ensuite être retirés. Pour cela les nanocristaux sont centrifugés. L'ajout de quelques gouttes de mauvais solvant (éthanol ou toluène) peut favoriser la précipitation des nanocristaux. Le surnageant est ensuite jeté et les nanocristaux dispersés dans la DMF à une concentration de 250 mg/mL. Cette concentration très élevée est nécessaire à la réalisation d'un film épais en une seule étape de dépôt (*127, 128*).

### ii. Paramètres de dépôt

Les films sont déposés par *spin-coating*. Pour obtenir des films épais, des vitesses de rotation faibles, de l'ordre de 400 rotations par minute (rpm) sont utilisées. L'accélération doit également être lente : on utilise donc une accélération de 100 à 200 rpm/s. La DMF étant un solvant difficile à évaporer, la rotation dure 5 à 6 minutes pour faciliter le séchage. Enfin, la vitesse de rotation est augmentée à 2000 rpm pendant 1 à 2 minutes (accélération de 2000 rpm/s) pour éjecter les dernières gouttes de solvant. Sur substrat de verre, les films obtenus ont une épaisseur d'environ 300 nm et une rugosité de l'ordre de 20 nm, soit une à deux monocouches de nanocristaux (Figure 72). En faisant varier les vitesses et les accélérations des différentes étapes du dépôt, il est possible d'atteindre des épaisseurs de couches de 500 nm en une seule fois.

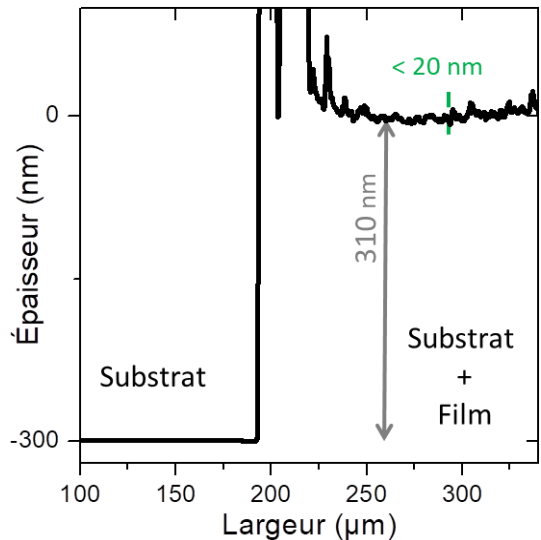

*Figure 72 : Épaisseur d'un film de nanocristaux HgTe 4000 cm⁻¹ obtenue par profilométrie mécanique. Entre 0 et 190 µm, le film a été retiré et la pointe scanne la surface du substrat nu. Au-delà de 190 µm, la pointe scanne la surface du film.*

## 2. Propriétés des films réalisés à base d'encre

Avant d'intégrer ces films de nanocristaux dans un détecteur infrarouge, il est nécessaire de vérifier leurs performances de photoconduction et leurs propriétés de stabilité dans le temps.

### a. Propriétés de photoconduction

Pour vérifier les propriétés de photoconduction de l'encre décrite au paragraphe précédent, quelques gouttes sont déposées par *spin-coating* sur des électrodes. Le courant dans l'obscurité et sous illumination est mesuré (Figure 73a) et une réponse de 1,1 mA/W est obtenue (Figure 73b). Cette valeur correspond à l'ordre de grandeur de réponse que l'on attend pour des films obtenus à partir d'échange de ligands EDT sur film, qui est la méthode utilisée par Amardeep Jagtap pour l'élaboration de sa diode (*123*). Le film obtenu à base d'encre est donc bien photoconducteur.

Pour valider la formulation choisie, j'ai comparé les performances du film obtenu avec l'encre utilisant les ligands MPOH et HgCl₂ avec celles de deux autres encres : une à base de ligands EDT, et une à base de ligands S²⁻. L'encre à base de ligands MPOH et HgCl₂ permet d'obtenir la conductivité la plus importante et la meilleure photoréponse (tableau de la Figure 73b).

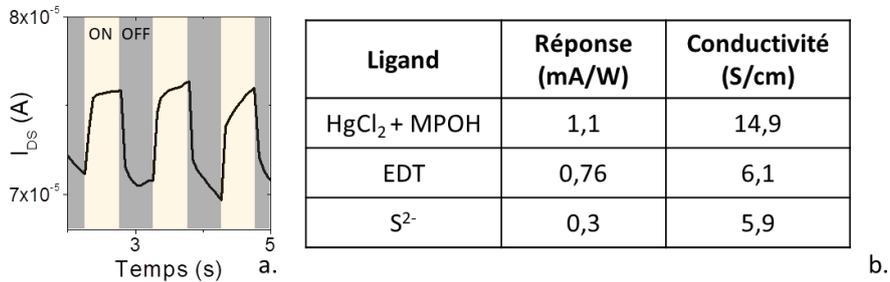

*Figure 73 : (a) Courant détecté par un dispositif photoconducteur à base de nanocristaux de HgTe 4000 cm⁻¹ préparés avec une encre HgTe MPOH – HgCl₂. Le laser utilisé pour l'illumination est un laser à 1,55 µm, à une puissance de 4 mW ; (b) tableau récapitulatif des propriétés de 3 encres : EDT, HgCl₂ MPOH et S²⁻. La réponse est mesurée à une tension de 1 V.*

Dans la suite du manuscrit, j'utiliserai le terme « encre » pour décrire l'encre à base de ligands MPOH et HgCl₂.

### b. Propriétés de transport – dopage

Avant d'intégrer les films à base d'encre dans une diode ESWIR, il faut également vérifier que l'énergie de Fermi est la même dans le cas des deux chimies de surface, échange de ligands EDT sur film et encre MPOH HgCl₂. En effet, les performances de la diode que je cherche à construire sont basées sur l'alignement des niveaux entre les différentes couches. Notamment, la couche de HgTe 6000 cm⁻¹ agit comme barrière pour les électrons tout en laissant passer les trous (Figure 68b et Figure 74a). Si l'énergie de Fermi de la couche de HgTe 4000 cm⁻¹ était modifiée, l'alignement des niveaux serait bouleversé et la barrière de HgTe 6000 cm⁻¹ ne jouerait plus son rôle (Figure 74b).

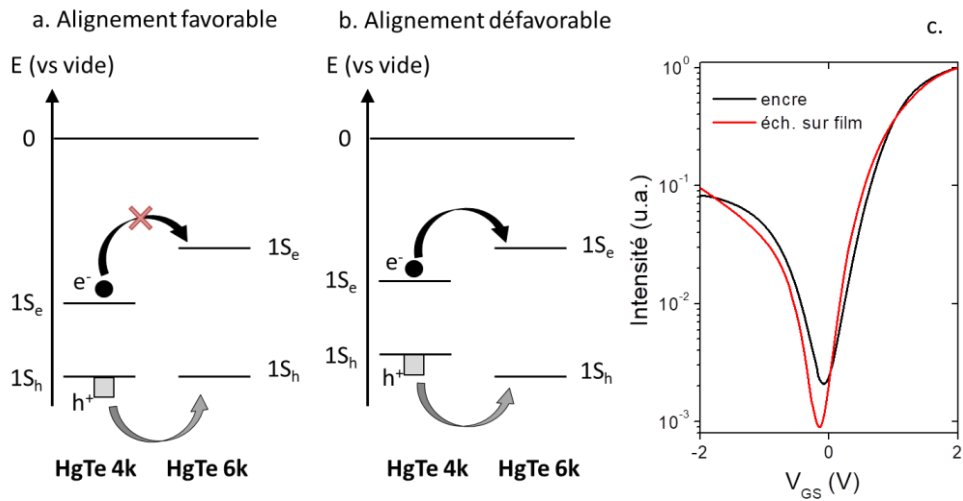

*Figure 74 : (a) Alignement favorable des niveaux électroniques entre les couches HgTe 4000 cm⁻¹ et 6000 cm⁻¹ (4k et 6k respectivement). Dans ce cas, la couche de HgTe 6k joue le rôle de barrière pour les électrons. (b) Alignement défavorable des niveaux électroniques. La couche de HgTe 6k ne joue plus son rôle de barrière. (c) Courbes de transfert de films de nanocristaux de HgTe 4000 cm⁻¹ préparés par échange de ligands EDT sur film (rouge) et sous forme d'encre (noir). Les courbes de transfert ont été normalisées par la valeur du courant à 1 V. $V_{DS} = 100$ mV dans le cas de l'échange sur film, $V_{DS} = 5$ mV dans le cas de l'encre.*

Afin de vérifier que l'alignement des niveaux reste inchangé entre les deux chimies de surface, j'effectue des mesures de transistor à effet de champ sur des films ayant été préparés par chacune des techniques (Figure 74c). Les courbes de transfert obtenues sont pratiquement superposables, ce qui confirme que les propriétés de transport, et donc les énergies des niveaux de Fermi, sont les mêmes dans les deux cas. La couche de HgTe 4k préparée sous forme d'encre peut donc être incluse dans la diode sans modifier les propriétés d'alignement des niveaux entre les différentes couches.

Les mesures de transistor à effet de champ sur des grilles Si/SiO$_2$ de capacité connue permettent également de remonter à la mobilité et donc à la longueur de diffusion des porteurs au sein du film. Alors que cette longueur de diffusion avait été estimée à une dizaine de nanomètres pour un film de nanocristaux de HgTe échangés avec des ligands EDT (échange sur film), elle est mesurée à environ 200 nm dans le cas de nanocristaux à base d'encre. L'amélioration de l'absorption des photons infrarouge obtenue en augmentant l'épaisseur du film à 500 nm sera donc efficace puisque les photo-porteurs présents dans la quasi-intégralité du film pourront rejoindre les électrodes.

### c. Stabilité des films à base d'encre

Les performances des films de nanocristaux de chalcogénures de mercure peuvent rapidement se dégrader dans le temps. Amardeep Jagtap a montré que le courant d'obscurité d'un dispositif photoconducteur à base de nanocristaux de HgTe 4000 cm$^{-1}$ pouvait augmenter d'un facteur 100 en à peine une dizaine de jours lorsque ce dernier est stocké à l'air (*123*).

J'ai donc étudié l'évolution de la résistance électrique de films de nanocristaux de HgTe. La résistance est directement proportionnelle au courant d'obscurité et donc au niveau de bruit dans le système : plus elle est faible, plus le niveau de bruit est important. J'ai comparé les performances de films de HgTe 4000 cm$^{-1}$ préparés par échange de ligands EDT sur film et de films préparés à base d'encre. Les résultats obtenus sont présentés sur la Figure 75a. Alors que la résistance du film HgTe EDT est divisée par mille en 8 jours, la résistance du film HgTe encre ne diminue que d'un facteur 3. Cela est probablement dû au fait que les ligands MPOH et HgCl$_2$ permettent une meilleure passivation de la surface. Les films préparés à base d'encre présentent donc une meilleure stabilité dans le temps.

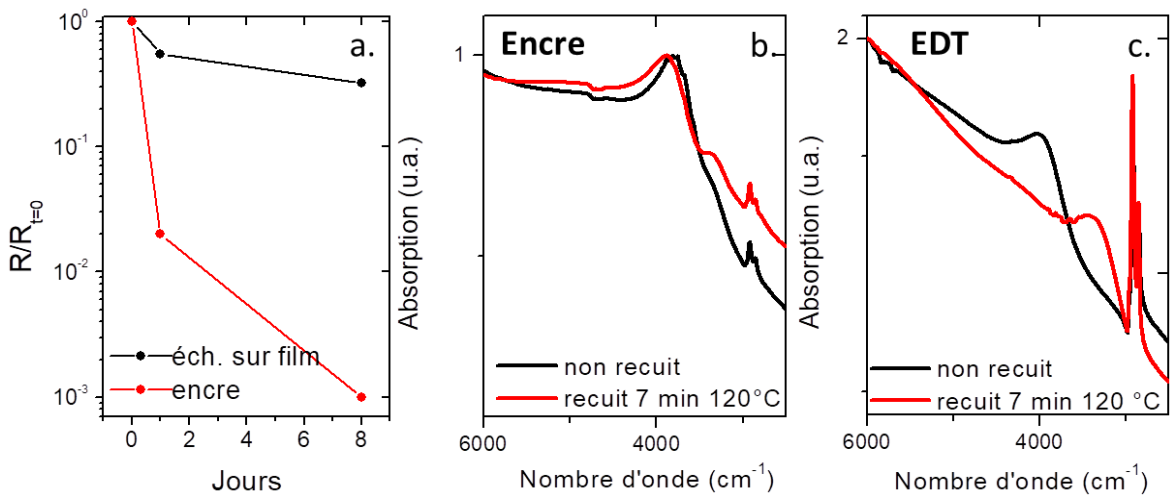

*Figure 75 : Influence du vieillissement et d'un recuit sur des films à base de nanocristaux HgTe 4000 cm$^{-1}$ : les ligands ont soit été échangés sur le film (HgTe EDT) soit en solution (HgTe encre). (a) Évolution de la résistance électrique de films de nanocristaux, en noir pour HgTe EDT, en rouge pour HgTe encre. Les résultats ont été normalisés par la valeur de la résistance du film à t = 0 jour ; (b) spectre infrarouge du film HgTe 4000 cm$^{-1}$ encre avant (noir) et après un recuit de 7 minutes à 120 °C (rouge) ; (c) spectre infrarouge d'un film HgTe EDT avant (noir) et après un recuit de 7 minutes à 120 °C (rouge).*

La stabilité en température des films de nanocristaux de chalcogénures de mercure est également un paramètre important à étudier. En effet, pendant la préparation de films HgTe par échange de ligands sur film, il est préférable d'éviter les étapes de recuit. Les nanocristaux étant synthétisés à basse température (entre 80 et 120 °C), les recuits peuvent entraîner une agrégation voire une fusion des nanocristaux entre eux. La transition optique se décale alors vers les plus basses énergies et la structure de l'exciton se détériore, signe que les nanocristaux sont de plus en plus polydisperses (*44, 123*). La diminution de l'énergie de bande interdite entraîne également une augmentation dramatique du courant d'obscurité, et réduit donc les performances du détecteur. Or, les étapes de recuit peuvent être utiles dans le processus de fabrication de détecteurs infrarouges à base de nanocristaux. Le recuit permet d'évaporer les solvants plus rapidement, notamment les solvants polaires à haut point d'ébullition, et donc d'accélérer la fabrication du détecteur. Dans l'optique de réaliser des détecteurs plus complexes, les recuits sont également nécessaires pour réaliser des motifs sur le film, par exemple des pixels (*129*), via des techniques de lithographie.

L'influence du recuit sur deux films de HgTe 4000 cm$^{-1}$, l'un réalisé à base d'encre et l'autre préparé par un échange de ligands EDT sur film, a donc été étudiée par spectroscopie infrarouge. Le spectre d'un film à base d'encre n'est pratiquement pas affecté par un recuit de 7 min à 120 °C (Figure 75b), tandis qu'un film préparé par un échange de ligands sur film se décale de plusieurs centaines de cm$^{-1}$ vers le rouge (Figure 75c). La stabilité en température des films à base d'encre est donc bien meilleure que celle des films réalisés par échange de ligands sur film.

### d. Bilan sur les propriétés de l'encre MPOH HgCl$_2$

L'encre à base de ligands MPOH et HgCl$_2$ permet donc d'obtenir des films photoconducteurs, moins sensibles au vieillissement et au recuit que les films obtenus par la méthode d'échange de ligands sur film. De plus, cette encre permet d'obtenir des films épais (plusieurs centaines de nanomètres) en une seule étape de dépôt, et donc d'améliorer l'absorption. Dans la suite de ce chapitre, je vais donc l'utiliser pour réaliser un dispositif de détection infrarouge photovoltaïque.

### III. Diode ESWIR à base d'encre

L'encre que j'ai optimisée et caractérisée dans la partie précédente est utilisée pour concevoir un dispositif de détection infrarouge dans la gamme ESWIR (4000 cm$^{-1}$, 2,5 µm, 500 meV). Je commencerai ce chapitre en présentant la structure de la diode étudiée. Je caractériserai ensuite la diode et déterminerai ses performances en tant que détecteur infrarouge. Une analyse du photocourant à plusieurs tensions permettra également de déterminer les différents régimes de fonctionnement de la diode.

Je démontrerai que cette diode peut être utilisée pour détecter de faibles puissances, de l'ordre de quelques dizaines de nanowatts. Je vérifierai en particulier que, contrairement aux détecteurs commerciaux QWIP que j'ai présentés dans le chapitre 1, aucune règle de sélection sur l'incidence des photons ne s'applique pour cette diode.

#### 1. Structure de la diode

Pour améliorer les performances de la diode ESWIR proposée par Amardeep Jagtap (*24*), je vais simplifier la structure en retirant la barrière de filtration des trous TiO$_2$, et augmenter l'épaisseur de la couche active de nanocristaux de HgTe 4000 cm$^{-1}$ en utilisant l'encre décrite dans la partie précédente.

La structure de diode que j'ai réalisée est comparée à celle proposée par Amardeep Jagtap (*24*) sur la Figure 76. La diode proposée par Amardeep Jagtap se réalise en une quinzaine d'étapes, il faut donc au minimum quatre heures pour la préparer. Celle que je propose peut se réaliser en moins de cinq étapes de dépôt, ce qui réduit le temps de fabrication à deux heures environ.

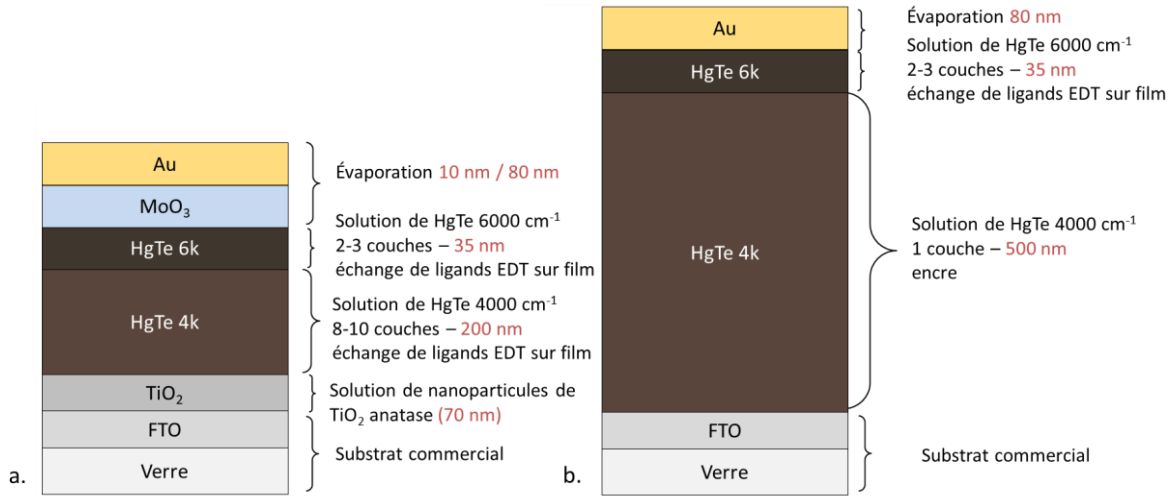

*Figure 76 : (a) Principe de fabrication de la diode ESWIR réalisée par Amardeep Jagtap (24). Pour réaliser le contact supérieur, 10 nm de MoO₃ puis 80 nm d'or sont déposés ; (b) principe de fabrication de la diode ESWIR présentée dans ce manuscrit.*

Expérimentalement, les substrats de FTO sur verre sont découpés en carrés de 15 mm × 15 mm. Ils sont ensuite gravés en utilisant une gravure acide catalysée par une poudre de zinc pour reproduire les motifs présentés sur la Figure 77 (gris foncé). Les couches de nanocristaux sont ensuite déposées, HgTe 4000 cm⁻¹ en encre et HgTe 6000 cm⁻¹ par un échange de ligands sur film. Le film est ensuite nettoyé sur la partie supérieure du substrat (partie 1 sur la Figure 77). Enfin, une électrode d'or est évaporée à travers un masque pour compléter la structure. Huit pixels avec une électrode d'or commune sont obtenus. Pour faire fonctionner la diode, l'électrode d'or (donc le contact 1) joue le rôle de drain et l'électrode de FTO (un des pixels 2 à 9) de source.

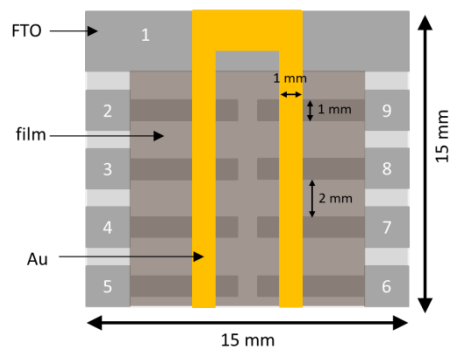

*Figure 77 : Schéma de la diode aux huit pixels obtenue. Le film de nanocristaux est nettoyé au niveau de la bande horizontale continue (en haut du schéma) pour que l'or soit en contact avec le FTO sur cette partie. Les contacts seront donc pris en 1 pour l'or, et sur un des autres pixels (2 – 9) pour le FTO.*

Pour vérifier l'épaisseur des couches de HgTe déposées, une des diodes est coupée en deux et la tranche est analysée par microscopie électronique à balayage (une des images obtenues est présentée sur la Figure 78). Les couches correspondants aux nanocristaux HgTe 4000 cm⁻¹ et 6000 cm⁻¹ ne sont pas différenciables sur le cliché, je peux donc uniquement mesurer l'épaisseur du bloc HgTe, qui est de 560 nm. Des calibrations précédentes dans le groupe ont montré que l'épaisseur d'un film de HgTe 6000 cm$^{-1}$ préparé dans les mêmes conditions que les miennes était de l'ordre de 50 nm. Je peux donc estimer l'épaisseur de la couche active de HgTe 4000 cm$^{-1}$ à environ 500 nm.

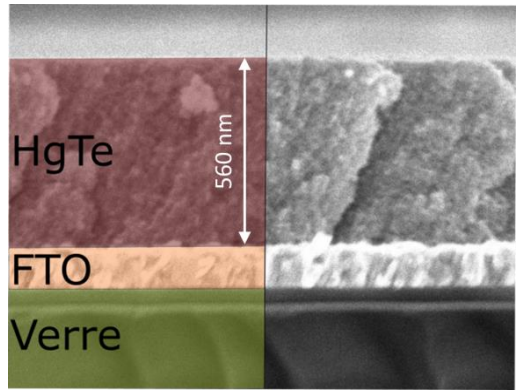

*Figure 78 : Image en coupe de la diode obtenue par microscopie électronique à balayage. Sur la partie gauche de l'image, des fausses couleurs ont été ajoutées pour différencier les différentes couches.*

### 2. Performances de la diode obtenue

Les performances de la diode seront caractérisées via trois figures de mérite usuelles pour les détecteurs infrarouges : la **réponse** (A/W), qui quantifie la capacité du détecteur à convertir les photons infrarouges en signal électrique, le **temps de réponse**, et la **détectivité** (Jones), qui permet de quantifier le rapport signal sur bruit du détecteur.

### a. Photoréponse

Le dispositif fabriqué est photovoltaïque. Il présente donc une caractéristique courant-tension asymétrique, comme présenté sur la Figure 79a. La gamme de tensions autour de 0 V est la plus intéressante pour des applications de détection infrarouge puisque c'est dans cette gamme que le courant d'obscurité est le plus faible tandis que le photocourant peut circuler grâce au champ électrique interne du dispositif (encart de la Figure 79a).

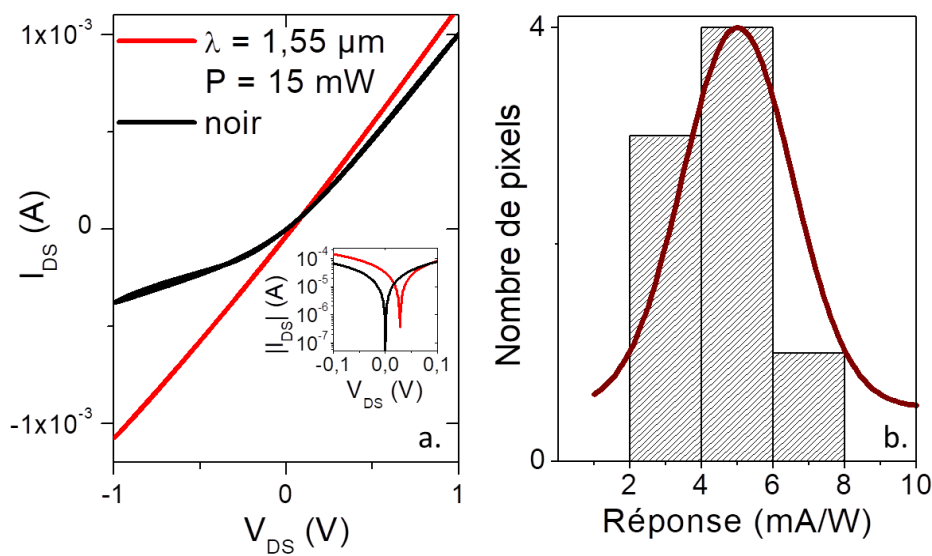

*Figure 79 : Caractéristique courant-tension obtenue pour un pixel de diode. L'illumination est effectuée par un laser à 1,55 μm, sous une puissance de 15 mW. Encart : Zoom sur la gamme [-0,1 V ; 0,1 V], en échelle logarithmique ; (b) histogramme des valeurs de réponse obtenues pour les huit pixels d'un même substrat.*

Une diode contenant 8 pixels, les caractéristiques courant-tension des huit pixels peuvent être mesurés et les amplitudes des réponses sont déterminées pour les huit pixels. Les huit pixels fonctionnent, c'est-à-dire que tous présentent une photoréponse sous illumination, et l'histogramme des réponses est présenté sur la Figure 79b. Tous les pixels présentent une réponse de l'ordre de quelques mA/W à 0 V : la moyenne est de 5 mA/W.

La gaussienne modélisant l'histogramme des réponses à 0 V (Figure 79b) présente une largeur à mi-hauteur de 3 mA/W, soit de l'ordre de 50 % de la valeur moyenne de réponse. Cette large distribution des réponses peut être due à la non-homogénéité du film. En effet, à cause de la faible vitesse de rotation pendant le dépôt, la solution de nanocristaux peut s'accumuler sur les bords du substrat sous forme de bourrelets, l'épaisseur du film sera donc plus importante sur les bords qu'au centre de l'échantillon. Lorsque plus de photons sont absorbés, la réponse peut être plus importante. Ainsi, sur l'échantillon utilisé pour effectuer les mesures de la Figure 79b, le pixel donnant la plus grande réponse est le pixel n°6 (au bord, donc plus épais), le pixel donnant la plus faible réponse est le pixel n°2 (au centre, donc plus fin). Les numérotations correspondent au schéma de la Figure 77.

Une fois l'amplitude de la réponse étudiée, il est ensuite nécessaire d'étudier la dynamique de la diode. Pour cela, j'étudie la réponse du détecteur à 0 V, sur une gamme de fréquences allant du continu à 1 kHz, grâce au montage schématisé sur la Figure 80a.

Un corps noir à 980 °C est utilisé comme source et un filtre en germanium permet de couper les photons à haute énergie ($\lambda < 1{,}9$ µm). Un modulateur optique (ou *optical chopper*) permet de moduler l'intensité lumineuse (0% ou 100%) à une fréquence choisie. Le photocourant généré par la diode est ensuite collecté par un amplificateur de transimpédance qui transmet le signal à un oscilloscope. Pour calculer la réponse, le photocourant mesuré sur l'oscilloscope (en A) est divisé par la puissance lumineuse émise, soit 65 µW dans le cas du montage utilisé. La source utilisée, le corps noir, a une puissance (65 µW) plus proche des conditions réelles d'utilisation que le laser infrarouge utilisé précédemment (4 mW). Or, dans les détecteurs à base de nanocristaux, la réponse augmente lorsque la puissance optique diminue. La photoréponse observée dans le cas où la source est un corps noir est souvent supérieure à celle observée avec un laser infrarouge.

Dans la gamme [1 Hz ; 1000 Hz], la réponse mesurée est quasiment constante, autour de 20 mA/W (Figure 80b). La dynamique de cette diode est donc compatible avec des applications de type imagerie. À noter que la réponse obtenue est 100 fois supérieure à celle obtenue sur la génération de diodes précédente (200 µA/W dans les mêmes conditions) (*24*), grâce à l'augmentation de l'absorption ($1\text{-}10^{-\alpha l} \approx 40\%$) et à la suppression de la barrière de $TiO_2$ filtrant les photoélectrons.

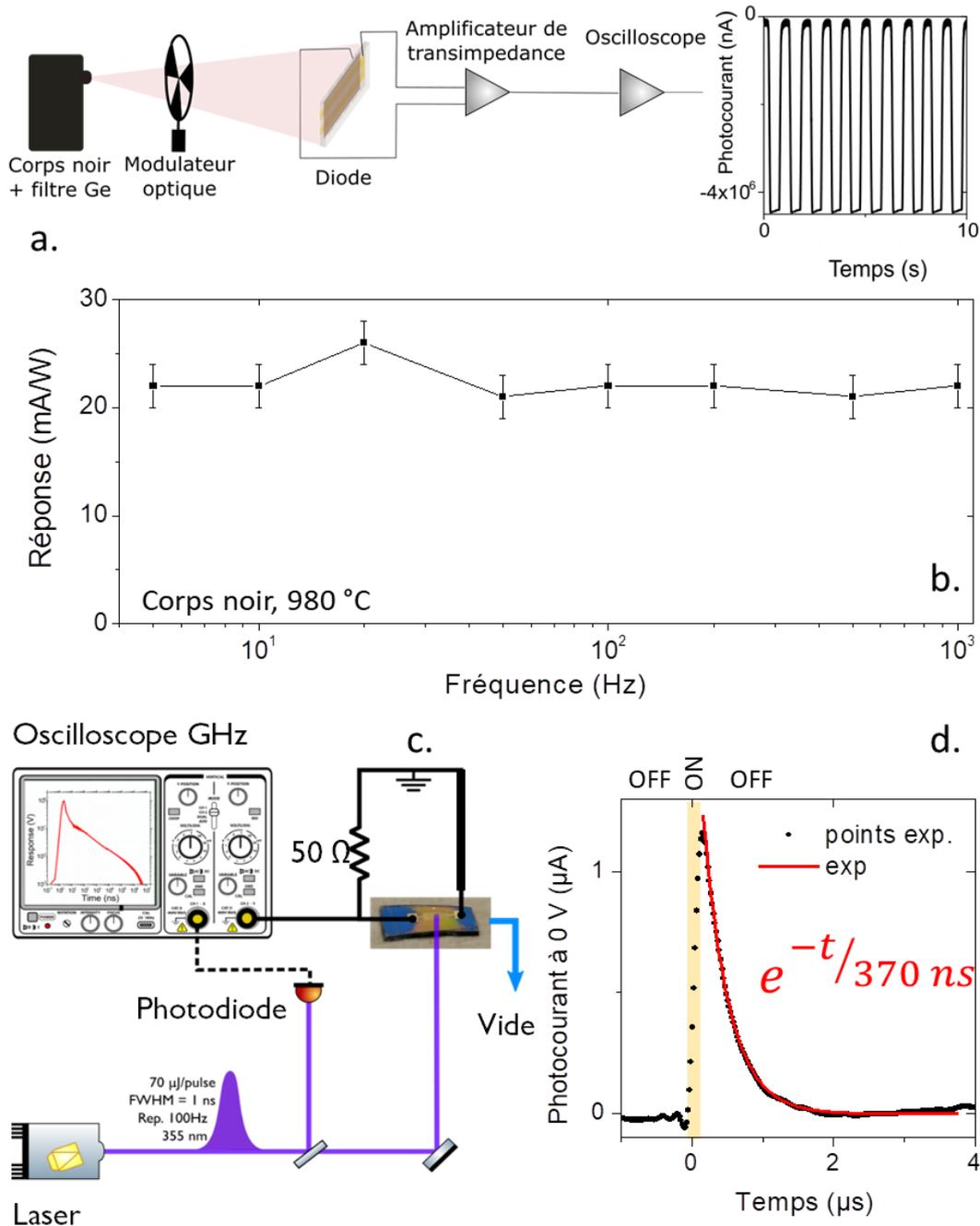

*Figure 80 : (a) Schéma du montage permettant la mesure de réponse fréquentielle de la diode ; (b) évolution de la réponse de la diode en fonction de la réponse ; (c) schéma du montage GHz utilisé pour déterminer le temps de réponse de la diode ; (d) évolution du photocourant à 0 V de la diode en réponse à une impulsion lumineuse (en jaune sur le graphe) de quelques nanosecondes.*

Afin de déterminer le temps de réponse de la diode, qui est donc bien inférieur à la milliseconde, il est nécessaire de modifier le montage, les modulateurs optiques utilisés n'étant pas compatibles avec des fréquences supérieures au kilohertz. J'utilise donc le montage que j'ai introduit dans le chapitre 2 (et rappelé sur la Figure 80c). Ce montage permet d'étudier la décroissance du photocourant après une impulsion lumineuse de quelques nanosecondes.

Le temps de réponse de la diode est modélisé grâce à une exponentielle décroissante et vaut environ 370 ns (Figure 80d).

### b. Détectivité

Pour déterminer la détectivité de la diode, il faut notamment déterminer la densité spectrale de bruit. Pour cela, j'utilise le montage décrit dans le schéma de la Figure 81a. Le courant d'obscurité, à 0 V, est collecté par un amplificateur de transimpédance et renvoyé sur un analyseur de spectres qui détermine la densité spectrale de bruit $S_I$, en $A.Hz^{-0,5}$. Dans le cas d'un film de nanocristaux, une dépendance du carré de la densité spectrale de bruit en 1/f est généralement observée (voir chapitre 1, p34). Dans le cas de la diode, on observe cette dépendance pour des fréquences inférieures à 100 Hz (droite en pointillés rouge sur le graphe de la Figure 81a). Pour des fréquences supérieures, on observe une déviation de la dépendance en 1/f, d'autres sources de bruit, probablement du bruit électronique lié à la mesure, limitent la densité spectrale de bruit.

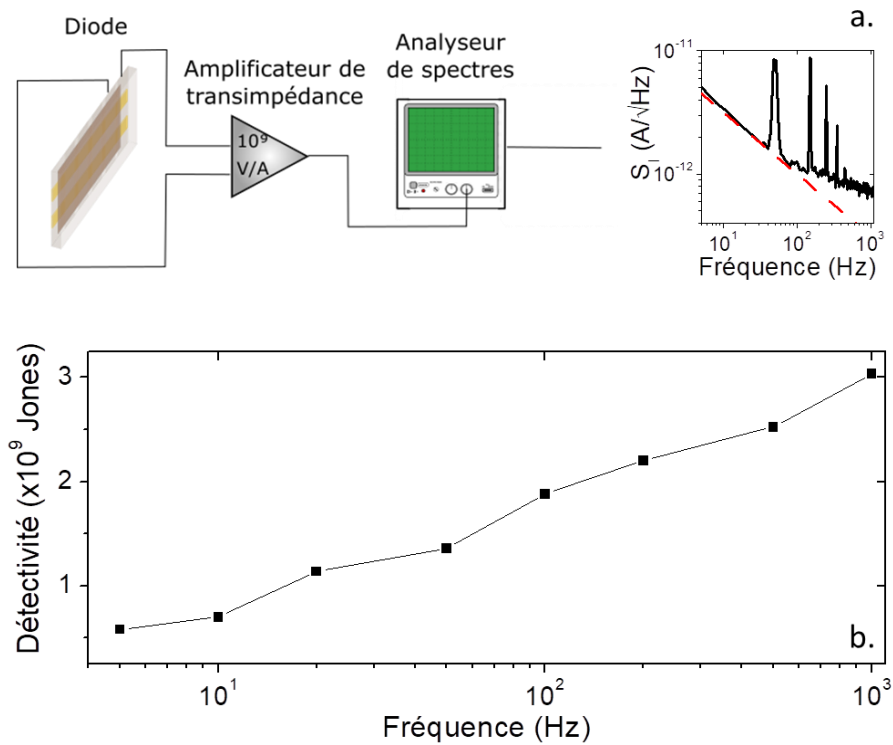

*Figure 81 : Mesures de détectivité sur la diode. (a) Schéma du montage utilisé ; (b) détectivité à 0 V de la diode à plusieurs fréquences*
.

Pour déterminer la détectivité, il faut ensuite multiplier la réponse $\mathcal{R}$ (Figure 80b) par le carré de l'aire du pixel (0,1 cm) et diviser par la densité spectrale de bruit (voir équation 4.1). On obtient alors les résultats de la Figure 81b.

$$D^* = \frac{\sqrt{A}.\,\mathcal{R}}{S_I} \qquad\qquad (4.1)$$

La réponse est constante entre 1 et 1000 Hz, et la densité spectrale de bruit diminue quand la fréquence augmente. Par conséquent, la détectivité maximale est obtenue à des fréquences élevées. Ainsi, une détectivité de $3 \times 10^9$ Jones à 1 kHz, à 0 V est obtenue, soit un ordre de grandeur au-dessus de la détectivité obtenue pour la génération de diodes précédentes.

La barrière de TiO₂ présente dans la génération précédente permettait de réduire le courant d'obscurité dû aux trous. Pour faciliter le transport des photoélectrons vers l'électrode FTO, cette barrière a été retirée dans la diode que j'ai étudiée. Avec cette diode, le courant d'obscurité, donc le bruit, est plus élevé. C'est une des raisons pour laquelle la détectivité n'est améliorée que d'un facteur 10 tandis que la réponse est améliorée d'un facteur 100 avec la structure de diode que je propose.

### 3. Analyse des différents régimes de fonctionnement de la diode

Pour aller plus loin dans la compréhension du fonctionnement de la diode, je vais étudier les différentes contributions du photocourant en réalisant des spectres de photocourant. Ces spectres indiquent l'intensité du photocourant en fonction du nombre d'onde de la lumière incidente. Par exemple, dans le cas de la diode étudiée, aucun matériau n'absorbe la lumière dont le nombre d'onde est inférieur à 3500 cm⁻¹. Le photocourant est donc nul dans cette gamme. Vers 4000 cm⁻¹, les HgTe 4000 cm⁻¹ commencent à absorber la lumière et génèrent un courant : le photocourant augmente. Au-delà de 6000 cm⁻¹, la contribution au photocourant des HgTe 6000 cm⁻¹ s'ajoute à celle des HgTe 4000 cm⁻¹ qui continuent d'absorber. Ces spectres permettent donc de connaître la contribution au photocourant global des différents matériaux constituants la diode.

Des spectres de photocourant à différentes tensions ont été enregistrés et normalisés par rapport à la contribution de la couche active de HgTe 4000 cm⁻¹ (Figure 82b). Pour faciliter la compréhension du lecteur, les caractéristiques courant-tension dans le noir et sous illumination sont rappelées sur la Figure 82a et les tensions auxquelles ont été mesurés les spectres de photocourant ont été indiquées.

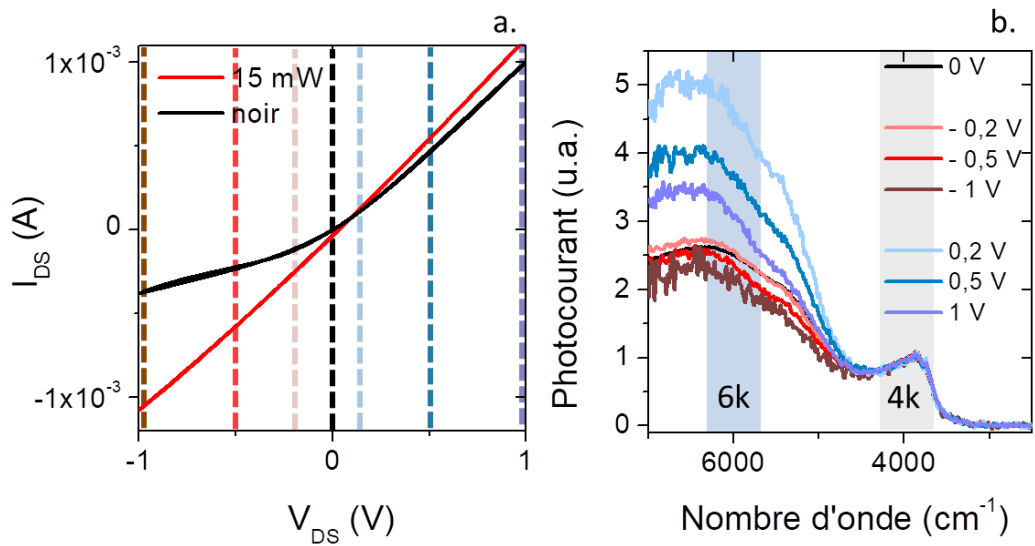

*Figure 82 : (a) Caractéristiques courant-tension d'un des pixels de la diode dans l'obscurité (courbe pleine en noir) et sous illumination (courbe pleine en rouge). Les lignes verticales en pointillés correspondent aux tensions utilisées pour la partie b. (b) Spectres de photocourant du même pixel enregistrés à différentes tensions.*

Les spectres de photocourant sont similaires pour des tensions drain-source négatives, tandis que la contribution des HgTe 6000 cm⁻¹ est plus importante (relativement à la contribution des HgTe 4000 cm⁻¹) pour des tensions drain-source positives.

Les ratios du photocourant obtenu pour une illumination à 4000 cm⁻¹ sur le photocourant obtenu pour une illumination à 6000 cm⁻¹ ont été tracés sur la Figure 83e. Les alignements des niveaux électroniques à différentes tensions ont également été reportés (Figure 83a, b, c et d). Pour des tensions drain-sources négatives, la barrière unipolaire de HgTe 6000 cm⁻¹ ne bloque pas le photocourant issu de la couche de HgTe 4000 cm⁻¹. Les contributions au photocourant de HgTe 4000 cm⁻¹ et de HgTe 6000 cm⁻¹ parviennent à l'électrode de FTO, quelle que soit la tension drain-source (négative) appliquée. Pour des tensions drain-source positives (> V$_{OC}$), les photoélectrons doivent rejoindre l'électrode d'or. À des faibles tensions drain-source, c'est-à-dire lorsque le produit e.F.L, où e est la charge élémentaire, F le champ électrique et L l'épaisseur de la diode, est faible devant la différence d'énergie entre la couche de HgTe 4000 cm$^{-1}$ et la barrière de 6000 cm$^{-1}$, les photoélectrons issus de la couche de HgTe 4000 cm$^{-1}$ sont filtrés par la barrière unipolaire. Le ratio HgTe 4000 cm$^{-1}$/6000 cm$^{-1}$ diminue. En augmentant la tension drain-source, la probabilité de passage tunnel des photoélectrons de la couche de HgTe 4000 cm$^{-1}$ à travers la barrière augmente, le ratio HgTe 4000 cm$^{-1}$/6000 cm$^{-1}$ augmente avec la tension drain-source.

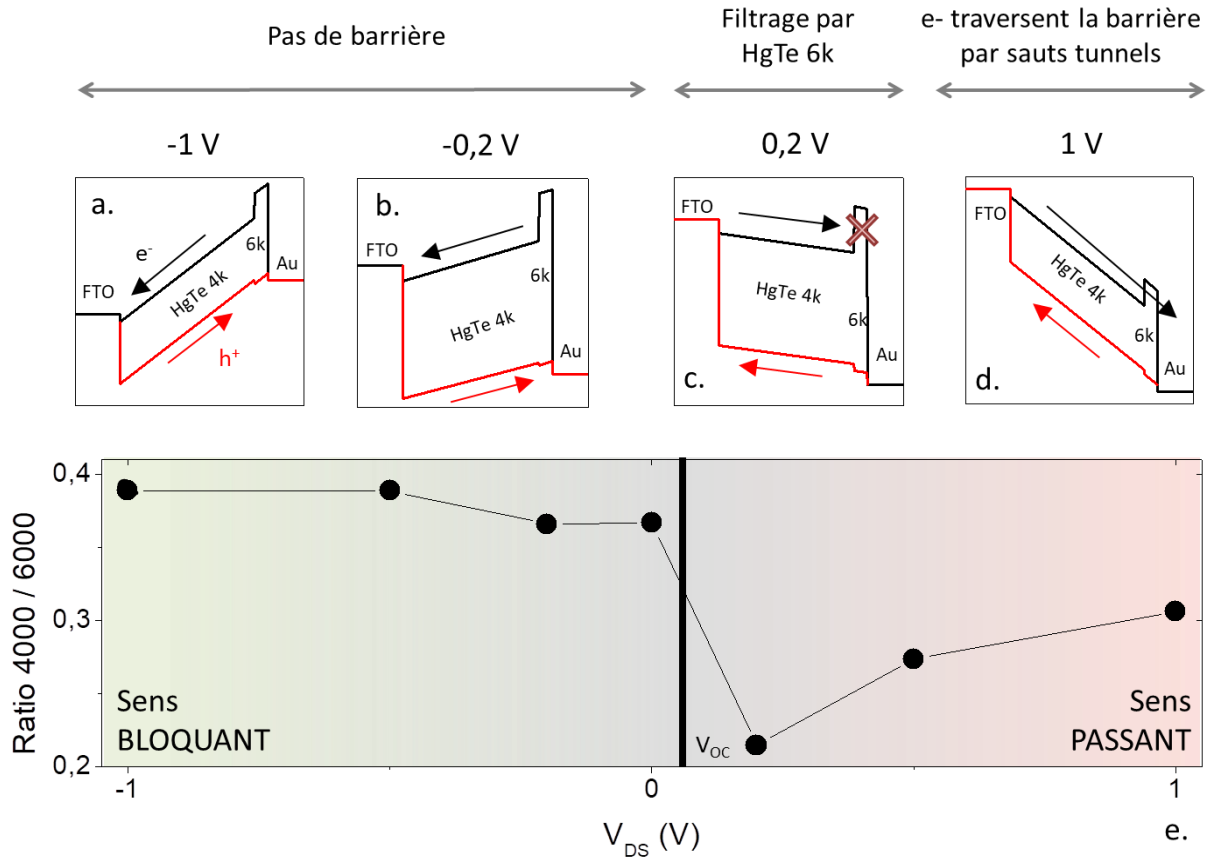

*Figure 83 : (a), (b), (c) et (d) sont des schémas correspondant à l'alignement des niveaux électroniques dans la diode pour des tensions de -1 V, -0,2 V, 0,2 V et 1 V respectivement. De gauche à droite, on retrouve les éléments FTO / HgTe 4000 cm$^{-1}$ (4k) / HgTe 6000 cm$^{-1}$ (6k) / Au. (e) Ratio des contributions 4000 cm$^{-1}$ /6000 cm$^{-1}$ en fonction de la tension drain-source appliquée. La barre noire verticale correspond à la tension de circuit ouvert : pour V$_{DS}$ > V$_{OC}$, les électrons rejoignent l'électrode d'or, pour V$_{DS}$ < V$_{OC}$, les électrons rejoignent l'électrode de FTO.*

### 4. Détection de faibles puissances et directionnalité

Dans cette dernière partie consacrée à l'étude de la diode ESWIR, je vais explorer les propriétés de détection dans des conditions plus difficiles : pour des faibles puissances lumineuses, de l'ordre quelques dizaines de nanowatts, et pour différents angles.

#### a. Détection de faibles puissances

Pour obtenir une source lumineuse de faible puissance, nous décidons d'utiliser un corps noir éloigné de quelques mètres de la diode. Pour une distance diode – corps noir de 3,8 m, la puissance émise est de 260 nW lorsque le corps noir est à 980 °C, elle est de 61 nW lorsque le corps noir est à 700 °C.

Le photocourant obtenu pour de faibles puissances est inférieur au nanoampère. Pour le détecter, le montage de mesure de la réponse que j'ai présenté plus tôt (p107) ne suffit pas. Ce montage ne comptait qu'une seule étape d'amplification avec un amplificateur de transimpédance. Pour détecter des signaux plus faibles, j'utilise une seconde étape d'amplification en ajoutant un détecteur synchrone et un modulateur optique, qui permet d'amplifier la composante du photocourant à 100 Hz. Enfin, pour pouvoir observer le courant d'obscurité et le courant sous illumination, un modulateur optique à 1 Hz est ajouté : le courant final observé sur l'oscilloscope est donc un créneau de fréquence 1 Hz. Le schéma de ce montage est présenté sur la Figure 84a, et une photo du montage est présentée sur la Figure 84b.

Les mesures de photocourant en fonction de la puissance reçue par la diode sont présentées sur la Figure 84c. **Des faibles puissances de 61 nW sont détectables par la diode**.

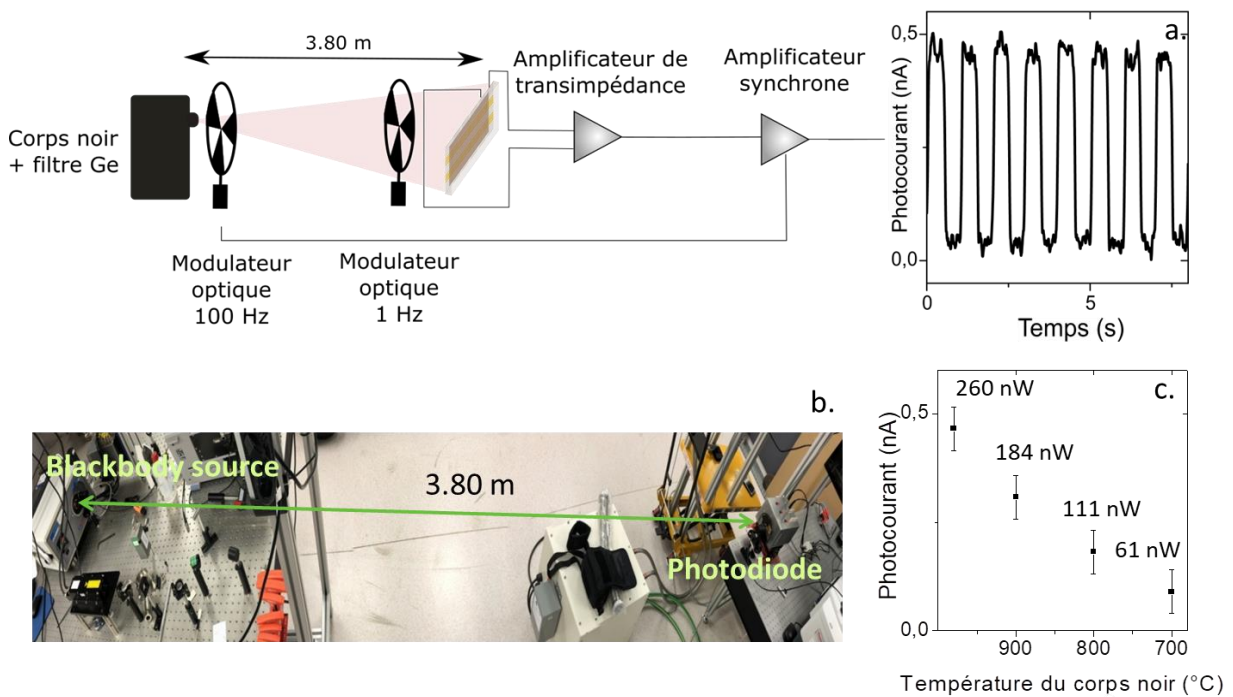

*Figure 84 : Détection longue distance. (a) Schéma du montage ; (b) photo du montage ; (c) mesures de photocourant en fonction de la température du corps noir. Les puissances reçues par le pixel ont été calculées et sont indiquées pour chaque température.*

### b. Directionnalité

Parmi les dispositifs de détection infrarouge commerciaux, certains présentent des règles de sélection sur l'angle d'incidence des photons. Dans les détecteurs à puits quantiques (QWIPs), les photons en incidence normale ne peuvent pas être absorbés. Des réseaux à la surface du matériau actif doivent donc être gravés pour optimiser l'absorption des photons (*30*).

Dans le cas de diodes à base de nanocristaux isotropes, comme c'est le cas de la diode ESWIR présentée dans ce chapitre, aucune direction n'est privilégiée donc l'angle d'incidence des photons ne devrait pas avoir d'influence. Pour le vérifier, la réponse angulaire de la diode est mesurée et les résultats sont reportés sur la Figure 85.

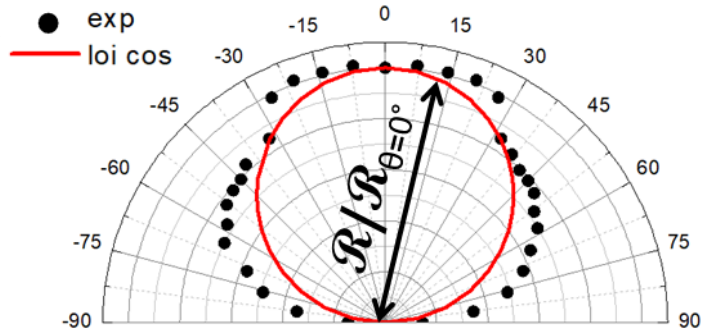

*Figure 85 : Mesure de la réponse de la diode en fonction de l'angle d'incidence de la source (la référence est prise par rapport à l'incidence normale). Le laser utilisé est un laser 1,55 µm à une puissance de 4 mW. Les points noirs correspondent à la réponse de la diode divisée par la réponse à incidence normale (0°). Le cercle rouge représente une évolution du cosinus en fonction de l'angle.*

La réponse décroit avec l'angle en suivant une loi de type cosinus, comme attendu dans le cas où aucune règle de sélection ne s'applique.

## 5. Diode ESWIR – bilan

Grâce à l'encre de nanocristaux de HgTe 4000 cm$^{-1}$ que j'ai développée, j'ai pu obtenir une diode dans l'ESWIR dont les performances ont été améliorées par rapport aux générations précédentes développées dans notre groupe, voir Figure 86. La réponse a été augmentée d'un facteur 10$^4$ et la détectivité d'un facteur 300 entre la première génération au début de l'année 2018 et la diode présentée dans ce manuscrit.

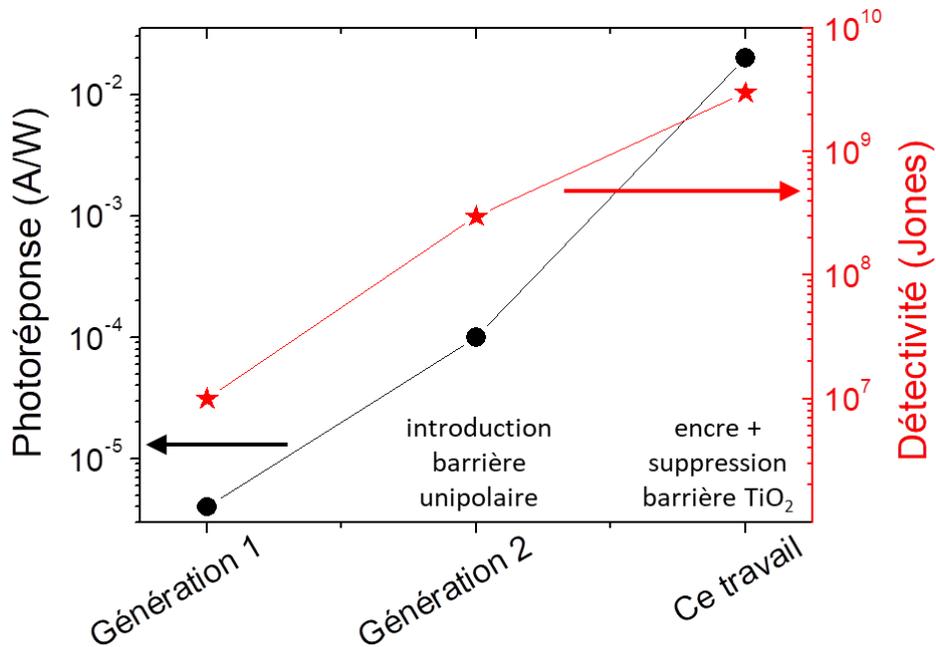

*Figure 86 : Évolution des performances de diode à base de HgTe 4000 cm$^{-1}$ dans notre groupe. Les cercles noirs correspondent à la photoréponse, les étoiles rouges à la détectivité.*

Pour encore augmenter la réponse et la détectivité de ces diodes. Une possibilité est d'utiliser des couches de transport d'électrons telles que $Bi_2Se_3$, qui permettent d'améliorer l'extraction des photoélectrons et donc d'augmenter encore la réponse de ces diodes (*130*).

## IV.     Fabrication de dispositifs multipixels

Dans le but de réaliser une caméra, et donc de reconstruire une image, des dispositifs multipixels sont nécessaires. Dans les dispositifs de détection infrarouge commerciaux, des systèmes de détection multipixels couplés à des circuits de lecture (*read-out integrated circuits*, ROIC) peuvent être utilisés. Les pixels font quelques dizaines de microns de large et un dispositif peut contenir quelques dizaines de milliers de pixels. L'intégration de nanocristaux dans ces dispositifs d'imagerie a déjà été testée avec succès pour des HgTe 2000 cm$^{-1}$ (*131*), voir l'image de la Figure 87a et pour des nanocristaux de PbS comme sur le schéma de la Figure 87b (*132–135*).

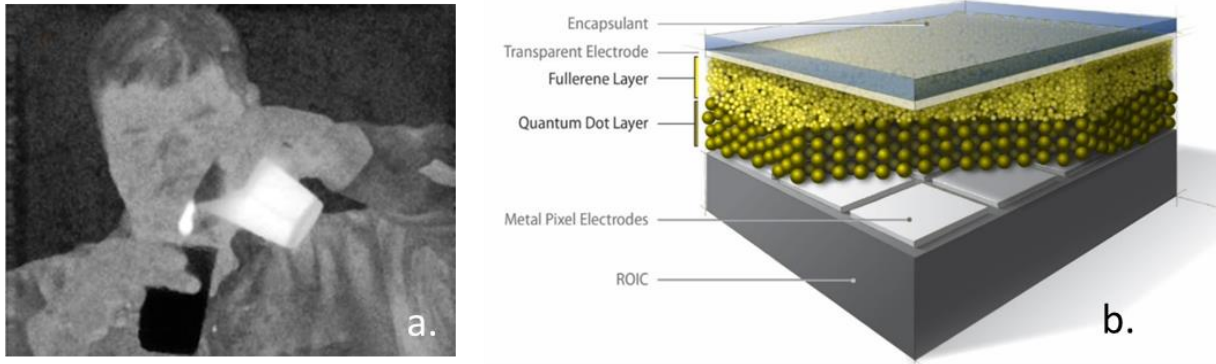

*Figure 87 : (a) Données issues de l'article « MWIR Imaging With Low Cost Colloidal Quantum Dot Films » par C. Buurma et al. en 2016 (131) - image prise avec un détecteur infrarouge. Le matériau actif est HgTe et l'image a été réalisée grâce à un circuit de lecture ROIC dont les pixels ont une largeur de 30 μm. (b) Données issues de l'article « PbS colloidal quantum dot photodiodes for low-cost SWIR sensing » de E. J. D. Klem et al. en 2015 (132) – schéma d'un détecteur photovoltaïque multipixels à base de PbS (quantum dot layer) sur un circuit de lecture ROIC.*

Cependant, ces dispositifs commerciaux sont coûteux (500 $/ pièce) et nécessitent une intégration dans un système de mesure du courant électrique complexe. Pour commencer à appréhender les contraintes dues aux composants multipixels (homogénéité, stabilité, couplage entre pixels …), tout en tenant compte de la fragilité des nanocristaux de tellure de mercure, nous avons décidé de construire notre propre matrice multipixels. Pendant mon doctorat, j'ai donc travaillé à l'élaboration d'une matrice de 100 pixels. L'homogénéité du signal obtenu après dépôt des nanocristaux est étudiée, et la matrice est utilisée pour réaliser une première image de faisceau laser infrarouge.

### 1.     Conception et fabrication de la matrice de pixels

Je propose de réaliser une matrice de 10 × 10 pixels, dont les pixels font une trentaine de μm de large et sont séparés de 6 μm, en utilisant la photolithographie UV. Cette taille de pixels est proche des tailles utilisées pour les pixels des circuits de lecture ROIC. Les étapes de préparation de cette matrice de pixels sont toutes réalisées dans la salle blanche de l'Institut des Nanosciences de Paris.

Le motif de matrice choisi est celui présenté sur la Figure 88a. Chaque pixel est composé d'une électrode centrale, en forme de H, et d'une électrode externe, comme indiqué sur l'image de microscopie électronique en fausses couleurs de la Figure 88b. Les électrodes centrales sont déposées en colonnes, les électrodes externes en ligne. Chaque pixel est donc adressé par une ligne et une colonne. Pour éviter que les électrodes externes soient en contact avec l'électrode centrale au niveau de l'intersection avec les colonnes, une couche de silice doit être déposée entre les deux dépôts d'or.

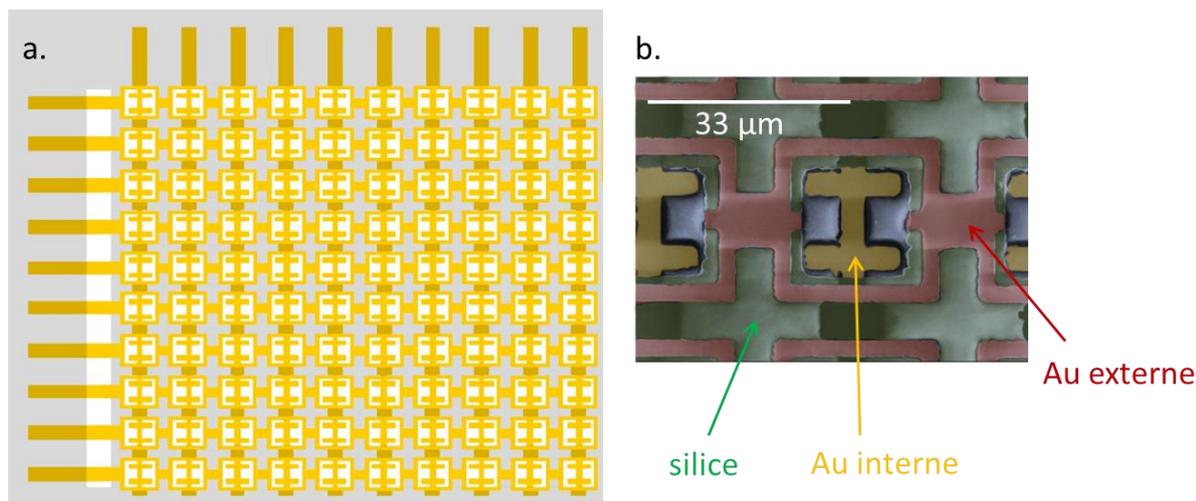

*Figure 88 : (a) Schéma des cent pixels de la matrice réalisée pendant cette étude. L'or est représenté en jaune, la silice en gris. Les fils d'or en haut et à gauche permettent de reporter les contacts ; (b) image de microscopie électronique à balayage en fausses couleurs d'un des pixels. L'or interne (jaune) correspond à une des électrodes (colonne), l'autre électrode est réalisée par le contact d'or externe (rouge, ligne). Pour ne pas court-circuiter les pistes, une couche de silice (vert) est déposée entre les deux couches d'or.*

Pour réaliser cette matrice, trois étapes de lithographie sont nécessaires. La première étape consiste à déposer les colonnes qui correspondent aux électrodes centrales. Cette première étape de lithographie est également utilisée pour effectuer tous les reports de contact de la matrice sur des contacts macroscopiques plus faciles à connecter. La deuxième étape consiste à déposer une couche de silice pour isoler les pistes électriques de la première étape et éviter les courts-circuits. Les seuls motifs à ne pas protéger sont les pixels sur lesquels il faut déposer les électrodes externes pendant la troisième étape et les contacts macroscopiques qui doivent encore être accessibles pour la mesure. Cette couche de silice permet également de s'assurer que le courant mesuré au niveau d'un pixel provient bien du pixel lui-même et pas de la piste de report de contact. Enfin, la troisième étape de lithographie permet de déposer les motifs correspondant aux électrodes externes.[11] Un schéma représentant les motifs après chaque étape de lithographie est présenté sur la Figure 89.

La résistance de chaque pixel est ensuite mesurée pour vérifier qu'il n'y a pas de court-circuit. Cette procédure nous permet d'obtenir des matrices de pixels, dont moins de 5% sont court-circuités.

Une fois cette matrice réalisée, l'homogénéité des résultats obtenus sur les différents pixels doit être vérifiée.

---

[11] Le protocole de préparation de cette matrice de pixels est décrit plus en détails dans l'Annexe 2.

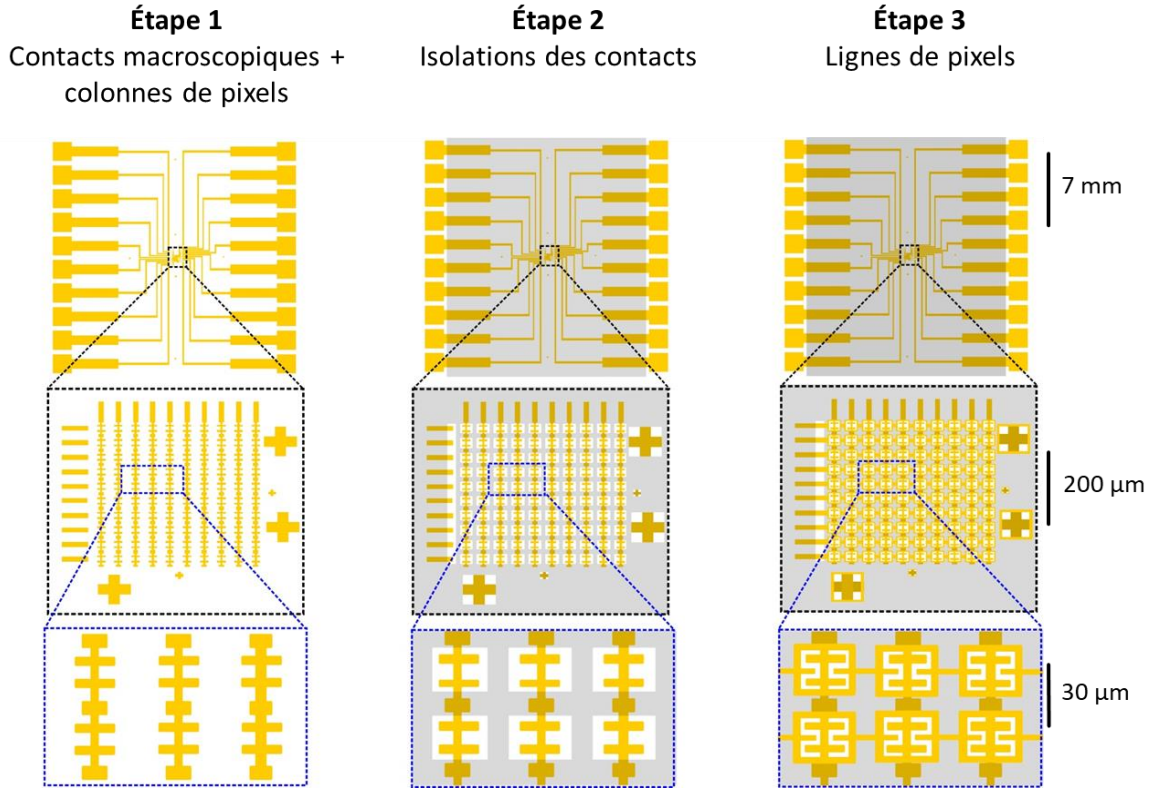

*Figure 89 : Schéma illustrant les différentes étapes permettant la fabrication de la matrice de pixels : la réalisation des colonnes de pixels, l'isolation des contacts par la silice et la réalisation des lignes de pixels. Chaque ligne correspond à un niveau de zoom différent. Les croix présentes sur les côtés de la matrice sont des croix d'alignement qui permettent d'aligner le masque de lithographie par rapport aux motifs déjà déposés.*

## 2. Homogénéité des pixels

Un film de nanocristaux à base d'encre (décrit plus tôt dans ce chapitre, p99) est déposé sur la matrice, et je mesure la réponse de chaque pixel dans des conditions identiques, en l'occurrence dans l'obscurité. Pour réaliser la cartographie de l'intensité du courant d'obscurité sur la matrice de pixels, j'utilise un multiplexeur Keithley DAQ6510 qui permet de commuter entre les différents pixels. La cartographie de la matrice est présentée sur la Figure 90a, et on constate que seuls trois pixels présentent un court-circuit. Les 97 restants présentent un courant d'obscurité inférieur à $10^{-7}$ A.

L'histogramme des valeurs de courants pour ces 97 pixels est présenté sur la Figure 90b. La largeur à mi-hauteur de la gaussienne qui modélise l'histogramme est assez élevée, de l'ordre de 50 % de la valeur du maximum. Les hétérogénéités dans les courants d'obscurité peuvent provenir d'une hétérogénéité des pixels ou d'une hétérogénéité dans l'épaisseur du film. Dans mon cas, la première option est privilégiée puisqu'une corrélation spatiale est observée entre les pixels présentant les courants d'intensité les plus élevés et les trois pixels court-circuités (ils sont situés sur les mêmes lignes ou les mêmes colonnes).

Pour finaliser l'étude du potentiel de cette matrice en imagerie infrarouge, je vais étudier sa réponse sous illumination. Les pixels ont été conçus selon une géométrie de photoconducteurs. La détectivité de cette matrice sera donc assez faible. La matrice devra donc être testée dans une application ou le flux de photons est intense, pour maximiser le rapport signal sur bruit. J'ai donc choisi de l'utiliser pour étudier le profil d'intensité d'un faisceau laser infrarouge.

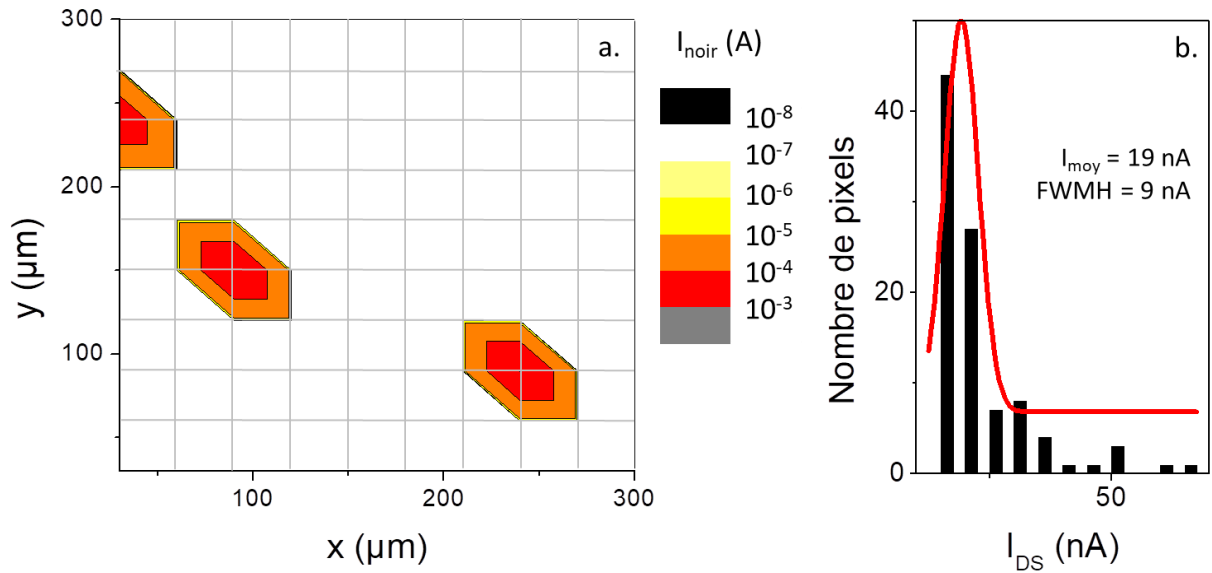

*Figure 90 : (a) Cartographie de l'intensité du courant d'obscurité après dépôt du film de nanocristaux. Chaque intersection de la grille correspond à la valeur mesurée pour un pixel. Entre les intersections, les valeurs indiquées sont des interpolations effectuées par le logiciel de traitement des données (Origin 2016) et ne correspondent donc pas à des mesures physiques. Les centres des hexagones déformés correspondent à des pixels court-circuités. (b) Histogrammes des courants d'obscurités des 97 pixels non court-circuités. Les pixels sont soumis à une tension de 100 mV.*

### 3. Détection d'un faisceau laser

Pour vérifier les propriétés d'imagerie dans l'infrarouge de la matrice de pixels couverte d'un film d'une encre de HgTe 4000 cm$^{-1}$, elle est utilisée pour détecter un faisceau laser infrarouge focalisé.

#### a. Focalisation du faisceau laser

La première étape pour réaliser une image du faisceau laser est de s'assurer que la taille de celui-ci est compatible avec la taille des pixels de la matrice. Je mesure donc la largeur d'un faisceau laser focalisé en utilisant un cache opaque aux infrarouges que je déplace par pas de 10 µm devant le faisceau et je mesure la puissance transmise. La taille du faisceau laser est obtenue en modélisant par une gaussienne la dérivée de la courbe obtenue, voir Figure 91.

Le faisceau fait 65 µm de large, ce qui correspond à deux pixels de la matrice. Sa largeur est donc bien adaptée pour qu'une image soit obtenue via la matrice de pixels.

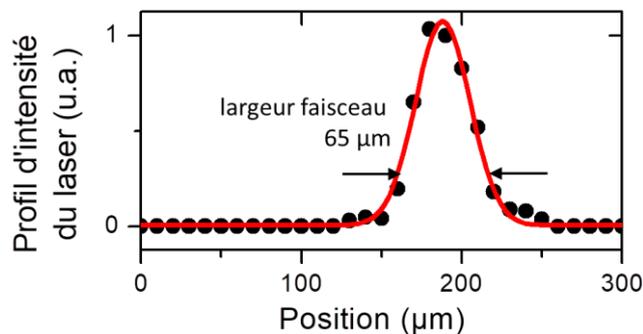

*Figure 91 : Profil d'intensité du laser utilisé pour tester la matrice de pixels. Le profil obtenu peut être modélisé par une gaussienne de largeur 65 µm.*

### b. Détection du faisceau

Le faisceau laser est donc dirigé sur la matrice de pixels et l'intensité du courant de chaque pixel est enregistrée. Une cartographie de l'intensité est ensuite réalisée et on obtient les résultats présentés sur la Figure 92. Un faisceau est effectivement détecté par la matrice de pixels, et l'image du faisceau a une largeur proche de celle du faisceau réel (75 µm contre 65 µm).

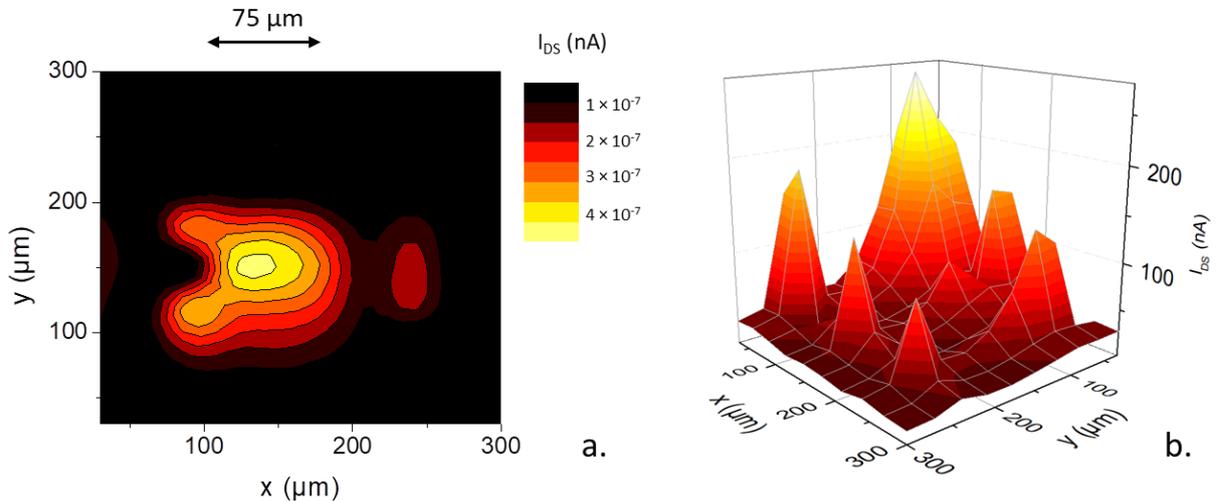

*Figure 92 : (a) Cartographie de l'intensité mesurée sous illumination (faisceau laser émettant à 1,55 µm, 10 mW, de largeur 65 µm). Le pic obtenu est modélisé par une gaussienne qui indique que la largeur vaut 75 µm. (b) Intensité du courant sur les différents pixels. Le faisceau laser éclaire un coin de la matrice, ce qui permet de voir sur le reste de la matrice des pics d'intensité correspondant aux ordres supérieurs du laser.*

En déplaçant le laser sur la matrice de pixels, la cartographie révèle également l'apparition de pics correspondant aux modes d'ordres supérieurs du laser.

La combinaison d'un film de HgTe 4000 cm[-1] déposé via une encre et de la matrice que j'ai développée ont donc permis d'obtenir une **preuve de concept de la capacité des nanocristaux colloïdaux pour imager le profil d'un laser dans l'infrarouge.**

### V. Conclusions et enjeux futurs

Dans ce chapitre, j'ai montré que les nanocristaux de chalcogénures de mercure pouvaient être intégrés dans des dispositifs de détection infrarouge complexes. En particulier, grâce au développement d'une encre, une diode dans l'ESWIR et un dispositif photoconducteur de 100 pixels ont été développés. Ces détecteurs ont montré des résultats encourageants pour l'utilisation de nanocristaux de tellure de mercure pour la détection vers 2,5 μm.

**Encre**

L'encre que j'ai développée permet d'obtenir des couches épaisses de nanocristaux chalcogénures de mercure en une seule étape de dépôt. Les films ainsi obtenus sont photoconducteurs et présentent une meilleure résistance au vieillissement et aux hautes températures que les films obtenus par des techniques d'échange de ligands sur film.

Cette encre et les paramètres de dépôt peuvent encore être améliorés pour réaliser des couches plus homogènes, et plus épaisses pour augmenter le pourcentage de lumière absorbée. En effet, avec les épaisseurs de film obtenues jusqu'à présent, 40 % de la lumière est absorbé par le film de nanocristaux.

**Diode**

En supprimant une barrière de filtration des trous non adaptée et en augmentant l'épaisseur de la couche active de matériaux grâce au développement d'une encre, j'ai obtenu une diode dans la gamme ESWIR présentant une photoréponse de l'ordre de 20 mA/W à 0 V et une détectivité de $3 \times 10^9$ Jones. Cette diode présente des temps de réponse faibles, de l'ordre de la centaine de nanosecondes, ce qui permet d'envisager son intégration dans des systèmes dynamiques.

Pour améliorer les performances des diodes dans cette gamme de longueurs d'onde, plusieurs aspects peuvent être envisagés. Pour améliorer l'absorption, on peut évidemment penser à l'élaboration de couches de matériau actif encore plus épaisses. Une autre idée est d'utiliser des résonateurs types Fabry-Pérot (*136*) ou encore d'intégrer des résonateurs plasmons type plots d'or (*137*) pour focaliser l'absorption dans la couche active. Enfin, l'extraction des porteurs n'étant pas optimale, des couches de transport HTL (*hole transport layer*, couche de transport de trous) et ETL (*electron transport layer*, couche de transport d'électron) peuvent être ajoutées (*130*). Le développement de jonctions p-i-n permettra également d'améliorer la collecte du photocourant.

**Matrice de pixels**

La conception d'une matrice de 100 pixels, entièrement réalisable en salle blanche par des techniques de lithographie optique, d'évaporation et de pulvérisation cathodique, a été démontrée. Ces matrices peuvent être utilisées pour prouver le potentiel d'imagerie des nanocristaux : dans ce manuscrit, j'ai montré que la détection d'un faisceau laser de 65 μm de large était possible en utilisant ce dispositif.

Pour optimiser ces matrices, une première option serait de diminuer encore la taille du pixel jusqu'à la limite de diffraction (10 μm), et d'en augmenter le nombre pour réaliser des images d'objets ayant des formes plus complexes. Une autre possibilité serait de graver le film de nanocristaux via une étape de lithographie supplémentaire pour éviter le couplage entre pixels et ainsi améliorer la qualité de l'image finale. Enfin, plutôt que de réaliser cette matrice en géométrie photoconducteur, il faudrait réaliser une photodiode en géométrie verticale afin d'améliorer la détectivité de tels dispositifs.

# Conclusion générale

Mes travaux de doctorat se sont concentrés sur les nanocristaux colloïdaux de semiconducteurs à faible bande interdite, en particulier le séléniure de mercure (HgSe) et le tellure de mercure (HgTe). Leurs transitions optiques modulables dans l'infrarouge en font des candidats potentiels pour être utilisés dans des dispositifs de détection infrarouge. Les objectifs de mes travaux de doctorat étaient de permettre une meilleure connaissance des propriétés optoélectroniques et de transport de ces matériaux, ainsi que de proposer une architecture de détecteur infrarouge adaptée.

Dans un premier temps, la détermination de l'énergie des niveaux de ces nanocristaux à différentes tailles, et donc à différents degrés de confinement quantique, a permis de mettre en avant l'influence de la taille sur les propriétés de dopage. Dans le cas des nanocristaux de HgTe, les matériaux sont de plus en plus n à mesure que la taille augmente. Pour les nanocristaux de HgSe, naturellement dopés à la fin de leur synthèse, le niveau de dopage augmente avec la taille des cristaux. Les plus petits ($\approx 5$ nm) sont des semiconducteurs dopés contenant 1 à 2 électron(s) et présentant des transitions intrabandes. Les plus gros ($\approx 15 - 20$ nm) peuvent contenir plus de 18 électrons et ont donc des propriétés plus proches de celles des métaux : transitions plasmoniques, transport... Pour compléter la reconstruction de la structure électronique, la présence de pièges dans la bande interdite a également été étudiée. Pour cela, un système de mesure indirecte de l'énergie d'Urbach via la mesure du photocourant a été mis en place. Les processus de piégeage-dépiégeage des porteurs avant de rejoindre l'électrode ont une dynamique dépendante de l'énergie d'Urbach, que nous pouvons mesurer en étudiant le photocourant transitoire. En réalisant cette étude sur des nanocristaux recouverts par différents ligands, nous pouvons déterminer que les ligands courts à base de thiols permettent de passiver au mieux les pièges dans la bande interdite induits par les états de surface.

**La connaissance de la structure électronique est particulièrement utile pour concevoir un système d'extraction des porteurs.** Elle permet de concevoir une architecture de détecteur où les barrières Schottky ainsi que le nombre de pièges dans la bande interdite sont réduits, ce qui améliore la réponse du détecteur.

Le contrôle du dopage dans les nanocristaux est essentiel puisque les propriétés optiques et de transport des nanocristaux en dépendent fortement. Tout d'abord, la modification de l'énergie de Fermi permet d'interdire ou d'autoriser certaines transitions : ainsi, dans des nanocristaux de HgSe de plus de 5 nm de diamètre, la transition interbande est interdite car le niveau de Fermi est au-dessus du niveau $1S_e$. Le niveau de dopage a également un impact crucial sur le courant d'obscurité d'un film à base de nanocristaux, et donc sur le bruit d'un détecteur à base de nanocristaux. Après avoir montré que la taille avait une influence sur le niveau de dopage des nanocristaux dans le chapitre 2, j'ai exploré d'autres pistes qui permettent de contrôler le dopage, en particulier les méthodes post-synthèse basées sur la chimie de surface. L'utilisation de ligands présentant un fort dipôle tels que $S^{2-}$ permet ainsi de diminuer le nombre d'électrons par nanocristal de HgSe de 2 à 0,2. Cette modification du niveau de dopage apparaît grâce à la courbure des niveaux électronique sous l'action du dipôle. Une autre approche basée sur l'utilisation de ligands oxydants est également étudiée. L'utilisation de polyoxométallates, capables de se greffer à la surface des nanocristaux de HgSe grâce à des fonctions thiols, a permis de diminuer le niveau de dopage sur des nanocristaux de HgSe, plus gros donc plus dopés initialement, de 5 à 0,03 électrons par cristal. Ce transfert d'électrons entre le cristal et le ligand oxydant est le plus important réalisé en utilisant des techniques d'oxydoréduction.

Pour compléter cette étude, j'ai étudié l'intégration de nanocristaux de HgTe, présentant une transition vers 0,5 eV, dans des systèmes de détection infrarouge. Pour augmenter le rapport signal sur bruit, ce qui est particulièrement important pour des dispositifs dans l'infrarouge où les transitions sont de l'ordre de quelques $k_BT$, une géométrie de détecteur de type photodiode est privilégiée. Dans ces photodiodes, un champ électrique interne permet d'extraire les porteurs générés sous illumination tandis que le bruit reste faible si les tensions appliquées restent proche de 0 V. Les résultats du chapitre 2 nous ont permis de construire une barrière unipolaire à base de nanocristaux HgTe plus petits, particulièrement adaptée pour bloquer le courant d'obscurité dû aux électrons tout en laissant passer les photo-trous. La diode proposée a alors une détectivité de $3 \times 10^8$ Jones, mais une réponse assez faible, de l'ordre de 100 µA/W. Pour augmenter la réponse, je propose d'augmenter l'épaisseur de la couche active pour augmenter l'absorption, en utilisant un échange de ligands en solution. En plus d'augmenter l'épaisseur de la couche active, cet échange de ligands en solution permet de simplifier le processus de dépôt ainsi que d'augmenter la mobilité des porteurs. Je propose également de retirer la barrière de dioxyde de titane ($TiO_2$), utilisée dans les cellules solaires à base de nanocristaux de sulfure de plomb, qui n'est pas adaptée aux nanocristaux à plus faible bande interdite et qui filtre donc une partie du photocourant. En combinant ces deux améliorations, la réponse est augmentée d'un facteur 200 (20 mA/W) et la détectivité d'un facteur 10, atteignant une valeur de $3 \times 10^9$ Jones.

Enfin, pour étudier la possibilité d'intégrer ces cristaux dans une caméra infrarouge, j'ai étudié un dispositif multipixels. Pour cela, j'ai construit une matrice de 100 pixels de 30 µm × 30 µm, proche de ce qui se fait pour les caméras commerciales à base d'InGaAs par exemple. Une fois les nanocristaux déposés sur cette matrice, j'ai montré qu'il était possible d'imager le profil d'un faisceau laser infrarouge de 60 µm de large.

## Perspectives

Les diodes dans l'ESWIR à base de nanocristaux de HgTe 4000 cm$^{-1}$ que j'ai fabriquées ont vu leur réponse et leur détectivité augmenter grâce à l'ingénierie d'alignement des niveaux électroniques que j'ai développée. Pour améliorer le signal de détection de cette diode, plusieurs stratégies peuvent être envisagées :

- Augmenter l'absorption de la lumière afin de générer plus de photocourant. Cela peut se faire par exemple en réalisant des couches de nanocristaux plus épaisses en développant le dépôt par encre. Des résonateurs de type Fabry-Pérot ou de structures plasmoniques permettant d'augmenter localement l'absorption peuvent également être utilisés pour augmenter l'absorption dans la couche active.
- Améliorer l'extraction des porteurs vers les électrodes pour récolter plus efficacement le photocourant, tout en gardant un niveau de bruit le plus faible possible. Pour cela, ajouter des couches de transport de trous et d'électrons, et renforcer le caractère p-n de la jonction peuvent être mises en place dans les diodes à base de nanocristaux.

Pour aller vers des longueurs d'onde plus grandes (domaines MWIR ou LWIR), des nanocristaux plus larges peuvent être utilisés. Ainsi, des diodes à base de HgTe 2000 cm$^{-1}$ existent déjà dans la littérature (*126*). Une autre solution est d'utiliser des diodes à base de nanocristaux dopés, tels que HgSe, basée sur des transitions intrabandes. Si HgSe seul est un matériau lent et introduisant un niveau de bruit élevé, des travaux sur des hétérostructures cœur-coquille HgSe-HgTe ont montré que ces matériaux présentaient des transitions intrabandes, caractéristiques de HgSe, tout en conservant la dynamique rapide des nanocristaux de HgTe (*97*). L'utilisation de mélanges de nanocristaux HgSe et HgTe est également une piste intéressante pour combiner transition intrabande et faibles temps de réponse (*138*). Le travail de conception de diode pour les HgTe 4000 cm$^{-1}$ constitue un point de départ pour la réalisation de ces diodes intrabandes.

Enfin, ces dernières années, des efforts ont été réalisés dans la communauté des nanocristaux colloïdaux pour développer des matériaux moins toxiques. Dans le domaine visible, les nanocristaux de phosphure d'indium (InP) remplacent à présent les nanocristaux de séléniure de cadmium (CdSe) dans les écrans de télévision QLED. Dans le domaine infrarouge, des alternatives aux matériaux à base de mercure doivent être trouvées. Par exemple, depuis le début des années 2010, de nombreux travaux ont porté sur Ag$_2$Se (*76, 139, 140*), présentant des transitions intrabandes dans l'infrarouge. Des réponses de l'ordre du mA/W ont été démontrées en 2019 (*139*). Le contrôle du dopage dans les oxydes dopés tels que les oxydes de zinc dopés au gallium ou les oxydes d'étain dopés à l'indium permet également d'atteindre des transitions dans l'infrarouge (*55, 75*) et peut constituer une piste prometteuse.

## Liste des publications

### I.    Articles publiés dans des journaux scientifiques à comité de lecture

**2017**
- H. Cruguel, C. Livache, **B. Martinez**, S. Pedetti, D. Pierucci, E. Izquierdo, M. Dufour, S. Ithurria, H. Aubin, A. Ouerghi, E. Lacaze, M. Silly, B. Dubertret, E. Lhuillier, Electronic structure of CdSe-ZnS 2D nanoplatelets, *Applied Physics Letters*, 110, 15, 152103
- C. Livache, E. Izquierdo, **B. Martinez**, M. Dufour, D. Pierucci, S. Keuleyan, H. Cruguel, L. Becerra, J.-L. Fave, H. Aubin, A. Ouerghi, E. Lacaze, M. Silly, B. Dubertret, S. Ithurria, E. Lhuillier, Charge dynamics and optoelectronic properties in HgTe colloidal quantum wells, *Nano Letters*, 17, 7, 4067-4074
- W. J. Mir, A. Assouline, C. Livache, **B. Martinez**, N. Goubet, X. Z. Xu, G. Patriarche, H. Aubin, E. Lhuillier, Electronic properties of (Sb ;Bi)$_2$Te$_3$ colloidal heterostructured nanoplates down to the single particle level, *Scientific Reports*, 7, 1, 9647
- **B. Martinez**, C. Livache, L. D. Notemgnou Mouafo, N. Goubet, S. Keuleyan, H. Cruguel, S. Ithurria, H. Aubin, A. Ouerghi, B. Doudin, E. Lacaze, B. Dubertret, M. Silly, R. P. S. M. Lobo, J.-F. Dayen, E. Lhuillier, HgSe self-doped nanocrystals as a platform to investigate the effects of vanishing confinement, *ACS Applied Materials and Interfaces*,  9, 41, 36173-36180

**2018**
- **B. Martinez**, C. Livache, N. Goubet, A. Jagtap, H. Cruguel, A. Ouerghi, E. Lacaze, M. Silly, E. Lhuillier, Probing charge carrier dynamics to unveil the role of surface ligands in HgTe narrow band gap nanocrystals, *the Journal of Physical Chemistry C*, 122, 1, 859-865
- W. J. Mir, C. Livache, N. Goubet, **B. Martinez**, A. Jagtap, A. Chu, N. Coutard, H. Cruguel, T. Barisien, S. Ithurria, A. Nag, B. Dubertret, A. Ouerghi, M. Silly, E. Lhuillier, Strategy to overcome recombination limited photocurrent generation in CsPbX$_3$ nanocrystal arrays, *Applied Physics Letters*, 112, 11, 113503
- C. Livache, N. Goubet, **B. Martinez**, A. Jagtap, J. Qu, S. Ithurria, M. Silly, B. Dubertret, E. Lhuillier, Band edge dynamics and multiexciton generation in narrow band gap HgTe nanocrystals, *ACS Applied Materials and Interfaces*, 10, 14, 11880-11887
- N. Goubet, A. Jagtap, C. Livache, **B. Martinez**, H. Portalès, X. Z. Xu, R. P. S. M. Lobo, B. Dubertret, E. Lhuillier, Terahertz nanocrystals: beyond confinement, *Journal of the American Chemical Society*, 140, 15, 5033-5036

- E. Izquierdo, M. Dufour, A. Chu, C. Livache, **B. Martinez**, D. Amelot, G. Patriarche, N. Lequeux, E. Lhuillier, S. Ithurria, Coupled HgSe colloidal quantum wells through a tunable barrier: a strategy to uncouple optical and transport band gap, *Chemistry of Materials*, 30, 12, 4065-4072

- N. Goubet, C. Livache, **B. Martinez**, X. Z. Xu, S. Ithurria, S. Royer, H. Cruguel, G. Patriarche, A. Ouerghi, M. Silly, B. Dubertret, E. Lhuillier, Wave-function engineering in HgSe/HgTe colloidal heterostructures to enhance mid-infrared photoconductive properties, *Nano Letters*, 18, 7, 4590-4597

- A. Jagtap, N. Goubet, C. Livache, A. Chu, **B. Martinez**, C. Gréboval, J. Qu, E. Dandeu, L. Becerra, N. Witkowski, S. Ithurria, F. Mathevet, M. Silly, B. Dubertret, E. Lhuillier, Short-wave infrared devices based on HgTe nanocrystals with air stable performances, *the Journal of Physical Chemistry C*, 122, 26, 14979-14985

- J. Qu, N. Goubet, C. Livache, **B. Martinez**, D. Amelot, C. Gréboval, A. Chu, J. Ramade, H. Cruguel, S. Ithurria, M. Silly, E. Lhuillier, Intraband mid-infrared transitions in $Ag_2Se$ nanocrystals: potential and limitations for Hg-free low-cost photodetection, *the Journal of Physical Chemistry C*, 122, 31, 18161-18167

- A. Jagtap, **B. Martinez**, N. Goubet, A. Chu, C. Livache, C. Gréboval, J. Ramade, D. Amelot, P. Trousset, A. Triboulin, S. Ithurria, M. Silly, B. Dubertret, E. Lhuillier, Design of a unipolar barrier for a nanocrystal-based short-wave infrared photodiode, *ACS Photonics*, 5, 11, 4569-4576

- **B. Martinez**, C. Livache, E. Meriggio, X. Z. Xu, H. Cruguel, E. Lacaze, A. Proust, S. Ithurria, M. Silly, G. Cabailh, F. Volatron, E. Lhuillier, Polyoxometalate as control agent for the doping in HgSe self-doped nanocrystals, *the Journal of Physical Chemistry C*, 122, 46, 26680-26685

**2019**

- C. Gréboval, E. Izquierdo, C. Livache, **B. Martinez**, M. Dufour, N. Goubet, N. Moghaddam, J. Qu, A. Chu, J. Ramade, H. Aubin, H. Cruguel, M. Silly, E. Lhuillier, S. Ithurria, Impact of dimensionality and confinement on the electronic properties of mercury chalcogenide nanocrystals, *Nanoscale*, 11, 9, 3905-3915

- M. Dufour, E. Izquierdo, C. Livache, **B. Martinez**, M. Silly, T. Pons, E. Lhuillier, C. Delerue, S. Ithurria, Doping as a strategy to tune color of 2D colloidal nanoplatelets, *ACS Applied Materials and Interfaces*, 11, 10, 10128-10134

- J. Qu, C. Livache, **B. Martinez**, C. Gréboval, A. Chu, E. Merigio, J. Ramade, H. Cruguel, X. Z. Xu, A. Proust, F. Volatron, G. Cabailh, N. Goubet, E. Lhuillier, Transport in ITO Nanocrystals with Short- to Long-Wave Infrared Absorption for Heavy-Metal-Free Infrared Photodetection, *ACS Applied Nano Materials*, 2, 3, 1621-1630

- C. Livache, **B. Martinez**, N. Goubet, C. Gréboval, J. Qu, A. Chu, S. Royer, S. Ithurria, M. Silly, B. Dubertret, E. Lhuillier, A random colloidal quantum dot infrared photodetector and its use for intraband detection, *Nature Communications*, 10, 1, 2125

- **B. Martinez**, J. Ramade, C. Livache, N. Goubet, A. Chu, C. Gréboval, J. Qu, L. Becerra, E. Dandeu, J.-L. Fave, E. Lacaze, E. Lhuillier, HgTe nanocrystal inks for extended short wave infrared detection, Effect of pressure on interband and intraband transition of mercury chalcogenides quantum dots, *Advanced Optical Materials,* 1900348

- C. Livache, N. Goubet, C. Gréboval, **B. Martinez**, J. Ramade, J. Qu, A. Triboulin, H. Cruguel, B. Baptiste, S. Klotz, G. Fishman, S. Sauvage, F. Capitani, E. Lhuillier, Effect of pressure on interband and intraband transition of mercury chalcogenides quantum dots, *the Journal of Physical Chemistry C*, 123, 20, 13122-13130

- C. Gréboval, U. Noumbe, N. Goubet, C. Livache, J. Qu, A. Chu, **B. Martinez**, Y. Prado, S. Ithurria, A. Ouerghi, H. Aubin, J.-F. Dayen, E. Lhuillier, Field effect transistor of narrow band gap nanocrystal arrays using ionic glasses, *Nano Letters,* 19, 6, 3981-3986

## II. Revues publiées dans des journaux scientifiques à comité de lecture

- A. Jagtap, C. Livache, **B. Martinez**, J. Qu, A. Chu, C. Gréboval, N. Goubet, E. Lhuillier, Emergence of intraband transitions in colloidal nanocrystals, *Optical Materials Express*, 8, 5, 1174-1183, 2018
- C. Livache, **B. Martinez**, N. Goubet, J. Ramade, E. Lhuillier, Road map for nanocrystal based infrared photodetectors, *Frontiers in chemistry*, 6, 2018

## III. Communications orales

- Transport in films of semiconductor colloidal nanocrystals – Application to the IR range, Rencontre des Jeunes Physiciens, 2017, Paris, France
- HgSe self-doped nanocrystals as a knob to probe the semiconductor to metal transition in nanocrystal solid, Nano TN, 2018, Marrakech, Maroc
- Extended short wave infrared photodiodes based on HgTe nanocrystals, Colloidal nanocrystals workshop organized at INSP, 2019, Paris, France
- Control of doping in infrared nanocrystals, EMRS, printemps 2019, Nice, France

## IV. Présentation de posters

- Transport with narrow gap nanocrystals, GDR Nacre, 2016, Aussois, France
- Controlling the intraband absorption in HgSe self-doped nanocrystals, *Exciting nanostructures: characterizing advances confined systems* summer school, 2017, Bad Honnef, Allemagne
- Probing charge carried dynamics to unveil the role of surface ligands on HgTe narrow band gap nanocrystals, Colloidal nanocrystals workshop organized at INSP, 2017, Paris, France
- HgSe self-doped nanocrystals as a knob to probe the semiconductor to metal transition in nanocrystal solid, GRC, 2018, Boston, États-Unis
- Narrow band gap nanocrystal based devices for IR photodetection, ONERA scientific days : Latest revolutions in detectors, from the visible to infrared, 2018, Palaiseau, France

# Annexe 1
# Synthèses

## I. Synthèses de nanocristaux

### 1. Précurseurs

Tous les produits chimiques sont utilisés tels quels, à l'exception de l'oleylamine qui est centrifugée avant utilisation.

#### a. Précurseurs métalliques et ioniques

Chlorure de mercure ($HgCl_2$, Strem Chemicals, 99%), acétate de mercure ($Hg(OAc)_2$, Sigma-Aldrich), tellurium (Te, Sigma-Aldrich, 99,99%), sélénium (Se, Strem Chemicals), disulfure de sélénium ($SeS_2$, Sigma-Aldrich).

#### b. Solvants et ligands utilisés pendant la synthèse

N-hexane (Carlo Erba), éthanol absolu anhydre (Carlo Erba, 99,9%), méthanol (Carlo Erba, 99,8%), chloroforme (Carlo Erba), acétone (Carlo Erba, 99,8%), n-octane (SDS, 99%), toluène (Carlo Erba, 99,3%), N-méthyl formamide (NMF, Alfa Aesar, 99%), N,N diméthylformamide (DMF, Sigma Aldrich), acide oléique (OA, Sigma Aldrich, 90%), oleylamine (OLA, Acros, 80-90%), trioctylphosphine (TOP, Sigma-Aldrich, 97%), dodécanethiol (DDT, Sigma-Aldrich, 98%).

#### c. Protocoles de synthèse des précurseurs TOPSe et TOPTe

**TOPSe** : 1,58 g de Se est mélangé à 20 mL de TOP dans un pilulier. La dissolution du sélénium est obtenue par sonication dans un bain à ultrasons après 1 h. La solution finale est transparente.

**TOPTe** : 2,54 g de Te sont mélangés à 20 mL de TOP dans un tricol. La solution est dégazée sous vide, à 80 °C pendant 1 h. La solution est ensuite placée sous atmosphère argon, à une température de 270 °C, jusqu'à ce que le tellurium soit dissout (environ 2 h). La solution devient orange, puis jaune quand la température est redescendue vers 20 °C.

### 2. Protocoles de synthèses des nanocristaux de chalcogénures de mercure

Pendant mon doctorat, j'ai eu la possibilité de synthétiser mes propres nanocristaux, grâce à une collaboration avec le Laboratoire de Physique et d'Étude des Matériaux (LPEM) de l'ESPCI, et notamment grâce à Sandrine Ithurria-Lhuillier. J'ai également pu bénéficier de l'aide d'autres personnes du groupe : Emmanuel Lhuillier, Audrey Chu, Junling Qu et surtout Nicolas Goubet qui ont préparé certaines des solutions de nanocristaux que j'ai utilisées pendant mes travaux.

Les nanocristaux dont les protocoles de synthèse sont décrits dans cette section seront systématiquement décrits par leur nature (HgTe ou HgSe) et par le nombre d'onde de l'exciton observé à la fin de la synthèse (en $cm^{-1}$).

#### a. Synthèses de nanocristaux de tellure de mercure (HgTe)

Les protocoles de synthèse proposés dans cette section ont été adaptés de la synthèse de Sean Keuleyan (*47*) par Nicolas Goubet, post-doctorant dans notre équipe.

##### i. HgTe 7000 $cm^{-1}$

Dans un tricol de 100 mL, 513 mg de $HgCl_2$ sont ajoutés à 60 mL d'oleylamine. Le mélange est dégazé pendant 1 h à 110 °C. La température est ensuite fixée à 45 °C et la solution placée sous atmosphère d'argon. 1,9 mL de TOPTe et 10 mL d'oleylamine sont ensuite ajoutés au mélange ; la couleur de la solution vire lentement au marron. Après 3 min, la réaction est stoppée en ajoutant un mélange de 1 mL de DDT et 9 mL de toluène, ainsi qu'en abaissant rapidement la température à 20 °C. Les nanocristaux obtenus sont ensuite précipités en utilisant de l'éthanol comme mauvais solvant, puis l'opération est répétée en utilisant cette fois du méthanol. Les cristaux sont ensuite re-dispersés dans quelques mL de CHCl$_3$ et précipités sans ajout de mauvais solvant afin de retirer le maximum de phase lamellaire. Le culot obtenu est ensuite re-dispersé dans 6 mL de CHCl$_3$ avant d'être filtré à travers un filtre PTFE (0,2 µm). Les cristaux sont ensuite précipités une dernière fois avec du méthanol avant d'être re-dispersés dans le toluène (50 mg/mL).

### ii.    HgTe 6000 cm$^{-1}$

Le protocole de synthèse est le même que celui utilisé pour les nanocristaux de HgTe 7000 cm$^{-1}$, mais en utilisant une température de synthèse de 60 °C.

### iii.    HgTe 4000 cm$^{-1}$

Le protocole de synthèse est le même que celui utilisé pour les nanocristaux de HgTe 7000 cm$^{-1}$, mais en utilisant une température de synthèse de 80 °C. Pour les synthèses réalisées à des températures supérieures à 80 °C, la coloration du mélange réactionnel apparaît plus rapidement (quelques secondes). De plus, ces réactions sont plus complètes, ce qui facilite les lavages. Les nanocristaux obtenus sont donc précipités en utilisant de l'éthanol et du méthanol comme mauvais solvants. Ce lavage est répété deux fois. Après centrifugation, les cristaux sont re-dispersés dans le toluène (50 mg/mL). Pour greffer les ligands POM-SH sur ces nanocristaux, la réaction doit être arrêtée sans ajouter de dodécanethiol (p83).

### iv.    HgTe 2000 cm$^{-1}$

Le protocole de synthèse est le même que celui utilisé pour les HgTe 4000 cm$^{-1}$, mais en utilisant une température de synthèse de 120 °C.

### v.    HgTe 700 cm$^{-1}$

Dans un tricol de 50 mL, on verse 20 mL d'oleylamine que l'on dégaze pendant 1 h à 110 °C. 171 mg de HgCl$_2$ sont mélangés à 4,3 mL d'oleylamine, puis 0,63 mL de TOPTe est ajouté à ce mélange. L'oleylamine est ensuite placée sous atmosphère d'argon, à 180 °C. Une fois la température stabilisée, le mélange contenant le chlorure de mercure et la TOPTe est injecté dans l'oleylamine à 180 °C. Après 3 min, la réaction est arrêtée en ajoutant 1 mL de DDT et 9 mL de toluène, ainsi qu'en abaissant rapidement la température. Les nanocristaux obtenus sont précipités en utilisant de l'éthanol comme mauvais solvant, avant d'être redispersés dans quelques millilitres de chloroforme.

### b.    Synthèses de nanocristaux de séléniure de mercure (HgSe)

Les protocoles de synthèse de nanocristaux de HgSe ont été adaptés des protocoles décrits dans l'article de Robin et al. (*48*) par Nicolas Goubet.

###### i. HgSe 2700 cm$^{-1}$

Dans un tricol de 50 mL, 0,5 g de Hg(OAc)$_2$ est ajouté à 10 mL d'acide oléique et 25 mL oleylamine. Le mélange obtenu est placé sous vide, à 85 °C pendant 1 h. La température est ensuite fixée à 80 °C et la solution placée sous atmosphère d'argon. 1,6 mL de TOPSe est ensuite rapidement injecté dans le tricol. Après 1 min, la réaction est stoppée en injectant 1 mL de DDT et en abaissant la température du mélange à 20 °C rapidement. Les nanocristaux sont précipités en utilisant de l'éthanol et du méthanol comme mauvais solvants. Ce lavage est répété deux fois. Après centrifugation, les cristaux sont re-dispersés dans le toluène (50 mg/mL).

###### ii. HgSe 2200 cm$^{-1}$

Le protocole de synthèse est le même que celui utilisé pour les nanocristaux de HgSe 2700 cm$^{-1}$, mais en utilisant un temps de réaction de 4 min.

###### iii. HgSe 1000 cm$^{-1}$

Dans un tricol de 50 mL, 0,1 g de Hg(OAc)$_2$ est ajouté à 4 mL d'acide oléique et 10 mL d'oleylamine. Le mélange obtenu est placé sous vide, à 85 °C pendant 30 min. Pendant le dégazage, 0,13 g de SeS$_2$ est ajouté à 2 mL d'oleylamine dans un pilulier et le tout est placé dans un bain à ultrasons jusqu'à ce que le mélange devienne marron. La solution contenant le mercure est ensuite placée sous atmosphère argon, à 90 °C. Le précurseur de Se est rapidement injecté dans le tricol. Après 6 min, la réaction est stoppée en injectant 1 mL de DDT et en abaissant la température du mélange à 20 °C rapidement. Les nanocristaux sont ensuite précipités en utilisant de l'éthanol et du méthanol. Ce lavage est répété deux fois. Après centrifugation, les cristaux sont re-dispersés dans le toluène (50 mg/mL).

###### iv. HgSe 400 cm$^{-1}$

Le protocole de synthèse est le même que celui utilisé pour les nanocristaux de HgSe 1000 cm$^{-1}$, mais en utilisant une température de synthèse de 120 °C et un temps de réaction de 20 min.

## II. Synthèses de polyoxométallates et de polyoxométallates fonctionnalisés

Les polyoxométallates et polyoxométallates fonctionnalisés utilisés dans ces travaux de doctorat ont été synthétisés par Florence Volatron, de l'Institut Parisien de Chimie Moléculaire (IPCM).

Les polyoxométallates utilisés dans cette étude ont pour formule brute : K$_7$[PW$_{11}$O$_{39}$]. Ils sont synthétisés selon la procédure décrite dans la référence (*88*). Pour ajouter les groupes thiolés fonctionnels, le protocole suivant est utilisé. K$_7$[PW$_{11}$O$_{39}$] (0,64 g, 0,2 mmol) est dissout dans un mélange eau/acétonitrile (30 mL, 1:1). Une solution aqueuse d'acide chloridrique (HCl, 1 M) est ajoutée goutte à goutte jusqu'à obtenir un pH apparent de 3. La solution est ensuite refroidie en utilisant un bain de glace et le précurseur (OCH$_3$)$_3$SiC$_3$H$_6$SH (0,157 mL, 0,8 mmol) est ajouté. Une solution d'acide chloridrique à 1 M est à nouveau ajoutée pour obtenir un pH apparent de 2. La réaction tourne toute la nuit. Du bromure de tétrabutylammonium (TBABr, 0,26 g, 0,8 mmol) est ensuite ajouté et le solvant est retiré en utilisant un évaporateur rotatif. Le solide est ensuite re-dissout dans un minimum d'acétonitrile et précipité à nouveau grâce à un excès de diéthyléther. Après une étape de centrifugation et un lavage supplémentaire au diéthyléther, on obtient le solide final sous forme de poudre blanche (0,55 g, 76%).

# Annexe 2
# Fabrication d'électrodes

## I.    Principe de la photolithographie optique

Les électrodes sont fabriquées en salle blanche, par photolithographie optique. Pour cela, on utilise une résine photosensible, exposée aux ultraviolets à travers un masque. Cette exposition aux UV peut soit renforcer la résine (par polymérisation par exemple) on parle alors de résine négative, soit la dégrader, on parle alors de résine positive. Pendant mon doctorat, j'ai utilisé la résine AZ5214 E, qui peut être positive ou négative selon le temps d'exposition aux UV et les temps de recuit imposés. Un schéma indiquant les différentes étapes de lithographie pour des résines négatives et positives est donné sur les Figure 93 et Figure 94.

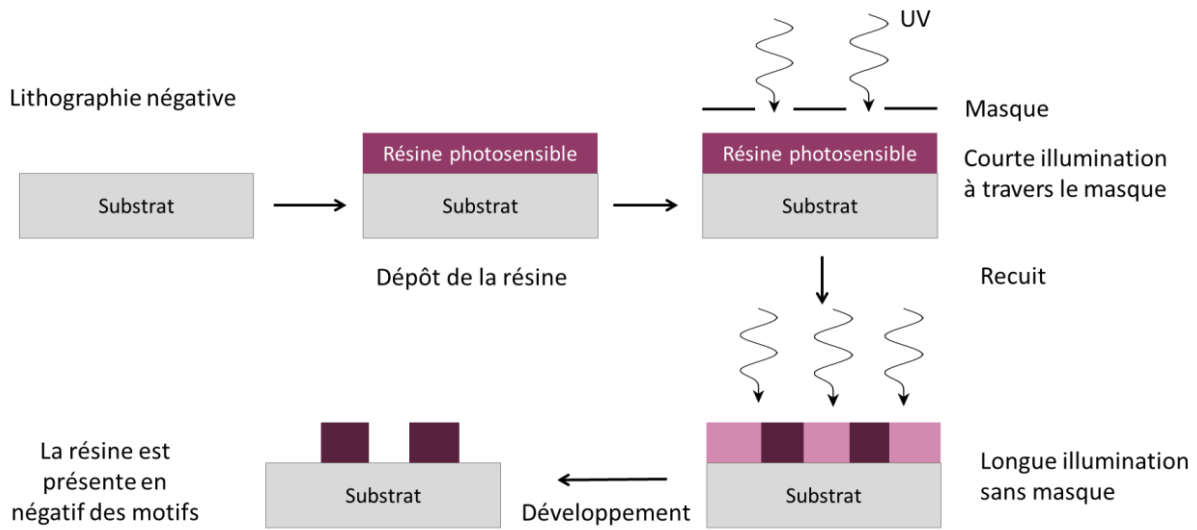

*Figure 93: Schéma de principe de la photolithographie en utilisant une résine négative. La résine négative polymérise après une courte exposition aux UV et un recuit (plus foncée sur le schéma). Au contraire, la résine qui n'a pas polymérisé est détériorée après une longue exposition aux UV (plus claire sur le schéma). À la fin du procédé, la résine est présente sur le négatif des motifs.*

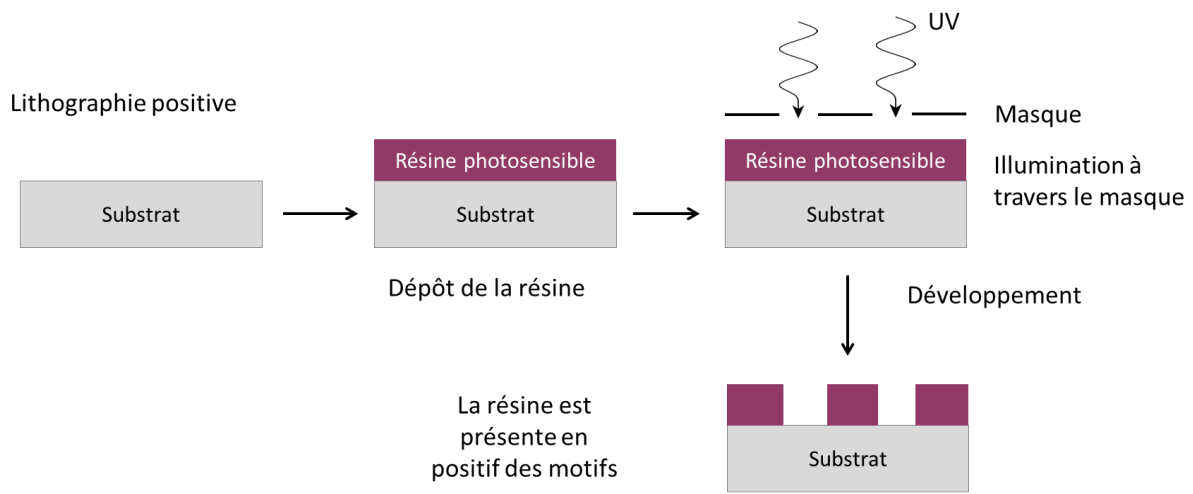

*Figure 94 : Schéma de principe de la photolithographie en utilisant une résine positive. La résine positive est détériorée après une exposition aux UV. À la fin du procédé, la résine est présente sur le positif des motifs.*

## II. Fabrication d'électrodes en salle blanche

### 1. Électrodes interdigitées Cr-Au sur substrat Si – SiO$_2$ (lithographie négative)

Les substrats Si/SiO$_2$ (400 nm d'oxyde) sont d'abord nettoyés abondamment à l'acétone, puis placés dans un cristallisoir rempli d'acétone dans un bain à ultrasons pendant 5 min. Ils sont ensuite rincés à l'acétone et à l'isopropanol avant d'être nettoyés avec un plasma d'oxygène pendant 5 min afin de retirer tous les résidus organiques sur le substrat.

Un promoteur d'adhésion (TI PRIME) est déposé à la tournette (en anglais *spin-coating*) selon les paramètres suivants : 4000 rotations par minute (rpm), 4000 rpm/s, 30 s. Le promoteur d'adhésion est ensuite recuit à 120 °C pendant 2 min afin d'évaporer le solvant. La résine photosensible, AZ 5214E, est à son tour déposée sur les substrats à la tournette (mêmes paramètres de dépôt), et recuite à 110 °C pendant 90 s.

Un masque en quartz, sur lequel les motifs d'électrodes voulus ont été reproduits en chrome par lithographie électronique, est utilisé pour cacher les futures parties du substrat correspondant aux électrodes pendant l'exposition. Cette dernière se fait dans l'ultraviolet (UV, entre 350 et 450 nm) et grâce à un aligneur MJB4 pendant 1,5 s. Cette étape permet d'initier la polymérisation de la résine en négatif du masque (donc sur tout la partie de l'échantillon ne correspondant pas aux motifs d'électrodes). La polymérisation se poursuit en recuisant les substrats à 125 °C pendant 2 min. Une deuxième illumination, sans masque, est réalisée sur tous les substrats pendant 40 s afin de rendre soluble dans le développeur la résine correspondant aux motifs d'électrodes.

Le développement des échantillons se fait dans le développeur AZ 726 MIF, pendant 20 s, et est stoppé par le rinçage à l'eau distillée. Afin d'éliminer toute trace de résine résiduelle sur les motifs, les échantillons sont nettoyés dans un plasma oxygène pendant 5 min.

5 nm de Cr et 80 nm d'Au sont évaporés (évaporateur VINCI) sur la surface des substrats, à une pression proche de $5.10^{-6}$ mbar. Les échantillons sont ensuite placés dans un bain d'acétone pendant 15 h minimum et sont éventuellement soniqués afin de retirer toute la résine. Les électrodes sont finalement rincées à l'acétone et à l'isopropanol.

Les motifs obtenus sont présentés sur la Figure 95. Afin de s'assurer que les électrodes sont fonctionnelles, les motifs sont vérifiés au microscope optique et toutes les paires d'électrodes sont testées électriquement pour vérifier l'absence de courts-circuits.

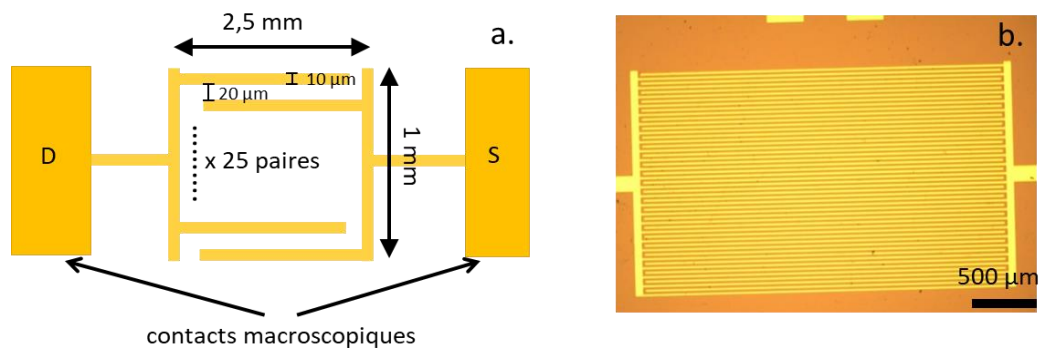

*Figure 95 : (a) Schéma des motifs d'électrodes interdigitées utilisées dans cette étude ; (b) photographie prise au microscope optique des électrodes.*

### 2. Électrodes interdigitées Cr-Au sur substrat verre (lithographie négative)

Ces électrodes sont notamment utilisées dans le cas où une illumination par la face arrière est nécessaire, ce qui n'est possible que pour des nombres d'onde supérieurs à 3000 cm$^{-1}$.

Des lames de microscope (2 mm d'épaisseur) sont coupées en deux dans le sens de la longueur. La suite du protocole de fabrication des électrodes est la même que pour les substrats Si/SiO$_2$.

Les motifs réalisés sont les mêmes que ceux obtenus sur substrat Si/SiO$_2$ (Figure 95).

### 3. Électrodes interdigitées Cr-Au sur substrat PET-ITO (lithographie positive)

Ces électrodes sont utilisées pour effectuer les mesures de photocourant transitoire avec le laser pulsé à 350 nm (p71). Elles permettent d'obtenir moins de réflexions parasites qui viennent perturber la mesure aux temps courts (quelques nanosecondes).

Les substrats utilisés (Sigma-Aldrich) sont des substrats de polytéréphtalate d'éthylène (*polyethylene terephtalate*, PET), recouverts d'une couche d'oxyde d'indium-étain (*indium tin oxide*, ITO) d'une épaisseur d'environ 80 nm et d'une résistance de l'ordre de 60 Ω/□.

Le nettoyage des substrats et le dépôt de la résine sont réalisés en suivant le même protocole que celui décrit pour les électrodes interdigitées sur substrat Si/SiO$_2$. Pour obtenir les motifs, on réalise une lithographie positive : c'est-à-dire que la résine est conservée au niveau des motifs. L'exposition des échantillons se fait au travers d'un masque chromé et dure 4 s. La résine illuminée est ainsi rendue soluble dans le développeur. Les échantillons sont ensuite développés dans un bain d'AZ 726 pendant 45 s avant d'être rincés dans l'eau distillée. Le négatif des motifs n'est plus protégé par la résine et est gravé par une solution d'HCl à 25 % pendant 15 s. L'ITO reste là où la surface est protégée par la résine tandis qu'il est gravé partout ailleurs.

Les échantillons sont ensuite rincés à l'acétone et à l'isopropanol pour retirer les restes de résine. Les motifs obtenus sont très semblables à ceux réalisés sur substrat de verre ou silicium. La seule différence est qu'ils font 1 mm de large et non pas 2,5 mm (Figure 95).

### 4. Électrodes 8 pixels ITO sur verre (lithographie positive)

Ces électrodes ont été utilisées pour réaliser la première génération de diode dans l'ESWIR à base de nanocristaux de HgTe 4000 cm$^{-1}$.

Les substrats de verre, couverts d'une fine couche de 30 nm d'ITO (Solem, 70 à 100 Ω/□) sont découpés en carré de 15 mm de côté. Les procédures de nettoyage du substrat et de dépôt de la résine photosensible sont les mêmes que celles utilisées pour la lithographie d'électrodes interdigitées Cr-Au sur substrat Si – SiO$_2$.

Les échantillons sont ensuite exposés à l'UV pendant 5 s à travers le masque chromé correspondant aux motifs. La résine illuminée est ainsi rendue soluble dans le développeur. Les échantillons sont ensuite développés dans un bain d'AZ 726 pendant 40 s avant d'être rincés dans l'eau distillée. Le négatif des motifs n'est plus protégé par la résine et est gravé par une solution d'HCl à 25 % pendant 12 min et rincés à l'eau distillée.

Les échantillons sont ensuite rincés à l'acétone et à l'isopropanol. Les motifs obtenus sont présentés sur la Figure 96.

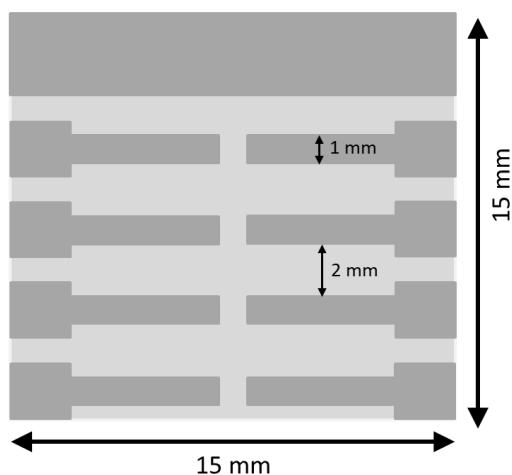

*Figure 96 : Schéma des électrodes d'ITO (gris foncé) sur verre (gris clair)*

5. **Électrodes 8 pixels FTO sur verre (lithographie positive)**

Les substrats de verre, couverts d'une couche d'oxyde d'étain dopé au fluor (*fluorine doped tin oxide*, FTO, Solem, 70 à 90 Ω/□) sont découpés en carré de 15 mm de côté. Les étapes correspondant au nettoyage des substrats, dépôt de résine, exposition à l'UV et développement dans l'AZ 726 sont les mêmes que celles utilisées pour la lithographie d'électrodes 8 pixels ITO sur verre. Après rinçage des échantillons à l'eau distillée, de la poudre de zinc (Zn, Sigma, 98%) est parsemée sur la surface et l'excès est retiré. Le procédé est répété deux fois et sert d'activation à la gravure. Les échantillons sont ensuite placés dans une solution d'HCl à 2 M (2 mL d'HCl à 37% et 10 mL d'eau distillée) pendant 15 min, et un coton tige est utilisé pour retirer la poudre de Zn de la surface.

Les échantillons sont ensuite nettoyés à l'eau distillée pour arrêter la gravure, puis plongés dans un bain d'acétone afin de retirer toute la résine. Ils sont finalement rincés l'acétone et à l'isopropanol. Les motifs réalisés sont les mêmes que ceux utilisés pour les électrodes ITO sur verre (Figure 96).

### III. Fabrication de la matrice de pixels

#### 1. Première étape – colonnes (lithographie négative)

Une demi-lame de microscope est d'abord nettoyée suivant le même protocole que pour les électrodes interdigitées (voir p132). Le promoteur d'adhésion est ensuite déposé à la tournette sur le substrat (4000 rpm, 1000 rpm/s, 30s) puis est recuit à 120 °C pendant 2 minutes. La résine photosensible AZ5214E est ensuite déposée à la tournette sur le substrat (4000 rpm, 1000 rpm/s, 30s) avant d'être recuite à 110 °C pendant 1 min 30 s.

L'échantillon est ensuite exposé à une lampe UV à travers un masque correspondant aux motifs des colonnes et aux reprises de contacts macroscopiques. L'exposition dure 1,5 s. La résine est ensuite inversée grâce à un recuit à 125 °C pendant 2 minutes, et l'échantillon est à nouveau exposé à la lampe UV, cette fois sans masque, pendant 40 secondes. Le développement est réalisé dans le développeur AZ726, pendant une durée allant de 16 à 25 secondes (les motifs au bord de la lame de verre peuvent être plus longs à développer). L'échantillon est alors rincé à l'eau distillée, puis nettoyé par un plasma oxygène pendant 5 min. Des évaporations de chrome (5 nm) et d'or (40 nm) sont réalisées, et la résine qui recouvrait le négatif des motifs est retirée dans un bain d'acétone. On obtient les motifs présentés sur la Figure 97.

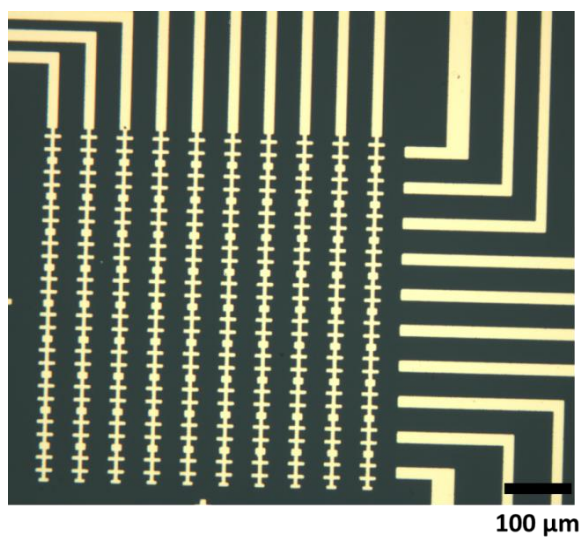

*Figure 97 : Motifs d'or obtenus après la première étape de lithographie. Les lignes d'or en haut et à droite de la matrice de pixels permettent de reporter les contacts sur les bords de la lame de verre.*

2. **Deuxième étape – isolation des contacts (lithographie positive)**

Cette étape consiste à déposer de la silice pour isoler les reports de contacts déposés dans l'étape 1. Pour cela, la silice doit recouvrir tout le substrat excepté le centre des pixels, sur lequel on viendra déposer les électrodes correspondant aux lignes, le report des contacts des lignes sur les contacts macroscopiques déposés pendant l'étape 1, et les contacts macroscopiques.

L'échantillon est nettoyé avec de l'acétone et de l'isopropanol avant d'être séché avec un flux d'azote. Les dépôts de TI-PRIME et de la résine photosensible AZ5214E sont déposés en suivant la même procédure que dans l'étape 1. L'échantillon est exposé aux rayons UV à travers un masque présentant les motifs à protéger pendant 8 secondes. Il est ensuite développé par le développeur AZ 726 pendant 45 secondes avant d'être rincé à l'eau distillée. L'échantillon est nettoyé par plasma d'oxygène pendant 5 minutes avant le dépôt de silice.

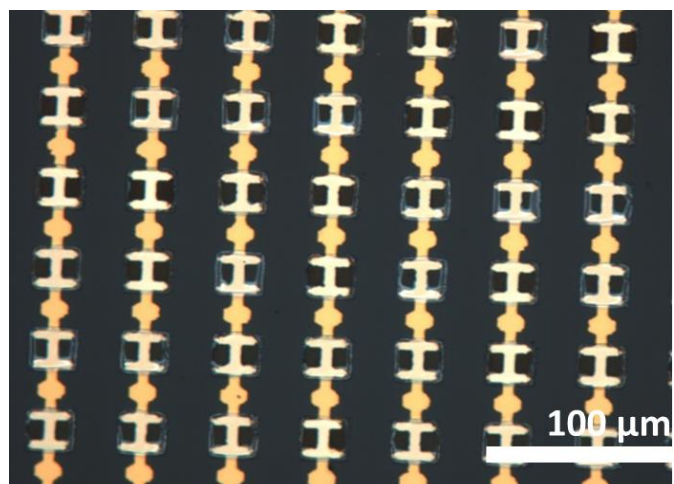

*Figure 98 : Motifs de silice déposés sur la matrice. Autour des pixels (en forme de H) se dessinent des carrés. Les pixels en jaune clair sont nus, tandis que le reste de la matrice est recouvert de silice.*

La silice est déposée par pulvérisation cathodique magnétron (Alcatel). 75 nm de silice sont déposés à une puissance de 120 W, pendant 27 minutes. À la fin de la pulvérisation, l'échantillon est plongé dans un bain d'acétone pendant une dizaine d'heures afin de retirer la résine restante. Les motifs obtenus sont présentés sur la Figure 98.

3. **Troisième étape – lignes (lithographie négative)**

Après avoir rincé l'échantillon à l'acétone et à l'isopropanol, le promoteur d'adhésion Ti-PRIME et la résine photosensible AZ5214E sont déposés en suivant le même protocole que dans les étapes 1 et 2. L'échantillon est ensuite exposé aux UV à travers un masque reproduisant les lignes de pixels et les reports de contact. Le processus d'inversion de la résine (recuit et exposition aux UV sans masque) et le même que celui décrit à l'étape 1. L'échantillon est finalement développé pendant 16 s dans le développeur AZ 726 et rincé à l'eau distillée.

Après un nettoyage au plasma d'oxygène pendant 5 minutes, 10 nm de chrome et 150 nm sont évaporés à la surface de l'échantillon. À la fin de l'évaporation, l'échantillon est placé dans un bain d'acétone pendant une dizaine d'heures minimum pour retirer la résine restante. Les motifs finalement obtenus sont présentés sur la Figure 99.

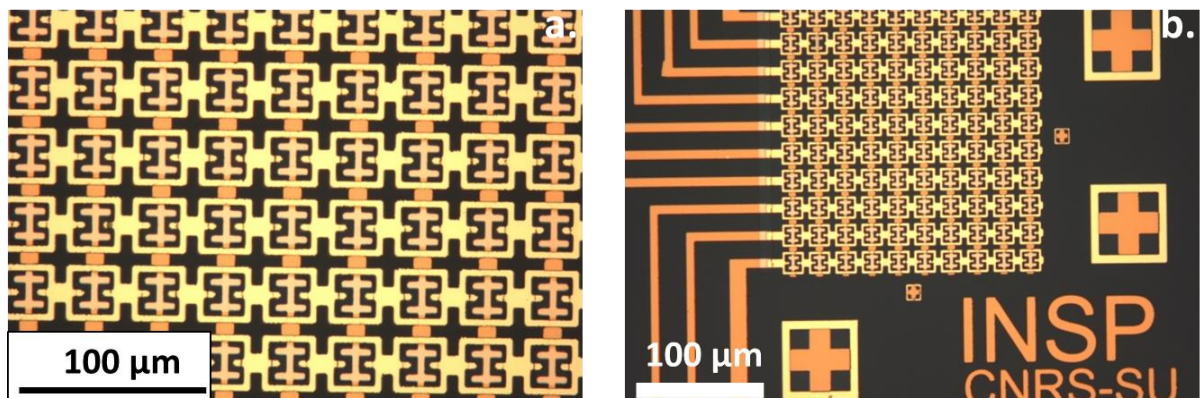

*Figure 99 : Motifs obtenus après la troisième étape de lithographie. (a) zoom sur les pixels. (b) schéma d'ensemble de la matrice avec reports de contact à gauche. Les croix sur la droite de l'image et en bas ont permis d'aligner les masques sur les motifs déjà présents.*

# Annexe 3
# Caractérisations

## I. Caractérisations structurales et spectroscopiques des nanocristaux colloïdaux

Après la synthèse, les nanocristaux peuvent être caractérisés en utilisant différentes techniques, donnant des informations complémentaires telles que la taille, la forme, la stœchiométrie, les énergies des transitions accessibles, la phase cristalline *etc* …

Dans cette annexe, j'indique les procédures utilisées pour réaliser ces caractérisations, ainsi que les appareils employés.

### 1. Microscopie électronique à transmission

Pour obtenir les clichés de microscopie électronique à transmission, une goutte de solution de nanocristaux diluée est déposée sur une grille en cuivre recouverte d'un film de carbone amorphe. La grille est dégazée sous vide secondaire pendant une nuit afin de retirer le solvant. Le microscope utilisé est un **JEOL 2010F**. Pour l'acquisition d'images, on applique une tension de 200 kV.

### 2. Analyse dispersive en énergie

L'analyse dispersive en énergie nous permet de remonter aux différents matériaux présents dans les nanocristaux et ce, de manière quantitative. Elle nous permet notamment de déterminer les ratios mercure/chalcogène au sein des nanocristaux.

Pour réaliser une analyse dispersive en énergie (*Energy Dispersive X-ray analysis* ou *EDX*), une goutte de solution colloïdale de nanocristaux est déposée sur un substrat conducteur. L'échantillon est ensuite placé dans un microscope électronique à balayage **FEI Magellan**, sous vide secondaire. Le courant du faisceau est ajusté à 0,8 nA, et la tension est fixée à 15 kV. L'analyse EDX est obtenue grâce à une sonde **Oxford**.

### 3. Absorption infrarouge

Le spectre infrarouge des nanocristaux est obtenu en utilisant un spectromètre infrarouge à transformée de Fourier, en configuration ATR (réflexion totale atténuée, ou *Attenuated Total Reflection*). Le modèle de spectromètre utilisé est un **Vertex 70** de **Bruker**. Pour cela, une goutte de solution de nanocristaux est déposée sur la platine ATR, et le spectre est acquis une fois le solvant évaporé. Le spectre infrarouge peut également être acquis sur des films de nanocristaux, en transmission.

Pour les mesures d'absorption à basse température, nous utilisons un spectromètre **Bruker IFS66v**, et un cryostat à hélium. Les mesures sont effectuées en transmission.

### 4. Diffraction des rayons X

La diffraction permet notamment de vérifier la phase cristalline des échantillons obtenus et de déterminer la taille des cristallites grâce à la formule de Scherrer.

Les échantillons étudiés en diffraction sont obtenus en déposant une goutte de solution de nanocristaux sur un substrat de silicium. Les mesures sont effectuées sur un diffractomètre **Philips X'pert**, sous une tension de 40 kV et un courant de 40 mA. L'illumination de l'échantillon est réalisée avec la raie Kα de l'anode de cuivre ($\lambda = 0{,}154$ nm).

5. **Photoémission**

La photoémission, utilisée pour mesurer l'énergie de la bande de valence par rapport à l'énergie de Fermi, le travail de sortie des nanocristaux, ou les états d'oxydation des différents éléments, est décrite en détails dans le corps du texte (p53).

## II. Échange de ligands

Pour améliorer le transport au sein d'un film de nanocristaux, les longs ligands utilisés pendant la synthèse peuvent être échangés pour des ligands plus courts qui faciliteront les sauts tunnels d'un nanocristal à l'autre.

### 1. Produits utilisés

Sulfure de sodium nonahydrate ($Na_2S$, $9H_2O$, Sigma-Aldrich, 98,0%), 1,2 éthanedithiol (EDT, Fluka, 98,0%), trisulfure d'arsenic ($As_2S_3$, Alfa Aesar, 99,9%), butylamine ($BuNH_2$, Sigma Aldrich), octanethiol (OSH, Fluka), butanethiol (BuSH, Sigma-Aldrich), 1,4 benzènedithiol (BeSH, Alfa Aesar), thiocyanate d'ammonium ($NH_4SCN$, Sigma Aldrich), chlorure d'ammonium ($NH_4Cl$, Sigma Aldrich), 2 mercaptoéthanol (mercaptoalcool ou MPOH, Merck > 99%), chlorure de mercure ($HgCl_2$, Strem Chemicals, 99%).

### 2. Protocoles

Les protocoles d'échanges de ligands sont décrits dans le corps du texte. L'échange de ligands sur film est décrit p42, l'échange de ligands par transfert de phase est décrit p42 et rappelé p99.

## III. Mesures de transistors

### a. Préparation de l'électrolyte

Matériaux utilisés : Perchlorate de lithium ($LiClO_4$, Sigma Aldrich, 98%), polyéthylène glycol (PEG 6k, $M_w$ = 6 kg.mol$^{-1}$, Fluka).

En boîte à gants, sous atmosphère d'argon, 500 mg de $LiClO_4$ sont mélangés à 2,3 g de PEG 6k, sur plaque chauffante, à 170 °C, jusqu'à ce que le mélange devienne transparent et homogène (environ 2 h). L'électrolyte est ensuite refroidi à température ambiante et est conservé sous atmosphère inerte dans la boîte à gants. Pour réaliser un transistor, un film de nanocristaux est déposé entre deux électrodes (le plus souvent des électrodes interdigitées sur un substrat Si/SiO$_2$ ou verre). Une goutte d'électrolyte à 100 °C, l'électrolyte se présente alors sous la forme d'un gel visqueux, est placée sur le film. Une fois l'électrolyte refroidi et durci, une grille de microscopie en cuivre est déposée sur la partie supérieure pour constituer l'électrode de grille.

### b. Obtention des courbes de transfert

L'échantillon est connecté à un **Keithley 2634b** qui impose une tension drain-source et fait varier la tension grille source. Il mesure les courants drain-source et grille-source pour chaque valeur de tension grille-source appliquée.

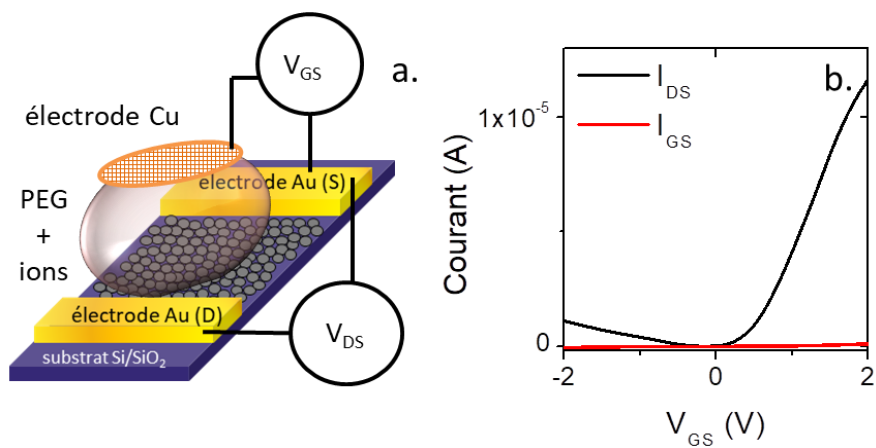

*Figure 100 : (a) Schéma de transistor à base de nanocristaux avec une grille électrolytique ; (b) exemple de courbe transfert.*

## IV. Caractérisations des dispositifs de détection infrarouge à base de nanocristaux

### 1. Caractéristiques courant-tension

Les propriétés de transport des films de nanocristaux peuvent être sondées via les caractéristiques courant-tension. Pour des photoconducteurs, elles sont souvent linéaires (si les contacts sont ohmiques), tandis qu'elles présentent un sens bloquant et un sens passant pour les photodiodes (p90).

Pour obtenir ces caractéristiques, l'échantillon est connecté à un **Keithley 2634b** qui permet d'imposer la gamme de tensions à balayer ainsi que le pas en tension, et mesure le courant qui circule.

Ces caractéristiques sont également mesurées sous illumination. Pour cela, on utilise un laser qui illumine la zone active. Les précisions sur la longueur d'onde ou la puissance utilisée sont données dans le corps du texte.

### 2. Mesures du transport à basses températures

L'évolution du courant d'obscurité avec la température renseigne sur l'énergie d'activation des porteurs, la densité de pièges … Dans le cadre de l'étude de la transition semiconducteur – métal observée au chapitre 2, l'étude du courant en fonction de la température nous a permis de mettre en avant le caractère métallique des nanocristaux de HgSe les plus larges.

Pour réaliser ces mesures, l'échantillon est connecté à un **Keithley 2634b** et est placé dans un cryostat dont la gamme de températures accessibles est [2 K ; 310 K].

### 3. Mesure de réponse à différentes fréquences

Pour obtenir des mesures de réponse, on utilise comme source de lumière un corps noir Omega BB-4A à 980 °C. Le flux de lumière est filtré par un filtre en germanium qui permet de couper les longueurs d'onde inférieures à 1,9 µm, et est modulé optiquement entre 1 Hz et 1 kHz. Le dispositif de détection infrarouge est placé en face du corps noir, et le photocourant, amplifié et converti en tension par un amplificateur de transimpédance **Femto DLPCA-200**, est mesuré grâce à détecteur synchrone **Zurich Instruments MFLI**, qui permet également d'imposer une tension drain-source. Le schéma du montage est présenté dans le corps du texte (p107, Figure 80a).

La puissance optique reçue par le détecteur se calcule comme :

$$P\,(W) = A_d\pi \cdot \cos(\theta) \cdot sin^2(\alpha) \int_{1,9\,\mu m}^{seuil\ QD} \frac{2hc^2}{\lambda^5} \frac{1}{e^{\frac{hc}{\lambda kT}} - 1}\,d\lambda$$

Où $A_d$ est l'aire du détecteur, $\theta$ est l'angle entre la normale du détecteur et le corps noir, fixé à 0° pendant les mesures, $\alpha$ est le demi angle de vue avec lequel le détecteur voit le corps noir, dépendant de la distance entre le corps noir et le détecteur, h est la constante de Planck, k est la constante de Boltzmann, T est les température du corps noir. L'intégrale est calculée entre 1,9 µm, qui correspond au filtre en germanium, et le seuil d'absorption des nanocristaux (ou QD) utilisés dans le détecteur.

### 4. Mesure de détectivité

Le courant d'obscurité du dispositif de détection infrarouge, à une tension fixée, est amplifié par un amplificateur **Femto DLPCA-200**, puis analysé par l'analyseur de spectres **SRS SR780**. Le schéma du montage est rappelé dans le corps du texte (p108, Figure 81a).

# Liste des figures

# Liste des tableaux

# Bibliographie

1. Faraday Michael, X. The Bakerian Lecture. —Experimental relations of gold (and other metals) to light. *Philos. Trans. R. Soc. Lond.* **147**, 145–181 (1857).
2. G. Mie, Beiträge zur Optik trüber Medien, speziell kolloidaler Metallösungen. *Ann. Phys.* **330**, 377–445 (1908).
3. C. Delerue, D. Vanmaekelbergh, Electronic band structure of zinc blende CdSe and rock salt PbSe semiconductors with silicene-type honeycomb geometry. *2D Mater.* **2**, 34008 (2015).
4. R. Benchamekh, N. A. Gippius, J. Even, M. O. Nestoklon, J.-M. Jancu, S. Ithurria, B. Dubertret, A. L. Efros, P. Voisin, Tight-binding calculations of image-charge effects in colloidal nanoscale platelets of CdSe. *Phys. Rev. B*. **89**, 35307 (2014).
5. C. Kittel, *Théorie quantique du solide* (Dunod, 1967).
6. R. Hill, Energy-gap variations in semiconductor alloys. *J. Phys. C Solid State Phys.* **7**, 521–526 (1974).
7. A. Sadao, *GaAs And Related Materials* (World Scientific, 1994).
8. Y. P. Varshni, Temperature dependence of the energy gap in semiconductors. *Physica*. **34**, 149–154 (1967).
9. V. I. Klimov, *Semiconductor and Metal Nanocrystals: Synthesis and Electronic and Optical Properties* (CRC Press, 2003).
10. V. Rinnerbauer, K. Hingerl, M. Kovalenko, W. Heiss, Effect of quantum confinement on higher transitions in HgTe nanocrystals. *Appl. Phys. Lett.* **89**, 193114 (2006).
11. N. Goubet, A. Jagtap, C. Livache, B. Martinez, H. Portalès, X. Z. Xu, R. P. S. M. Lobo, B. Dubertret, E. Lhuillier, Terahertz HgTe Nanocrystals: Beyond Confinement. *J. Am. Chem. Soc.* **140**, 5033–5036 (2018).
12. J. M. Pietryga, Y.-S. Park, J. Lim, A. F. Fidler, W. K. Bae, S. Brovelli, V. I. Klimov, Spectroscopic and Device Aspects of Nanocrystal Quantum Dots. *Chem. Rev.* **116**, 10513–10622 (2016).
13. C. B. Murray, D. J. Norris, M. G. Bawendi, Synthesis and characterization of nearly monodisperse CdE (E = sulfur, selenium, tellurium) semiconductor nanocrystallites. *J. Am. Chem. Soc.* **115**, 8706–8715 (1993).
14. J. Jasieniak, L. Smith, J. van Embden, P. Mulvaney, M. Califano, Re-examination of the Size-Dependent Absorption Properties of CdSe Quantum Dots. *J. Phys. Chem. C*. **113**, 19468–19474 (2009).
15. M. A. Hines, P. Guyot-Sionnest, Synthesis and Characterization of Strongly Luminescing ZnS-Capped CdSe Nanocrystals. *J. Phys. Chem.* **100**, 468–471 (1996).
16. S. Tamang, C. Lincheneau, Y. Hermans, S. Jeong, P. Reiss, Chemistry of InP Nanocrystal Syntheses. *Chem. Mater.* **28**, 2491–2506 (2016).
17. J. L. Stein, E. A. Mader, B. M. Cossairt, Luminescent InP Quantum Dots with Tunable Emission by Post-Synthetic Modification with Lewis Acids. *J. Phys. Chem. Lett.* **7**, 1315–1320 (2016).
18. G. Nedelcu, L. Protesescu, S. Yakunin, M. I. Bodnarchuk, M. J. Grotevent, M. V. Kovalenko, Fast Anion-Exchange in Highly Luminescent Nanocrystals of Cesium Lead Halide Perovskites (CsPbX3, X = Cl, Br, I). *Nano Lett.* **15**, 5635–5640 (2015).
19. B. G. Streetman, S. K. Banerjee, *Solid State Electronic Devices* (2014), *Pearson*.
20. M. A. Hines, G. D. Scholes, Colloidal PbS Nanocrystals with Size-Tunable Near-Infrared Emission: Observation of Post-Synthesis Self-Narrowing of the Particle Size Distribution. *Adv. Mater.* **15**, 1844–1849 (2003).
21. Y. Pan, Y. R. Li, Y. Zhao, D. L. Akins, Synthesis and Characterization of Quantum Dots: A Case Study Using PbS. *J. Chem. Educ.* **92**, 1860–1865 (2015).
22. B. Martinez, C. Livache, L. D. Notemgnou Mouafo, N. Goubet, S. Keuleyan, H. Cruguel, S. Ithurria, H. Aubin, A. Ouerghi, B. Doudin, E. Lacaze, B. Dubertret, M. G. Silly, R. P. S. M.

Lobo, J.-F. Dayen, E. Lhuillier, HgSe Self-Doped Nanocrystals as a Platform to Investigate the Effects of Vanishing Confinement. *ACS Appl. Mater. Interfaces*. **9**, 36173–36180 (2017).

23. G. Allan, C. Delerue, Tight-binding calculations of the optical properties of HgTe nanocrystals. *Phys. Rev. B*. **86**, 165437 (2012).

24. A. Jagtap, B. Martinez, N. Goubet, A. Chu, C. Livache, C. Gréboval, J. Ramade, D. Amelot, P. Trousset, A. Triboulin, S. Ithurria, M. G. Silly, B. Dubertret, E. Lhuillier, Design of a Unipolar Barrier for a Nanocrystal-Based Short-Wave Infrared Photodiode. *ACS Photonics*. **5**, 4569–4576 (2018).

25. A. McWorther, 1/f noise nd related surface effects in Germanium (1955).

26. F. N. Hooge, A. M. H. Hoppenbrouwers, 1/ƒ noise in continuous thin gold films. *Physica*. **45**, 386–392 (1969).

27. Y. Lai, H. Li, D. K. Kim, B. T. Diroll, C. B. Murray, C. R. Kagan, Low-Frequency (1/f) Noise in Nanocrystal Field-Effect Transistors. *ACS Nano*. **8**, 9664–9672 (2014).

28. J. F. Stephany, Origin of 1/f noise. *J. Appl. Phys.* **46**, 665–667 (1975).

29. M. Jiang, X. Cao, S. Bao, H. Zhou, P. Jin, Regulation of the phase transition temperature of VO2 thin films deposited by reactive magnetron sputtering without doping. *Thin Solid Films*. **562**, 314–318 (2014).

30. A. Rogalski, Infrared detectors: an overview. *Infrared Phys. Technol.* **43**, 187–210 (2002).

31. D. W. Palmer, *www.semiconductors.co.uk* (2003), (available at www.semiconductors.co.uk).

32. A. Nag, M. V. Kovalenko, J.-S. Lee, W. Liu, B. Spokoyny, D. V. Talapin, Metal-free Inorganic Ligands for Colloidal Nanocrystals: S2–, HS–, Se2–, HSe–, Te2–, HTe–, TeS32–, OH–, and NH2– as Surface Ligands. *J. Am. Chem. Soc.* **133**, 10612–10620 (2011).

33. S. Keuleyan, E. Lhuillier, V. Brajuskovic, P. Guyot-Sionnest, Mid-infrared HgTe colloidal quantum dot photodetectors. *Nat. Photonics*. **5**, 489–493 (2011).

34. G. M. Dalpian, J. R. Chelikowsky, Self-Purification in Semiconductor Nanocrystals. *Phys. Rev. Lett.* **96**, 226802 (2006).

35. M. Amelia, C. Lincheneau, S. Silvi, A. Credi, Electrochemical properties of CdSe and CdTe quantum dots. *Chem. Soc. Rev.* **41**, 5728–5743 (2012).

36. N. P. Osipovich, S. K. Poznyak, V. Lesnyak, N. Gaponik, Cyclic voltammetry as a sensitive method for in situ probing of chemical transformations in quantum dots. *Phys. Chem. Chem. Phys.* **18**, 10355–10361 (2016).

37. S. K. Poznyak, N. P. Osipovich, A. Shavel, D. V. Talapin, M. Gao, A. Eychmüller, N. Gaponik, Size-Dependent Electrochemical Behavior of Thiol-Capped CdTe Nanocrystals in Aqueous Solution. *J. Phys. Chem. B*. **109**, 1094–1100 (2005).

38. M. Chen, P. Guyot-Sionnest, Reversible Electrochemistry of Mercury Chalcogenide Colloidal Quantum Dot Films. *ACS Nano*. **11**, 4165–4173 (2017).

39. C. Wang, M. Shim, P. Guyot-Sionnest, Electrochromic Nanocrystal Quantum Dots. *Science*. **291**, 2390–2392 (2001).

40. J. Robertson, High dielectric constant gate oxides for metal oxide Si transistors. *Rep. Prog. Phys.* **69**, 327–396 (2005).

41. E. Lhuillier, S. Ithurria, A. Descamps-Mandine, T. Douillard, R. Castaing, X. Z. Xu, P.-L. Taberna, P. Simon, H. Aubin, B. Dubertret, Investigating the n- and p-Type Electrolytic Charging of Colloidal Nanoplatelets. *J. Phys. Chem. C*. **119**, 21795–21799 (2015).

42. B. Martinez, C. Livache, N. Goubet, A. Jagtap, H. Cruguel, A. Ouerghi, E. Lacaze, M. G. Silly, E. Lhuillier, Probing Charge Carrier Dynamics to Unveil the Role of Surface Ligands in HgTe Narrow Band Gap Nanocrystals. *J. Phys. Chem. C*. **122**, 859–865 (2018).

43. H. Henck, D. Pierucci, J. Chaste, C. H. Naylor, J. Avila, A. Balan, M. G. Silly, M. C. Asensio, F. Sirotti, A. T. C. Johnson, E. Lhuillier, A. Ouerghi, Electrolytic phototransistor based on graphene-MoS2 van der Waals p-n heterojunction with tunable photoresponse. *Appl. Phys. Lett.* **109**, 113103 (2016).

44. E. Lhuillier, A. Robin, S. Ithurria, H. Aubin, B. Dubertret, Electrolyte-Gated Colloidal Nanoplatelets-Based Phototransistor and Its Use for Bicolor Detection. *Nano Lett.* **14**, 2715–2719 (2014).

45. E. Lhuillier, M. Scarafagio, P. Hease, B. Nadal, H. Aubin, X. Z. Xu, N. Lequeux, G. Patriarche, S. Ithurria, B. Dubertret, Infrared Photodetection Based on Colloidal Quantum-Dot Films with High Mobility and Optical Absorption up to THz. *Nano Lett.* **16**, 1282–1286 (2016).

46. A. C. Thompson, D. T. Attwood, E. M. Gullikson, M. R. Howells, J. B. Kortright, A. L. Robinson, J. H. Underwood, Z. J. Kim, J. Kirz, I. Lindau, P. Pianetta, H. Winick, G. P. Williams, J. H. Scofields, *X-Ray Data Booklet* (2009).

47. S. Keuleyan, E. Lhuillier, P. Guyot-Sionnest, Synthesis of Colloidal HgTe Quantum Dots for Narrow Mid-IR Emission and Detection. *J. Am. Chem. Soc.* **133**, 16422–16424 (2011).

48. A. Robin, C. Livache, S. Ithurria, E. Lacaze, B. Dubertret, E. Lhuillier, Surface Control of Doping in Self-Doped Nanocrystals. *ACS Appl. Mater. Interfaces.* **8**, 27122–27128 (2016).

49. J. Jeong, B. Yoon, Y.-W. Kwon, D. Choi, K. S. Jeong, Singly and Doubly Occupied Higher Quantum States in Nanocrystals. *Nano Lett.* **17**, 1187–1193 (2017).

50. E. Lhuillier, S. Keuleyan, P. Guyot-Sionnest, Optical properties of HgTe colloidal quantum dots. *Nanotechnology.* **23**, 175705 (2012).

51. Z. Deng, K. S. Jeong, P. Guyot-Sionnest, Colloidal Quantum Dots Intraband Photodetectors. *ACS Nano.* **8 (11)**, 11707–11714 (2014).

52. N. F. Mott, Metal-Insulator Transition. *Rev. Mod. Phys.* **40**, 677–683 (1968).

53. H. Fu, K. V. Reich, B. I. Shklovskii, Hopping conductivity and insulator-metal transition in films of touching semiconductor nanocrystals. *Phys. Rev. B.* **93**, 125430 (2016).

54. T. Chen, K. V. Reich, N. J. Kramer, H. Fu, U. R. Kortshagen, B. I. Shklovskii, Metal-insulator transition in films of doped semiconductor nanocrystals. *Nat. Mater.* **15**, 299–303 (2016).

55. J. Qu, C. Livache, B. Martinez, C. Greboval, A. Chu, E. Merrigio, J. Ramade, H. Cruguel, X. Xu, A. Proust, F. Volatron, G. Cabailh, N. Goubet, E. Lhuillier, Transport in ITO Nanocrystals with Short- to Long-Wave Infrared Absorption for Heavy Metal-Free Infrared Photodetection. *ACS Appl. Nano Mater.* **2**, 1621–1630 (2019).

56. M. K. H. Taha, O. Boisron, B. Canut, P. Melinon, J. Penuelas, M. Gendry, B. Masenelli, Control of the compensating defects in Al-doped and Ga-doped ZnO nanocrystals for MIR plasmonics. *RSC Adv.* **7**, 28677–28683 (2017).

57. S. Einfeldt, F. Goschenhofer, C. R. Becker, G. Landwehr, Optical properties of HgSe. *Phys. Rev. B.* **51**, 4915–4925 (1995).

58. A. M. Schimpf, N. Thakkar, C. E. Gunthardt, D. J. Masiello, D. R. Gamelin, Charge-Tunable Quantum Plasmons in Colloidal Semiconductor Nanocrystals. *ACS Nano.* **8**, 1065–1072 (2014).

59. G. Shen, P. Guyot-Sionnest, HgS and HgS/CdS Colloidal Quantum Dots with Infrared Intraband Transitions and Emergence of a Surface Plasmon. *J. Phys. Chem. C.* **120**, 11744–11753 (2016).

60. T. B. Tran, I. S. Beloborodov, X. M. Lin, T. P. Bigioni, V. M. Vinokur, H. M. Jaeger, Multiple Cotunneling in Large Quantum Dot Arrays. *Phys. Rev. Lett.* **95**, 76806 (2005).

61. H. Moreira, Q. Yu, B. Nadal, B. Bresson, M. Rosticher, N. Lequeux, A. Zimmers, H. Aubin, Electron Cotunneling Transport in Gold Nanocrystal Arrays. *Phys. Rev. Lett.* **107**, 176803 (2011).

62. P. Guyot-Sionnest, Electrical Transport in Colloidal Quantum Dot Films. *J. Phys. Chem. Lett.* **3**, 1169–1175 (2012).

63. C.-H. M. Chuang, A. Maurano, R. E. Brandt, G. W. Hwang, J. Jean, T. Buonassisi, V. Bulović, M. G. Bawendi, Open-Circuit Voltage Deficit, Radiative Sub-Bandgap States, and Prospects in Quantum Dot Solar Cells. *Nano Lett.* **15**, 3286–3294 (2015).

64. P. Guyot-Sionnest, E. Lhuillier, H. Liu, A mirage study of CdSe colloidal quantum dot films, Urbach tail, and surface states. *J. Chem. Phys.* **137**, 154704 (2012).

65. C. Main, D. Nesheva, Transient Photocurrent Techniques as a Means of Characterising Amorphous Semiconductors. *J. Optoelectron. Adv. Mater.* **3**, 655–664 (2001).

66. J. Orenstein, M. A. Kastner, V. Vaninov, Transient photoconductivity and photo-induced optical absorption in amorphous semiconductors. *Philos. Mag. Part B.* **46**, 23–62 (1982).

67. J. Orenstein, M. Kastner, Photocurrent Transient Spectroscopy: Measurement of the Density of Localized States in -As2Se3. *Phys. Rev. Lett.* **46**, 1421–1424 (1981).

68. J. Tang, K. W. Kemp, S. Hoogland, K. S. Jeong, H. Liu, L. Levina, M. Furukawa, X. Wang, R. Debnath, D. Cha, K. W. Chou, A. Fischer, A. Amassian, J. B. Asbury, E. H. Sargent, Colloidal-quantum-dot photovoltaics using atomic-ligand passivation. *Nat. Mater.* **10**, 765–771 (2011).

69. A. T. Fafarman, W. Koh, B. T. Diroll, D. K. Kim, D.-K. Ko, S. J. Oh, X. Ye, V. Doan-Nguyen, M. R. Crump, D. C. Reifsnyder, C. B. Murray, C. R. Kagan, Thiocyanate-Capped Nanocrystal Colloids: Vibrational Reporter of Surface Chemistry and Solution-Based Route to Enhanced Coupling in Nanocrystal Solids. *J. Am. Chem. Soc.* **133**, 15753–15761 (2011).

70. E. Lhuillier, S. Keuleyan, P. Zolotavin, P. Guyot-Sionnest, Mid-Infrared HgTe/As2S3 Field Effect Transistors and Photodetectors. *Adv. Mater.* **25**, 137–141 (2013).

71. M. Dufour, E. Izquierdo, C. Livache, B. Martinez, M. G. Silly, T. Pons, E. Lhuillier, C. Delerue, S. Ithurria, Doping as a Strategy to Tune Color of 2D Colloidal Nanoplatelets. *ACS Appl. Mater. Interfaces*. **11**, 10128–10134 (2019).

72. S. C. Erwin, L. Zu, M. I. Haftel, A. L. Efros, T. A. Kennedy, D. J. Norris, Doping semiconductor nanocrystals. *Nature*. **436**, 91–94 (2005).

73. A. Sahu, M. S. Kang, A. Kompch, C. Notthoff, A. W. Wills, D. Deng, M. Winterer, C. D. Frisbie, D. J. Norris, Electronic Impurity Doping in CdSe Nanocrystals. *Nano Lett.* **12**, 2587–2594 (2012).

74. D. J. Norris, A. L. Efros, S. C. Erwin, Doped Nanocrystals. *Science*. **319**, 1776–1779 (2008).

75. M. K. Hamza, J.-M. Bluet, K. Masenelli-Varlot, B. Canut, O. Boisron, P. Melinon, B. Masenelli, Tunable mid IR plasmon in GZO nanocrystals. *Nanoscale*. **7**, 12030–12037 (2015).

76. J. Qu, N. Goubet, C. Livache, B. Martinez, D. Amelot, C. Gréboval, A. Chu, J. Ramade, H. Cruguel, S. Ithurria, M. G. Silly, E. Lhuillier, Intraband Mid-Infrared Transitions in Ag2Se Nanocrystals: Potential and Limitations for Hg-Free Low-Cost Photodetection. *J. Phys. Chem. C*. **122**, 18161–18167 (2018).

77. A. H. Nethercot, Prediction of Fermi Energies and Photoelectric Thresholds Based on Electronegativity Concepts. *Phys. Rev. Lett.* **33**, 1088–1091 (1974).

78. C. Livache, E. Izquierdo, B. Martinez, M. Dufour, D. Pierucci, S. Keuleyan, H. Cruguel, L. Becerra, J. L. Fave, H. Aubin, E. Lacaze, M. G. Silly, B. Dubertret, S. Ithurria, E. Lhuillier, Charge Dynamics and Optolectronic Properties in HgTe Colloidal Quantum Wells. *Nano Lett.* **17**, 4067–4074 (2017).

79. P. Guyot-Sionnest, M. A. Hines, Intraband transitions in semiconductor nanocrystals. *Appl. Phys. Lett.* **72**, 686–688 (1998).

80. P. R. Brown, D. Kim, R. R. Lunt, N. Zhao, M. G. Bawendi, J. C. Grossman, V. Bulović, Energy Level Modification in Lead Sulfide Quantum Dot Thin Films through Ligand Exchange. *ACS Nano*. **8**, 5863–5872 (2014).

81. D. M. Kroupa, M. Vörös, N. P. Brawand, B. W. McNichols, E. M. Miller, J. Gu, A. J. Nozik, A. Sellinger, G. Galli, M. C. Beard, Tuning colloidal quantum dot band edge positions through solution-phase surface chemistry modification. *Nat. Commun.* **8**, 15257 (2017).

82. C.-H. M. Chuang, P. R. Brown, V. Bulović, M. G. Bawendi, Improved performance and stability in quantum dot solar cells through band alignment engineering. *Nat. Mater.* **13**, 796–801 (2014).

83. K. S. Jeong, Z. Deng, S. Keuleyan, H. Liu, P. Guyot-Sionnest, Air-Stable n-Doped Colloidal HgS Quantum Dots. *J. Phys. Chem. Lett.* **5**, 1139–1143 (2014).

84. S. Yang, D. Prendergast, J. B. Neaton, Tuning Semiconductor Band Edge Energies for Solar Photocatalysis via Surface Ligand Passivation. *Nano Lett.* **12**, 383–388 (2012).

85. B. de Boer, A. Hadipour, M. M. Mandoc, T. van Woudenbergh, P. W. M. Blom, Tuning of Metal Work Functions with Self-Assembled Monolayers. *Adv. Mater.* **17**, 621–625 (2005).

86. Z. Deng, P. Guyot-Sionnest, Intraband Luminescence from HgSe/CdS Core/Shell Quantum Dots. *ACS Nano*. **10**, 2121–2127 (2016).

87. C. Rocchiccioli-Deltcheff, M. Fournier, R. Franck, R. Thouvenot, Vibrational investigations of polyoxometalates. 2. Evidence for anion-anion interactions in Molybdenum(VI) and Tungsten(VI) compounds related to the Keggin structure. *Inorg. Chem.* **22**, 207–216 (1983).

88. P. Souchay, *Polyanions and polycations* (1963), *Gauthier-Villars, Paris*.

89. L. Huder, C. Rinfray, D. Rouchon, A. Benayad, M. Baraket, G. Izzet, F. Lipp-Bregolin, G. Lapertot, L. Dubois, A. Proust, L. Jansen, F. Duclairoir, Evidence for Charge Transfer at the Interface between Hybrid Phosphomolybdate and Epitaxial Graphene. *Langmuir*. **32**, 4774–4783 (2016).

90. M. Sadakane, E. Steckhan, Electrochemical Properties of Polyoxometalates as Electrocatalysts. *Chem. Rev.* **98**, 219–238 (1998).

91. J. Huang, W. Liu, D. S. Dolzhnikov, L. Protesescu, M. V. Kovalenko, B. Koo, S. Chattopadhyay, E. V. Shenchenko, D. V. Talapin, Surface Functionalization of Semiconductor and Oxide Nanocrystals with Small Inorganic Oxoanions (PO43– , MoO42–) and Polyoxometalate Ligands. *ACS Nano*. **8**, 9388–9402 (2014).

92. K. T. Ng, D. M. Hercules, XPS studies of oxides of row transition metals of W. *J. Phys. Chem. C*. **80**, 2095 (1976).

93. G. E. McGuire, G. K. Schweitzer, T. A. Carlson, Core electron binding energies in some Group IIIA, VB, and VIB compounds. *Inorg. Chem.* **12**, 2450–2453 (1973).

94. S. Pedetti, B. Nadal, E. Lhuillier, B. Mahler, C. Bouet, B. Abécassis, X. Xu, B. Dubertret, Optimized Synthesis of CdTe Nanoplatelets and Photoresponse of CdTe Nanoplatelets Films. *Chem. Mater.* **25**, 2455–2462 (2013).

95. M. Shim, P. Guyot-Sionnest, n-type colloidal semiconductor nanocrystals. *Nature*. **407**, 981–983 (2000).

96. W. Koh, A. Y. Koposov, J. T. Stewart, B. N. Pal, I. Robel, J. M. Pietryga, V. I. Klimov, Heavily doped *n*-type PbSe and PbS nanocrystals using ground-state charge transfer from cobaltocene. *Sci. Rep.* **3**, 2004 (2013).

97. N. Goubet, C. Livache, B. Martinez, X. Z. Xu, S. Ithurria, S. Royer, H. Cruguel, G. Patriarche, A. Ouerghi, M. Silly, B. Dubertret, E. Lhuillier, Wave-Function Engineering in HgSe/HgTe Colloidal Heterostructures To Enhance Mid-infrared Photoconductive Properties. *Nano Lett.* **18**, 4590–4597 (2018).

98. M. E. Cryer, J. E. Halpert, 300 nm Spectral Resolution in the Mid-Infrared with Robust, High Responsivity Flexible Colloidal Quantum Dot Devices at Room Temperature. *ACS Photonics*. **5**, 3009–3015 (2018).

99. M. Böberl, M. V. Kovalenko, S. Gamerith, E. J. W. List, W. Heiss, Inkjet-Printed Nanocrystal Photodetectors Operating up to 3 μm Wavelengths. *Adv. Mater.* **19**, 3574–3578 (2007).

100. J. P. Clifford, K. W. Johnston, L. Levina, E. H. Sargent, Schottky barriers to colloidal quantum dot films. *Appl. Phys. Lett.* **91**, 253117 (2007).

101. J. M. Luther, M. Law, M. C. Beard, Q. Song, M. O. Reese, R. J. Ellingson, A. J. Nozik, Schottky Solar Cells Based on Colloidal Nanocrystal Films. *Nano Lett.* **8**, 3488–3492 (2008).

102. L.-Y. Chang, R. R. Lunt, P. R. Brown, V. Bulović, M. G. Bawendi, Low-Temperature Solution-Processed Solar Cells Based on PbS Colloidal Quantum Dot/CdS Heterojunctions. *Nano Lett.* **13**, 994–999 (2013).

103. J. Tang, H. Liu, D. Zhitomirsky, S. Hoogland, X. Wang, M. Furukawa, L. Levina, E. H. Sargent, Quantum Junction Solar Cells. *Nano Lett.* **12**, 4889–4894 (2012).

104. A. G. Pattantyus-Abraham, I. J. Kramer, A. R. Barkhouse, X. Wang, G. Konstantatos, R. Debnath, L. Levina, I. Raabe, M. K. Nazeeruddin, M. Grätzel, E. H. Sargent, Depleted-Heterojunction Colloidal Quantum Dot Solar Cells. *ACS Nano*. **4**, 3374–3380 (2010).

105. M. Law, M. C. Beard, S. Choi, J. M. Luther, M. C. Hanna, A. J. Nozik, Determining the Internal Quantum Efficiency of PbSe Nanocrystal Solar Cells with the Aid of an Optical Model. *Nano Lett.* **8**, 3904–3910 (2008).

106. J. Gao, J. M. Luther, O. E. Semonin, R. J. Ellingson, A. J. Nozik, M. C. Beard, Quantum Dot Size Dependent J−V Characteristics in Heterojunction ZnO/PbS Quantum Dot Solar Cells. *Nano Lett.* **11**, 1002–1008 (2011).

107. K. S. Leschkies, T. J. Beatty, M. S. Kang, D. J. Norris, E. S. Aydil, Solar Cells Based on Junctions between Colloidal PbSe Nanocrystals and Thin ZnO Films. *ACS Nano*. **3**, 3638–3648 (2009).

108. E. M. Sanehira, A. R. Marshall, J. A. Christians, S. P. Harvey, P. N. Ciesielski, L. M. Wheeler, P. Schulz, L. Y. Lin, M. C. Beard, J. M. Luther, Enhanced mobility CsPbI3 quantum dot arrays for record-efficiency, high-voltage photovoltaic cells. *Sci. Adv.* **3**, 4204 (2017).

109. B. Li, Y. Zhang, L. Fu, T. Yu, S. Zhou, L. Zhang, L. Yin, Surface passivation engineering strategy to fully-inorganic cubic CsPbI 3 perovskites for high-performance solar cells. *Nat. Commun.* **9**, 1076 (2018).

110. M. A. Green, A. Ho-Baillie, H. J. Snaith, The emergence of perovskite solar cells. *Nat. Photonics.* **8**, 506–514 (2014).

111. H. Chen, S. Xiang, W. Li, H. Liu, L. Zhu, S. Yang, Inorganic Perovskite Solar Cells: A Rapidly Growing Field. *Sol. RRL.* **2**, 1700188 (2018).

112. E. H. Sargent, Solar Cells, Photodetectors, and Optical Sources from Infrared Colloidal Quantum Dots. *Adv. Mater.* **20**, 3958–3964 (2008).

113. O. E. Semonin, J. M. Luther, S. Choi, H.-Y. Chen, J. Gao, A. J. Nozik, M. C. Beard, Peak External Photocurrent Quantum Efficiency Exceeding 100% via MEG in a Quantum Dot Solar Cell. *Science.* **334**, 1530–1533 (2011).

114. J. M. Pietryga, R. D. Schaller, D. Werder, M. H. Stewart, V. I. Klimov, J. A. Hollingsworth, Pushing the Band Gap Envelope: Mid-Infrared Emitting Colloidal PbSe Quantum Dots. *J. Am. Chem. Soc.* **126**, 11752–11753 (2004).

115. P. Guyot-Sionnest, J. A. Roberts, Background limited mid-infrared photodetection with photovoltaic HgTe colloidal quantum dots. *Appl. Phys. Lett.* **107**, 253104 (2015).

116. M. V. Kovalenko, M. Scheele, D. V. Talapin, Colloidal Nanocrystals with Molecular Metal Chalcogenide Surface Ligands. *Science.* **324**, 1417–1420 (2009).

117. Z. Ning, H. Dong, Q. Zhang, O. Voznyy, E. H. Sargent, Solar Cells Based on Inks of n-Type Colloidal Quantum Dots. *ACS Nano.* **8**, 10321–10327 (2014).

118. A. A. Bessonov, M. Allen, Y. Liu, S. Malik, J. Bottomley, A. Rushton, I. Medina-Salazar, M. Voutilainen, S. Kallioinen, A. Colli, C. Bower, P. Andrew, T. Ryhänen, Compound Quantum Dot–Perovskite Optical Absorbers on Graphene Enhancing Short-Wave Infrared Photodetection. *ACS Nano.* **11**, 5547–5557 (2017).

119. Y. Zhang, G. Wu, C. Ding, F. Liu, Y. Yao, Y. Zhou, C. Wu, N. Nakazawa, Q. Huang, T. Toyoda, R. Wang, S. Hayase, Z. Zou, Q. Shen, Lead Selenide Colloidal Quantum Dot Solar Cells Achieving High Open-Circuit Voltage with One-Step Deposition Strategy. *J. Phys. Chem. Lett.*, 3598–3603 (2018).

120. M. Yuan, M. Liu, E. H. Sargent, Colloidal quantum dot solids for solution-processed solar cells. *Nat. Energy.* **1**, 16016 (2016).

121. D. M. Balazs, N. Rizkia, H.-H. Fang, D. N. Dirin, J. Momand, B. J. Kooi, M. V. Kovalenko, M. A. Loi, Colloidal Quantum Dot Inks for Single-Step-Fabricated Field-Effect Transistors: The Importance of Postdeposition Ligand Removal. *ACS Appl. Mater. Interfaces.* **10**, 5626–5632 (2018).

122. J. Kim, O. Ouellette, O. Voznyy, M. Wei, J. Choi, M.-J. Choi, J. W. Jo, S.-W. Baek, J. Fan, M. I. Saidaminov, B. Sun, P. Li, D.-H. Nam, S. Hoogland, Z.-H. Lu, F. P. G. de Arquer, E. H. Sargent, Butylamine-Catalyzed Synthesis of Nanocrystal Inks Enables Efficient Infrared CQD Solar Cells. *Adv. Mater.*, 1803830 (2018).

123. A. Jagtap, N. Goubet, C. Livache, A. Chu, B. Martinez, C. Greboval, J. Qu, E. Dandeu, L. Becerra, N. Witkowski, S. Ithurria, F. Mathevet, M. G. Silly, B. Dubertret, E. Lhuillier, Short Wave Infrared Devices Based on HgTe Nanocrystals with Air Stable Performances. *J. Phys. Chem. C.* **122**, 14979–14985 (2018).

124. D. Zhitomirsky, O. Voznyy, L. Levina, S. Hoogland, K. W. Kemp, A. H. Ip, S. M. Thon, E. H. Sargent, Engineering colloidal quantum dot solids within and beyond the mobility-invariant regime. *Nat. Commun.* **5**, 3803 (2014).

125. S. Ghosh, L. Manna, The Many "Facets" of Halide Ions in the Chemistry of Colloidal Inorganic Nanocrystals. *Chem. Rev.* **118**, 7804–7864 (2018).

126. M. M. Ackerman, X. Tang, P. Guyot-Sionnest, Fast and Sensitive Colloidal Quantum Dot Mid-Wave Infrared Photodetectors. *ACS Nano.* **12**, 7264–7271 (2018).

127. T. Fukuda, A. Takahashi, K. Takahira, H. Wang, T. Kubo, H. Segawa, Limiting factor of performance for solution-phase ligand-exchanged PbS quantum dot solar cell. *Sol. Energy Mater. Sol. Cells.* **195**, 220–227 (2019).

128. K. Qiao, Y. Cao, X. Yang, J. Khan, H. Deng, J. Zhang, U. Farooq, S. Yuan, H. Song, Efficient interface and bulk passivation of PbS quantum dot infrared photodetectors by PbI 2 incorporation. *RSC Adv.* **7**, 52947–52954 (2017).

129. X. Tang, X. Tang, K. W. C. Lai, Scalable Fabrication of Infrared Detectors with Multispectral Photoresponse Based on Patterned Colloidal Quantum Dot Films. *ACS Photonics*. **3**, 2396–2404 (2016).

130. X. Tang, M. M. Ackerman, M. Chen, P. Guyot-Sionnest, Dual-band infrared imaging using stacked colloidal quantum dot photodiodes. *Nat. Photonics*, 1 (2019).

131. C. Buurma, R. E. Pimpinella, A. J. Ciani, J. S. Feldman, C. H. Grein, P. Guyot-Sionnest, MWIR imaging with low cost colloidal quantum dot films. *Opt. Sens. Imaging Photon Count. Nanostructured Devices Appl. 2016*. **9933**, 993303 (2016).

132. E. J. D. Klem, C. Gregory, D. Temple, J. Lewis, PbS colloidal quantum dot photodiodes for low-cost SWIR sensing. *Proc. SPIE*. **9451**, 945104 (2015).

133. T. Rauch, M. Böberl, S. F. Tedde, J. Fürst, M. V. Kovalenko, G. Hesser, U. Lemmer, W. Heiss, O. Hayden, Near-infrared imaging with quantum-dot-sensitized organic photodiodes. *Nat. Photonics*. **3**, 332–336 (2009).

134. E. Heves, C. Ozturk, V. Ozguz, Y. Gurbuz, Solution-Based PbS Photodiodes, Integrable on ROIC, for SWIR Detector Applications. *IEEE Electron Device Lett.* **34**, 662–664 (2013).

135. E. Georgitzikis, P. E. Malinowski, L. M. Hagclsieb, V. Pejovic, G. Uytterhoeven, S. Guerrieri, A. Süss, C. Cavaco, K. Chatzinis, J. Maes, Z. Hens, P. Heremans, D. Cheyns, in *2018 IEEE SENSORS* (2018), pp. 1–4.

136. X. Tang, M. M. Ackerman, G. Shen, P. Guyot-Sionnest, Towards Infrared Electronic Eyes: Flexible Colloidal Quantum Dot Photovoltaic Detectors Enhanced by Resonant Cavity. *Small*. **15**, 1804920 (2019).

137. X. Tang, M. M. Ackerman, P. Guyot-Sionnest, Thermal Imaging with Plasmon Resonance Enhanced HgTe Colloidal Quantum Dot Photovoltaic Devices. *ACS Nano*. **12**, 7362–7370 (2018).

138. C. Livache, B. Martinez, N. Goubet, C. Gréboval, J. Qu, A. Chu, S. Royer, S. Ithurria, M. G. Silly, B. Dubertret, E. Lhuillier, A colloidal quantum dot infrared photodetector and its use for intraband detection. *Nat. Commun.* **10**, 2125 (2019).

139. S. B. Hafiz, M. R. Scimeca, P. Zhao, I. J. Paredes, A. Sahu, D.-K. Ko, Silver Selenide Colloidal Quantum Dots for Mid-Wavelength Infrared Photodetection. *ACS Appl. Nano Mater.* **2**, 1631–1636 (2019).

140. A. Sahu, A. Khare, D. D. Deng, D. J. Norris, Quantum confinement in silver selenide semiconductor nanocrystals. *Chem. Commun.* **48**, 5458–5460 (2012).

# ÉTUDE DES PROPRIÉTÉS OPTOÉLECTRONIQUES DE NANOCRISTAUX COLLOÏDAUX À FAIBLE BANDE INTERDITE : APPLICATION À LA DÉTECTION INFRAROUGE

**Résumé** : Les nanocristaux colloïdaux de semiconducteurs sont des nanomatériaux synthétisés en solution. En deçà d'une certaine taille, ils acquièrent des propriétés de confinement quantique : leurs propriétés optiques et électroniques deviennent dépendantes de leur taille. Le développement de tels nanocristaux a atteint une grande maturité dans le visible ; l'enjeu est maintenant d'étendre la gamme accessible et d'obtenir des nanocristaux ayant des propriétés dans l'infrarouge. Parmi les candidats potentiels, on trouve les nanocristaux de tellure de mercure (HgTe) et de séléniure de mercure (HgSe). L'objectif de ce doctorat est d'approfondir la connaissance des propriétés optiques, optoélectroniques et de transport de ces nanocristaux dans l'optique de concevoir un système de détection infrarouge.

Pour y parvenir, la structure électronique de ces matériaux a été déterminée expérimentalement pour différentes tailles et différentes chimies de surface. Nous pouvons en déduire les énergies des différents niveaux électroniques, l'énergie de Fermi, le type de dopage (et le quantifier). Nous montrons que la taille des cristaux a une forte influence sur le dopage, qui devient de plus en plus n quand les cristaux sont de plus en larges, allant jusqu'à observer une transition semiconducteur-métal pour des cristaux de HgSe. Le contrôle du dopage est ensuite étudié en fonction de la chimie de surface. En utilisant des effets dipolaires ou des transferts d'électrons via des ligands fortement oxydants, nous montrons un contrôle du dopage sur plusieurs ordres de grandeur en modifiant l'environnement à la surface des nanocristaux.

Ces études nous permettent de proposer un détecteur infrarouge à base de nanocristaux de HgTe, fonctionnant à 2,5 µm, dont la structure permet de convertir efficacement les photons absorbés en courant électrique, tout en obtenant un bon rapport signal sur bruit. Nous obtenons ainsi une réponse de l'ordre de 20 mA/W et une détectivité de $3 \times 10^9$ Jones.

**Mots clés** : Photodétection, infrarouge, nanocristaux, HgSe, HgTe, dopage, optoélectronique

# OPTOELECTRONIC PROPERTIES OF NARROW BAND GAP NANOCRYSTALS: APPLICATION TO INFRARED DETECTION

**Abstract**: Colloidal semiconductor nanocrystals are nanomaterials synthesized in solution. Below a certain size, these nanocrystals acquire quantum confinement properties: their optoelectronic properties depend on the nanoparticle size. In the visible range, colloidal nanocrystals are quite mature. The next objective in this field is to get infrared colloidal nanocrystals. Mercury selenide (HgSe) and mercury telluride (HgTe) are potential candidates. The goal of this PhD work is to strengthen our knowledge on optical, optoelectronic and transport properties of these nanocrystals, in order to design an infrared detector.

To do so, we studied the electronic structure of HgSe and HgTe for different sizes and surface chemistries. We can then determine the energies of the electronic levels and the Fermi energy, quantify doping level … We show that the nanocrystal size has an influence on doping level, which gets more and more n-type as the nanocrystal size gets larger. We even observe a semiconductor-metal transition in HgSe nanocrystals as the size is increased. The doping control with surface chemistry is then investigated. By using dipolar effects or oxidizing ligands, we show a doping control over several orders of magnitude.

Thanks to these studies, we are able to propose a HgTe based device for detection at 2,5 µm, which structure allows to convert effectively the absorbed photons into an electrical current and to get a high signal over noise ratio. We get a photoresponse of 20 mA/W and a detectivity of $3 \times 10^9$ Jones.

**Key words**: Photodetection, infrared, nanocrystals, HgSe, HgTe, doping, optoelectronics